\documentclass[rmp,aps,nofootinbib]{revtex4} 
\pdfoutput=1
\usepackage{amsmath}
\usepackage{amssymb,graphicx}
\usepackage{epsfig}

\def\inbar{\,\vrule height1.5ex width.4pt depth0pt}
\def\IR{\relax{\rm I\kern-.18em R}}
\def\IC{\relax\hbox{$\inbar\kern-.3em{\rm C}$}}

\def\beq{\begin{eqnarray}}
\def\eeq{\end{eqnarray}}

\def\largeimagesize{0.6\hsize}
\def\imagesize{0.5\hsize}
\def\smallimagesize{0.4\hsize}

\begin{document}
\title{Alternatives to an Elementary Higgs}

\author{Csaba Cs\'aki}
\email{ccsaki@gmail.com}
\affiliation{Department of Physics, LEPP, Cornell
University, Ithaca, NY 14853}
\author{Christophe Grojean\footnote{On leave from ICREA and IFAE, Barcelona Institute of Science and Technology (BIST)
Campus UAB,  E-08193 Bellaterra, Spain}}
\email{christophe.grojean@desy.de}
\affiliation{
DESY, Notkestrasse 85, D-22607 Hamburg, Germany}
\author{John Terning}
\email{jterning@gmail.com}
\affiliation{Department of Physics, University of California, Davis, CA
95616}

\begin{abstract}  
We review strongly coupled and extra dimensional models of
electroweak symmetry breaking.  Models examined include
warped extra dimensions, bulk Higgs,  ``little" Higgs, dilaton Higgs, composite Higgs, twin Higgs, quantum critical Higgs, and ``fat" SUSY Higgs.
We also discuss current bounds and future LHC searches for this class of models.
\end{abstract}                                                                 

\date{November 2015}
\maketitle
\tableofcontents

\section{INTRODUCTION}
\label{sec:intro}

A key ingredient of the standard model (SM) of elementary particle physics is the electroweak sector that explains the relation between weak interactions and ordinary electromagnetism. This sector requires a mechanism that forces the vacuum to distinguish the weak gauge bosons, the $W$ and $Z$, from the photon, thus breaking the symmetry of the electroweak  gauge interactions. The $W$ and $Z$  have large masses while the photon is exactly massless.  This symmetry breaking goes by the name ``Higgs mechanism". In addition to fields corresponding to the spin-1  gauge bosons (force carriers), the SM includes another boson field with zero spin, called the Higgs field, which is arranged to have  a non-zero vacuum expectation value (VEV). In the absence of electroweak gauge couplings, three components of the Higgs field would have massless Goldstone excitations, that is there would be three exact Goldstone bosons \cite{Goldstone:1961eq}.  With the electroweak gauge interactions turned on, these excitations provide the missing longitudinal modes for the $W^\pm$ and $Z$.  Using an elementary Higgs field \cite{Englert:1964et,Higgs:1964ia,Higgs:1964pj,Guralnik:1964eu,Weinberg:1967tq} to arrange for the spontaneous breaking of electroweak gauge symmetry is certainly the simplest model that makes sense of the electroweak interactions; at least at tree-level that is. At loop-level one encounters large radiative corrections (aka quadratic divergences) that tend to drive the renormalized Higgs mass parameter, and hence the $W$ and $Z$ masses, up to the highest scale in the theory.  In other words, in the absence of some incredible fine-tuning there should be some new physics beyond the SM near the TeV scale \footnote{It does not necessarily mean that TeV scale new physics stabilizing the Higgs mass will be easily accessible at the LHC, see for instance the discussion of neutral naturalness models \cite{Chacko:2005pe, Craig:2015pha} in section~\ref{sec:twinHiggs}. A recent proposal \cite{Graham:2015cka, Espinosa:2015eda} relies on the cosmological evolution of the Universe to drive it neear a critical point for electroweak symmetry breaking and thus alleviates  the hierarchy problem without the need for TeV scale new physics at all.}. The existence of these quadratic divergences is a basic feature of elementary scalar fields, first noted by \textcite{Weisskopf:1939zz}, and is  referred to as the hierarchy problem. In fact Weisskopf argued that this was the explanation of why no one had ever discovered an elementary scalar field. Weisskopf's argument essentially still holds; the well-known loop holes are supersymmetry and compositeness.  The quadratic divergences can be tamed with the introduction of supersymmetry, and the fine-tuning eliminated with superpartners below the TeV scale that cut off the divergence. On the other hand, if the scalar is composite rather than 
elementary, the new interactions that produce the composite can serve to cut off the divergence.

Now that the LHC has found a Higgs-like resonance \cite{Aad:2012tfa,Chatrchyan:2012ufa} the next questions are: ``Is there also supersymmetry?" and ``Is the Higgs elementary?" Superpartners below the TeV scale can be uncovered by the LHC, but the question of whether the Higgs is elementary or not is much more subtle.  There are a large variety of scenarios that cover a continuum of possibilities  from elementary to composite (it was even suggested that a composite scalar acquires a small VEV that then induces the VEV of the elementary Higgs \cite{Samuel:1990dq,Chang:2014ida}). There are even SUSY theories with a composite Higgs.
In this review we will attempt to survey these possibilities, pointing out when their phenomenologies overlap and where there are unique signals. 

Since the idea of a composite Higgs relies on having new strong interactions, we will begin our review with a brief discussion of the prototypical model for breaking electroweak
symmetry with strong interactions: technicolor.  Although these models have been discarded,  they will set the stage for the more successful models that avoid the pitfalls of technicolor. An alternative approach to strong interactions is to use a weakly coupled ``dual" description in terms of an extra dimension.  Whether the extra dimension or the dual strong interactions are the fundamental description can only be answered at even higher energies,  but the extra dimensional description is certainly easier to calculate with near the TeV scale.  An alternative to canceling divergences with superpartners is to cancel them with partners of the same spin. Such ``little" Higgs theories can in principle  raise the compositeness/SUSY scale from 1\,TeV to 10\,TeV. Extending the space-time symmetries to include conformal symmetry rather than supersymmetry is another possibility for protecting the Higgs mass from divergences.  We will also examine a more bottom-up approach that relies on constructing a general low-energy effective field theory for the Higgs. It will be useful to see how this effective theory can be matched on to some interesting composite models.

\subsection{An Instructive Failure: Technicolor}
\label{subsec:technicolor}

Technicolor\footnote{For a thorough review see \cite{Hill2003235}.} was the first alternative to an elementary SM Higgs; it is a beautiful idea that seems not to have lived up
to its potential.  In some ways technicolor is analogous to superconductivity, and somewhat ironically, before the Higgs mechanism
was discovered by particle physicists, \textcite{PhysRev.130.439} had emphasized how composite degrees of freedom lead to an effective photon mass in a Bardeen--Cooper--Schrieffer superconductor \cite{PhysRev.108.1175}, and 
Nambu had developed the field theory version of the composite Higgs mechanism in his work with Jona-Lasinio \cite{Nambu:1961tp,Nambu:1961fr}.
In a superconductor, electrons can attract each other (very weakly) by exchanging phonons (quanta of lattice vibrations) forming Cooper pairs. If the charged Cooper pairs undergo Bose condensation, then the lowest energy state of the system has an arbitrarily large charge (limited only by the number of electrons in the superconductor). Photons moving through this charged medium are effectively massive, as can be seen by the fact that magnetic fields cannot penetrate a superconductor (the Meissner effect).  In a superconductor the analog of the Higgs is the Cooper pair, and the analog of the massive $W$ and $Z$ is the photon which has a penetration depth inversely proportional to its effective mass.  
However a Cooper pair is so weakly bound that its physical size\footnote{The electron pairs in high $T_c$ superconductors are much smaller, so they look much more like an analog of the SM Higgs.} (100's of nanometers) is much larger than its Compton wavelength, so there is no sense in which we can describe this system using an effective field theory with a Higgs-like field  standing-in for the Cooper pair. Thus the first known implementation of the Higgs mechanism does not have an elementary field but rather a  loosely bound composite.

\textcite{Weinberg:1976gm,PhysRevD.19.1277} and  \textcite{Susskind:1978ms} independently proposed that composite particles formed by a new strong interaction could replace the SM Higgs boson. The new interactions were supposed to be similar to those of quantum chromodynamics (QCD), and these theories were hence dubbed technicolor theories. Susskind showed that if the Higgs boson was absent from the Standard Model, QCD would provide electroweak symmetry breaking through quark composites (although it would give masses for the $W$ and $Z$ that are about a factor of 2,600 too small). Technicolor theories thus harkened back to superconductivity where a gauge symmetry is broken by a composite of two fermions, a crucial difference being that the interactions responsible for superconductivity are quite weak, whereas the technicolor interactions must remain strong. Technicolor theories essentially resolve the fine-tuning problem by lowering the effective cut-off scale to 1\,TeV. Remarkably, technicolor theories predicted the correct ratio for the $W$ and $Z$ masses. This is due to the global symmetry breaking pattern of QCD being $SU(2)\times SU(2)/SU(2)$ rather than simply the minimal breaking $SU(2)\times U(1)/U(1)$ required by electroweak gauge symmetry breaking. This enhanced symmetry is referred to as the custodial symmetry \cite{Sikivie:1980hm}, although authors differ on whether they use this term to refer to the full $SU(2)\times SU(2)$ symmetry or the unbroken diagonal $SU(2)$ subgroup.  

The Achilles' heel of technicolor has always been producing masses for the quarks and leptons. This requires several complicated extensions of the model (e.g. extended technicolor \cite{Eichten:1979ah,Dimopoulos:1979es}, and often extra pseudo-Goldstone bosons), and even then one finds problems with flavor changing neutral currents (FCNCs).  The FCNC problem could be resolved for the first two generations \cite{PhysRevD.24.1441,Holdom1985301} by assuming approximately conformal behavior above 1\,TeV (aka a ``walking" rather than running coupling constant \cite{Holdom:1981rm,Yamawaki:1986zg,Appelquist:1986an,Appelquist:1987tr}). However this is not enough to explain the top quark, and further model building gymnastics are required. A subtler problem with technicolor was revealed by the comparison with precision electroweak measurements \cite{Golden:1990ig,Holdom:1990tc,Peskin:1992sw,Peskin:1990zt,Altarelli:1990zd}. Following the idea of scaling up QCD to obtain the correct $W$ and $Z$ masses, it was possible to scale up QCD data (essentially using QCD as an ``analog computer") to predict the deviations of a technicolor theory from the SM. These deviations were not seen at SLAC or LEP. It remained logically possible that there was some version of technicolor that does not behave like QCD, but in the absence of an explicit, workable model interest in technicolor waned during the 1990s. 

For the remainder of this review we will be interested in models that have a light composite scalar boson, unlike the QCD-like models discussed above.

\subsection{Classifying the Alternatives}
\label{subsec:classifying}
While there have been many proposals for alternative models of electroweak symmetry breaking there are some basic concepts that can be applied to all the models
that can impose some structure that allows us to easily compare and contrast them.
One way to think about  how to classify the range of models is to first consider the scaling dimension of the operator that breaks electroweak symmetry.  For an elementary Higgs, the scaling dimension is obviously one, up to perturbative corrections.  In contrast, in an alternate Universe where technicolor breaks electroweak symmetry, the operator that does the breaking is a fermion bilinear.  At tree-level this operator has scaling dimension three, but at strong coupling it should be significantly smaller.  In walking technicolor the scaling dimension was assumed to be two. Thus the scaling dimension of the operator is an essential ingredient to understanding the UV completion of the Higgs sector. In theories with a light composite Higgs, it is especially important to know the scaling dimension of the square of the Higgs field, i.e. the mass term.  If the scaling dimension of the mass term is less than four, then the operator is relevant and can receive divergent corrections; if the scaling dimension of the mass term is greater than four, then the operator is irrelevant and the hierarchy problem is solved! In general the scaling dimension of the mass term is not just twice the scaling dimension of the Higgs field, but if the composite Higgs is weakly coupled this should be a good approximation. 
 
  In the next section we will discuss extra dimensional models in anti-de Sitter (AdS) space, and we will see that there is a direct connection between how the Higgs is localized in the extra dimension and the effective scaling dimension. In the original Randall--Sundrum (RS) model, where the Higgs is localized at one end of the extra dimension, the effective scaling dimension is infinite.  When this model is generalized to allow the Higgs to extend into the extra dimension (aka the bulk Higgs model) one finds that the scaling dimension can vary from infinity all the way down to one. For a fixed scaling dimension one is still free to vary how much of the $W$ and $Z$ masses come from the Higgs VEV and how much comes from mixing with higher resonances (i.e. from AdS curvature effects), so there is a two dimensional parameter space of models.
  
  An interesting way to keep a composite Higgs light is for it to be a pseudo-Goldstone boson corresponding to a global symmetry broken  at a scale, $f$, that is much higher than the Higgs VEV, $v$.  This can occur with an elementary Higgs, which therefore has scaling dimension near one (as in ``little" Higgs models), or via strong coupling/extra dimensions.  In the explicit case of the minimal composite pseudo-Goldstone boson Higgs the scaling dimension turns out to be two.  In these types of models there are also additional mixing corrections to the $W$ and $Z$ masses which are parametrically of the order of $v^2/f^2$. The ``Fat" SUSY Higgs is another type of composite where the scaling dimension can range between one and two.  In it's minimal form, the the Minimal Composite Supersymmetric Standard Model (MCSSM), we will see that the scaling dimension is fixed to be close to one, with a small amount of mixing for the $W$ and $Z$. 

\begin{figure}[htb]
\includegraphics[width=\imagesize]{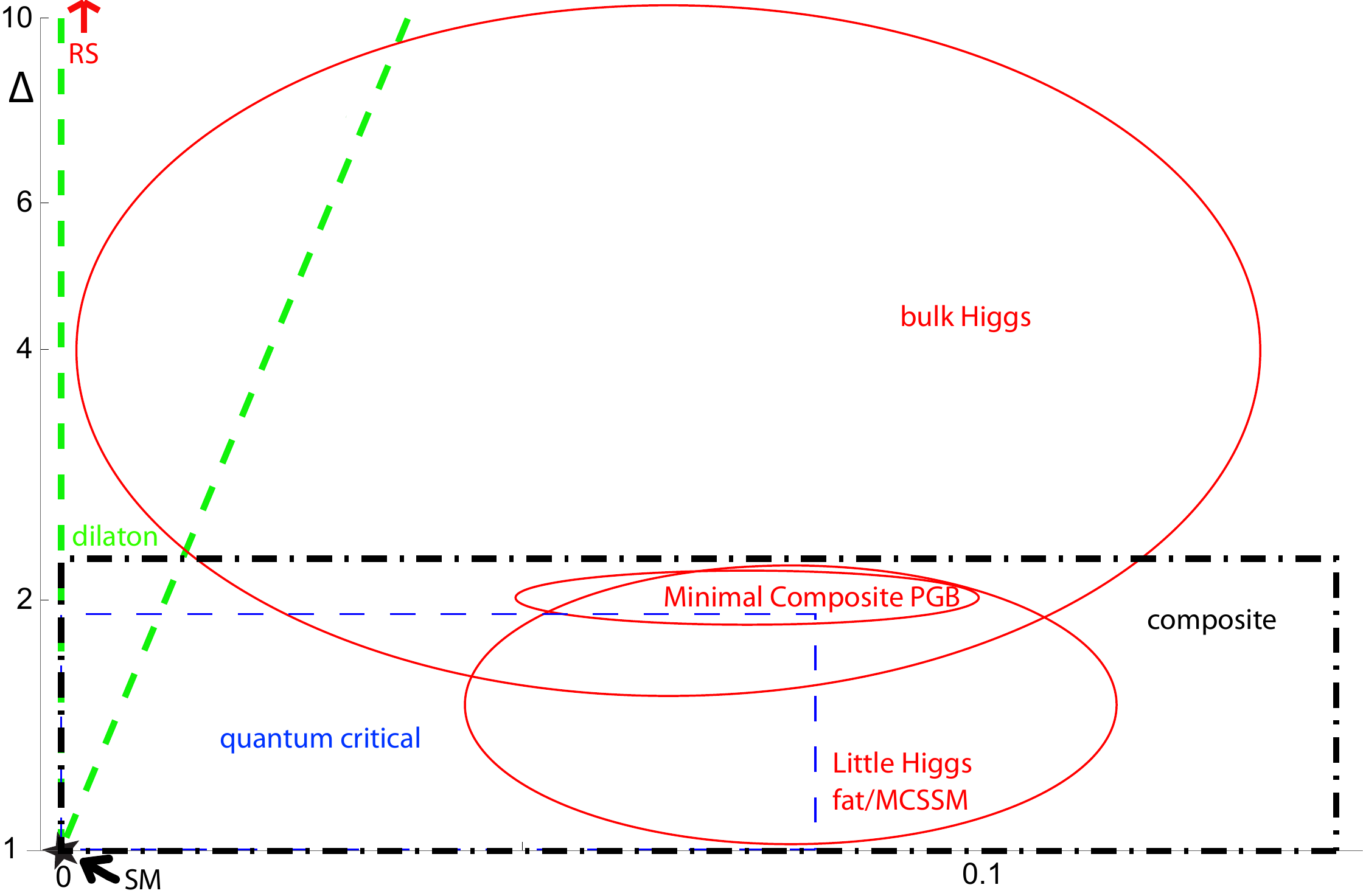}
\caption[]{Parameter space of alternative Higgs models, the vertical axis is the scaling dimension, $\Delta$, of the operator that condenses and breaks the electroweak gauge symmetry, the horizontal axis is the fraction of the $W$ and $Z$ masses that arise from mixing. Precision Electroweak measurements rule out models that lie too far to the right. The location of the SM, at the origin, is marked by a star. The dash-dot lines denote the approximate range of composite Higgs models, while the thin blue dashes denote the approximate range of quantum critical Higgs models. The green dashed lines enclosing a wedge-shaped area denote the approximate region for dilaton models, with $\Delta$ denoting the dimension of the scale breaking condensate. Color online.} 
\label{fig:classifying}
\end{figure}

Several types of models use conformal invariance to try to keep the Higgs light. A dilaton Higgs is another example of a pseudo-Goldstone boson, this time from broken conformal invariance. Conformal Technicolor \cite{Luty:2004ye} was an interesting attempt to find theories where the scaling dimension of the Higgs is close to one, while the scaling dimension of the mass term is greater than four. Generically if there is a parameter in the theory that can be adjusted to move from the phase where electroweak symmetry is broken to the symmetric phase, and the phase transition is continuous, we will find a light scalar in the broken phase that is a fluctuation of the order parameter. This is what happens in the SM, where the Higgs mass parameter in the Lagrangian has to be delicately adjusted to be close to the critical point with a light Higgs. In condensed matter parlance such transitions are called quantum critical points, since at zero temperature it is quantum fluctuations that dominate rather than thermal fluctuations.  Experimentally we seem to be near such a critical point; this means that there is potentially a very long renormalization group (RG) flow, which usually results in approaching an IR fixed point.  This fixed point could be trivial (i.e. free) as in the SM, or non-trivial, that is an interacting conformal field theory (CFT). The quantum critical Higgs model assumes a quantum critical point where the light composites are weakly interacting; from the extra dimensional point of view it is a special case of the bulk Higgs where the scaling dimension is between one and two.
A rough sketch of the range of models we will consider is shown in Fig.~\ref{fig:classifying}.

\section{EXTRA DIMENSIONS}
\label{sec:extradim}
In the late 1990s extra dimensions became a popular framework for extensions of the SM. Many early attempts \cite{ArkaniHamed:1998rs,Antoniadis:1998ig,Appelquist:2000nn} did not really address the hierarchy problem until the advent of the warped extra dimensional model of \textcite{Randall:1999ee}, generally referred to as RS.
This type of model is a five dimensional AdS space where the warped extra dimension is truncated at two 4D boundaries (aka 3-branes).  The warping allows us to associate
one end of the space with low-energies, the infrared (IR),  and the other end with high-energies, the ultraviolet (UV).  Without the IR cutoff the theory resembles a CFT, a result that can be understood through the AdS/CFT correspondence. 

\subsection{The AdS/CFT correspondence}
\label{subsec:adscft}

A major breakthrough of the late 90's was the discovery of the AdS/CFT correspondence by \textcite{Maldacena:1997re}, who conjectured (based on several independent consistency checks) that there is an exact equivalence between type IIB string theory on an AdS$_5\times S^5$ background and non-gravitational 4D ${\cal N}=4$ supersymmetric SU(N) gauge theories in the large N limit. In this correspondence operators of the CFT are associated with bulk fields in AdS$_5$, and the value of the field on the boundary of AdS$_5$ acts as a source for the CFT operator. While the initial excitement was mainly confined to the string community, soon it was realized that AdS/CFT has wide reaching consequences and applications in many branches of physics. For example, the presence of supersymmetry does not appear to be essential, and it was conjectured by \textcite{ArkaniHamed:2000ds,Rattazzi:2000hs,PerezVictoria:2001pa} that the proper interpretation of the original Randall--Sundrum models is in terms of a non-supersymmetric version of the correspondence, whereby the bulk of an extra dimension with anti-de Sitter background corresponds to a large N limit of a non-supersymmetric 4D CFT. It is actually not too hard to understand the underlying reason for this: the metric of 5D AdS space (in so-called conformal coordinates) is given by 
\begin{equation}
ds^2=  \left( \frac {R}{z} \right)^2   \Big( \eta_{\mu \nu} dx^\mu
dx^\nu - dz^2 \Big)
\label{eq:AdSbackground}
\end{equation}
which has an isometry $z\to e^\alpha z, x\to e^\alpha x$. The physical meaning is that a motion along the fifth dimension $z$ is equivalent to a rescaling of the 4D coordinates, implying that movement along the fifth direction is actually a RG transformation. Since this leaves the metric invariant,  one expects the corresponding 4D scaling to be a symmetry, hence a 4D CFT must be the underlying structure. 

The RS model does not however have a full AdS space $0\leq z\leq \infty$, but rather only coincides with a slice of it: the so-called UV brane (or Planck-brane) at $z=R$ forming one of the boundaries, while the IR brane (TeV brane) at $z=R'$ forming the other boundary. The effect of the UV brane is to render the graviton zero mode  normalizable (while gravity decouples in full AdS due to a non-normalizable graviton zero mode). Thus the presence of the UV brane will recouple gravity to the 4D CFT. Since gravity provides an explicit scale (the Planck scale), the correct interpretation of the UV brane is that it provides an explicit breaking at scale $1/z_{UV}$ for the CFT. 

The interpretation of the IR brane is more subtle: introducing the IR brane will provide a mass gap of order $1/z_{IR}$ into the Kaluza--Klein expansion of all types of fields, calling for an interpretation different from that of the UV brane. The most natural interpretation is to assume that the presence of the IR brane signals a spontaneous breaking of the conformal symmetry: the theory was perturbed away from the exact fixed point, as a result the coupling became strong and generated a condensate of scale $1/z_{IR}$ resulting in the mass gap. 

The next question is how to deal with global symmetries, which may or may not be weakly gauged. It is quite clear from the formulation of the correspondence that global symmetries in the CFT require bulk gauge fields in AdS: the conserved global current $J_\mu$ of the CFT must have a bulk vector field $A_\mu$ that it can couple to on the boundary. The question we need to answer next is what determines if this will be a global or a weakly gauged symmetry. Clearly this will again be set by the presence (or absence) of a 4D gauge field zero mode. Carefully examining the 5D Maxwell equation $\partial_\mu [ \sqrt{g} g^{\mu\nu} F_{\nu\rho}] =0$ in the background (\ref{eq:AdSbackground}) shows that the profile, $f(z)$, of a potential gauge zero-mode $A_\mu (z,x) = f(z) A_\mu (x)$ is flat along the extra dimension. Imposing flat (Neumann) boundary conditions (BC's) for the gauge field profile 
\begin{equation}
\partial_z f(z)|_{z=R,R'} =0 
\end{equation}
will allow the zero mode in the spectrum, and thus corresponds to a weakly gauged symmetry. However imposing a Dirichlet BC $f(R)=0$ on the UV brane will remove the gauge zero mode  and hence the combination with a Dirichlet BC in the UV and Neumann BC in the IR corresponds to a global symmetry. Keeping a Neumann BC on the UV but imposing a Dirichlet BC on the IR will have the effect of raising the zero mode to a mass of order $1/z_{IR}$: this is the case expected for a spontaneous breaking of the gauge symmetry: the CFT condensate that  broke  the CFT will also contribute to breaking the weakly gauged symmetry, similar to the technicolor models discussed in the previous section. Thus 5D models of technicolor can be built by imposing Dirichlet BC's for the appropriate combinations of bulk gauge field, yielding the so called higgsless models \cite{Csaki:2003dt,CGPT}. 

The final possibility of Dirichlet BC's on both UV and IR branes presents another important possibility with wide applications. A Dirichlet BC on the UV brane will render that symmetry global, and the additional breaking by a Dirichlet BC on the IR brane will produce a broken global symmetry, which should have the appropriate Goldstone bosons. Indeed it turns out that in this case the fifth component of the gauge field, $A_5$, will have a zero mode, peaked on the IR brane. This is the mode often used for a holographic implementation of composite Higgs models where the Higgs arises as a Goldstone boson. A short summary of the AdS/CFT dictionary used by BSM model builders is presented Table~\ref{tab:dictionary}.

\begin{table}
\[
\begin{tabular}{lll}
 {\bf Bulk of AdS} \hspace*{4cm}& $\leftrightarrow$ & \hspace*{1cm}{\bf CFT} \\[.25cm]
 Inverse coordinate ($1/z$) along AdS & $\leftrightarrow$  & \hspace*{1cm}Energy scale in CFT \\[.25cm]
 UV brane & $\leftrightarrow$ & \hspace*{1cm}CFT has a cutoff \\[.25cm]
 IR brane & $\leftrightarrow$ &
\hspace*{1cm}conformal symmetry broken
spontaneously by CFT\\[.35cm]
 KK modes localized near IR brane & $\leftrightarrow$ & \hspace*{1cm}composites of CFT \\[.25cm]
 Modes on the UV brane & $\leftrightarrow$ & \hspace*{1cm}Elementary fields
coupled
to CFT \\[.25cm]
 Gauge fields in bulk & $\leftrightarrow$ & \hspace*{1cm}CFT has a global symmetry  \\[.25cm]
Bulk gauge symmetry broken on UV brane
& $\leftrightarrow$ & \hspace*{1cm}Global symmetry not gauged \\[.4cm]
Bulk gauge symmetry unbroken on UV brane
& $\leftrightarrow$ &
\hspace*{1cm}Global
symmetry weakly gauged \\[.35cm]
 Higgs on IR brane & $\leftrightarrow$ &
\hspace*{1cm}Strong CFT produces
composite Higgs  \\[.35cm]
BC breaking on IR brane & $\leftrightarrow$ &
\hspace*{1cm}CFT condensate breaks gauge symmetry \\[.35cm]
BC breaking on both branes & $\leftrightarrow$ &
\hspace*{1cm}Broken global symmetry with $A_5$ Goldstones \\
\end{tabular}
\]
\caption{Summary of the AdS/CFT dictionary used in BSM model building from \cite{Csaki:2005vy}.}
\label{tab:dictionary}
\end{table}

\subsection{Realistic RS}
\label{subsection:realisticRS}

\begin{figure}[htb]
\includegraphics[width=\imagesize]{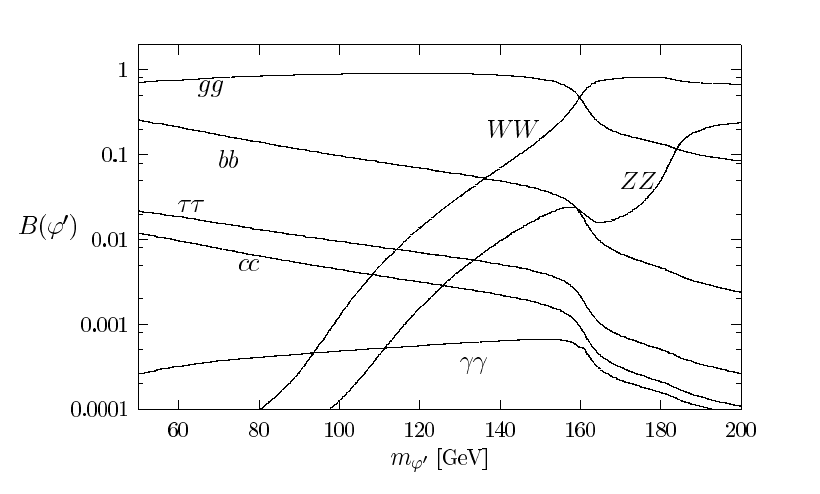}
\caption[]{Light radion decay fractions: the decay to gluons is enhanced while the decays to $b{\overline b}$  and $\gamma \gamma$ are suppressed, from \cite{Giudice:2000av}.} \label{radiondecays}
\end{figure}

There are many variations on RS models (for reviews see \textcite{Rattazzi:2003ea,Kribs:2006mq,Cheng:2010pt,Gherghetta:2006ha,Csaki:2005vy,Sundrum:2005jf,Ponton:2012bi,Gherghetta:2010cj}). The first iterations had all the SM fields localized on the IR brane (aka the TeV brane).  
Through the AdS/CFT correspondence one sees that states localized on the IR brane are the analogs of strongly bound composites of the CFT, while states localized on the UV brane are external spectators of the strongly coupled CFT that have been added with a weak coupling. 
As such, the novel phenomenology of the early RS models is entirely due to the KK modes of 5D gravity, which includes spin-2 modes as well as a scalar radion
\cite{Davoudiasl:1999jd,Csaki:2000zn}.
The distance between the IR and UV branes is arbitrary, corresponding to the massless radion.  In fact the size of the extra dimension is unstable to small perturbations \cite{Csaki:1999jh} and must be stabilized \cite{Goldberger:1999uk}, and this stabilization results in a mass for the radion \cite{Goldberger:1999un,Csaki:1999mp}.
It can be shown that a light radion decays predominantly to gluons due to the trace anomaly \cite{Giudice:2000av} as show in Fig. \ref{radiondecays}.

 The AdS/CFT correspondence suggests that the standard model states localized on the IR brane
should have a plethora of higher dimension operators coupling them and that these operators should be suppressed by powers of the TeV scale.  Since there is no experimental evidence of such operators, more realistic RS models were developed where the gauge bosons and light fermions live throughout the bulk of the extra dimension while the Higgs
and the top quark are localized near the IR brane~\cite{Huber:2000fh,Carena:2003fx,Davoudiasl:1999tf,Davoudiasl:2000wi,Csaki:2002gy,Grossman:1999ra,Pomarol:1999ad,Gherghetta:2000qt}.  Since the Higgs is effectively a composite state with an inverse TeV size there are no quadratic divergences in its mass. In these models the top needs to be localized near the IR brane so that it can have a large coupling to the Higgs.

\begin{figure}[htb]
\includegraphics[width=\imagesize]{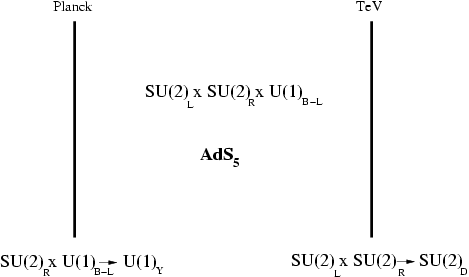}
\caption[]{Realistic RS models: the gauged $SU(2)_L \times SU(2)_R$ symmetry in the bulk 
corresponds to a custodial global symmetry in the hypothetical dual CFT.
The gauge symmetry is broken to the SM gauge group on the Planck (UV) brane
and is broken by the Higgs VEV down to the diagonal subgroup on the TeV (IR) brane. } 
\label{fig:custodial bulk}
\end{figure}

We will now focus on such realistic RS models.  Using the rules of the AdS/CFT correspondence discussed in the previous section, we can 
relatively easily find the type of model  that we are after. We want a
theory that has an $SU(2)_L\times SU(2)_R\times U(1)_{B-L}$ global
symmetry, with the $SU(2)_L\times U(1)_Y$ subgroup weakly gauged
and broken by a Higgs VEV on the IR brane. To have the full global
symmetry, we need to have an $SU(2)_L\times SU(2)_R\times U(1)_{B-L}$
gauge symmetry 
in the bulk of AdS$_5$. To make sure that we do not get unwanted
gauge fields at low energies, we need to break $SU(2)_R\times
U(1)_{B-L}$ to $U(1)_Y$ on the UV brane, which we can do with another Higgs-like field, or with BC's. 
Finally, a Higgs localized on the TeV brane gets a VEV and
breaks $SU(2)_L\times SU(2)_R$ to $SU(2)_D$. 
This setup 
 implements the necessary custodial symmetry by means of a bulk $SU(2)_L \times SU(2)_R$ gauge symmetry \cite{Agashe:2003zs}, as shown in Fig. \ref{fig:custodial bulk}. As in the SM and technicolor models, the custodial symmetry ensures the correct ratio of $W$ and $Z$ masses.

\subsection{Realistic RS LHC Searches}
\label{subsection:RSsearch}

\begin{figure}[htb]
\includegraphics[width=\imagesize]{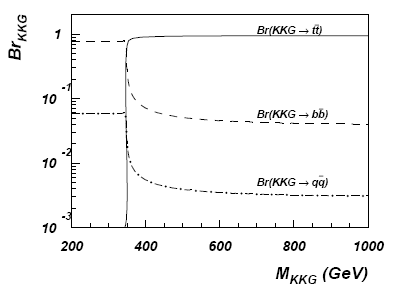}
\caption[]{KK gluons decay mainly to $t {\overline t}$ pairs,
from \cite{Agashe:2006hk}. } \label{fig:KKgluondecay}
\end{figure}

\begin{figure}[htb]
\includegraphics[width=\imagesize]{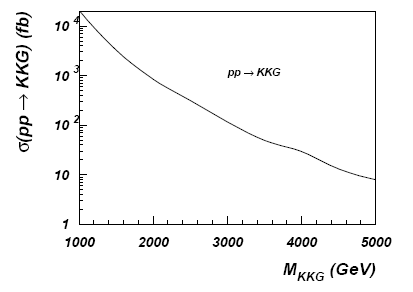}
\caption[]{KK gluons production cross section at the 14\,TeV LHC,
from \cite{Agashe:2006hk}. } \label{fig:KKgluonprod}
\end{figure}

\begin{figure}[htb]
\centerline{\includegraphics[width=\imagesize]{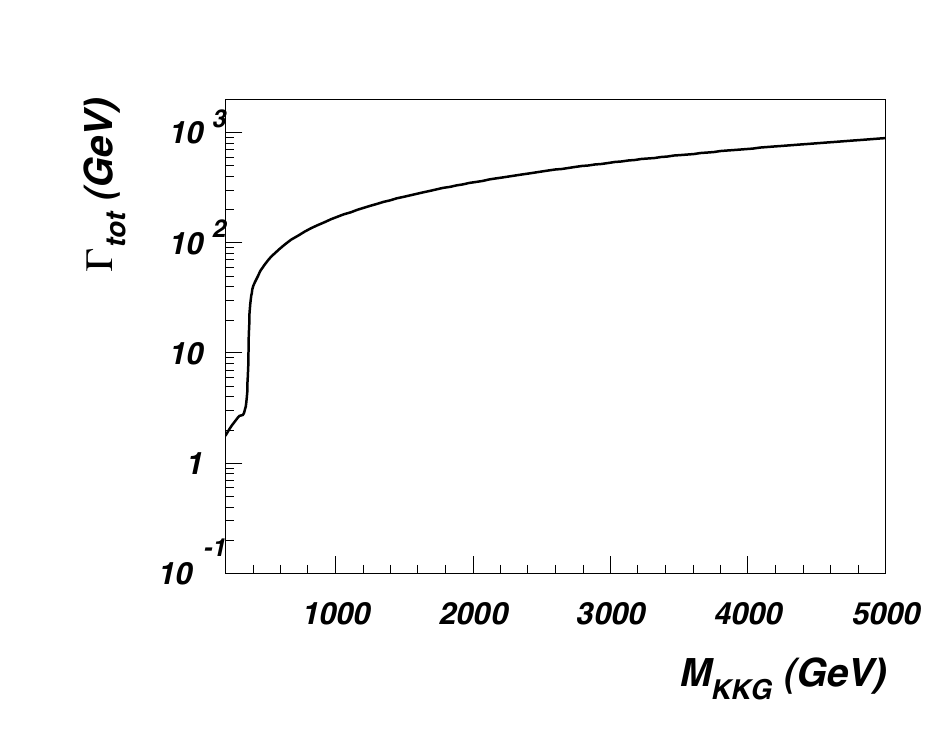}}
\caption[]{KK gluons typically have a large width, from \cite{Agashe:2006hk}. }
 \label{fig:KKgluonwidth}
\end{figure}

\begin{figure}[htb]
\includegraphics[width=\imagesize]{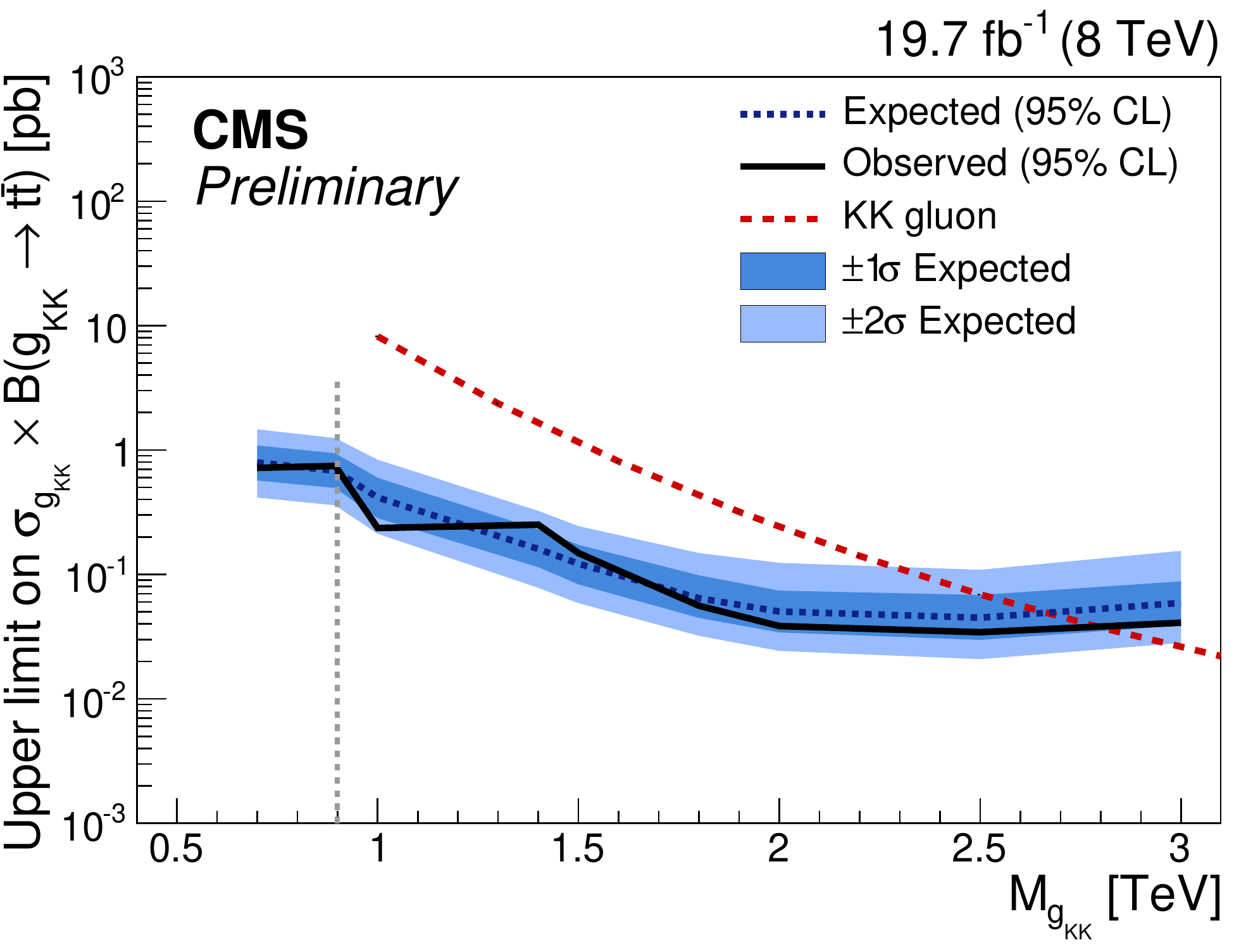}
\caption[]{CMS bound on KK gluon production, from \cite{CMS:2015nza}. Color online.}
 \label{fig:CMSKKgluon}
\end{figure}

While deviations in Higgs couplings are expected in RS models their values are not uniquely predicted, so most of the search strategies focus on the extra associated particles.
As mentioned above the top quark must be localized near the TeV brane in order to get a large enough Yukawa coupling to the Higgs.  This implies that
the color $SU(3)$ gauge group is also in the bulk.  All the bulk gauge fields will have a tower of KK modes.
Since they are strongly interacting, KK gluons are a primary target for LHC searches \cite{Lillie:2007yh}.
As seen in Fig. \ref{fig:KKgluondecay}, KK gluons decay almost exclusively to $t {\overline t}$ pairs.
\cite{Agashe:2006hk,Lillie:2007ve}. 
Figure \ref{fig:KKgluonprod} shows the production cross section for KK gluons at the 14\,TeV LHC.
A 3\,TeV KK gluon has a 0.1 pb production cross section, so it is fairly easy to produce.
KK gluons typically have a large width of the order of 100's of GeV \cite{Agashe:2006hk} as seen in Fig. \ref{fig:KKgluonwidth}. 
To search for these broad resonances decaying to tops, effectively one needs to take advantage of jet substructure \cite{Seymour:1993mx,Larkoski:2013eya,Ellis:2012sn}
to develop a top-tagger \cite{Kaplan:2008ie}, see \textcite{Plehn:2011tg,Shelton:2013an} for reviews.
A recent CMS analysis excludes KK gluons below 2.8\,TeV \cite{Chatrchyan:2012ku}, 
see Fig. \ref{fig:CMSKKgluon}.

\begin{figure}[htb]
\includegraphics[width=\smallimagesize]{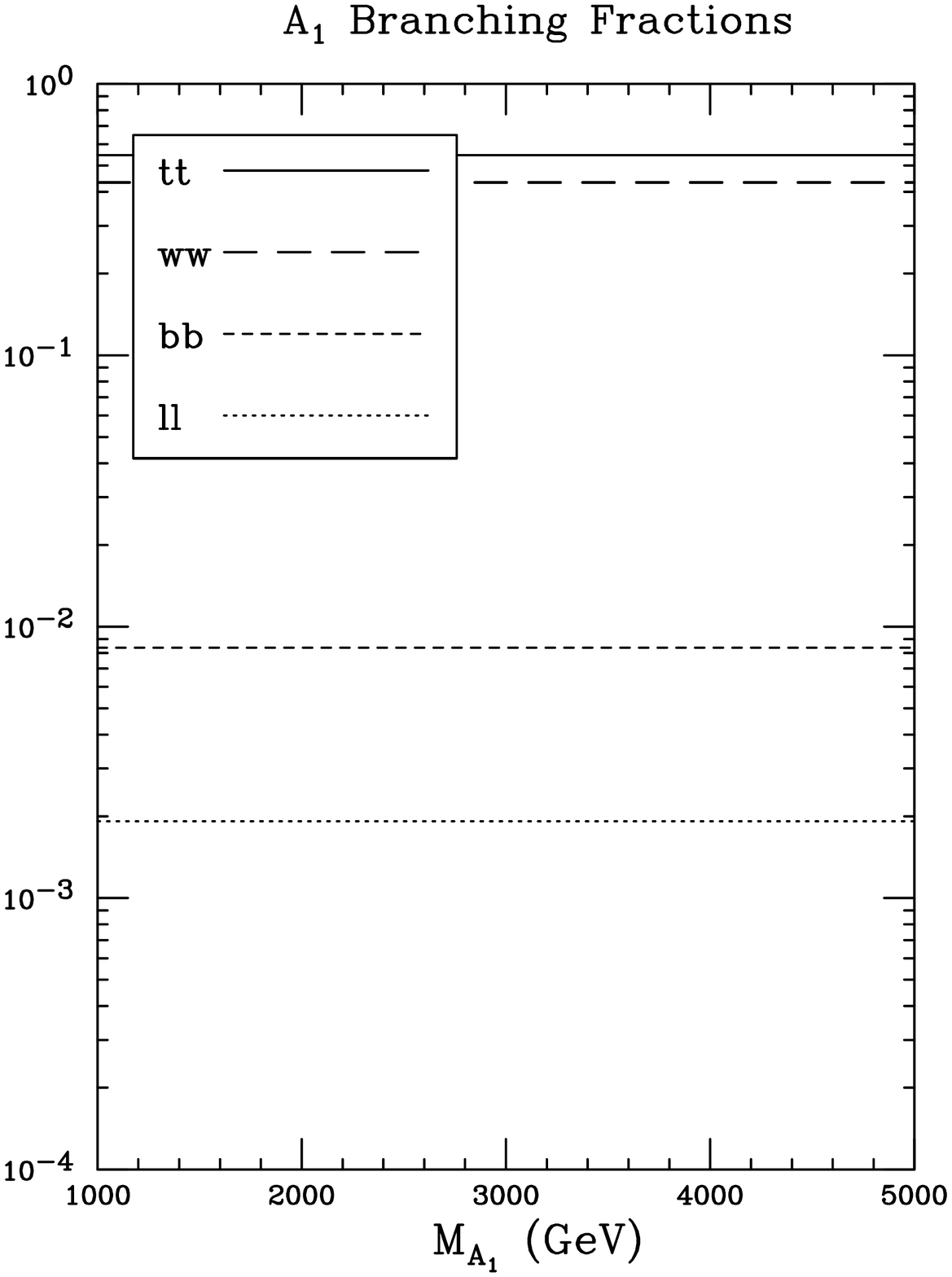}
\caption[]{Decay branching fractions for the first KK photon, from \cite{Agashe:2007ki}. }
 \label{fig:AgashebrA1}
\end{figure}

\begin{figure}[htb]
\includegraphics[width=\smallimagesize]{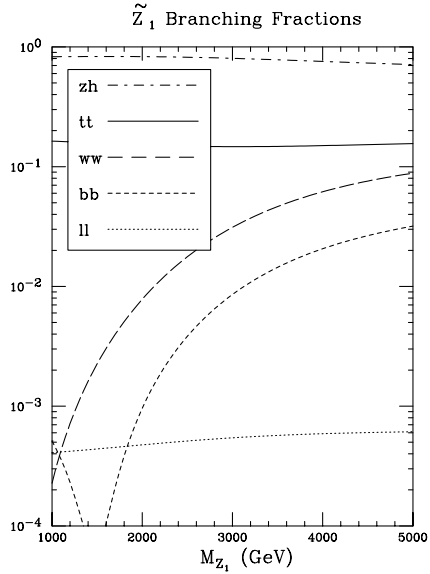}
\caption[]{Decay branching fractions for the first KK $Z$ mode, from \cite{Agashe:2007ki}. }
 \label{fig:AgashebrZ1}
\end{figure}

\begin{figure}[htb]
\includegraphics[width=\smallimagesize]{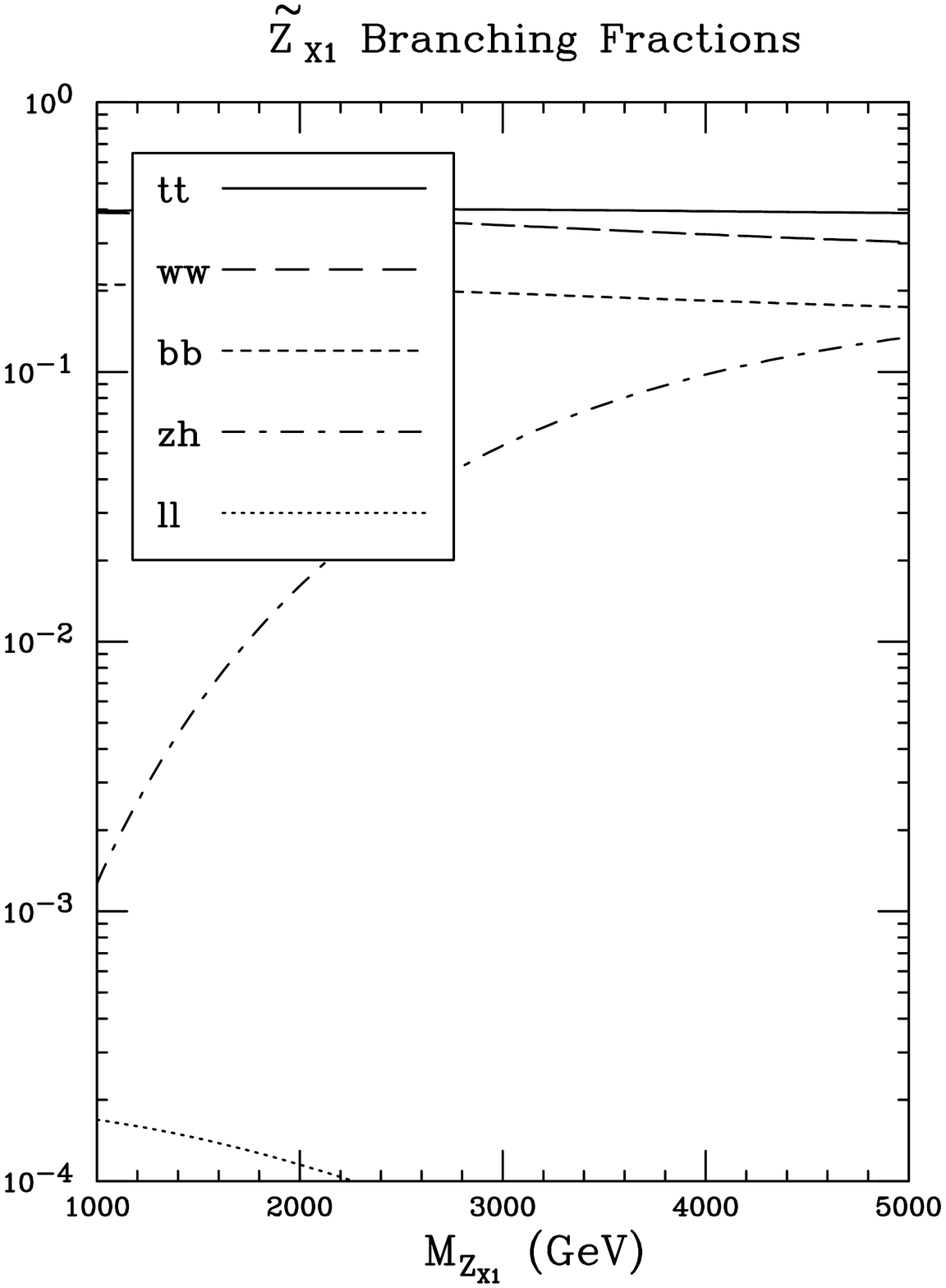}
\caption[]{Decay branching fractions for the KK $Z_x$ mode, from \cite{Agashe:2007ki}. }
 \label{fig:AgashebrZx1}
\end{figure}

Other KK gauge boson states are much harder to find. For example
production of the KK excitations of the $Z$ (aka the $Z^\prime$) is suppressed since it is localized on the
IR brane while the light quarks are localized on the UV brane. The KK excitations can also 
decay to difficult final states \cite{Agashe:2007ki} as seen in Figs. \ref{fig:AgashebrA1}-\ref{fig:AgashebrZx1}.
It would take about 100\,fb$^{-1}$ of integrated luminosity to uncover a 2\,TeV 
$Z^\prime$, and 1 ab$^{-1}$ for a 3\,TeV $Z^\prime$ \cite{Agashe:2007ki}.

\begin{figure}[htb]
\includegraphics[width=\imagesize]{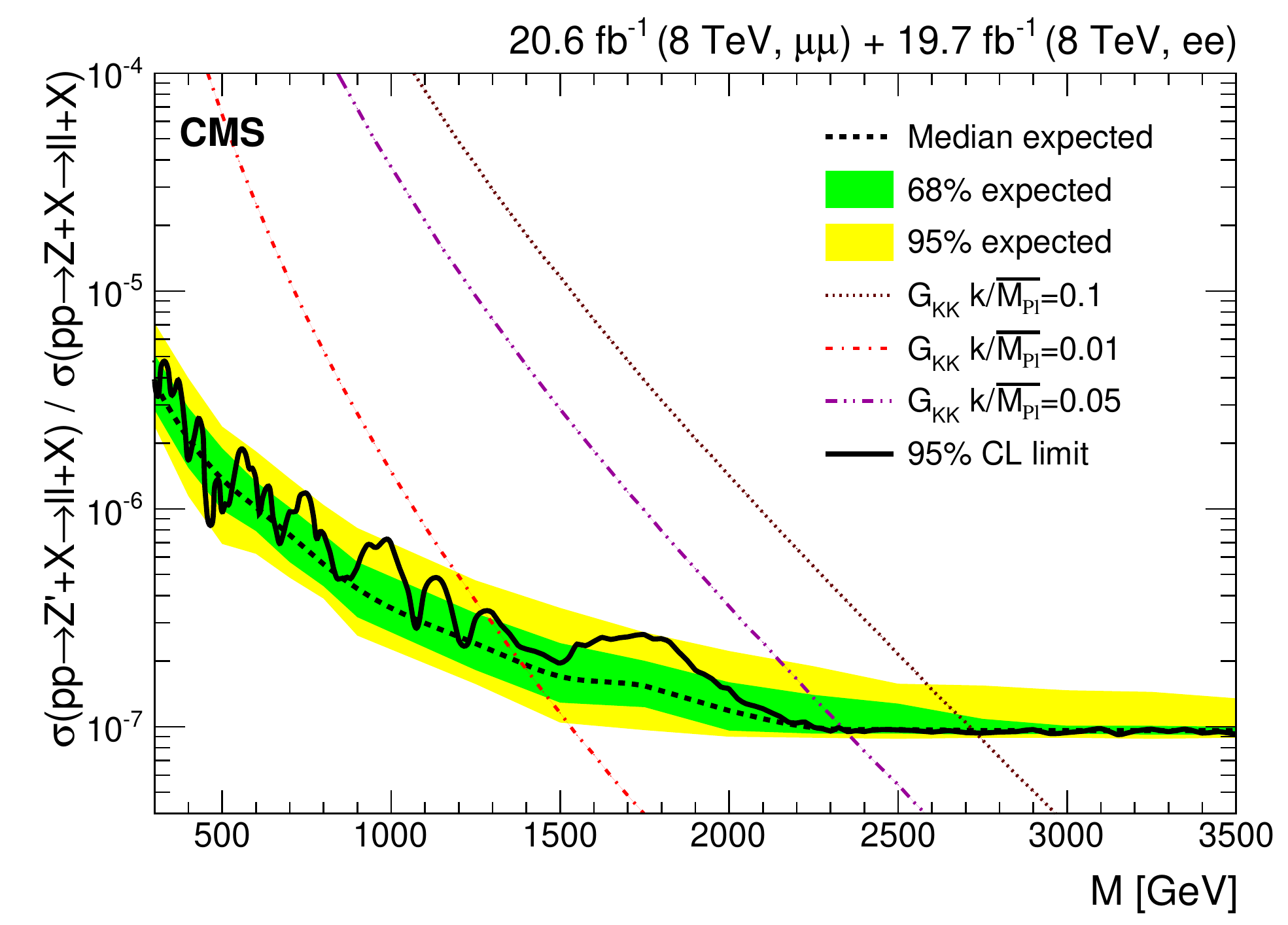}
\caption[]{Bound on the KK graviton from the dilepton channel in the RS model for different coupling strengths $k/{\bar M}_{\rm Pl}$, from \cite{Khachatryan:2014fba}. Color online.}
 \label{fig:RSgravitondilepton}
\end{figure}

\begin{figure}[htb]
\includegraphics[width=\smallimagesize]{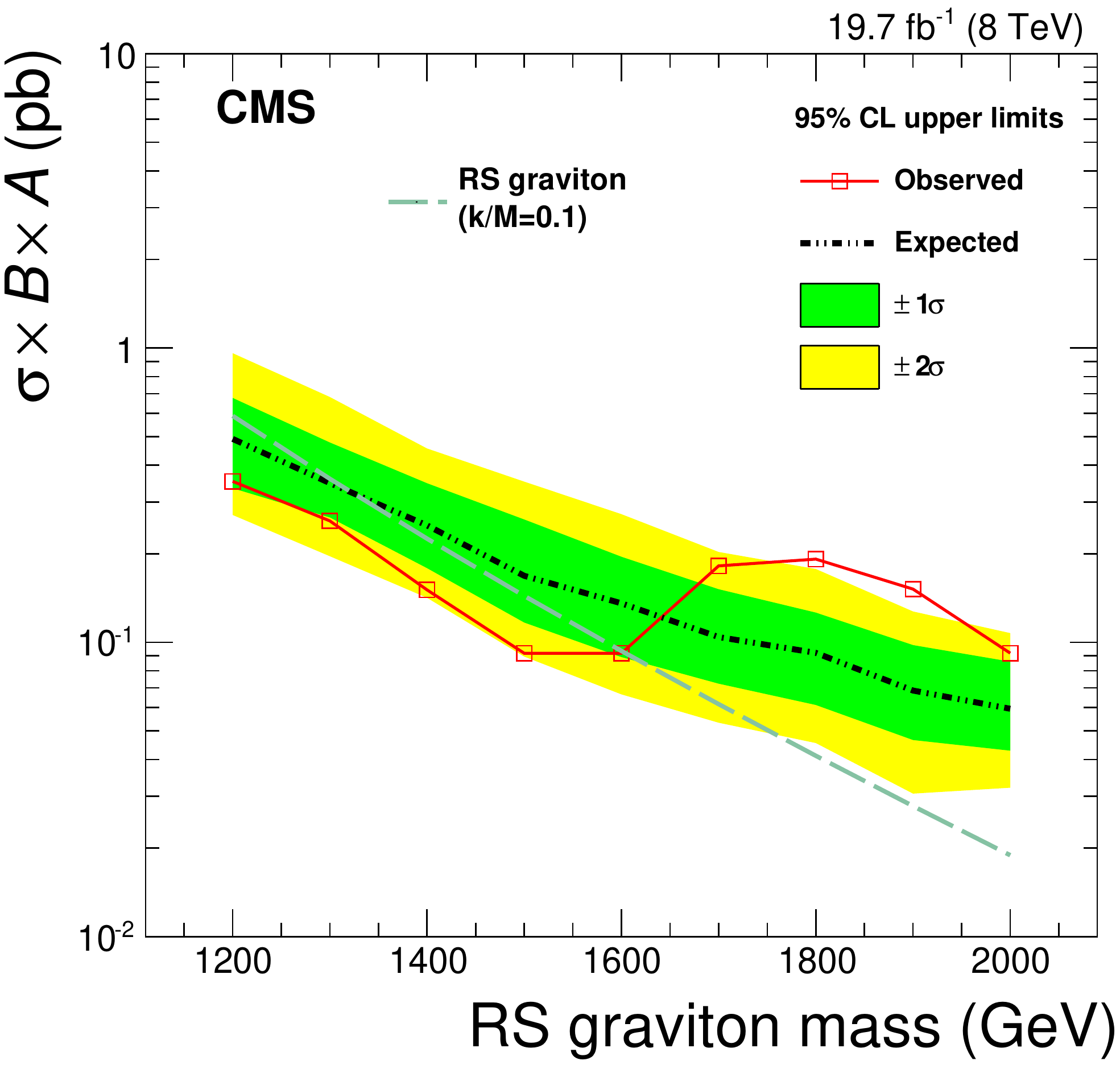}
\caption[]{Bound on the KK graviton from the dijet channel in the RS model, from \cite{Khachatryan:2015sja}. Color online.}
 \label{fig:RSgravitondijet}
\end{figure}

\begin{figure}[htb]
\includegraphics[width=\imagesize]{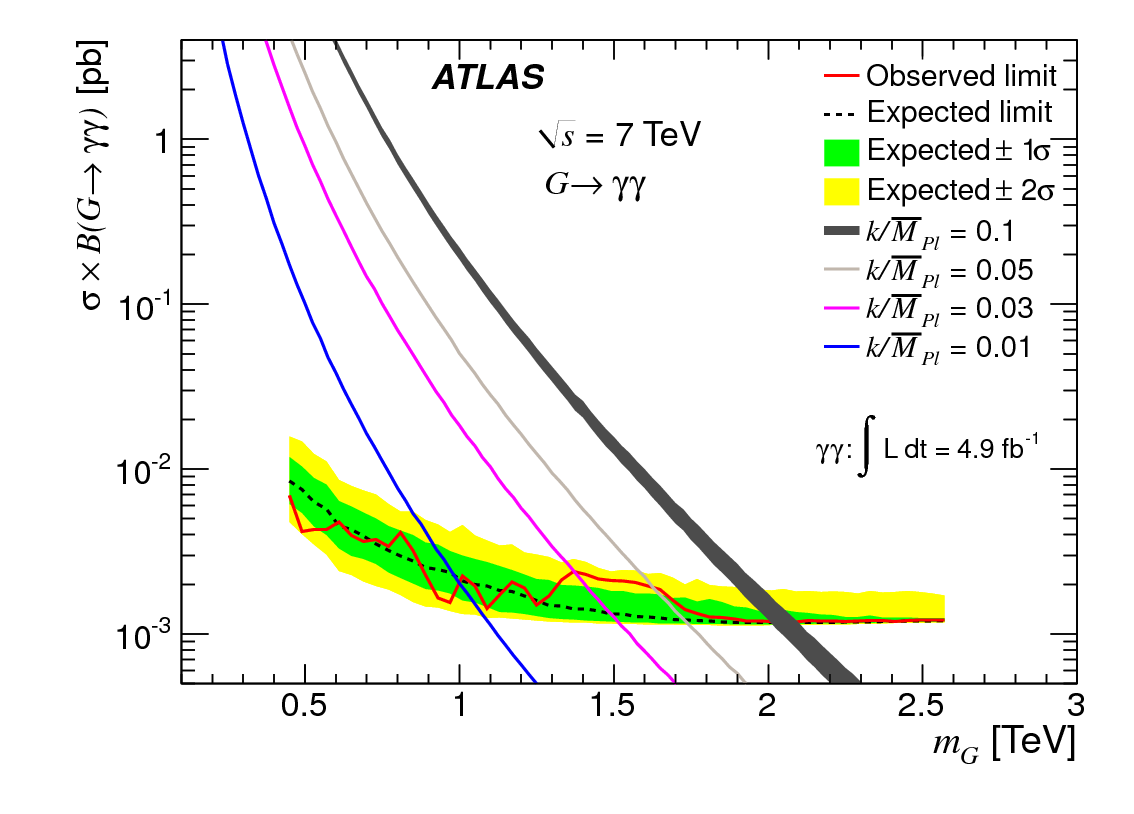}
\caption[]{Bound on KK graviton from the diphoton channel in the RS model for different coupling strengths $k/{\bar M}_{\rm Pl}$, from \cite{Aad:2012cy}. Color online.}
 \label{fig:RSgravitondiphoton}
\end{figure}

\begin{figure}[htb]
\includegraphics[width=\largeimagesize]{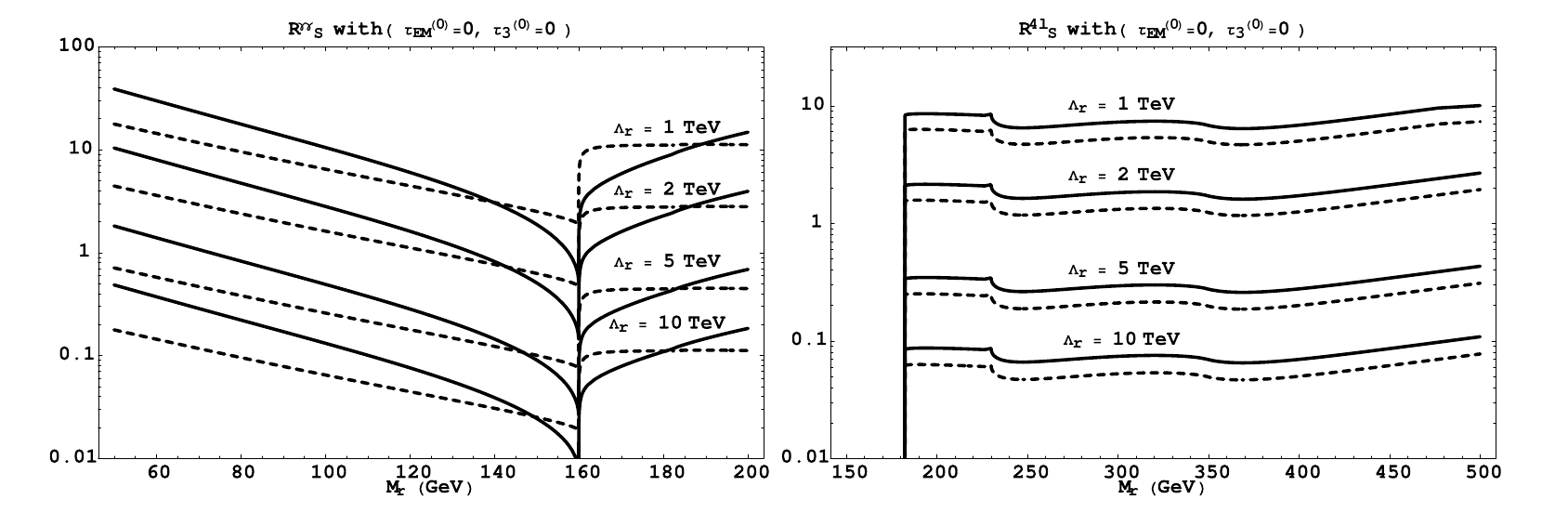}
\caption[]{Discovery significance for the radion in the realistic RS model (solid), and 
RS1 model (dashed), from \cite{Csaki:2007ns}. }
 \label{fig:realisticradion}
\end{figure}

\begin{figure}[htb]
\includegraphics[width=\imagesize]{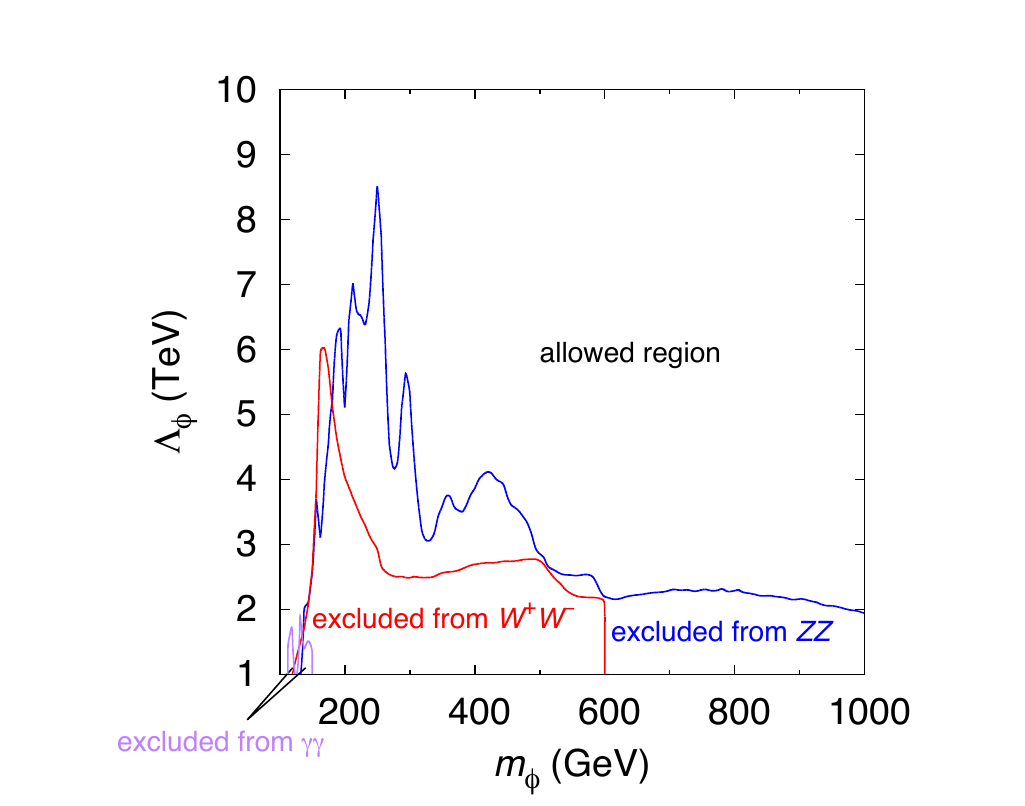}
\caption[]{Bound on the mass of the RS radion ($m_\phi$) as a function of its coupling ($\Lambda_\Phi$), where the coupling is given by $\frac{\Phi}{\Lambda_\Phi} T^\mu_\mu$, from \cite{Cho:2013mva}. Color online.}
 \label{fig:RSradionbound}
\end{figure}

Currently LHC data puts bounds around 1.3-2.7\,TeV on KK gravitons in RS models from dileptons (see Fig.~\ref{fig:RSgravitondilepton}), and 
dijets (see Fig.~\ref{fig:RSgravitondijet}). The diphoton channel currently gives KK graviton bounds around 1-2\,TeV as shown in Fig.~\ref{fig:RSgravitondiphoton}.
Taking into account the trace anomaly, one sees \cite{Csaki:2007ns}
that loops on the brane generate significant couplings for the radion.
In fact the discovery significance for the radion can be comparable to the Higgs \cite{Csaki:2007ns,Grzadkowski:2012ng},
as shown in Fig. \ref{fig:realisticradion}. The Higgs searches at the LHC can also be used to set bounds on the RS radion, see for example~\cite{Cho:2013mva,Bhattacharya:2014wha,Desai:2013pga}. The bound from decays to $WW$ and $ZZ$ (assuming a brane localized SM and as a function of its coupling and mass) are displayed in Fig.~\ref{fig:RSradionbound}. 


\subsection{Bulk Higgs}

\label{subsection:gaugephobic:bulkaction}

Extra dimensions allow us to modify the realistic RS models by lifting the Higgs itself into the bulk as well \cite{Davoudiasl:2005uu,gaugephobic,Dey:2009gf,Vecchi:2010em,Cabrer:2011vu,Das:2011fb,Archer:2012qa,Frank:2013un,Chang:2000zh}. In order to  solve the hierarchy problem, the profile should be peaked close to the IR brane, but does not necessarily have to be exactly localized (as we have assumed till now). In addition, the magnitude of the Higgs VEV does not have to exactly reproduce the value in the SM in order to obtain the correct $W$ and $Z$ masses. This can be understood in the following way. Raising the Higgs VEV will deform the wave functions of the gauge bosons, as a consequence even in the $v\to \infty$ limit one does not send the gauge boson masses to infinity. Instead one obtains Higgsless models \cite{Csaki:2003dt}, where the only role played by the Higgs is to enforce the IR brane boundary condition 
\begin{equation}
 g_{5L}A^{L\,a}_\mu-g_{5R}A^{R\,a}_\mu=0
\end{equation}
where $g_{5L,R}$ are the bulk gauge couplings of the $SU(2)_{L,R}$ groups, while the orthogonal combination has a Neumann BC. In this case the masses of the $W$ and $Z$  bosons are set entirely by the size and geometry of the extra dimension, and one can think of such higgsless models as extra dimensional versions of technicolor, with the important significant difference that the Higgsless models could be weakly coupled. In this case one still expects perturbative unitarization of the $WW, WZ$ scattering amplitudes, however since there is no state analogous to the SM Higgs, the unitarization happens \cite{Chivukula:2002ej,SekharChivukula:2001hz,Csaki:2003dt} via the exchange of the gauge boson KK modes $W', Z', W'', Z'', \ldots ~$ The generic large energy expansion of the $WW$ scattering amplitude is given by  
 \begin{equation}
\mathcal{A}= A^{(4)} \frac{E^4}{M_W^4} + A^{(2)} \frac{E^2}{M_W^2}+
A^{(0)}+{\cal O}\left(\frac{M_W^2}{E^2}\right).
 \end{equation}
The requirement of unitarity that the $A^{(4)}, A^{(2)}$ amplitudes must vanish imposes sum rules among the masses and couplings of the various KK modes of the $W$ and $Z$ given by \cite{Csaki:2003dt}
\begin{eqnarray}
g_{WWWW}^2 &=& \sum_n g_{WWZ^{(n)}}^2 \nonumber \\
g_{WWWW}^2 M_W^2 & = & \frac{3}{4} \sum_n g_{WWZ^{(n)}}^2 M_{Z^{(n)}}^2
\end{eqnarray}
for the case of $WW$ scattering, where $g_{WWZ^{(n)}}$ is the cubic coupling between the ordinary $W$ and the various KK modes of the $Z$, and $M_{Z^{(n)}}$ is the mass of the KK $Z$'s. A similar sum rule applies for $WZ$ scattering. 

Since a Higgs-like particle has been discovered pure Higgsless models are excluded. Bulk Higgs models can however lead to situations where some of the features (e.g. the contribution of KK modes to unitarization) persists, albeit at a sub-leading level. By increasing the value of the Higgs VEV one merely fixes the size of the extra dimension by making sure the observed $W$ and $Z$ masses are still reproduced correctly. Such models have been referred to as gaugephobic Higgs models \cite{gaugephobic} but we will refer to them here as bulk Higgs models.

If we simply follow the AdS bulk gauge setup of the realistic RS  models
then we can take the 
Higgs field to be a bidoublet of $SU(2)_L \times SU(2)_R$ with $U(1)_{B-L}$ charge zero.
The Lagrangian  for the Higgs includes a bulk mass ($\mu^2$ in units of the
inverse curvature radius $k=R^{-1}$), and brane potentials.
The brane potentials gives us the freedom to choose the Higgs BCs.  To ensure that the Higgs has a VEV 
\beq
\langle \mathcal{H}  \rangle = \mathcal{H} = \left( \begin{array}{cc}
v(z) & 0 \\
0 & v(z) \end{array} \right) 
\eeq
we can take
\begin{equation} \label{eq:Vtev}
V_{\rm TeV} = \left( \frac{R}{R'} \right)^4 \frac{\lambda R^2}{2} \left(  {\rm Tr}  \left| \mathcal{H} \right|^2 - \frac{v_{\rm TeV}^2}{2} \right)^2\,.
\end{equation}

The bulk equations of motion give us power law solutions for the Higgs VEV.
A convenient way to parameterize the bulk profile of the Higgs VEV is to write
\beq
v(z) = a \left(\frac{z}{R}\right)^{2 +\beta}~,
\quad\quad
\beta = \sqrt{4 +\mu^2}~.
\eeq
The AdS/CFT correspondence suggests that in a CFT description the scaling dimension of the operator that breaks 
electroweak symmetry is $d=2+\beta$.  For a weakly coupled Higgs the dimension of its mass operator $|\mathcal{H} |^2$ is
$2d= 4+2\beta$, so as long a $\beta >0$ the mass operator is irrelevant, which is to say that it does not suffer from  a quadratic divergence, and
the hierarchy problem is solved. It is also possible to arrange for $\beta < 0$ \cite{Cacciapaglia:2008ns,Klebanov:1999tb}, in which case one can only solve
the little hierarchy problem (cf. section \ref{subsec:littlest}).

A normalization, $V$, can be chosen \cite{gaugephobic} so that 
\beq\label{eq:VEVProfile}
v(z) = \sqrt{ \frac{2 (1+\beta) \log R'/R}{1-(R/R')^{2+2\beta}}} \frac{g V}{g_5} \frac{R'}{R} \left(\frac{z}{R'}\right)^{2+\beta}
\eeq
This is useful for comparing with the SM limit where the elementary Higgs has scaling dimension $d=1$ (and thus $\beta=-1$) and $V=246$\,GeV.
As discussed in Section \ref{subsec:classifying}, these models have a two dimensional parameter space ($V$,$\beta$).  We are already familiar with some of the limiting cases
of the parameter space.  The corner with large scaling dimension and small VEV is the realistic RS model; the large scaling dimension means that the Higgs is 
localized on the TeV brane.  The corner with a large VEV and large scaling dimension corresponds to the Higgsless limit.  The corner with $\beta=-1$ and $V=246$\,GeV is  the SM. As we have mentioned, the fact that the SM has $\beta=-1$ is indicative of its hierarchy problem. In this section we will restrict ourselves to $\beta\ge 0$, and defer discussion of $-1<\beta<0$ to section~\ref{sec:Unhiggs}. The relation of various models of electroweak symmetry breaking is shown in Table~\ref{table:GaugephobicModels}.

\begin{table}[htb]
    \begin{center}
\begin{tabular}{l|cc}
                                   & V& $\beta$  \\
    \hline
    Higgsless   &     $\infty$     & $\infty$  \\
     RS1    &  246\,GeV& $\infty$ \\
    Bulk Higgs benchmark
         & 300\,GeV   & 2   \\
    Minimal PGB Higgs &  246\,GeV & 0 \\
    quantum critical benchmark &    246\,GeV &   -0.3\\
    Standard Model                    & 246\,GeV   &      -1
\end{tabular}
\end{center}
\caption{Bulk Higgs parameters for benchmark points studied, and limits for other models. The minimal pseudo-Goldstone boson Higgs model is discussed in 
section~\ref{sec:minimalpseudo-goldstoneHiggs}, the quantum critical Higgs model is discussed in section~\ref{sec:Unhiggs}.  }\label{table:GaugephobicModels}
\end{table}

Unitarization proceeds differently in the bulk Higgs model compared to the SM.  The contributions to terms in the $WW$ scattering amplitude that grow like energy squared have
the following form
\begin{equation}
   \mathcal{A}^{(2)} \sim g_{WWWW}^2 - \frac{3}{4} \sum_{k} \frac{M_{Z^k}^2}{M_W^2} g_{WWZ^k}^2 - \frac{1}{4} \sum_{k} g_{WWH^k}^2\,,
\label{sumrule}
\end{equation}
where the first term is the four $W$ contact term, the first sum is over KK modes of the $Z$ and the final sum is over KK modes of the Higgs (cf. \textcite{Falkowski:2012vh,Bellazzini:2012tv}).  In the SM
the sums reduce to single terms.  In Higgsless toy-models the final sum vanishes but unitarization is maintained by the sum of $Z$ KK modes, which is usually dominated
by the lightest modes.  In a bulk Higgs model there are contributions from all three terms, again usually dominated by the lightest modes.  In every case 
$\mathcal{A}^{(2)} =0$, but different modes are responsible for the unitarization.

\begin{figure}[htb]
\includegraphics[width=0.8\hsize]{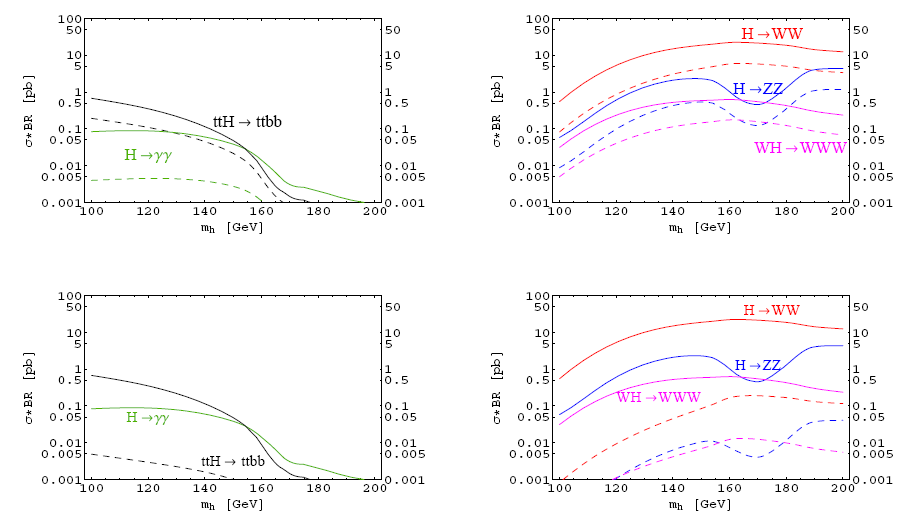}
\caption[]{Cross sections times branching ratios for various Higgs production and decay channels for the SM (solid lines) and bulk Higgs (dashed lines) for $\beta =2$ with $V=300$ (top) and $V=500$ (bottom), from \cite{gaugephobic}. Color online.} 
\label{fig:gaugephobiccrossections}
\end{figure}

It is useful to define the following parameter in order to quantify how
``Higgsless'' the model is:
\begin{equation}
   \xi_{BH} \equiv \frac{ \sum_{k} g_{WWH^k}^2}{g_{WWH}^2 (SM)} \,.
   \label{eq.xiGP}
\end{equation}
With the LHC data for the Higgs-like boson we are far away from the Higgsless limit $\xi_{BH} =0$, and are roughly constrained to  models with $\xi_{BH} >0.9$.
The production cross section times branching fractions of a bulk Higgs and an elementary Higgs are shown in Fig. \ref{fig:gaugephobiccrossections}. This suggests that the current data constrains us to $V <300$\,GeV.

As in any model with KK gauge bosons, care must be taken not to mess up the $Zb{\bar b}$ coupling.  An interesting way of doing this was proposed by \textcite{Agashe:2006at}
who suggested that the top-bottom quark doublet should have a left-right interchange symmetry to protect the $Zb{\bar b}$ coupling from corrections.  This has many implications \cite{Carena:2007ua,Carena:2006bn,Agashe:2006hk} including top-partner fermions with exotic electric charges \cite{Contino:2008hi,Mrazek:2009yu,Dissertori:2010ug}, cf. section \ref{sec:minimalpseudo-goldstoneHiggs}.

\subsection{Bulk Higgs LHC Searches}
\label{subsection:gaugephobic:searches}

\begin{figure}[htb]
\includegraphics[width=\smallimagesize]{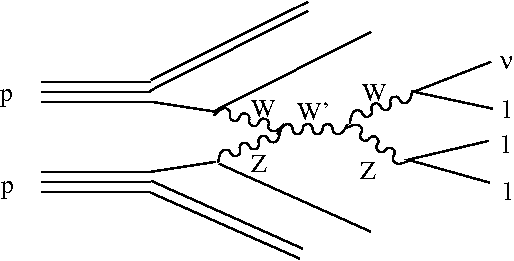}
\caption[]{$WZ$ scattering in $pp$ collisions.} 
\label{fig:WZscattering}
\end{figure}

\begin{figure}[htb]
\includegraphics[width=\smallimagesize]{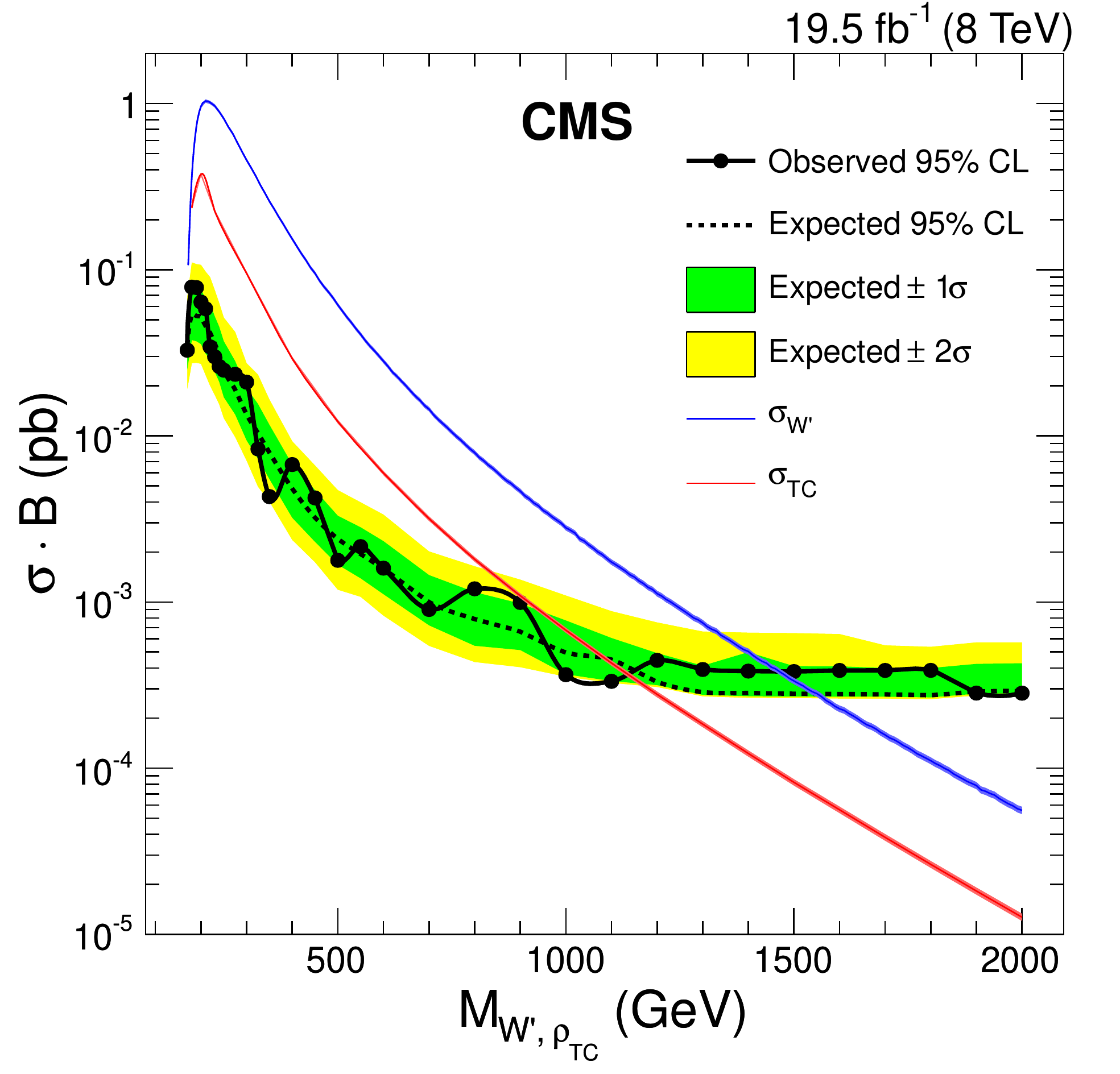}
\caption[]{CMS bound on the $W^\prime$ mass, from \cite{Khachatryan:2014xja}. Color online.} 
\label{fig:CMSWprime}
\end{figure}

\begin{figure}[htb]
\includegraphics[width=\imagesize]{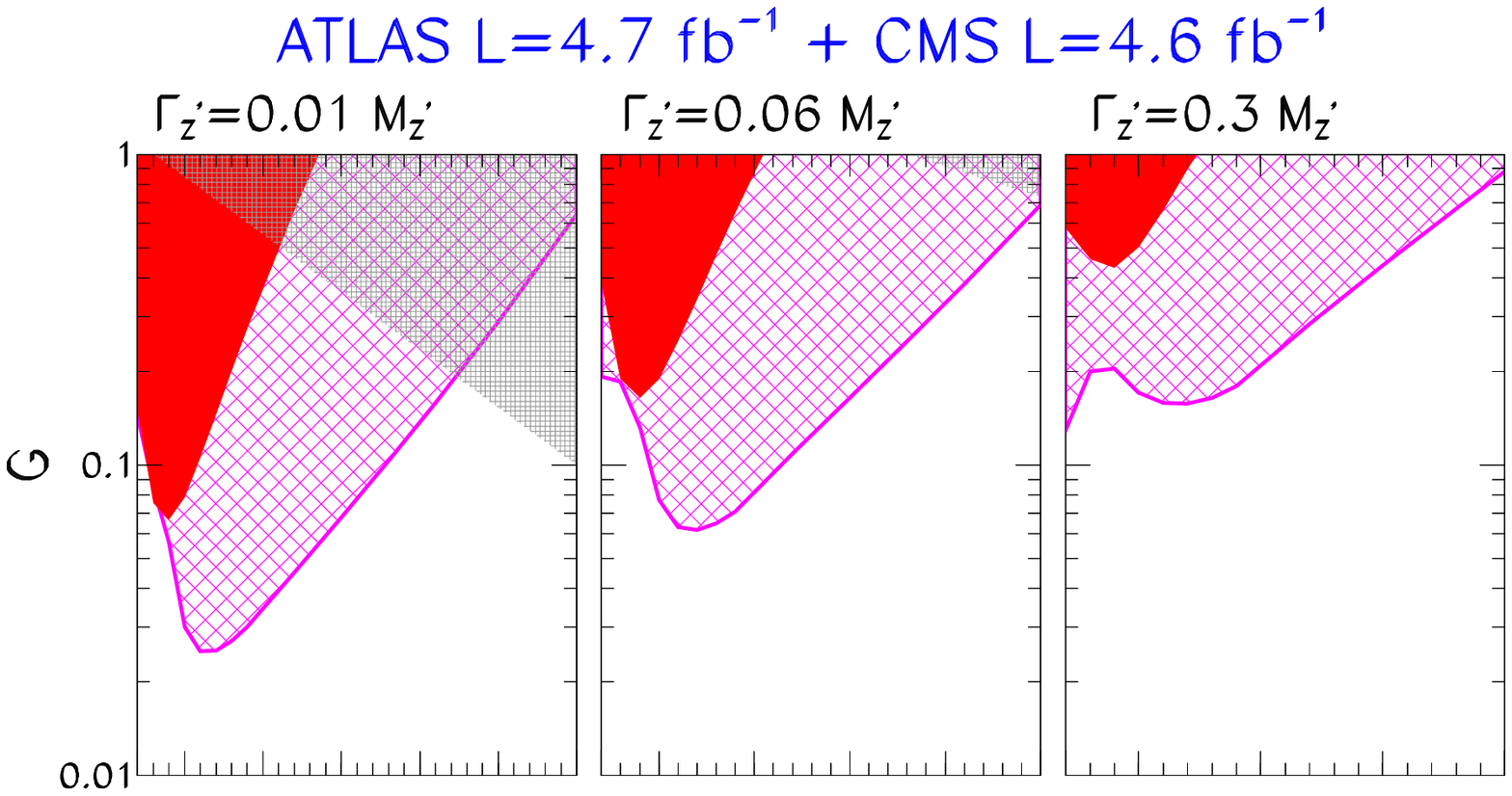}\\
\includegraphics[width=\imagesize]{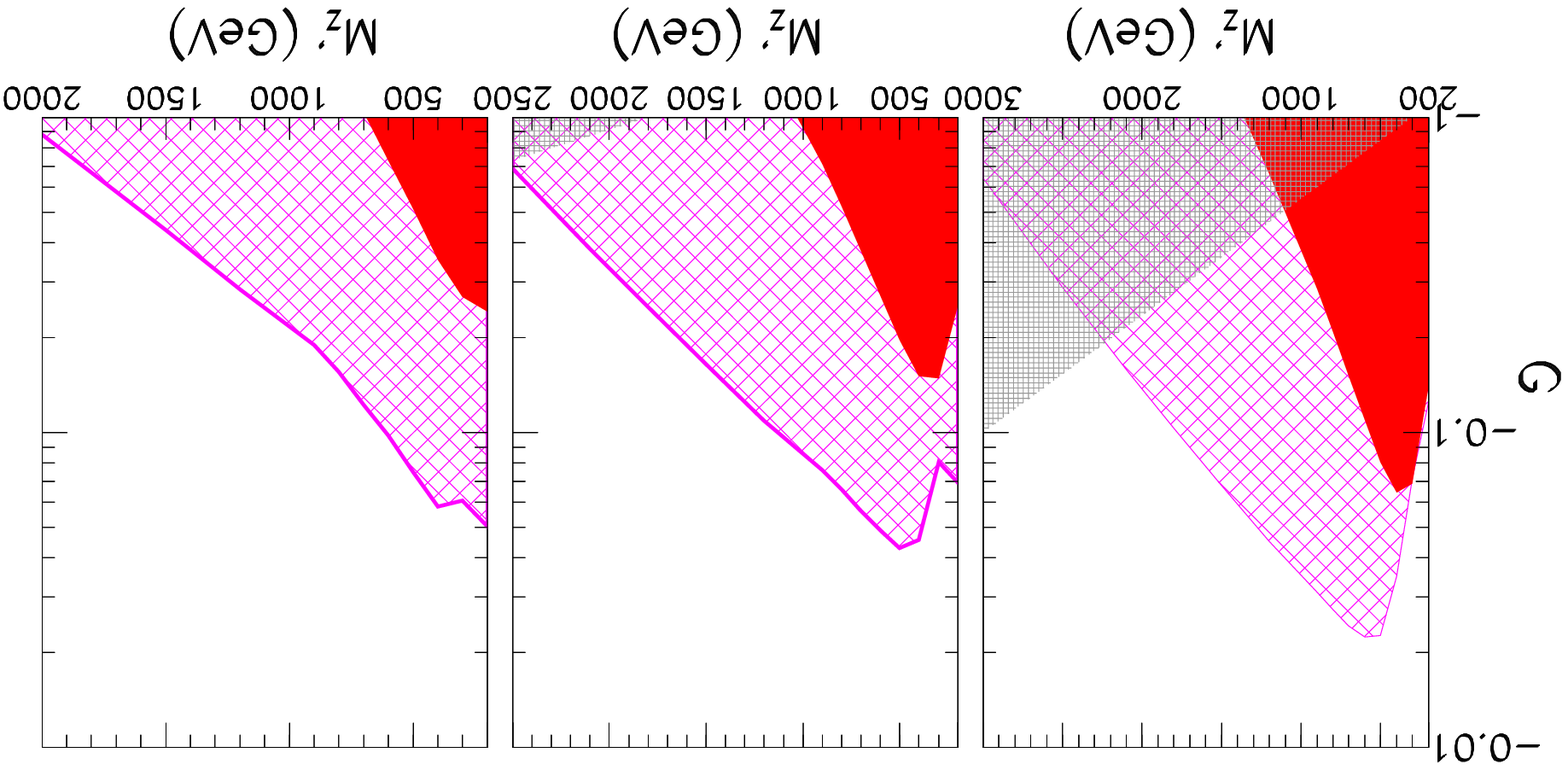}
\caption[]{Model independent bounds on neutral spin-1 resonances in the mass-coupling plane for different $Z^\prime$widths, from  \cite{GonzalezFraile:2012fq}. Color online.} 
\label{fig:eboli}
\end{figure}

The most promising channel for searching for these models is in $WZ$ scattering (see Fig. \ref{fig:WZscattering}) where one can observe the $W^\prime$ resonance
\cite{Birkedal:2005yg,Agashe:2008jb} with 300\,fb$^{-1}$ of integrated luminosity. 
Further studies of the $W^\prime$ were performed in \textcite{Davoudiasl:2000wi} and \textcite{Hewett:2004dv}.
CMS has already performed a preliminary search in this channel \cite{Khachatryan:2014xja}, the 
results are shown in Fig. \ref{fig:CMSWprime}. CMS finds a lower bound of 1500\,GeV on the mass of a sequential $W^\prime$ decaying to $WZ$. Of course KK $W$s have suppressed couplings and the bound is weaker for smaller couplings.
Model independent analyses  \cite{Eboli:2011ye,Andreev:2014fwa} of ATLAS  and CMS bounds on neutral spin-1 resonances (i.e. the $Z^\prime$ search \cite{Langacker:2008yv,Agashe:2007ki}) can also be applied to these models.
 Figure \ref{fig:eboli} shows the bounds in the mass-coupling plane for different $Z^\prime$ widths. A  model-independent parametrizaton of $WZ$ resonances has been advocated in \cite{Pappadopulo:2014qza} and has been used in experimental analyses \cite{Khachatryan:2015bma,Aad:2015yza}.

\begin{figure}[htb]
\includegraphics[width=\smallimagesize]{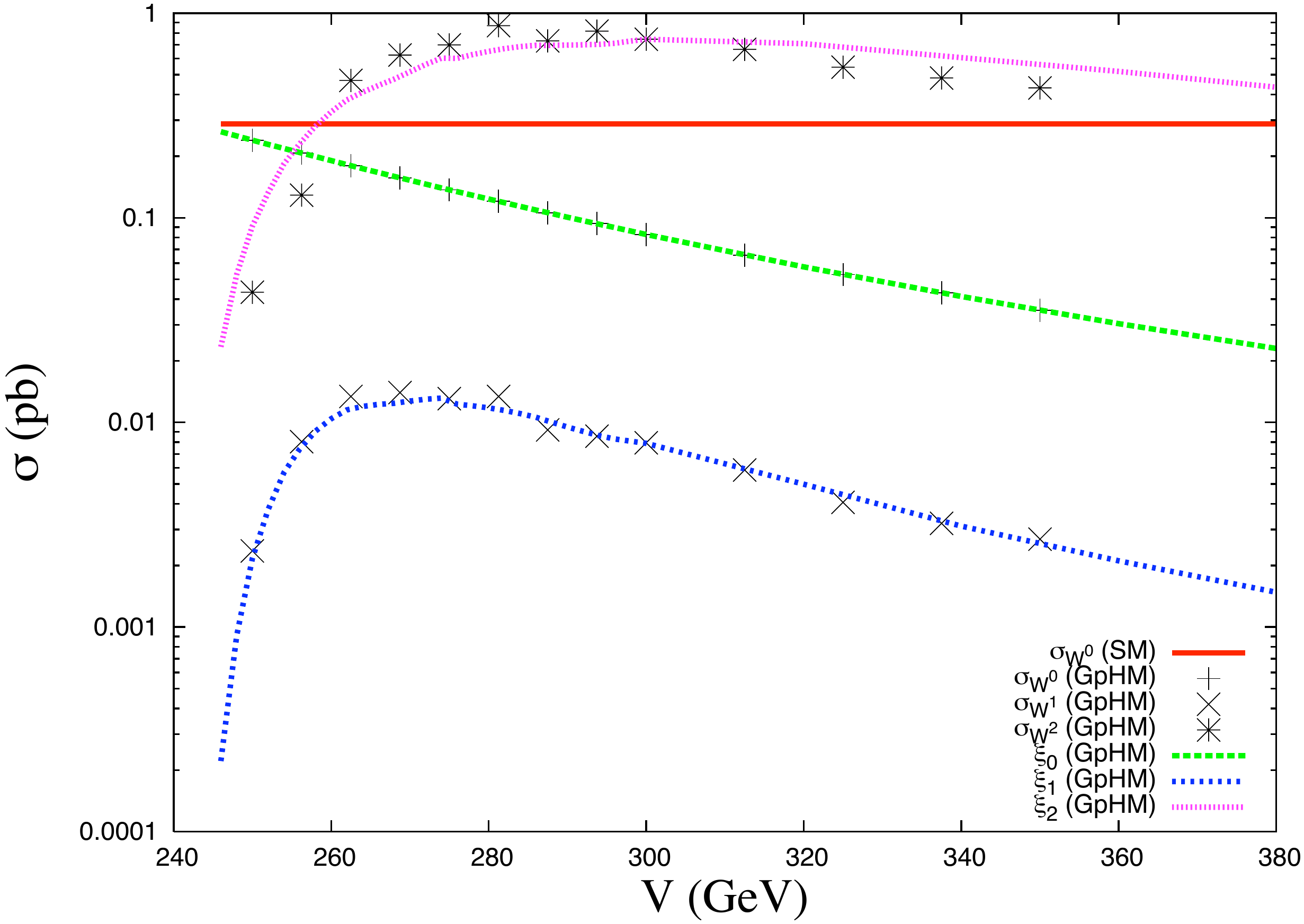}
\caption[]{Approximate contribution to $pp \to W H$ from an intermediate $W'$, from \cite{Galloway:2009xn}. Color online.  }
\label{fig:WixsvsV}
\end{figure}

\begin{figure}[htb]
\includegraphics[width=\smallimagesize]{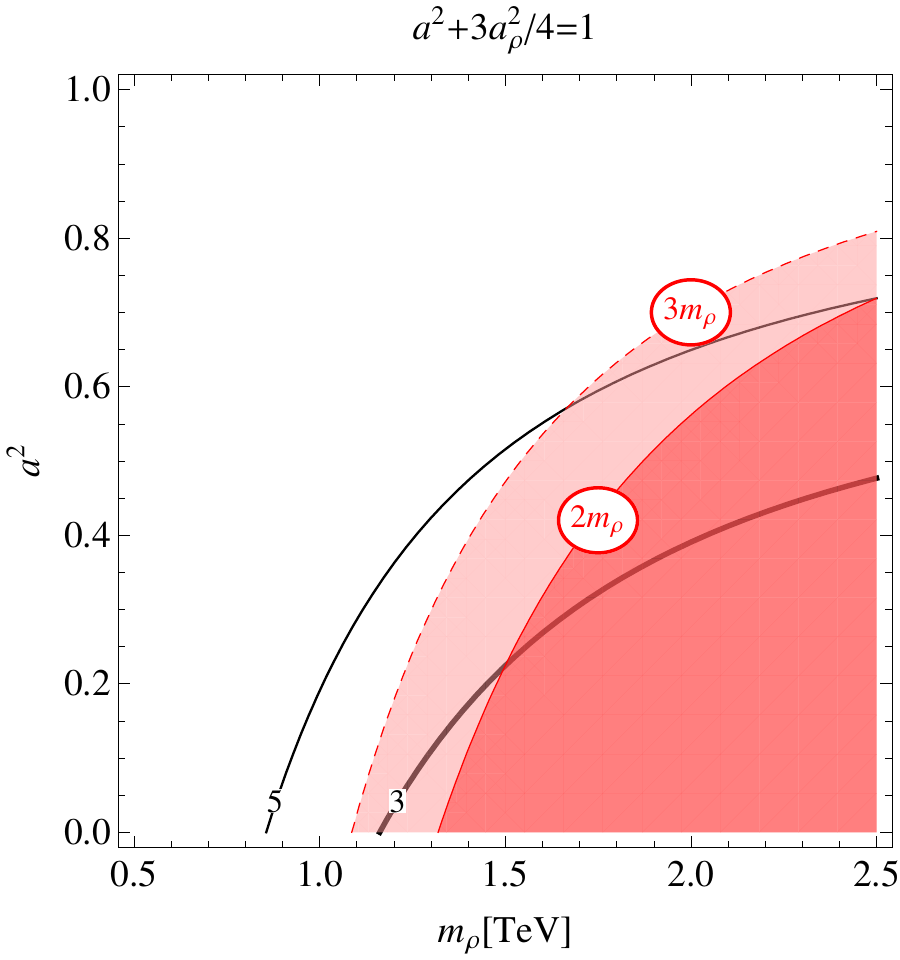}
\caption[]{Higgs and unitarity bounds combined to put an upper bound  on the spin-1 resonance, $\rho$, for different cutoffs $\Lambda= 3$\,TeV, 5\,TeV, $2 m_\rho$, $3m_\rho$, from  \cite{Bellazzini:2012tv}. Here $a$ is the strength of the Higgs
coupling to $WW$  relative to the SM. Color online.} 
\label{fig:bellazzini3}
\end{figure}

A promising channel for a bulk  Higgs search is $pp \to W H$ through an intermediate $W'$
\cite{Galloway:2009xn}.   In contrast to the similar process in a ``little" Higgs model (see section \ref{subsection:pseudoLHCsearch}), the bulk  Higgs has an enhanced coupling to $WW'$.
As shown in Fig.~\ref{fig:WixsvsV}, one finds an enhanced cross-section relative to the SM for a broad range of $V$.  The bulk Higgs signal also has very different kinematics from the SM background so it is relatively easy to introduce cuts that significantly reduce the background \cite{Galloway:2009xn}.
Another  interesting way to get bounds on these models is to use bounds from the Higgs $\rightarrow WW$ search.  Assuming that $WW$ scattering is unitary, then if
the Higgs coupling is not large enough, we need another resonance to unitarize the scattering, and this resonance cannot be too heavy or it cannot perform its job.  
With the observed Higgs mass  this bound can be translated into an upper bound  on the resonance mass as shown in Fig. \ref{fig:bellazzini3}

\section{COMPOSITE PSEUDO-GOLDSTONE BOSON HIGGS}

\subsection{The ``Littlest" Higgs Model}
\label{subsec:littlest}

\begin{figure}[htb]
\includegraphics[width=\imagesize]{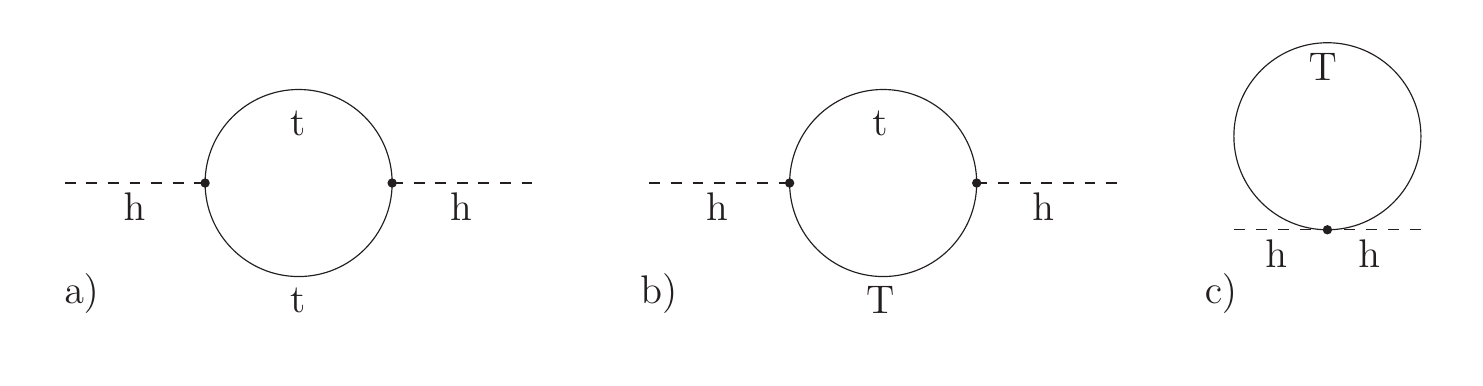}
\caption[]{Cancellation of the quadratic divergences in the top loop for ``little" Higgs models,  from \cite{Perelstein:2005ka}. }
 \label{fig:littleHiggs}
\end{figure}

\begin{figure}[htb]
\includegraphics[width=\smallimagesize]{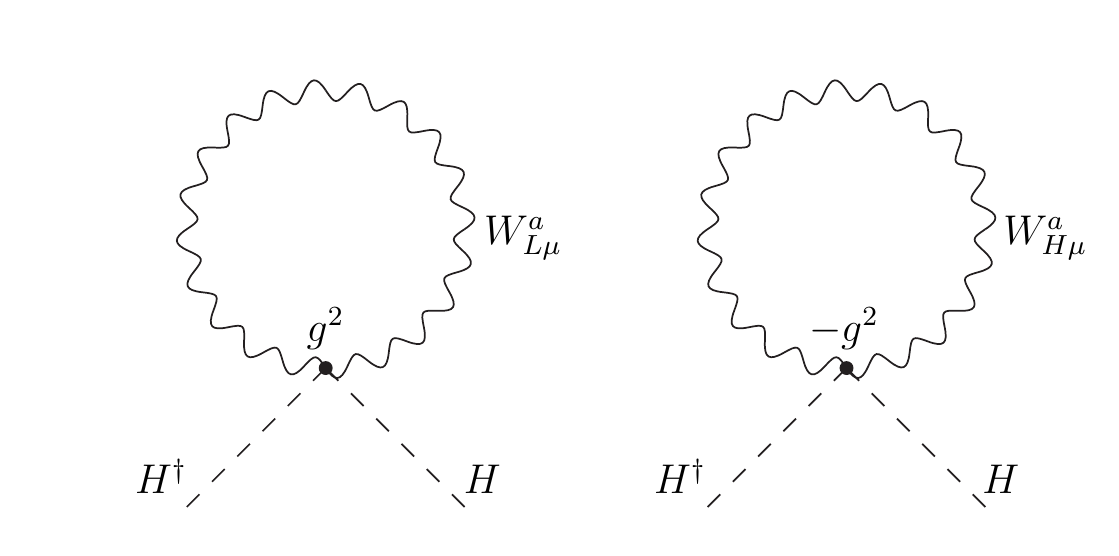}
\caption[]{Cancellation of the quadratic divergences in gauge loops for ``little" Higgs models,  from \cite{Perelstein:2005ka}. }
 \label{fig:littleHiggsgauge}
\end{figure}

Solving the hierarchy problem resolves why the weak scale is small compared to the Planck scale (or some other very high scale). The little hierarchy problem refers to a smaller problem: why were there no new effects seen at LEP suppressed by just a few TeV. In other words, why is the weak scale small compared to 10\,TeV? In order to address this we
will turn to an alternative where the Higgs is a pseudo-Goldstone Boson.
Models with pseudo-Goldstone boson Higgs bosons were proposed in the 1970s \cite{Georgi:1974yw,Georgi:1975tz} and thoroughly considered in the 1980s \cite{Georgi:1984ef,Kaplan:1983fs,Kaplan:1983sm,Georgi:1984af,Dugan:1984hq}.
However one-loop corrections to the Higgs mass meant that the cutoff scale of the effective theory could not be much above 1\,TeV, which is no better than the SM.
In the 21st century, the technique of dimensional deconstruction \cite{Hill:2000mu,ArkaniHamed:2001ca,ArkaniHamed:2001nc,ArkaniHamed:2002pa}, which is essentially latticizing an extra dimension, led to the idea of collective breaking.  If a global 
symmetry becomes exact when two different interactions go to zero separately, then the pseudo-Goldstone bosons that arise when this symmetry is spontaneously broken are doubly
protected.  If there are only loop corrections to the pseudo-Goldstone boson mass that raise it from being massless, then the leading contribution will come at two loop order, since
is must be proportional to both of the couplings rather than just one as is the usual case.  The way this actually comes about in specific models is that quadratic divergences
of the SM are cancelled by new `little partners' with the same spin as the SM particle, in contrast to SUSY where the superpartners have a different spin with the opposite statistics. The cancellation  of the quadratic divergence is shown in Fig.~\ref{fig:littleHiggs} and \ref{fig:littleHiggsgauge}.
This extra suppression is enough to move the naturalness cutoff from 1\,TeV,
as it is in the SM, to 10\,TeV. In this sense ``little" Higgs theories solve the little hierarchy problem.
What happens in such models at 10\,TeV is usually left as an open question for a future UV completion.
There are many different ``little" Higgs models that have been proposed, for detailed reviews of ``little" Higgs theories see \textcite{Schmaltz:2005ky,Perelstein:2005ka,Chen:2006dy,Cheng:2007bu,Bhattacharyya:2009gw}. Here we will limit ourselves mostly to the generic features of such models.
The weak point of these models is that while the mass terms only appear through loops, the Higgs quartic coupling can be generated at tree level, which results in to large
a quartic coupling and a physical Higgs mass that is generically above 125\,GeV.  Further tuning has to be introduced to get around this problem.

The so-called ``littlest Higgs" model \cite{ArkaniHamed:2002qy} is based on a non-linear $\sigma$ model describing
the breaking of 
$SU(5)$ to $SO(5)$~\cite{ArkaniHamed:2002qy}.
This can be arranged by giving a VEV to a symmetric tensor of the $SU(5)$ global symmetry, $\Sigma_0$.The Goldstone bosons are the fluctuations around this VEV, and can be parameterized by $\Pi =
\pi^a X^a$, where the $X^a$ are the broken generators of $SU(5)$. The
non-linear sigma model field is then
\beq
\Sigma (x) = e^{i\Pi/f} \Sigma_0 e^{i \Pi^T/f}=e^{2i\Pi/f} \Sigma_0.
\eeq
where $f$ is the analog of the pion decay constant that sets the scale of the symmetry breaking VEV.
Gauging an $SU(2)_1\times U(1)_1 \times SU(2)_2\times U(1)_2$ subgroup of $SU(5)$
completes the model. The model assumes that the quarks and leptons of the first two generations have their usual quantum numbers
under $SU(2)_L$
$\times$ U(1)$_Y$ assigned under the first $SU(2)_1 \times
U(1)_1$. The generators of  $SU(2)_1$, called $Q^a_1$, correspond (in a convenient basis) to the upper-left 2 $\times$ 2 block of the $SU(5)$ generator, similarly the
generators of  $SU(2)_2$ correspond to the lower-right 2 $\times$ 2 block.
The uneaten Goldstone fields include a ``little" Higgs doublet $(h^0,h^+)$ and $\phi$, a
complex triplet. One must ensure that $\phi$ does not get a large VEV, otherwise it will
give a large contribution to the isospin breaking $T$-parameter.

As in the chiral Lagrangian of QCD, the kinetic energy term in the low-energy effective theory for the Goldstone field is
\beq
\frac{f^2}{8} {\rm Tr} D_\mu \Sigma (D^\mu \Sigma)^\dagger 
\label{littleHiggseffective}
\eeq
where the gauge covariant derivative is given by
\beq D_\mu \Sigma = \partial_\mu \Sigma - i \sum_j \left[
g_j W_j^a (Q_j^a \Sigma + \Sigma Q_j^{aT} ) + g_j^\prime B_j( Y_j
\Sigma + \Sigma Y_j)\right]~. 
\eeq

At the scale of symmetry breaking $f$
(neglecting the Higgs VEV for the moment) the gauge group is broken to the diagonal subgroup.
The gauge bosons of the four groups mix to form the 
light electroweak gauge bosons and heavy partners.
In the ($W_1^a$, $W_2^a$) basis 
the mass matrix, which can be read off from Eq.~(\ref{littleHiggseffective}), is:
\beq
\frac{f^2}{4} \left(\begin{array}{cc} g_1^2  &  -g_1 g_2  \\
-g_1 g_2  &g_2^2   \end{array}\right)
\eeq
Diagonalizing we find that the light (actually massless until we include the Higgs VEV) and heavy mass eigenstates  are:
\beq
W_L^a = s W_1^a + c W_2^a, \quad 
W_H^a = -c W_1^a + s W_2^a,
\eeq
where we have the usual result for the mixing angles
\beq
s= \frac{g_2}{\sqrt{g_1^2+g_2^{2}}},\quad
c=\frac{g_1}{\sqrt{g_1^2+g_2^{2}}} ~.
\eeq
There are analogous results for the $U(1)$ mass eigenstates.
We can identify the light gauge bosons with the SM gauge bosons of $SU(2)_L\times U(1)_Y$,
and quarks and leptons of the first two generations have their SM couplings.
They also couple to the heavy gauge bosons ($W^a_H$,$B_H$) with strength
$(- g_1 c$ , $- g_1^\prime  c^\prime)$.

The kinetic term of
the ``little" Higgs field contains the coupling\footnote{For simplicity we will work in a unitary gauge and
only keep track of the $h \equiv {\rm Re}\ h^0$ component of the Higgs field.}
\beq
{\cal L}_{W^2 h^{2}}=
\frac{g_1 g_2}{4}  W^a_{1 \mu} W_2^{a \mu} h^{2}~.
\eeq
Expressing $W_1$ and $W_2$ in terms of the mass eigenstates we obtain a
\beq
{\cal L}_{W^2 h^{2}}= c\,s\, \left(W^a_{\mu L}W^{a \mu}_{ L}-W^{a \mu}_H W^{a}_{ \mu H}\right) h^2   +(s^2-c^2)  W^a_{\mu L}W^{a \mu}_H {h}^2~.
\eeq
Now we can see exactly how the cancellation of the quadratic divergence comes about, the symmetry forces equal but opposite couplings to $W_L^2$ and $W_H^2$ so their
one-loop contributions to the Higgs mass cancel. The mixed term does not contribute at one-loop because we cannot close off the $W_L$ propagator with $W_H$.
A similar analysis shows that the top Yukawa coupling $\lambda$ is related to the $h^2$  coupling to the top partner, $T$:
\beq
{\cal L}_{yuk} = \lambda\, h \,t_R^\dagger t_L - \frac{\lambda}{2 f} h^2 \,T_R^\dagger T_L + h.c.
\eeq
Again the one-loop contributions cancel, see Fig.\ref{fig:littleHiggs}.

Integrating out $W^a_H$ and $B_H$ induces  additional
operators in the effective theory,  which are quadratic in the light gauge fields
 and
quartic in Higgs fields or quadratic in $\phi$:
These operators 
give corrections to the light
gauge boson masses once the Higgs  gets a VEV:
\beq
\langle h \rangle = \frac{v}{\sqrt{2}}~,\quad
\langle \phi \rangle = v^\prime~,
\eeq
and including the effects of these higher dimension operators we find
that the masses of the $W$ and $Z$ receive corrections of order $  v^2/ f^2$ and $v^{\prime 2}/v^2$, which are potentially 
dangerous without some addition suppression mechanism\footnote{See the discussion of $T$-parity in section \ref{subsec:littlestvar}.}.

\begin{figure}[htb]
\includegraphics[width=\smallimagesize]{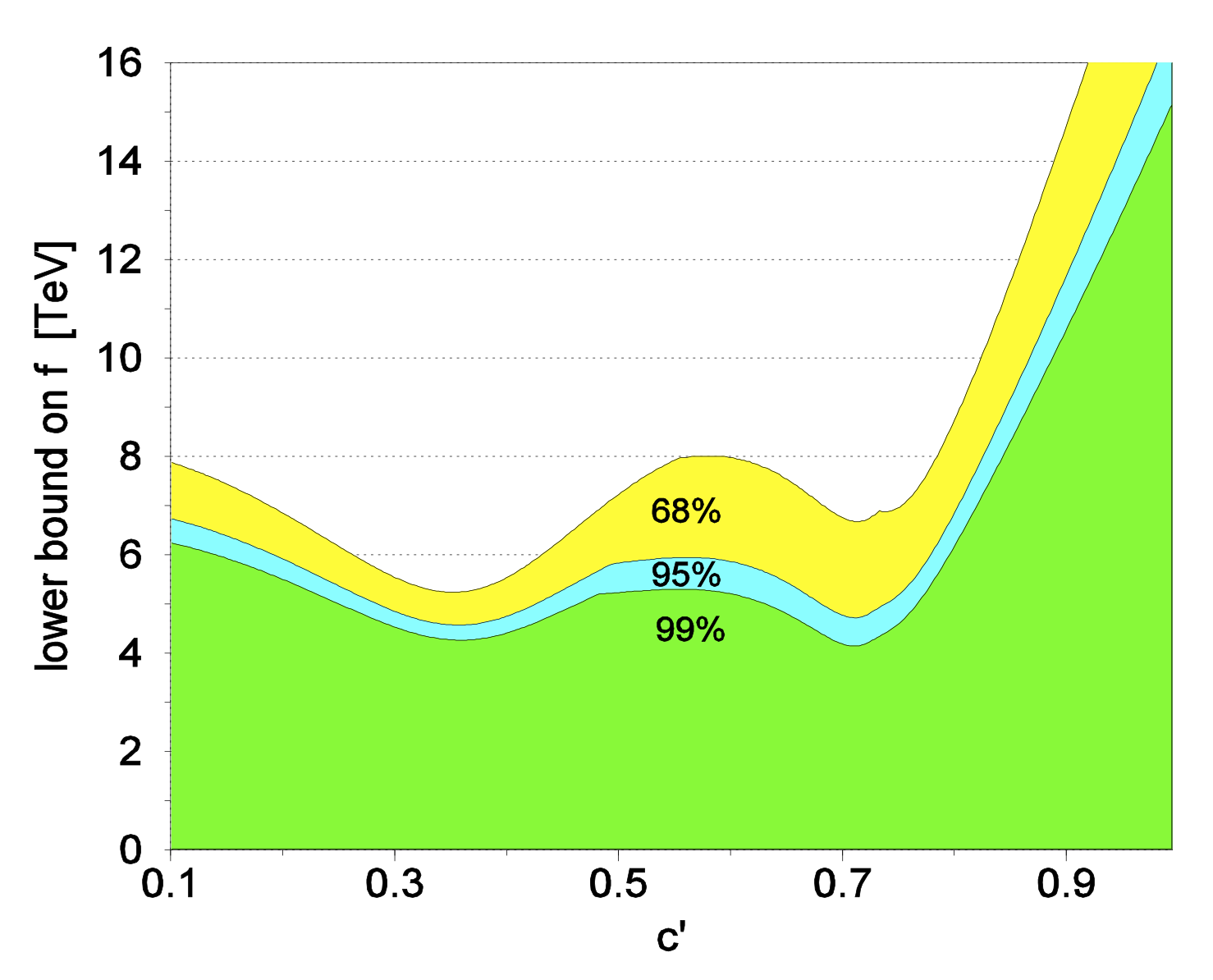}
\caption[]{Lower bound on the ``littlest" Higgs scale $f$,  from \cite{Csaki:2002qg}. }
 \label{fig:littleHiggsbound}
\end{figure}

Exchanges of $W^a_H$ and $B_H$ also give
corrections to the coupling of the $SU(2)_L\times U(1)_Y$ gauge
bosons  and additional four-fermion operators. So we see that ``little" Higgs models can give big corrections to precision electroweak observables which results in a lower
bound on the scale $f$.  Fig.~\ref{fig:littleHiggsbound} shows the bound in the case of the ``littlest" Higgs model \cite{Csaki:2002qg}.
To satisfy the bound the top partner must be heavy, and thus we find that there is still fine-tuning at the per cent level in this model.

\subsection{Variations on ``Little" Higgs}
\label{subsec:littlestvar}

\begin{figure}[htb]
\includegraphics[width=0.2\hsize]{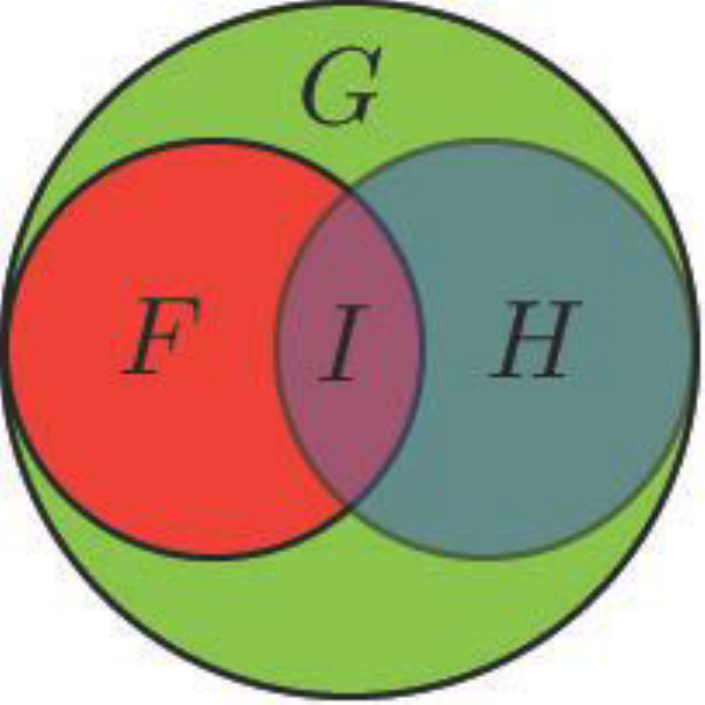}
\caption[]{Coset space for global symmetry group $G$ spontaneously broken to $H$. If $G$ has a weakly gauged subgroup $F$
then $I$ will be the unbroken gauge group, from \cite{Cheng:2007bu}. Color online.   }
 \label{fig:coset}
\end{figure}

This type of scenario can obviously be generalized to different breaking patterns. The general situation is shown in Fig.~\ref{fig:coset}
Consider a global symmetry group $G$ that is spontaneously broken to $H$, if $G$ has a weakly gauged subgroup $F$
then $I = F \cap H$ will be the unbroken gauge group. For a ``little" Higgs model we want to have $I = SU(2)_L \times U(1)_Y$.  The number
 of uneaten pseudo-Goldstone bosons is given by counting the number of broken generators, $N(G)-N(H)$, and subtracting the number eaten by gauge symmetries, $N(F)-N(I)$.
A few examples of such models are:
\begin{itemize}
\item minimal moose~\cite{ArkaniHamed:2002qx}: $G/H= SU(3)^2/SU(3), \; F= [SU(2)\times U(1)]^2$;
\item ``littlest" Higgs~\cite{ArkaniHamed:2002qy}: $G/H= SU(5)/SO(5),\; F=[SU(2)\times U(1)]^2$;
\item simple group ``little" Higgs~\cite{Kaplan:2003uc}: $ G/H=[SU(3)/SU(2)]^2$, $F=SU(3)\times U(1)$;
\item ``bestest little" Higgs \cite{Schmaltz:2010ac} $ G/H=SO(6)\times SO(6)/SO(6)$, $F=SU(2)^4\times U(1)$.
\end{itemize}

For a general approach to constructing ``little" Higgs models using ``moose" diagrams see \textcite{ArkaniHamed:2002qx} and \textcite{Gregoire:2002ra}.
Other ``little" Higgs models are discussed in \textcite{Schmaltz:2002wx,Low:2002ws,Chang:2003un,Skiba:2003yf,Chang:2003zn,Contino:2003ve,Cheng:2003ju,
Katz:2003sn,Kaplan:2004cr,Schmaltz:2004de,Low:2004xc,Agashe:2004rs,Batra:2004ah,Roy:2005hg,Schmaltz:2008vd,Kearney:2013cca}.
There have even been models that incorporate a ``little" Higgs mechanism with SUSY\cite{Birkedal:2004xi,Roy:2005hg,Csaki:2005fc}.

Generically all of these ``little" Higgs models give large tree-level corrections to precision electroweak constraints \cite{Csaki:2002qg,Csaki:2003si,Hewett:2002px},
which implies the existence of fine-tuning, see also \cite{Grinstein:2008kt,Bazzocchi:2005gs,Casas:2005ev}.
Loop-level electroweak corrections \cite{Gregoire:2003kr} have also been considered in the $SU(6)/Sp(6)$ ``little" Higgs model, and in some regions of the parameter space the tuning can be weaker. Further studies of indirect constraints are given in \textcite{Chivukula:2002ww,Huo:2003vd,Kilic:2003mq,Casalbuoni:2003ft,Kilian:2003xt,
Yue:2004xt,Lee:2004me,Buras:2004kq,Choudhury:2004bh,Marandella:2005wd}.

The most stringent constraints typically come from isospin breaking corrections, aka contributions to the $T$-parameter, that show up in the differences between $M_W$ and $M_Z$. Isospin violation can be suppressed in models that incorporate a custodial symmetry  \cite{Chang:2003un,Chang:2003zn}.
Further improvements in suppressing all precision electroweak corrections can be made in models that incorporate a new symmetry: $T$-parity \cite{Cheng:2003ju,Cheng:2004yc,Cheng:2005as}.
$T$-parity is a $Z_2$ symmetry, reminiscent of R-parity in the MSSM, where ordinary particles are even and the new partners are odd.
This implies that the new partners have to be pair produced and that the lightest $T$-parity odd particle (the LTP) is stable, a possible dark matter candidate,
and a source of missing energy signatures. $T$-parity significantly weakens the constraints from precision electroweak measurements \cite{Hubisz:2005tx,Han:2005dz} since then
$T$-odd particles can only contribute at loop level to precision electroweak observables. The requirement of $T$-parity further increases the difficulty of finding a
consistent UV completion however \cite{Hill:2007zv}.

\subsection{Effective Theory of a Pseudo-Goldstone Boson Higgs}
\label{subsection:effective}

Without knowing the underlying theory one can always parameterize the effects of new physics with a low-energy, effective theory.
At low energies the compositeness of the Higgs reveals itself in the deviations of the couplings in the effective theory (as compared to the SM couplings). Assuming that the Higgs boson is a CP-even weak doublet, one has \cite{SILH,Contino:2013kra}
\begin{eqnarray}
\label{eq:silh}
&& 
\Delta {\cal L}_{\rm eff} = {\cal L}_{\rm SM}+
 \frac{\bar c_H}{2v^2}\, \partial^\mu\!\left( H^\dagger H \right) \partial_\mu \!\left( H^\dagger H \right) 
+ \frac{\bar c_T}{2v^2}\left (H^\dagger {\overleftrightarrow { D^\mu}} H \right) \!\left(   H^\dagger{\overleftrightarrow D}_\mu H\right)  
- \frac{\bar c_6\, \lambda}{v^2}\left( H^\dagger H \right)^3 
\nonumber\\
&&
+ \left( \left( \frac{\bar c_u}{v^2}\,  y_{u}\, H^\dagger H\,   {\bar q}_L H^c u_R +  \frac{\bar c_d}{v^2}\,  y_{d}\, H^\dagger H\,   {\bar q}_L H d_R 
+ \frac{\bar c_l}{v^2}\,  y_{l}\, H^\dagger H\,   {\bar L}_L H l_R \right)  + {\it h.c.} \right)
\nonumber \\
&& 
+\frac{i\bar c_W\, g}{2m_W^2}\left( H^\dagger  \sigma^i \overleftrightarrow {D^\mu} H \right )( D^\nu  W_{\mu \nu})^i
+\frac{i\bar c_B\, g'}{2m_W^2}\left( H^\dagger  \overleftrightarrow {D^\mu} H \right )( \partial^\nu  B_{\mu \nu})   
\nonumber\\
&&
+\frac{i \bar c_{HW} \, g}{m_W^2}\, (D^\mu H)^\dagger \sigma^i (D^\nu H)W_{\mu \nu}^i
+\frac{i\bar c_{HB}\, g^\prime}{m_W^2}\, (D^\mu H)^\dagger (D^\nu H)B_{\mu \nu} 
\nonumber\\
&&+\frac{\bar c_\gamma\,  {g'}^2}{m_W^2}\, H^\dagger H B_{\mu\nu}B^{\mu\nu}
   +\frac{\bar c_g \, g_S^2}{m_W^2}\, H^\dagger H G_{\mu\nu}^a G^{a\mu\nu}
\, . 
\label{lsilh}
\end{eqnarray}


The effects of composite nature of the Higgs boson and of resonances of the $W$ and $Z$ on precision electroweak measurements  have been studied  \cite{Pich:2012dv,Barbieri:2007bh,Orgogozo:2012ct,Ciuchini:2013pca},
with the conclusion that the main constraint is that the Higgs $WW$ coupling is close to the SM value: $ 0.99< 1-\bar{c}_H /2 <1.06$ (light fermion resonances can relax this constraint~\cite{Grojean:2013qca}). Deviations in the Higgs couplings of course mean that $WW$ scattering is only partially unitarized by the Higgs (cf. section \ref{subsection:gaugephobic:bulkaction}). 
A fairly general analysis of unitarization in this case is given in \cite{Falkowski:2011ua,Bellazzini:2012tv}.

When the new physics sector is characterized by a single scale $M$ and a coupling $g_*\equiv M/f$, and assuming that the classical action including the heavy fields involves at most two derivatives, the Wilson coefficients of the effective Lagrangian obey the simple scaling:
\begin{equation}
\label{eq:NDAestimates}
\bar c_{H}, \bar c_{T}, \bar c_{6}, \bar c_{\psi} \sim O\!\left(\frac{v^2}{f^2}\right) , \quad 
\bar c_{W}, \bar c_{B} \sim O\!\left(\frac{m_W^2}{M^2} \right)\, , \quad
\bar c_{HW}, \bar c_{HB}, \bar c_{\gamma}, \bar c_{g} \sim  O\!\left(\frac{m_W^2}{16\pi^2 f^2}\right). 
\end{equation}
When the Higgs doublet is a composite pseudo-Goldstone boson of a 
spontaneously-broken symmetry ${\cal G} \to {\cal H}$ of the strong dynamics~\cite{Contino:2003ve,Agashe:2004rs,SILH}, a further suppression of the contact operators to photon and gluons holds:
\begin{equation}
\label{eq:cgcga}
\bar c_{\gamma}, \bar c_g \sim O\!\left(\frac{m_W^2}{16\pi^2 f^2}\right) \!\times \frac{g_{\not G}^2}{g_*^2}\, ,
\end{equation}
where $g_{\not G}$ denotes any weak coupling that breaks the Goldstone symmetry (in minimal models the SM gauge couplings or Yukawa couplings).
It should be stressed that these estimates are valid at the UV scale $M$, at which the effective Lagrangian is matched onto explicit models. 
Renormalization effects between $M$ and the electroweak scale mix  operators with the same quantum numbers~\cite{Alonso:2013hga},  and give in general
subdominant corrections to the coefficients~\cite{Grojean:2013kd, Elias-Miro:2013gya, Elias-Miro:2013eta, Elias-Miro:2014eia, Cheung:2015aba}. The estimates of ${\bar c}_{W,B}$ and ${\bar c}_{T}$ apply when  these coefficients
are generated at tree-level.
However, specific  symmetry protections which might be at work in the UV theory, for example $R$-parity in SUSY theories, 
can force the leading corrections to arise at the 1-loop level.

\subsection{Pseudo-Goldstone Boson Higgs}
\label{sec:pseudo-goldstoneHiggs}

As we mentioned earlier, the idea of the Higgs as a composite pseudo-Goldstone boson goes back to the 1980's \cite{Georgi:1984ef,Kaplan:1983fs,Kaplan:1983sm,Georgi:1984af,Dugan:1984hq}.
As in ``little" Higgs models, the electroweak gauge symmetry is embedded in a larger global symmetry which is broken at a strong interaction scale $f \sim 1$\,TeV, producing a Goldstone boson that is identified with the SM Higgs, for reviews see \textcite{Grojean:2009fd}, \textcite{Contino:2010rs}, \textcite{Rychkov:2011br}, \textcite{Espinosa:2012qj}, \textcite{Bellazzini:2014yua} and \textcite{Panico:2015jxa} . In an extra dimensional setting\footnote{Often these models use more than 5 dimensions in order to get enough Higgs components.} these models are also referred to as gauge-Higgs unification \cite{Panico:2005dh,Serone:2009kf,Csaki:2002ur,Medina:2007hz} since the Goldstone boson shows up as the fifth component of a bulk gauge field. The effective theory has a cutoff $\Lambda \sim 4 \pi f\sim 5\textrm{--}10$\,TeV. The difference here is that there is no tree-level quartic allowed and the entire Higgs potential is loop-generated.  Top partners (as in ``little" Higgs models, cf section \ref{subsec:littlest}) and spin-1 partners are still needed to cancel the quadratic divergences. However, these models still must be tuned at the 1\% level; as discussed below the origin of the tuning is the requirement of keeping $v$ much smaller than $f$. Since the quartic is loop-generated the Higgs is naturally light, around 100 GeV, in agreement with the observed value. Generically the deviations from SM Higgs couplings at the 10\%--20\% level.

There are numerous ways to construct models where the Higgs can appear as a pseudo-Goldstone boson, we will restrict ourselves to those with an unbroken custodial $SU(2)\times SU(2)$ symmetry, as usual. A list of examples is given in Table \ref{table:NGBcosets} (see also \cite{Bellazzini:2014yua}).
\begin{table}[h!]
	\begin{center}
\begin{tabular}{lcccccc}
\hline
& $G$ & $H$ & $N_G$ &  $\textrm{rep.}[GB] = \textrm{rep.}[\textrm{SU}(2) \times \textrm{SU}(2)]$ \\

minimal & $SO(5)$ & $SO(4)$ & 4 & $\mathbf{4} = (\mathbf{2},\mathbf{2})$ \\

next to minimal & $SO(6)$ & $SO(5)$ & 5 & $\mathbf{5} = (\mathbf{1},\mathbf{1}) + (\mathbf{2},\mathbf{2})$ \\

& $SO(6)$ & $SO(4)$ $\times$ $SO(2)$ & 8 & $\mathbf{4_{2}} + \mathbf{\bar{4}_{-2}} = 2 \cdot (\mathbf{2},\mathbf{2})$ \\

& $SO(7)$ & $SO(6)$ & 6 & $\mathbf{6} = 2 \cdot (\mathbf{1},\mathbf{1}) + (\mathbf{2},\mathbf{2})$ \\

& $SO(7)$ & $\textrm{G}_2$ & 7 & $\mathbf{7} = (\mathbf{1},\mathbf{3})+(\mathbf{2},\mathbf{2})$ \\

& $SO(7)$ & $SO(5)$ $\times$ SO(2) & 10 & $\mathbf{10_0} = (\mathbf{3},\mathbf{1})+(\mathbf{1},\mathbf{3})+(\mathbf{2},\mathbf{2})$ \\

& $SO(7)$ & $[\textrm{SO}(3)]^3$ & 12 & $(\mathbf{2},\mathbf{2},\mathbf{3}) = 3 \cdot (\mathbf{2},\mathbf{2})$ \\

composite two-Higgs & $Sp(6)$ & $Sp(4)$ $\times$ SU(2) & 8 & $(\mathbf{4},\mathbf{2}) = 2 \cdot (\mathbf{2},\mathbf{2})$ or $ (\mathbf{2},\mathbf{2}) + 2 \cdot (\mathbf{2},\mathbf{1})$ \\

& $SU(5)$ & $SU(4)$ $\times$ U(1) & 8 & $\mathbf{4}_{-5} + \mathbf{\bar{4}_{5}} = 2 \cdot (\mathbf{2},\mathbf{2})$ \\

& $SU(5)$ & $SO(5)$ & 14 & $\mathbf{14} = (\mathbf{3},\mathbf{3}) + (\mathbf{2},\mathbf{2}) + (\mathbf{1},\mathbf{1})$ \\

\hline
\end{tabular} 
	\caption{Cosets $G/H$ from simple Lie groups, 
	with maximal subgroups. For each coset, 
	the number of Goldstone bosons, $N_G$, and the Goldstone bosons representation under $H$ and 
	$SO(4)\simeq SU(2)_L \times SU(2)_R$ are given, \cite{Mrazek:2011iu}}
	\label{cosets}
	\end{center}
	\label{table:NGBcosets}
\end{table} 
Some of these models have been investigated in the literature: the minimal model \cite{Contino:2003ve,Agashe:2004rs} (discussed in section \ref{sec:minimalpseudo-goldstoneHiggs}), the next to minimal model \cite{Gripaios:2009pe}, and the composite two-Higgs doublet model  \cite{Mrazek:2011iu}.

\begin{figure}[htb]
\includegraphics[width=15cm]{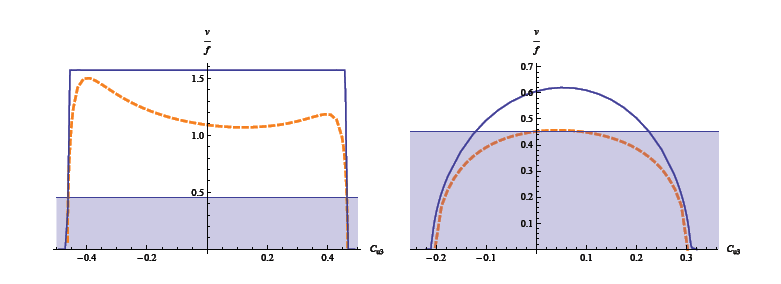}
\caption[]{The amount of tuning required to suppress $v/f$ in pseudo-Goldstone boson Higgs models. Generically a tuning at the per cent level is required to be in the acceptable shaded, blue, region, as shown on the left.
By carefully choosing other input parameters the apparent fine-tuning in $c_u$ is reduced, shown on the right, at the price of fine-tuning other parameters \cite{Csaki:2008zd}. Color online.} 
\label{fig:pGBHiggstuning}
\end{figure}

At lowest order in the weak gauge couplings, the effective Lagrangian is just a chiral Lagrangian with only derivative interactions. Thus the Higgs potential in these models is entirely generated by loop corrections \`a la Coleman--Weinberg \cite{Coleman:1973jx}. Generically the potential takes the form
\beq
V(H)=\frac{g^2}{16 \pi^2} f^4 \cos^n\left( \frac{H}{f}\right)~.
\eeq
 This means that there is no natural separation between the weak scale VEV $v$ and the strong breaking scale $f$.  Qualitatively the mass term and quartic term can be estimated as
\beq
m^2\sim \frac{g^2 \Lambda^2}{16 \pi^2}\sim g^2 f^2~,\quad\quad \lambda \sim  \frac{g^2 \Lambda^2}{16 \pi^2 f^2} \sim g^2~.
\eeq
Consistency with precision electroweak measurements (especially the $S$-parameter) \cite{Csaki:2008zd,Carena:2007ua} requires that approximately $v< 0.45 f$.
Detailed calculations show that generically achieving this suppression in these models requires  a per cent level tuning \cite{Csaki:2008zd}, as shown in Fig. \ref{fig:pGBHiggstuning}. This bound can be relaxed thanks to the contributions of the fermionic and bosonic resonances to the EW oblique parameters, see for instance \cite{Ciuchini:2013pca,Grojean:2013qca}.

\subsection{Minimal Composite Pseudo-Goldstone Boson Higgs}
\label{sec:minimalpseudo-goldstoneHiggs}

The simplest implementation of the composite pseudo-Goldstone boson scenario is a 5D model \cite{Contino:2003ve,Agashe:2004rs,Agashe:2005dk,SILH} known as the ``minimal composite pseudo-Goldstone boson Higgs", usually shortened to 
just the ``minimal composite Higgs". The model could be viewed as simply the low-energy effective theory for a particular pattern of symmetry breaking, without reference to a 5D theory. In the 5D description the Higgs is the fifth component of a gauge boson, and from its profile in the extra dimension, we can read off (via the AdS/CFT correspondence) that its scaling is 2. This model has a global $SO(5)\times U(1)_X$ symmetry in the bulk which is broken to $SU(2) \times U(1)_Y$ on the UV brane and $SO(4)\times U(1)_X$ on the IR (TeV) brane. The Goldstone boson of the $SO(5)\rightarrow SO(4)$ breaking is a vector of $SO(4)$ which is equivalent to a $(2,2)$ of 
$SU(2)_L\times SU(2)_R$ which has the correct quantum numbers to be the Higgs.  Gauging the SM $SU(2)_L \times U(1)_Y$ subgroup explicitly breaks the global symmetry and gives the Higgs a mass and quartic interactions.  This can be thought of as an extra dimensional version of a ``little" Higgs model with custodial symmetry \cite{Chang:2003un}. 

As in a ``little" Higgs model we construct the low-energy effective theory in terms of a field that provides a non-linear realization of the symmetry.  This field 
encodes the four Goldstone boson fluctuations around the VEV:
\begin{equation}
\Sigma(x) = \Sigma_0 e^{\Pi(x)/f}  \qquad 
\begin{array}{ll}
& \Sigma_0 = (0,0,0,0,1) \\[0.15cm]
& \Pi(x) = -i X^{a} h^{a}(x) \sqrt{2}\, ,
\end{array} 
\end{equation}
where $X^a$ are the broken $SO(5)$ generators. 
This is equivalent to 
\begin{equation}
\label{eq:sigma}
\Sigma = \frac{\sin(h/f)}{h}\, \big(  h^1,  h^2,  h^3,  h^4,  h \cot(h/f) \big) \, , \qquad h \equiv \sqrt{(h^{a})^2}\, .
\end{equation}
Following the classic paper of Callan, Coleman, Wess, and Zumino \cite{Coleman:1969sm,Callan:1969sn} we find the gauge field terms in the leading order Lagrangian (in momentum space) to be:
\begin{equation}
\begin{split}
{\cal L} =  K^{\mu\nu} \bigg\{
   & \frac{1}{2} \left( \frac{f^2 \sin^2(h /f)}{4} \right) \left( B_\mu B_\nu + W^3_\mu W^3_\nu -2 W^3_\mu B_\nu \right) \\
   & + \left( \frac{f^2 \sin^2( h /f)}{4} \right)  W^+_\mu W^-_\nu 
    - \frac{ q^2}{2} \Big[ \frac{1}{g^2}\, W^{a_L}_\mu W^{a_L}_\nu + \frac{1}{g^{\prime 2}} B_\mu B_\nu \Big]
 + \dots \bigg\}
\end{split}
\end{equation}where the transverse factor $K^{\mu\nu}$ is just $K^{\mu\nu}=\eta^{\mu\nu}-q^\mu q^\nu$, 
$W^a_\mu$ are the $SU(2)_L$ gauge bosons and $B_\mu$ is the  $U(1)_Y$ gauge boson.
Comparing with the gauge boson masses in the SM we can identify
\begin{equation} \label{eq:defxi}
v = f \,\sin \frac{\langle h\rangle}{f}~.
\end{equation}
It is useful to write expressions in terms of the ratios of the VEVs so we  define
\beq
\xi \equiv \frac{v^2}{f^2} = \sin^2 \frac{\langle h\rangle}{f}~.
\eeq
Expanding around the Higgs VEV, $h^{a} = (0 , 0, \langle h \rangle + h^0 , 0 )$,
we obtain the deviations of the couplings of the Higgs to the $W$ and $Z$:
\begin{equation}
\label{eq:ab}
\kappa_W\equiv \frac{g_{hWW}}{g_{hWW}^{\rm SM}}=1-\frac{{\bar c}_H}{2}=\sqrt{1-\xi}\, , \qquad \kappa_Z\equiv \frac{g_{hZZ}}{g_{hZZ}^{\rm SM}}=1-\frac{{\bar c}_H}{2}-2 {\bar c}_T=\sqrt{1-\xi}~.
\end{equation}
These deviations in the Higgs coupling mean that unitarity breaks down at a scale 
\beq
\Lambda \approx \frac{4 \pi v}{\sqrt{\xi}}~,
\eeq
so below this scale additional resonances of the strongly coupled sector must start to contribute to $WW$ scattering (cf. section \ref{subsection:gaugephobic:bulkaction}).

As $\xi\rightarrow 0$ the Higgs couplings approach the SM values, so for sufficiently small $\xi$ the model can pass most precision electroweak tests.
A left-right symmetry to protect the $Zb{\bar b}$ coupling can also be implemented in this type of model \cite{Agashe:2005dk,Contino:2006qr}, resulting in exotic top-partners,
cf Section~\ref{subsection:pseudoLHCsearch}.

\subsection{Composite Pseudo-Goldstone Boson Higgs LHC Searches}
\label{subsection:pseudoLHCsearch}

\begin{figure}[htb]
\includegraphics[width=\imagesize]{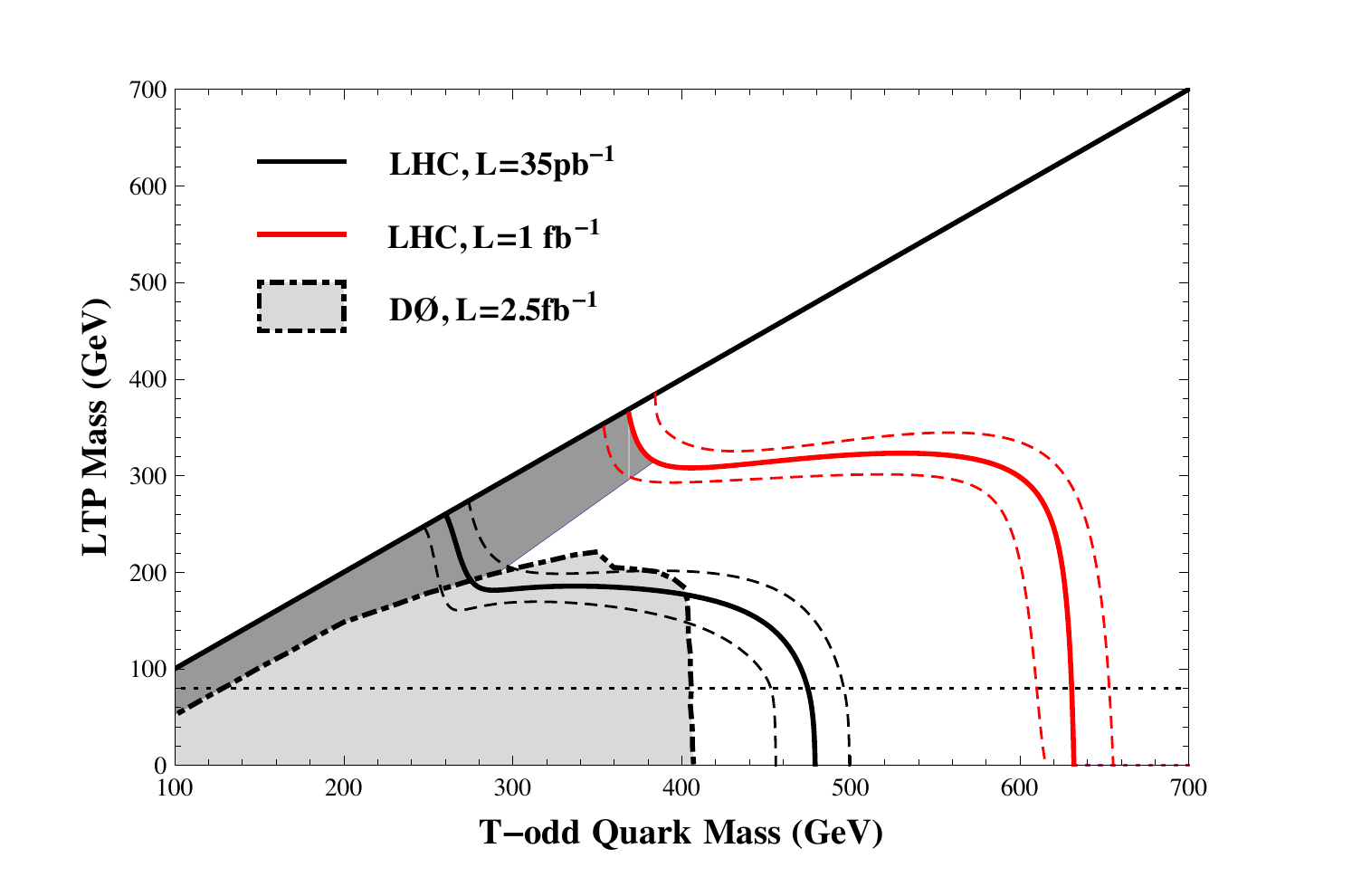}
\caption[]{Lower bound on the ``little" Higgs scale $f$,  from \cite{Perelstein:2011ds}. Color online.}
 \label{fig:perelstein}
\end{figure}

The signature of ``little" Higgs models is a fermionic top partner that cancels the quadratic divergence in the top loop \cite{Matsumoto:2006ws,Belyaev:2006jh,Meade:2006dw,Han:2005ru}.  For the model to address the little hierarchy problem this top partner has to be significantly below 1\,TeV.  The single lepton final state is often the optimal channel for finding such top partners \cite{AguilarSaavedra:2009es}. The gauge partners can be heavier since they have a smaller coupling to the Higgs, and this is easily arranged by having additional VEVs that contribute to the heavy gauge boson masses but not to the top partner mass \cite{Schmaltz:2010ac}. 
There have been many phenomenological studies \cite{Burdman:2002ns,Han:2003wu,Perelstein:2003wd,Hubisz:2004ft,Han:2003gf,Dib:2003zj,Sullivan:2003xy,Yue:2003yk,
Azuelos:2004dm,GonzalezSprinberg:2004bb,Kilian:2004pp,Yue:2004fv,Park:2003sq,BirkedalHansen:2003mpa,Hektor:2007uu,Barducci:2014ila,Yang:2014uza,Cacciapaglia:2013wha,Reuter:2012sd,Reuter:2013iya}. A model independent (effective Lagrangian) approach was advocated by \textcite{Buchkremer:2013bha}.
As shown by  \textcite{Perelstein:2011ds}, the SUSY search for jets and missing energy can be reinterpreted at a bound on the T-odd quark partner mass. The first 35 pb$^{-1}$ of data~\cite{Khachatryan:2011tk} led to a bound around 450\,GeV, while it was anticipated that a 1\,fb$^{-1}$ analysis should raise the bound to 600\,GeV, as shown in Fig.~\ref{fig:perelstein}. 

\begin{figure}[htb]
\includegraphics[width=\imagesize]{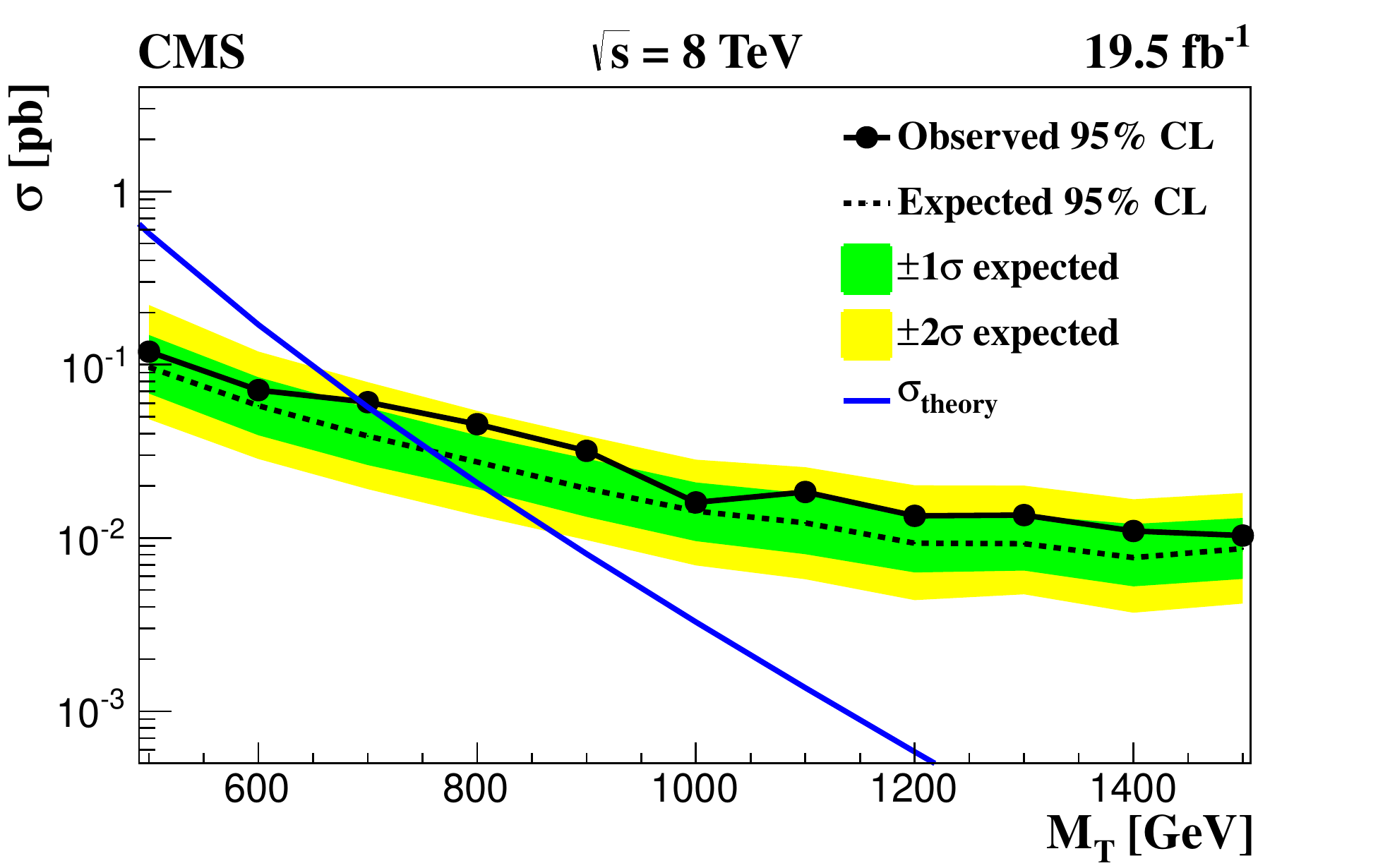}
\caption[]{CMS bound on the T-even top partner production cross section for branching fractions into bW, tH, tZ of 50\%, 25\%, and 25\% respectively,  from \cite{Chatrchyan:2013uxa}. Color online. }
 \label{fig:Tevenpartner}
\end{figure}

\begin{figure}[htb]
\includegraphics[width=\smallimagesize]{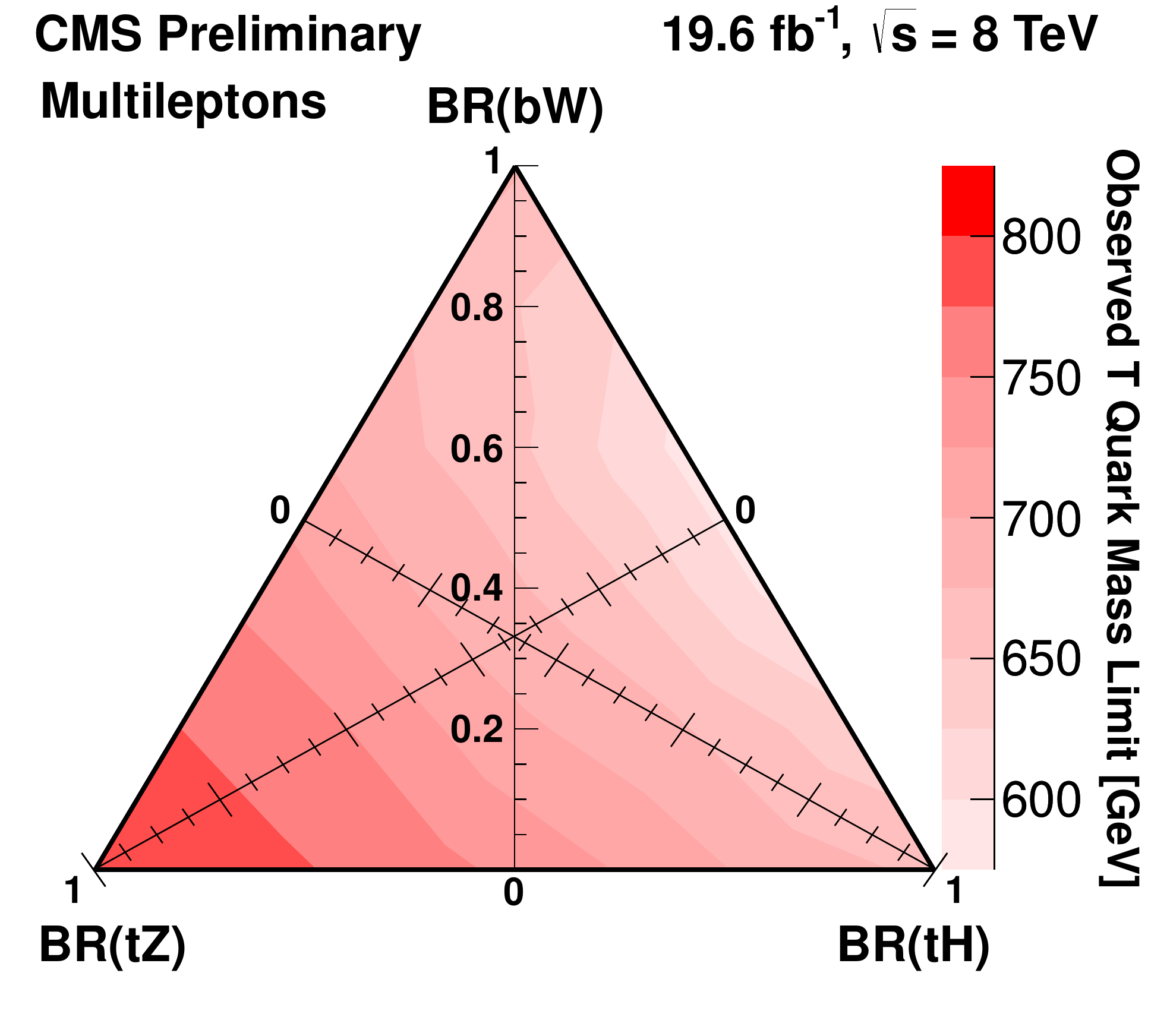}
\caption[]{CMS mass bound on the T-even top partner as a function of branching fractions into bW, tH, tZ,  from \cite{Chatrchyan:2013uxa}. Color online. }
 \label{fig:Tevenpartnerbfplane}
\end{figure}

CMS has performed a search \cite{Chatrchyan:2013uxa} for the T-even top partner (the partner that cancels the divergence) which gives a bound at 475\,GeV, assuming a branching fraction for $T\rightarrow t+Z$ of 100\%, as shown in Fig.~\ref{fig:Tevenpartner}~and~\ref{fig:Tevenpartnerbfplane}.  In the ``little" Higgs models the $T\rightarrow t+Z$  branching fraction is more like 25\%, so the actual limit is around 375\,GeV. One can also search for $T\rightarrow b+W$. Naturalness would suggest that the top partner mass should be around 250\,GeV in the absence of tuning, while a mass of  860\,GeV would require a 10\% tuning, and a mass of 3\,TeV would require a 1 \% tuning. The combined LHC bound \cite{Berger:2012ec}  for 5\,fb$^{-1}$, assuming appropriate branching fractions is about 450\,GeV, which implies at least a 20 \% fine tuning. A CMS analysis \cite{CMSprelimTpartner} with 8~TeV data (allowing for decays into  $bW$, $tZ$, and $tH$) found a lower bound of 687\,GeV for vector-like top partners using 19.6\,fb$^{-1}$  of data. More exotic top partners, for example with an $SO(10)/SO(5)^2$ coset, can have final states with b-jets and a large number of electroweak gauge bosons \cite{Kearney:2013cca}. Model-indepent parametrization of top partners have been advocated in \textcite{DeSimone:2012fs} and \textcite{Panico:2015jxa} using an effective Lagrangian approach.
 
\textcite{Godfrey:2012tf} studied the ``bestest" model in some detail, focusing on top partner pair production decaying to $b\bar{b}W^+W^-$ and $t\bar{t}ZZ$ and singly produced top partners from $W$ exchange.  With just 1.1\,fb$^{-1}$ of data there were already bounds on the top partner mass of about 400\,GeV.

\begin{figure}[htb]
\includegraphics[width=\imagesize]{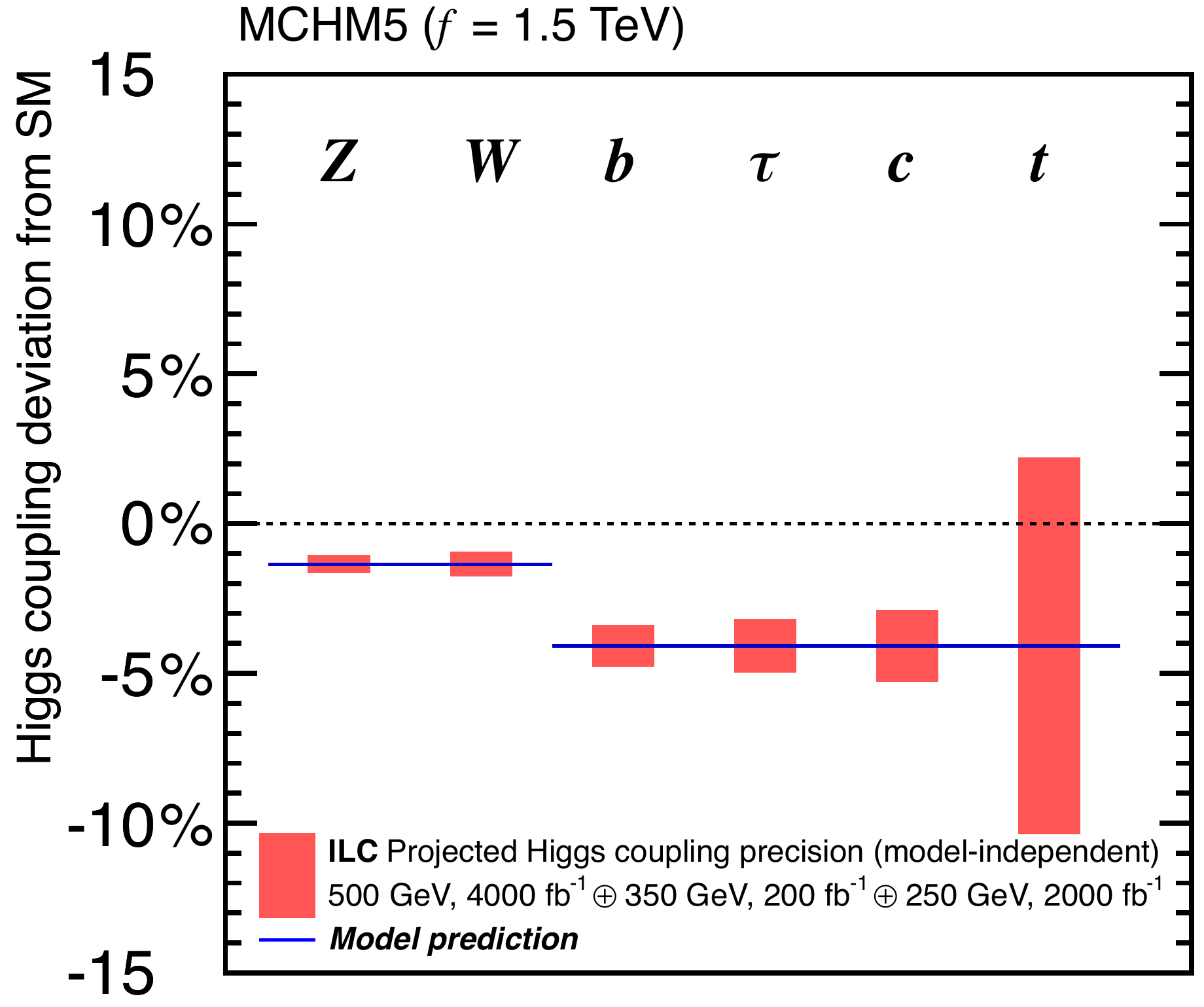}
\caption[]{Examples of deviations from the SM predictions of the Higgs couplings in the minimal composite models \cite{Fujii:2015jha}. The red bars denote the sensivity of these couplings at the ILC.} 
\label{fig:pGBHiggscouplings}
\end{figure}

\begin{figure}[htb]
\includegraphics[width=12cm]{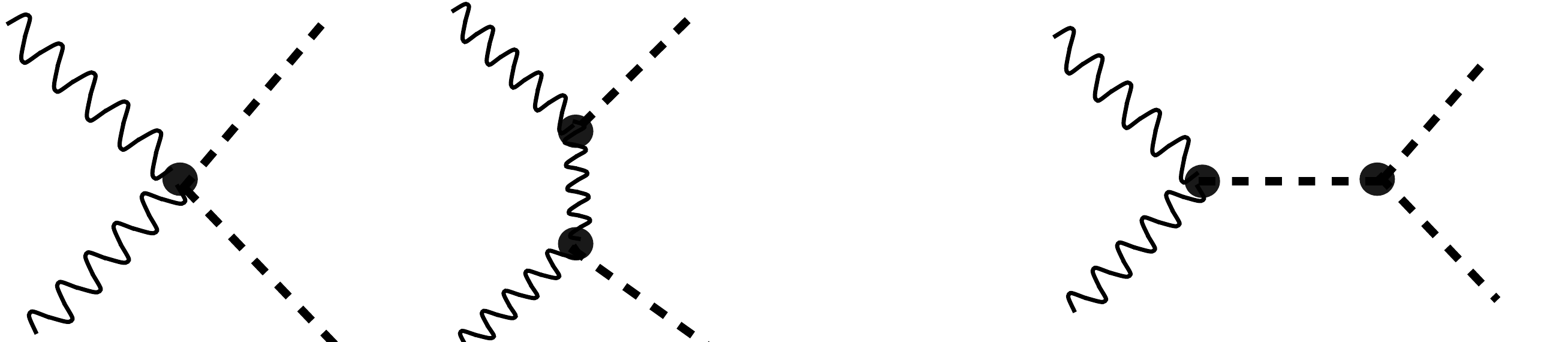}
\caption[]{Double Higgs production in composite pseudo-Goldstone boson  Higgs models have contributions that grow with the squre of the center-of-mass energy, \cite{Contino:2010mh}.} 
\label{fig:doubleHiggs}
\end{figure}

\begin{figure}[htb]
\includegraphics[width=\smallimagesize]{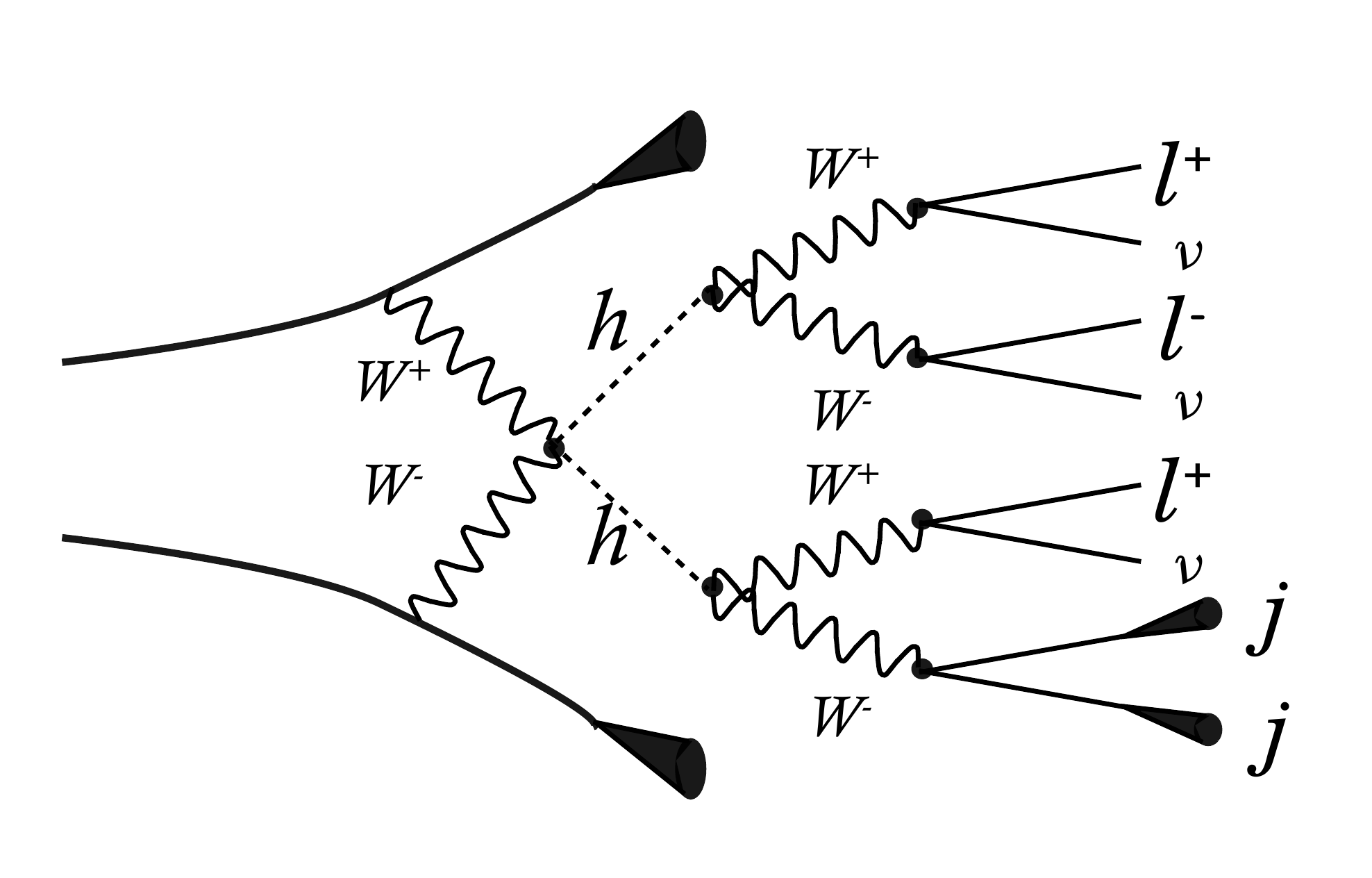}
\caption[]{Difficult 4 $W$ search channel for a pseudo-Goldstone boson  Higgs from strong double Higgs production by vector boson fusion, \cite{Contino:2010mh}.} 
\label{fig:pGBcompHiggssearch}
\end{figure}

The deviations of the Higgs couplings can provide interesting probes of composite pseudo-Goldstone boson models at the LHC \cite{Espinosa:2010vn,Duhrssen:2004cv,Azatov:2012qz,Azatov:2012rd,Azatov:2012bz,Cacciapaglia:2012wb,Espinosa:2012ir}.
Typical deviations of the Higgs couplings in  the minimal composite pseudo-Goldstone boson Higgs are shown in Fig.~\ref{fig:pGBHiggscouplings}.
One sees that generically the current LHC data are starting to set tight constraints on these models.
A future stringent test will come from double Higgs production \cite{Contino:2010mh,Gouzevitch:2013qca,Dolan:2012ac,Noble:2007kk} which can have contributions to the amplitude, shown in Fig.~\ref{fig:doubleHiggs}, that grow with energy squared.
A search involving 4 $W$'s (see Fig.~\ref{fig:pGBcompHiggssearch}) has been proposed \cite{Contino:2010mh} which would take between 450\,fb$^{-1}$ to 3500\,fb$^{-1}$ of luminosity for a 5$\sigma$ signal for reasonable values of $v/f$.
Composite Higgs models often have top-partner fermions with exotic electric charges \cite{Contino:2008hi,Mrazek:2009yu,Dissertori:2010ug} similar to the bulk Higgs models discussed in section \ref{subsection:gaugephobic:searches}.
Top-partners play a crucial role in the dynamics of EWSB since the top sector is a dominant breaking of the global symmetry structure and contributes in an instrumental way to the generation of the pseudo-Goldstone boson Higgs potential (see \textcite{Contino:2010rs} and \textcite{Panico:2015jxa} for a general discussion). It has even been shown~\cite{Marzocca:2012zn,Pomarol:2012qf} using general arguments based on the Weinberg sum rules that the mass of the Higgs around 125~GeV calls for top-partners below 1~TeV. The best way to search for exotic top-partners is in pair production $gg\to T_{5/3}\bar T_{5/3} \to t\bar {t} W^+ W^-$ leading to same-sign dilepton final states. ATLAS and CMS started have started probing the interesting mass range with a lower bound on their mass close to 800~GeV~\cite{Chatrchyan:2013wfa,TheATLAScollaboration:2013jha}.  These bounds can be further improved using single production 
channels~\cite{DeSimone:2012fs}. An idea of the potential reach of these LHC searches is shown in Table~\ref{table:toppartnerreachearch}.
\begin{table}[htb]
\includegraphics[width=\imagesize]{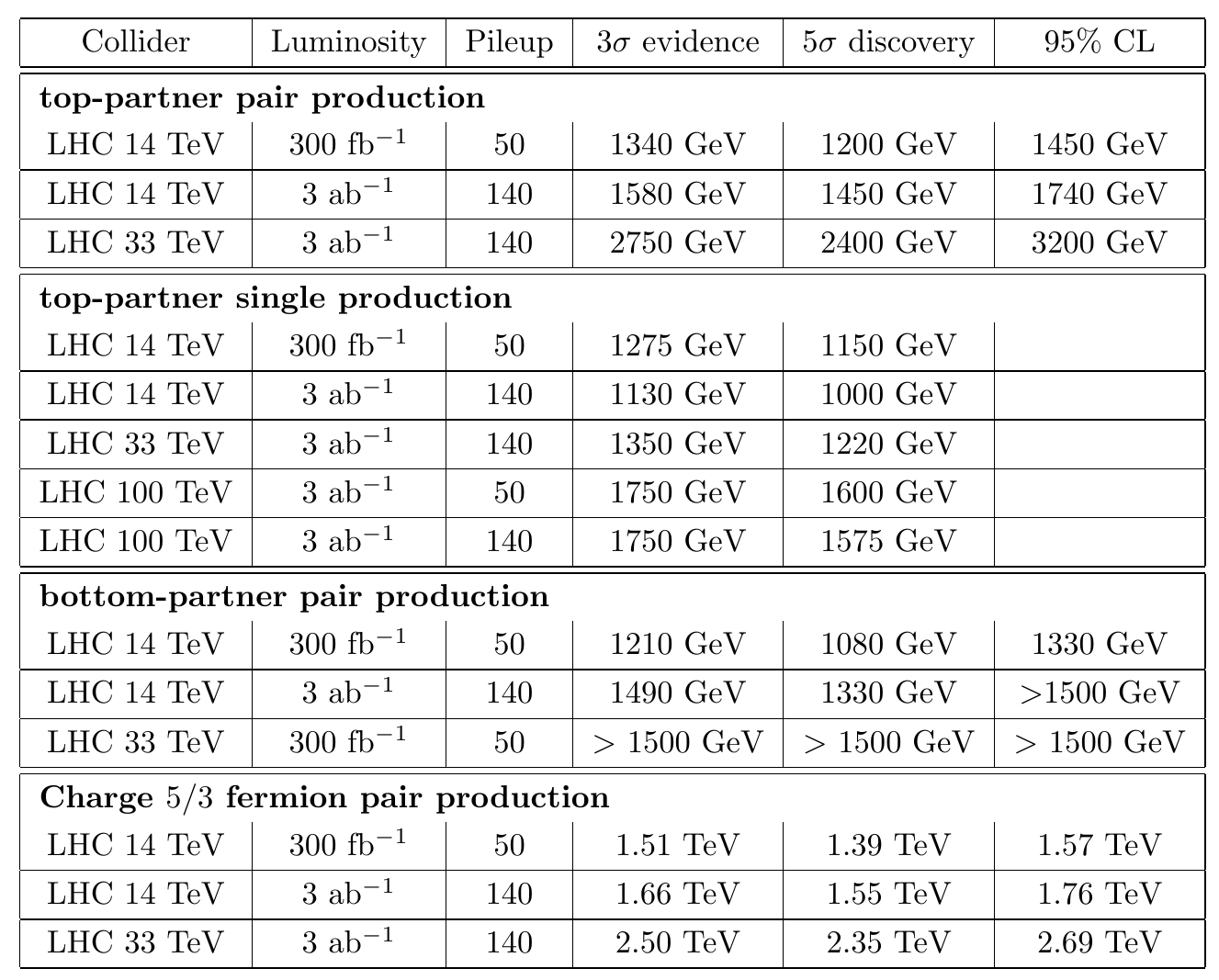}
\caption[]{Potential reach of top (and bottom) partner searches for different charge states, from \textcite{Agashe:2013hma}.} 
\label{table:toppartnerreachearch}
\end{table}

Unusual Higgs couplings may require unusual strategies to enhance particular signals \cite{Englert:2011us,Fox:2011qc}.
Separating out the vector boson fusion production channel is especially useful since it probes the couplings at the heart of electroweak symmetry breaking \cite{Chang:2012tb}.

\subsection{``Fat" SUSY Higgs}

Supersymmetric composite (aka ``fat") Higgs theories have an advantage over non-supersymmetric theories in that many details of the composite sector are under
theoretical control due to Seiberg duality~\cite{Seiberg:1994pq}. In the infrared limit dual description often reduce to a weakly coupled gauge theory with a Yukawa coupling, and the size of the Yukawa
coupling is set by ratios of strong interaction scales. This control allows for a much more detailed predictions in such theories. A further advantage over more conventional supersymmetric theories is that electroweak symmetry breaking can occur in the SUSY limit, thus avoiding having to tune SUSY breaking parameters against SUSY preserving parameters as happens in the Minimal Supersymmetric Standard Model (MSSM).

Models with composite\footnote{There have also been studies of models where the Higgs is mostly elementary but mixed with the composites of a strongly coupled SUSY sector \cite{Heckman:2012nt}.}  Higgs fields \cite{Luty:2000fj,Harnik:2003rs,Fukushima:2010pm} need to address the problem of fitting the electroweak precision measurements.  First of all the $S$-parameter tends to grow with the size of the electroweak sector, but extra contributions to the $S$-parameter from a composite Higgs are suppressed by the square of the VEV over the compositeness scale $v^2/f^2$, so these contributions are not necessarily troublesome here \cite{Galloway:2010bp,Pich:2012dv,Gripaios:2006nn}. The precision electroweak fits of the SM prefer a small $S$, but, as is well known~\cite{Peskin:2001rw}, 
it is possible to have additional, correlated, contributions to $S$ and $T$ and still be consistent with precision electroweak fits. 

An interesting class of   ``fat" SUSY  Higgs models  is where
the particles that contribute in loops to the Higgs mass are a least partially composite \cite{Delgado:2005fq}. In order for all the leading one-loop divergences to be determined by
the strong dynamics we need, the Higgses, the electroweak gauge fields, and (some of) the SM fermions to be composite.
To get a realistic theory, the composite $W$ and $Z$ need to be mixed
with elementary $W$ and $Z$ gauge bosons that couple to the elementary
quarks and leptons. The electric theory of the simplest such model, the Minimal Composite Supersymmetric Standard Model (MCSSM) model \cite{Csaki:2011xn,Csaki:2012fh},
has a strongly coupled $SU(4)$ group and 6 flavors of quarks, ${\cal Q}$, ${\cal \bar Q}$. The model also allows small tree-level masses for the electric quarks ${\cal Q}$, ${\cal \bar Q}$.

The IR behavior of this strongly coupled theory is given by the Seiberg
dual~\cite{Seiberg:1994pq} with a dual gauge group $SU(2)_{\rm mag}$, 6 flavors of dual quarks $q$, ${\bar q}$, a gauge singlet meson $M$, and with the additional dynamical superpotential term that couples the meson to the dual quarks, with a Yukawa coupling $y$.
The $SU(2)_{mag}\times SU(2)_{el}$ is eventually
broken to the diagonal subgroup: the SM
$SU(2)_L$ at a large scale $\cal{F}$.

The dual quarks contain the left-handed third generation quark
doublet, two Higgses $H_{u,d}$, and two bifundamentals ${\cal H, \bar{H}}$ that will be responsible for breaking the $SU(2)_{mag}\times SU(2)_{el}$ to the diagonal and generating the partially composite $W$ and $Z$. The embedding into the dual quarks is:
\beq
q= Q_{3} , {\cal H},  H_d \quad \quad
\bar q=X ,{\cal \bar{H}} ,  H_u~,
\eeq
where $X$ is an exotic. From the $q$, $\bar q$ charge assignments it follows that the meson $M$
contains the right-handed $t$, two singlets , two additional Higgses  transforming under the elementary $SU(2)_{el}$,   and some exotics. All the exotic extra fields can be removed in an anomaly-free way through effective mass terms.

From the point of view of electroweak symmetry breaking and the light fermion masses this final model is basically identical at low energies to the usual ``fat" Higgs models \cite{Luty:2000fj,Harnik:2003rs}. The Higgs potential, for ${\cal F}\gg f$, involves and additional singlet, $S$, and is given by 
\beq
&& V= y^2|H_u H_d  -f^2|^2   +y^2|S|^2 (|H_u|^2
+|H_d|^2) +m_S^2 |S|^2 + m_{H_u}^2 |H_u|^2 +m_{H_d}^2 |H_d|^2
\nonumber \\ &&\quad\quad+  (A S H_u H_d   + T S+ h.c.)+
\frac{g^2+g'^2}{8} (|H_u|^2 -|H_d|^2 )^2 \label{Higgspot}
\eeq
where $m_{S,H_u,H_d}^2$, $A$, and $T$ are soft supersymmetry breaking parameters, and the last term is the usual MSSM $D$-term. 
With the standard parametrization of the Higgs fields we have define the usual $\tan \beta$ as the ratio of the up and down-type VEVs.
The  interaction with the singlet provides a sizable
quartic, so $\tan \beta$
can be close to or  less than one.  Minimizing the potential with respect to the scalar
$S$ we find that $S$ develops a VEV.

Using holomorphic techniques one can can map the effects of soft-SUSY breaking
masses, $m_{UV}^2$, in SUSY QCD over to the dual description
\cite{Cheng:1998xg,ArkaniHamed:1998wc,Luty:1999qc}. At the edge of the
conformal window, where the MCSSM model sits
 ($F=4,N=6$), one has a hierarchy of the soft breaking terms, which takes
the form
\beq
A, m_{\tilde{q},\tilde{g}} \sim
\frac{m_{UV}^2}{\Lambda} \quad
T  \sim   f^2 m_{UV}~.
\eeq
This leads to parameters of order
\beq
\begin{array}{ll}
 m_{UV} \sim M_3 \sim
{\rm few}\cdot {\rm TeV} \nonumber \\ 
 \Lambda \sim 5-10\,  {\rm TeV}
\nonumber \\ 
m_H \sim m_{\tilde t} \sim M_1 \sim M_2 \sim A
\sim {\rm few}\cdot {\rm 100\,GeV} \hphantom{asdfsdf}  \nonumber \\ 
 f\sim 100\  {\rm
GeV} \end{array}
\begin{array}{ll}
 T\sim f^2 m_{UV}\sim {\rm few} \cdot
10^{-2}\, {\rm TeV}^3 \nonumber \\ 
 {\cal F} \sim {\rm few} \cdot\  {\rm
TeV} \nonumber \\ 
 \mu_{\rm eff} = y \langle S \rangle \sim A \nonumber  \\
 \tan \beta \sim {\cal O}(1)~. \nonumber 
\end{array}
\eeq
In particular the stop can remain light, even with a heavy gluino \cite{Cleary:2015koc}.

\subsection{``Fat" SUSY Higgs LHC searches}
\label{subsection:fatLHCsearch}

If the Higgs compositeness scale is at 10\,TeV or above then it is very difficult to observe the small deviations in the Higgs couplings without some type of Higgs factory to do precision studies.  As in the RS and ``little" Higgs models we can look for the new associated particles that the composite model predicts. For SUSY composite models these are the usual superpartners, but they can have unusual spectra due to their composite nature.
Some sample spectra have been generated \cite{Csaki:2012fh};
often the light stop, ${\tilde t_1}$, is lighter than the lightest neutralino, $N_1$, it is  assumed that  a low-scale for supersymmetry breaking
gives an LSP gravitino with a mass  of a
few GeV or less. 
This type of spectrum corresponds to a stealthy stop
scenario \cite{Fan:2011yu}, with the $\tilde{t}_1$ nearly degenerate with the $t$.
The largest SUSY production process at the LHC in this scenario is
$pp\rightarrow {\tilde t}_1 {\tilde t}_1^*$.  For a light gravitino, if the ${\tilde t}_1$ decays promptly 
to a top quark and a gravitino there is little  missing  energy and these events are very difficult to uncover, but novel search techniques using spin correlations are closing in on these processes \cite{Aad:2014mfk}.

The next largest SUSY production cross-sections are production of the lightest sbottom and the heavier stop: $pp\rightarrow {\tilde b}_1
{\tilde b}_1^*$ and $pp\rightarrow {\tilde t}_2 {\tilde t}_2^*$.  The
${\tilde b_1}$ decays  to $\tilde t_1 W$, giving rise to $t {\overline
t} WW$ final states and in principle, a small amount of missing energy.  The ${\tilde t_2}$ decays mostly to ${\tilde t_1}Z$  and again the  ${\tilde t_1}$ will further decay to $t$ + gravitino. The final state for $pp\rightarrow {\tilde t}_2 {\tilde t}_2^*$ thus mostly contains $t{\overline t}ZZ$ plus a small amount of missing energy.
 The $\tilde{b_1}$  production process would be an interesting channel, however all of these events will have very
little missing transverse energy, so generic SUSY searches could miss it.  Even though the light ${\tilde t_1}$ will be
boosted,  this will usually not bring the
missing energy above the standard cuts.  
Other benchmark spectra  have neutralino (N)LSP's, thus  the usual missing energy signals of supersymmetry are expected.  Due to the heaviness of the  gluino and first two generations of squarks, the production rates are reduced from those of a vanilla MSSM. These  spectra fall in the class of models considered by~\textcite{Papucci:2011wy}. 

\subsection{Twin Higgs models and their UV completions}
\label{sec:twinHiggs}

All models presented until now have the common feature that the particles responsible for canceling the quadratic divergences in the Higgs mass are charged under the SM gauge symmetries. In particular, the top partner carries color charge, implying a reasonably large minimal production cross section at the LHC. An alternative scenario (which is experimentally quite challenging) is the case referred to as ``neutral naturalness" \cite{Craig:2015pha,Craig:2014aea,Craig:2013fga,Burdman:2014zta,Craig:2014roa,Cohen:2015gaa,Burdman:2006tz,Chang:2006ra,Chacko:2005pe}, where the particles canceling the 1-loop quadratic divergences are neutral under the SM. The canonical example for such theories is the Twin Higgs model of~\textcite{Chacko:2005pe}. This is an example of a pseudo-Goldstone boson Higgs theory, with an approximate global $SU(4)$ symmetry broken to $SU(3)$. This breaking can be parametrized by the 4-component field
\begin{equation}
H = e^{i \pi^a T^a/f} \left( \begin{array}{c} 0 \\ 0 \\ 0 \\ f \end{array} \right) \ ,
\end{equation}
where the $T^a$'s are the seven broken generators of $SU(4)$ and the $\pi^a$'s are the corresponding Goldstone bosons. The Twin Higgs model is obtained by gauging the $SU(2)_A\times SU(2)_B$ subgroup of $SU(4)$, where $SU(2)_A$ is identified with the SM $SU(2)_L$, while $SU(2)_B$ is the twin $SU(2)$ group. Gauging this subgroup breaks the $SU(4)$ symmetry explicitly, but the form of the corrections that are quadratically divergent are given by
\begin{equation}
\frac{\Lambda^2}{64\pi^2} (g_A^2 |H_A|^2+g_B^2 |H_B|^2) \ .
\label{eq:twinHiggsgaugeloop}
\end{equation}
For $g_A=g_B$ this term is $SU(4)$ symmetric and does not contribute to the Goldstone masses. The $SU(4)\to SU(3)$ breaking will also result in the breaking of the twin $SU(2)_B$ group and as a result three of the seven Goldstone bosons will be eaten, leaving 4 Goldstone bosons corresponding to the SM Higgs doublet $h$. The leading expression for $H_A,H_B$ 
\begin{equation} 
H_A \sim h, \ \ H_B \sim \left( \begin{array}{c} 0 \\ f- \frac{h^2}{2f} \end{array} \right) 
\end{equation}
show nicely the cancellation of the quadratic divergences of the $h$-dependent terms in (\ref{eq:twinHiggsgaugeloop}). In fact imposing the $Z_2$ symmetry on the full model will ensure the cancellation of all 1-loop quadratic divergences to the Higgs mass, since the two independent quadratic invariants $|H_A|^2$ and $|H_B|^2$ always combine into $|H|^2$ by the $Z_2$ symmetry. Logarithmically divergent terms can however arise for example from gauge loops, and will be of the form $\kappa (|H_A|^4+|H_B|^4)$, with $\kappa = {\cal O} (\frac{g^4}{16\pi^2} \log \frac{\Lambda}{g f})$, leading to a Higgs mass of order $g^2 f /4\pi$, which is of the order of the physical Higgs mass for $f \sim 1$ TeV. The quadratic divergences from the top sector can be eliminated if the $Z_2$ protecting the Higgs mass remains unbroken by the couplings that result in the top Yukawa coupling. This can be achieved by introducing top partners charged under a twin $SU(3)_c$ via the Lagrangian
\begin{equation}
{\cal L}_{top} = - y_t H_A \bar{t}_L^A t_R^A - y_t H_B \bar{t}_L^B t_R^B + h.c.
\end{equation}
In this case the quadratic divergences are cancelled by top partners that are neutral under the SM gauge symmetries. One remaining question is how to obtain a $Z_2$ breaking VEV $\langle H_B\rangle =f \gg v = \langle H_A \rangle$. This can be achieved by adding a soft-breaking $\mu$-term for the $Z_2$ symmetry in the form $\mu |H_A|^2$ without the corresponding $|H_B|^2$ term. 

Twin Higgs models are low-energy effective theories valid up to a cutoff scale of order $\Lambda \sim 4\pi f \sim 5-10$ TeV, beyond which a UV completion has to be specified. The simplest such possibility is to also make the Higgs composite, and UV complete the twin-Higgs model via gauge and top partners at masses of the order of a few TeV. A concrete implementation is the holographic twin Higgs model~\cite{Geller:2014kta}, which also incorporates a custodial symmetry to protect the $T$-parameter from large corrections. It is based on a warped extra dimensional theory with a bulk SO(8) gauge group, which incorporates the $SU(4)$ global symmetry discussed above enlarged to contain the SU(2)$_L\times$SU(2)$_R$ custodial symmetry. In addition the bulk contains either a full $SU(7)$ group or an $SU(3)\times SU(3)\times U(1) \times U(1) \times Z_2$ subgroup of it to incorporate QCD, its twin, and hypercharge.  The breaking on the UV brane is to the standard model and the twin standard model, while on the IR brane $SO(8) \to SO(7)$, giving rise to the 7 Goldstone bosons, three of which will be again eaten by the twin $W,Z$. The main difference compared to ordinary composite Higgs models is that in composite twin Higgs models the cancellation of the one-loop quadratic divergences is achieved by the twin partners of order 700 GeV - TeV, which are uncharged under the SM gauge group. This allows the IR scale of the warped extra dimension to be raised to the multi-TeV range without reintroducing the hierarchy problem. The role of the composite partners is to UV complete the theory, rather than the cancellation of the one-loop quadratic divergences. For more details about the composite twin Higgs models see~\cite{Batra:2008jy,Geller:2014kta,Barbieri:2015lqa,Low:2015nqa}.

Another interesting variation of twin Higgs models is the so-called fraternal twin Higgs \cite{Craig:2015pha}, where only the third generation fermions are endowed with twin partners, those needed to cancel the quadratic divergence from the top loop and the states related to them by gauge symmetries.   This may change the expected collider signals of the model significantly.

\subsection{Twin Higgs LHC Searches}
\label{sec:twinHiggspheno}

\begin{figure}[htb]
\includegraphics[width=0.35\hsize]{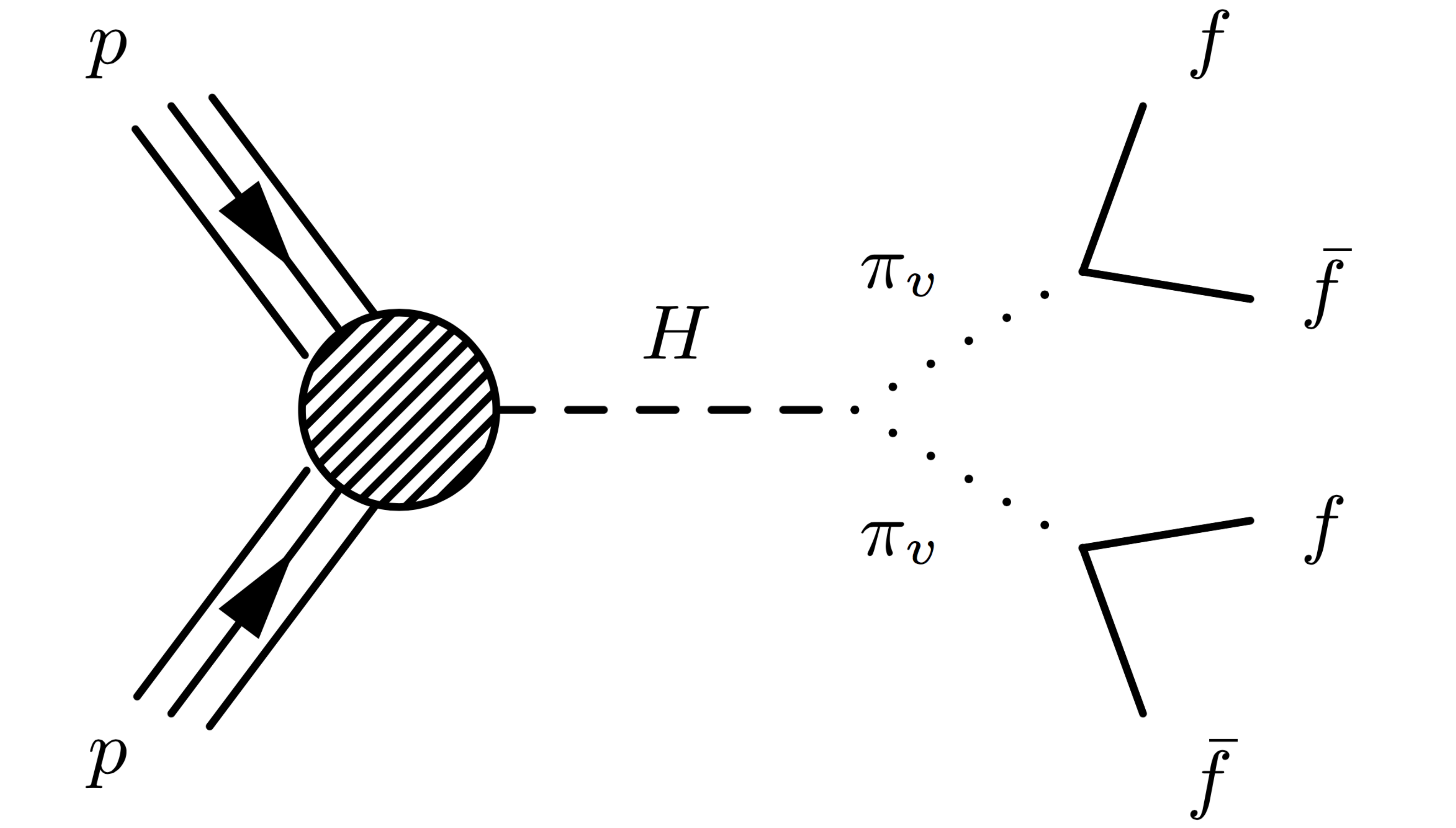}
\caption[]{Displaced decay of the SM Higgs due to decays to long-lived mirror glueballs which eventually decay to SM particles via mixing with the SM Higgs, from \textcite{Csaki:2015fba}.}
\label{fig:displacedHiggs}
\end{figure}

\begin{figure}[htb]
\includegraphics[width=0.35\hsize]{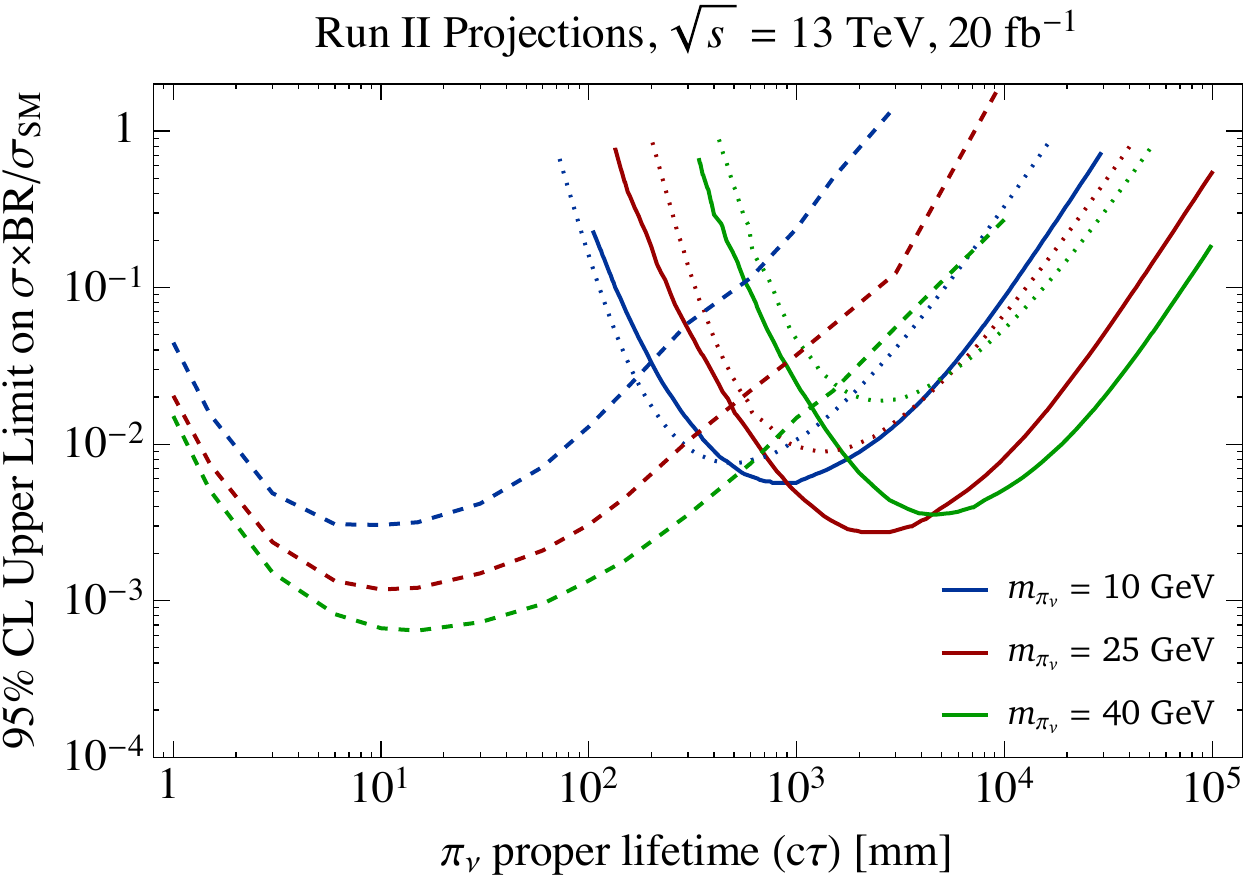}
\caption[]{Expected sensitivity of Run II displaced searches for displaced Higgs decays. The vertical axis is the branching fraction of the Higgs into the displaced decays, while the horizontal is the lifetime of the metastable decay products, for three different values of the mass of these intermediate states, from \textcite{Csaki:2015fba}. (Color online.)}
\label{fig:displacedHiggssensitivity}
\end{figure}

In twin Higgs models the new particles have low production cross sections or are very heavy, so it is difficult to directly test them at the LHC. The only direct connection between the visible and the mirror sector is via the Higgs sector, thus one expects any direct signals to appear in Higgs physics. Due to the mixing with the twin Higgs, one expects in all scenarios deviations of ${\cal O}(v^2/f^2) \lesssim 10$ \% in the Higgs couplings. Precision Higgs measurements should eventually be able to observe such deviations. More striking direct signals might also be expected depending on the details of the structure of the mirror sector, and in particular on the mirror QCD sector. The basic classes are~\cite{Curtin:2015fna} models with mirror QCD with long-lived glueballs (due to the absence of light mirror quarks), models with mirror QCD without long-lived glueballs (in the presence of light mirror quarks), and models without mirror QCD. For the case of mirror QCD with no light matter the glueballs of mirror QCD will be long lived and decay to the SM only via the mixing with the ordinary Higgs. Since the ordinary Higgs itself can decay to such glueballs one eventually ends up~\cite{Craig:2015pha,Curtin:2015fna} with displaced Higgs decays as depicted in Fig.~\ref{fig:displacedHiggs}.  The expected reach from the Run II of the LHC for such displaced decays has recently been estimated in~\cite{Csaki:2015fba} and the main results are summarized in Fig.~\ref{fig:displacedHiggssensitivity}. The case of mirror QCD with light quark matter is expected to result in Higgs to invisible decays (where the invisibles are mirror jets). Finally in the case of no mirror QCD the only expected signal would be the deviation of the Higgs couplings from their SM values. 

\section{BROKEN CONFORMAL SYMMETRY}
\label{sec:conformal}

\subsection{Dilatons}
\label{subsec:dilatons}

Technicolor was one of the most appealing ideas for a natural electroweak symmetry breaking scenario. With the discovery of the Higgs-like particle pure technicolor models have been excluded. A natural question to ask is whether models of dynamical electroweak symmetry breaking could nevertheless produce a Higgs-like particle \cite{Bando:1986bg,Yamawaki:1986zg}. Spontaneously broken conformal symmetry provides for this possibility since it produces a light scalar dilaton, the Goldstone boson of broken conformal  symmetry.  This is not too far-fetched since the  SM itself has a limit where the Higgs can be considered a dilaton: if the entire Higgs potential of the SM is turned off, the Higgs will be a classical flat direction, and the SM classically scale invariant. A Higgs VEV will break the scale invariance in addition to breaking the electroweak symmetry, and the physical Higgs boson can be identified with the dilaton of the spontaneously broken scale invariance. The couplings of the Higgs to other SM fields will be determined in this limit: the Higgs will couple to all the sources of conformality breaking suppressed by the Higgs VEV, including couplings to masses and to $\beta$-functions in the case of massless gauge bosons. Thus the tree level coupling of a dilaton to gauge bosons and fermions the same as the SM Higgs couplings (as guaranteed by low energy theorems \cite{Ellis:1975ap,Shifman:1979eb})  if the VEV that breaks conformal symmetry is the same as the VEV that breaks electroweak symmetry.  The loop level couplings to photons and gluons are model dependent.  For example if the gluons are composites of the conformal sector then then dilaton-gluon-gluon coupling is very different from the Higgs-gluon-gluon coupling \cite{Low:2012rj}.  If however the gluons merely weakly gauge a global symmetry of the conformal sector then it is possible for this scenario to work \cite{Bellazzini:2012vz,Chacko:2012vm,Goldberger:2007zk,Ryskin:2009kw,Foot:2008tz,Abe:2012eu}.

While QCD-like technicolor does not have a light dilaton \cite{Holdom:1987yu} there has been much work on trying to find technicolor-like  models that do have a light dilaton  \cite{Bando:1986bg,Yamawaki:1986zg,Kozlov:2012cr,Matsuzaki:2012xx,Fodor:2012ty,Elander:2012fk,Lawrance:2012cg,Anguelova:2012ka,Matsuzaki:2012gd,Elander:2011aa,Matsuzaki:2011ie,Hashimoto:2011cw,Elander:2010wd,Vecchi:2010aj,Appelquist:2010gy,Vecchi:2010gj,Sannino:2009za,Dietrich:2005jn}.
There have also been several other approaches to producing models with a Higgs-like dilaton \cite{Coriano:2012nm,Campbell:2011iw,Ryskin:2009kw,Jora:2009dh,Foot:2008tz}, here we focus on the generic properties.

Scale transformations  are part of the conformal group \cite{Coleman}. The scale transformation of an operator $ {\cal O}$  is given by (for $x\to x' = e^{-\alpha} x$)
\begin{equation}
 {\cal O} (x)\to {\cal O}'(x)=e^{\alpha \Delta} {\cal O} (e^\alpha x) \, ,
\end{equation}
where $\Delta$ is the scaling dimension of  $ {\cal O}$. If all the operators in the action have scaling dimension four, then the action is invariant.
If scale invariance is broken spontaneously by the VEV of an operator $\langle \mathcal{O}\rangle=f^n$ then there must be a Goldstone boson that transforms as
\begin{equation}
\sigma(x)\to \sigma(e^\alpha x)+ \alpha f \, .
\end{equation}
The low-energy effective theory can simply be obtained by replacing the VEV with the non-linear realization
\begin{equation}
f\to f\, \chi \equiv f \, e^{\sigma /f}~.
\label{eq:replacement}
\end{equation}
In the spirit of Callan, Coleman, Wess, and Zumino \cite{Coleman:1969sm,Callan:1969sn} we can find the low-energy effective action by requiring that it is invariant under scale transformations:
\beq
{\cal L}_{eff}&=&  - a_{0,0} \, (4\pi)^2 f^4 \chi^4 +\frac{f^2}{2}(\partial_\mu \chi)^2  +  \frac{a_{2,4}}{(4 \pi)^{2} }  \frac{(\partial \chi )^{4} } {\chi^4} + \ldots
\eeq
It is quite unusual that there is a non-derivative term in the potential, but it is a simple consequence of the fact that we are dealing with a space-time
symmetry and its transformation is cancelled by the change in the volume measure. This effective theory is somewhat of an embarrassment since if $a_{0,0} \ne 0$
it does not describe spontaneous conformal breaking.  If $a_{0,0} >  0$ then the vacuum will be at $f \chi =0$ and there is no spontaneous breaking, if  $a_{0,0} <  0$ then the vacuum will be at $f \chi \rightarrow \infty$ so there was no conformal theory to start with. Assuming that we actually can find theories that spontaneously break conformal symmetry (with a flat-direction since $a_{0,0} = 0$) then we still need to be very careful that any perturbations do not reintroduce $a_{0,0} \ne 0$.  If we add a perturbation $\lambda {\cal O}$ to the theory then the conditions for $a_{0,0}$ to remain small are that the scaling dimension of ${\cal O}$ is close to, but slightly below, four and the $\beta$ function for $\lambda$ is very small all the way from the UV down to the breaking scale $f \chi$ \cite{Bellazzini:2012vz,Chacko:2012sy,Chacko:2013dra,Evans:2013vca,Grinstein:2011dq}. If these conditions are satisfied $a_{0,0}$  will remain small and develop a weak dependence on $\lambda$ through the running, which will pick out a unique minimum for $\chi$ and give the dilaton a small mass.

A spurion analysis is sufficient to determine the couplings of the dilaton in the low-energy theory \cite{Bellazzini:2012vz}. For SM particles the leading couplings are mass$/f$ for fermions and mass$^2/f$ for bosons, so the they are very close to the SM Higgs couplings when $f=v$. The spurion analysis also gives the loop suppressed couplings to gluons and photons, but this can be seen even more simply through  dilaton dependence of the breaking scale $f$.
The IR gauge coupling is given by
\beq
\frac{1}{g^2(\mu)}= \frac{1}{g^2(\mu_0)} - \frac{b_{UV}}{8 \pi^2} \ln \frac{\mu_0}{f} - \frac{b_{IR}}{8 \pi^2} \ln \frac{f }{\mu}~.
\eeq
where $b_{IR}$ includes all the particles with masses below $f$ and $b_{UV}$ includes all the particles with masses below the UV cutoff $\mu_0$.
The coefficient of $F^{\mu \nu}F_{\mu\nu}$ in  the low-energy theory is 
$ -1/4g^2(\mu)$ and replacing $f$ by $f e^{\sigma/f}$ we find a linear dilaton coupling
\beq
\frac{g^2}{32 \pi^2} \left( b_{IR}-b_{UV}\right) F^{\mu \nu}F_{\mu\nu} \frac{\sigma}{f}~.
\label{eq:sigmaF}
\eeq
Thus if the SM particles are spectators of the conformal sector there contributions cancel out, and the low-energy coupling depends on the conformal sector contribution to $b_{UV}$, which is completely model dependent. Of course the SM particles still contribute in the loops of the low-energy theory, and since they have SM-like couplings these loops will precisely mimic the SM Higgs couplings to photons and gluons.  So, for example if the conformal sector has no colored degrees of freedom, we expect the dilaton coupling to $gg$ to be equal to the SM Higgs coupling to $gg$.

\subsection{Dilaton Higgs LHC Searches}
\label{subsec:dilatonsearch}

LHC searches have been proposed for general dilatons \cite{Barger:2011nu,Barger:2011hu,Coleppa:2011zx} but these studies tend to make model dependent assumptions about the dilaton couplings that are incompatible with identifying it with the Higgs-like boson discovered at the LHC. The most direct way to constrain the dilaton Higgs models is to look for deviations in the couplings of the Higgs-like boson from the SM prediction; this will be discussed further in section \ref{subsection:effective}. A substantive difference between the SM Higgs and a dilaton Higgs is in the three- and four-point self couplings. This can start to be addressed by double Higgs production searches \cite{Coriano:2012dg,Gouzevitch:2013qca,Dolan:2012ac,Noble:2007kk,Goldberger:2007zk} (cf. section \ref{subsection:pseudoLHCsearch}).

\subsection{Conformal Technicolor}
\label{subsec:conformaltechnicolor}

As we have discussed, technicolor theories solve the hierarchy problem by replacing the Higgs  by an operator  that has a dimension greater than two.  Naively a solution to the hierarchy problem would be that the analog of the quadratically divergent Higgs mass term (that leads to the hierarchy problem in the SM)  is replaced by an operator with a dimension greater than four, so that it cannot have a divergence.  An alternative scenario known as ``conformal technicolor" \cite{Luty:2004ye} posits a more subtle solution: what if the Higgs had a scaling dimension near one, but the mass operator was still irrelevant. This certainly cannot happen at weak coupling, but if the Higgs is a composite of some strong dynamics this may be possible.  This leads to a wealth of phenomenological possibilities \cite{Luty:2008vs,Morrissey:2009tf,Evans:2010ed,Galloway:2010bp,Fukushima:2010pm,Azatov:2011ht,Gherghetta:2011na,Orgogozo:2011kq}.
Valiant efforts have been made to extend the exact results on conformal theories to exclude this possibility with the result that the dimension of the Higgs must be larger than about 1.5 for its mass operator to be irrelevant \cite{Rychkov:2009ij,Rattazzi:2008pe,Poland:2010wg,Poland:2011ey,Fitzpatrick:2011hh,Green:2012nq}.
Thus the Higgs would not look much like a free particle (cf. the next section).  An additional problem is why would such a theory have a Higgs that is light compared to the compositeness scale?  The only known answer is that the Higgs-like scalar must be a pseudo-Goldstone boson and there must be some tuning involved to adjust its properties, in which case we have something very much like the minimal composite Higgs (see section \ref{sec:minimalpseudo-goldstoneHiggs}).

\subsection{Quantum Critical Points and the Higgs}
\label{sec:Unhiggs}

A continuous phase transition that is tuned near the critical point has a light scalar degree of freedom that is the fluctuation of the order parameter. In the SM a tiny change in the Higgs mass parameter moves us from the broken phase to the symmetric phase, so it must be tuned to be very close the the critical point. Experimentally we have already seen a Higgs-like resonance, so if there is new physics beyond the standard model it also should be tuned to be close to a critical point. This type of transition at zero temperature is call a quantum critical point  in order to emphasize that it is quantum fluctuations that dominate rather than thermal fluctuations. If we assume we are near a quantum critical point then there is a very long RG flow which generically either approaches a trivial IR fixed point or a non-trivial fixed. The Higgs sector of the SM is an example of a theory that approaches a trivial fixed point. If we want to allow for the possibility that the correct extension of the SM involves a non-trivial fixed point we need to study general CFT's with a variety  of possible IR breakings of scale invariance. This is a very formidable task, but we can greatly restrict the range of theories to be investigated since we know the Higgs resonance of such a theory must be weakly coupled since the observed Higgs is very similar to the SM Higgs. We will see that the class of quantum critical Higgs (aka Unhiggs) models provide particular, concrete realizations of this scenario.


\textcite{Georgi:2007ek} introduced a useful technique for studying conformal sectors that couple to the standard model using two-point functions of operators\footnote{See also \cite{Georgi:2007si}. The properties of higher $n$-point functions of unparticles are discussed in \textcite{Georgi:2009xq}. } with scaling dimension between one and two.  Formally the phase space corresponding to the spectral density of this two-point function resembles the phase space for a fractional number of particles, hence the name ``unparticles." 
For electroweak symmetry breaking these ideas have been applied in models where the Higgs couples to an approximately conformal sector and can mix with an unparticle \cite{Fox:2007sy,unother4,unother5,unother6,Lee:2008ph}.  For a non-elementary Higgs we need to study the case where the Higgs is a composite of an approximately conformal sector, or in other words the Higgs field itself includes a continuum in its spectrum in addition to a pole (see \textcite{unother1,unother2,unother3,unother4,unother5,unother6,Lee:2008vp,Lee:2008ph} for work on related ideas). This is called the quantum critical Higgs scenario. 
It can be formulated in 4D \cite{Stancato:2008mp} or 5D \cite{Falkowski:2008yr}. The 4D description is simpler to implement and will be described here, but the 5D 
 description has broader implications for form factors \cite{Bellazzini:2015cgj} and is very useful for calculating precision electroweak observables \cite{Falkowski:2009uy}.
 
Clearly a phenomenologically acceptable quantum critical Higgs cannot have a continuous spectrum that starts at zero energy, we must have a pole significantly below any continuum. Fortunately there is a simple way to parameterize the case with a finite threshold \cite{Fox:2007sy,coloredunparticles}.  Typically the introduction of a threshold introduces a single discrete state as well, which could play the role of the Higgs-like particle discovered at the LHC. An important part of seeing that the model is consistent  is showing that an quantum critical Higgs still unitarizes $WW$ scattering.  This only happens due to the subtle behavior of the non-standard Feynman vertices of unparticles coupled to gauge bosons. Such vertices are also crucial in the cancelation of anomalies for unfermions \cite{Galloway:2008sq}.
The quantum critical Higgs is taken to have a scaling dimension, $\Delta$,  between 1 and 2, and thus is a continuation of the bulk Higgs to smaller scaling dimensions.  In the notation of 
section~\ref{subsection:gaugephobic:bulkaction} this is the  range $-1<\beta<0$. For understanding the transition from $\Delta <2$ to $\Delta >2$ the 5D description \cite{Falkowski:2009uy}
is essential.

For a quantum critical Higgs we crucially need couplings to the electroweak gauge bosons, so we need gauge covariant derivatives ($D_\mu=\partial_\mu - i g A^a T^a$) in the kinetic term, and Yukawa couplings to 
the fermions, the largest being to the top quark.  Thus, neglecting couplings to light quarks and leptons the minimal effective action \cite{coloredunparticles,Stancato:2008mp} for a quantum critical Higgs is:
\beq
S= \int d^4 x  \,-H^\dagger  \left(D^2+\mu^2\right)^{2-\Delta}
H- V(H)-\lambda_t \,{\overline t}_R \frac{H^\dagger}{\Lambda^{\Delta-1}} \left( \begin{array}{c} t \\ b \end{array} \right)_L + \textrm{ h.c.} ~.
 \label{action}
\eeq
Note that if the scaling dimension of the quantum critical Higgs, $\Delta$, is larger than one, then the Yukawa coupling has a dimension larger than four, and is suppressed by a power of the 
cutoff.  Calculating loop corrections in this theory generically leads to potential terms which allow for 
the possibility of a minimum at a non-zero VEV of the quantum critical Higgs field.
To make further progress we need the couplings of the quantum critical Higgs to the gauge fields.
The general result  \cite{coloredunparticles} using Eq.  (\ref{action}) and the Mandelstam path-ordered exponential \cite{Mandelstam:1962mi,Terning:1991yt} is 
\beq
i g \Gamma^{a\alpha}(p,q)&=&
-i g T^a  \frac{2p^\alpha+q^\alpha}{2p \cdot q+q^2}\left[ \left(\mu^2-(p+q)^2\right)^{2-\Delta}
-\left(\mu^2-p^2\right)^{2-\Delta} \right]~.
\label{svertex}
\eeq

As usual we expand the quantum critical Higgs field in small fluctuations around the VEV and the resulting Goldstone bosons will be eaten by the gauge bosons, leaving behind a physical Higgs mode. It is convenient to remove the mixing terms by including gauge fixing terms, then
the propagators for the gauge bosons are very different from in the SM:
\beq
\Delta_W(q) &=& \frac{-i}{q^2-M_W^2+i\epsilon} 
 \left(g_{\alpha\beta}+ \left( \frac{\xi \left(q^2 -  M_W^2\right)\mu^{2-2\Delta}}{K(q^2)\left(q^2-\frac{\xi M_W^2}{2-\Delta}\right)} -\frac{1}{q^2}\right)q_\alpha q_\beta \right)~, \nonumber
\eeq
For practical calculations it seems best to use Landau gauge, $\xi=0$, where the gauge boson propagators are the same as the SM Landau gauge propagators.
The physical quantum critical Higgs propagator is 
\beq
\Delta_{h}(q)&=&
-\frac{i}{m^{4-2\Delta}-\mu^{4-2\Delta}+\left( \mu^2-q^2-i \epsilon \right)^{2-\Delta} }~,
   \label{scalarprop}
\eeq
where we explicitly see the occurrence of a pole (to be identified with the resonance at 125\,GeV) at 
\beq
q^2=\mu^2 -\left(\mu^{4-2\Delta}-m^{4-2\Delta}\right)^\frac{1}{2-\Delta}~,
\eeq
and a continuum for $q>\mu$.

\begin{figure}[htb]
\includegraphics[width=0.2\hsize]{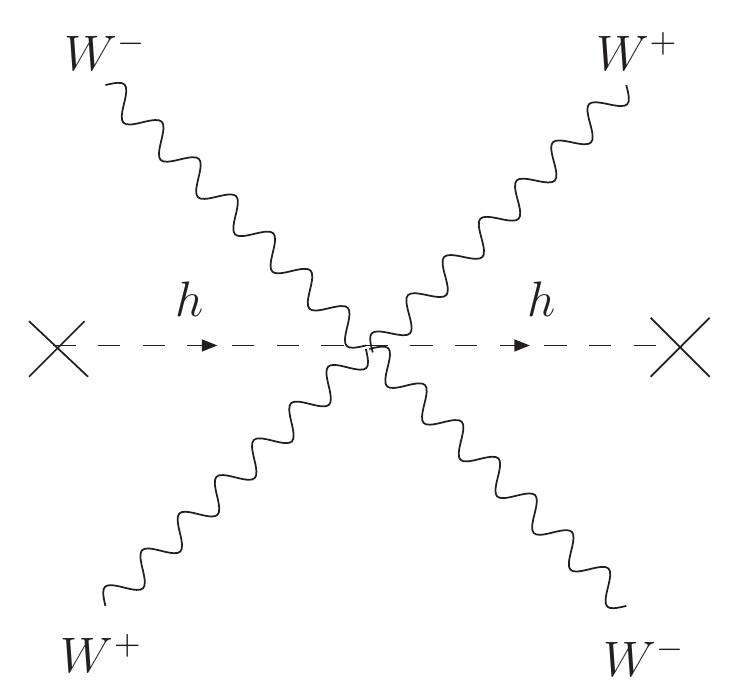}
\caption[]{The four gauge boson two (quantum critical) Higgs contribution to $WW$ scattering, from \textcite{Stancato:2008mp}.}
\label{fig:WW2UnH}
\end{figure}

The effects of unparticles on unitarity  have been studied for $WW$ scattering \cite{UnWWscattering} and Higgs-Higgs scattering \cite{UnHHscattering}, 
 assuming that the Higgs boson was an ordinary particle, and that the unparticle belonged to a non-SM sector.
The non-standard behavior of the quantum critical Higgs propagator  (\ref{scalarprop}) means that unitarization works differently in this model compared to the usual picture of heavy vector resonances discussed in section \ref{subsection:gaugephobic:bulkaction}. Taking the SM Higgs exchange diagrams and replacing the Higgs propagator by an quantum critical Higgs propagator gives softer growth with energy that is not sufficient for Unitarization.  However the full amplitude also has a new contribution from vertices with multiple gauge couplings that arise from expanding the path-ordered exponential.  The requisite vertex is shown in Fig.~\ref{fig:WW2UnH}.
Once all the contributions are included there are no positive powers of energy in the scattering amplitude. Partial wave unitarity also puts mild constraints on the
parameter values \cite{Stancato:2008mp}.  The unitarity of fermion scattering is discussed in \cite{Englert:2012dq}, where special care is needed for $d>1.5$.

From the usual top loop correction to the quadratic (quantum critical) Higgs term in the  action we find
\beq
\delta m_h^{4-2\Delta} = \frac{3|\lambda_t|^2}{8\pi^2} \Lambda^{4-2\Delta}~.
\eeq
So larger values of $\Delta$ lead to less sensitivity to the cutoff, smoothly matching onto the bulk Higgs (subsection \ref{subsection:gaugephobic:bulkaction}) case ($\Delta>2$) where the hierarchy problem is
completely solved. Thus while we cannot expect to solve the full hierarchy we can resolve the little hierarchy problem. For $\Delta \sim 1.7$ the cutoff can be near 10\,TeV. \textcite{Beneke:2011jk} provide an alternative interpretation of the little hierarchy in a theory with non-minimal gauge couplings.

\subsection{Quantum Critical Higgs LHC Searches}
\label{sec:UnhiggsSignals}

The quantum critical Higgs reduces to the bulk Higgs model \cite{gaugephobic} when its scaling dimension is near 2  or larger \cite{Cacciapaglia:2008ns}. We can quantify the suppression of the gauge boson coupling with the definition
\beq
\label{eqn:Xi^2}
\xi^2 \equiv \frac{\sigma_{Unh} (e^+ e^- \rightarrow HZ)}{\sigma_{SM} (e^+ e^- \rightarrow HZ)}
\eeq
\textcite{Stancato:2008mp} showed that $\xi^2$ falls as $\Delta$ gets larger and is approximately zero, as expected,  for $\Delta \rightarrow 2$. 
The suppression of gauge couplings was studied in \textcite{Englert:2012dq}  using $pp$ collisions at the LHC, where we again it was seen that for $\Delta \sim1$ there is only a small suppression of the cross section compared to the SM, but a severe suppression already for values of $\Delta$ around 1.2. This sensitivity study was done before the Higgs discovery, and focussed on resonances above the $ZZ$ threshold, so this type of analysis needs to be be redone by experimentalists using the actual data with the resonance set to 125\,GeV.


Another signature is that the fraction of top decays with a longitudinal $W$ emitted is different than in the SM  \cite{Stancato:2010ay}.
To compare with data, we must calculate the fraction of the top decays with a longitudinally produced $W$ boson: 
\beq
\mathcal{F}_0 \equiv \frac{\Gamma(t\rightarrow W^+_L b)}{\Gamma(t\rightarrow W^+_L b)+\Gamma(t\rightarrow W^+_T b)}~.
\eeq
The current top quark data from ATLAS with 1.04\,fb$^{-1}$ yields the following value for $\mathcal{F}_0$ \cite{Aad:2012ky}:
\beq
\label{eqn:F0CDF}
\mathcal{F}_0 = 0.67 \pm 0.07~.
\eeq
while CMS reports \cite{Khachatryan:2014vma} with 19.1\,fb$^{-1}$
\beq
\mathcal{F}_0=0.720+/- 0.039 \,({\rm stat.}) +/- 0.037\, ( {\rm syst.})~,
 \eeq
 compared with the SM prediction $\mathcal{F}_0=0.699$. 
 With more data the LHC promises to make a more accurate determination of this quantity, this process however can only significantly constrain models where the 
 threshold $\mu$ is below the top mass.

\begin{figure}[htb]
\begin{center}
\includegraphics[width=\smallimagesize]{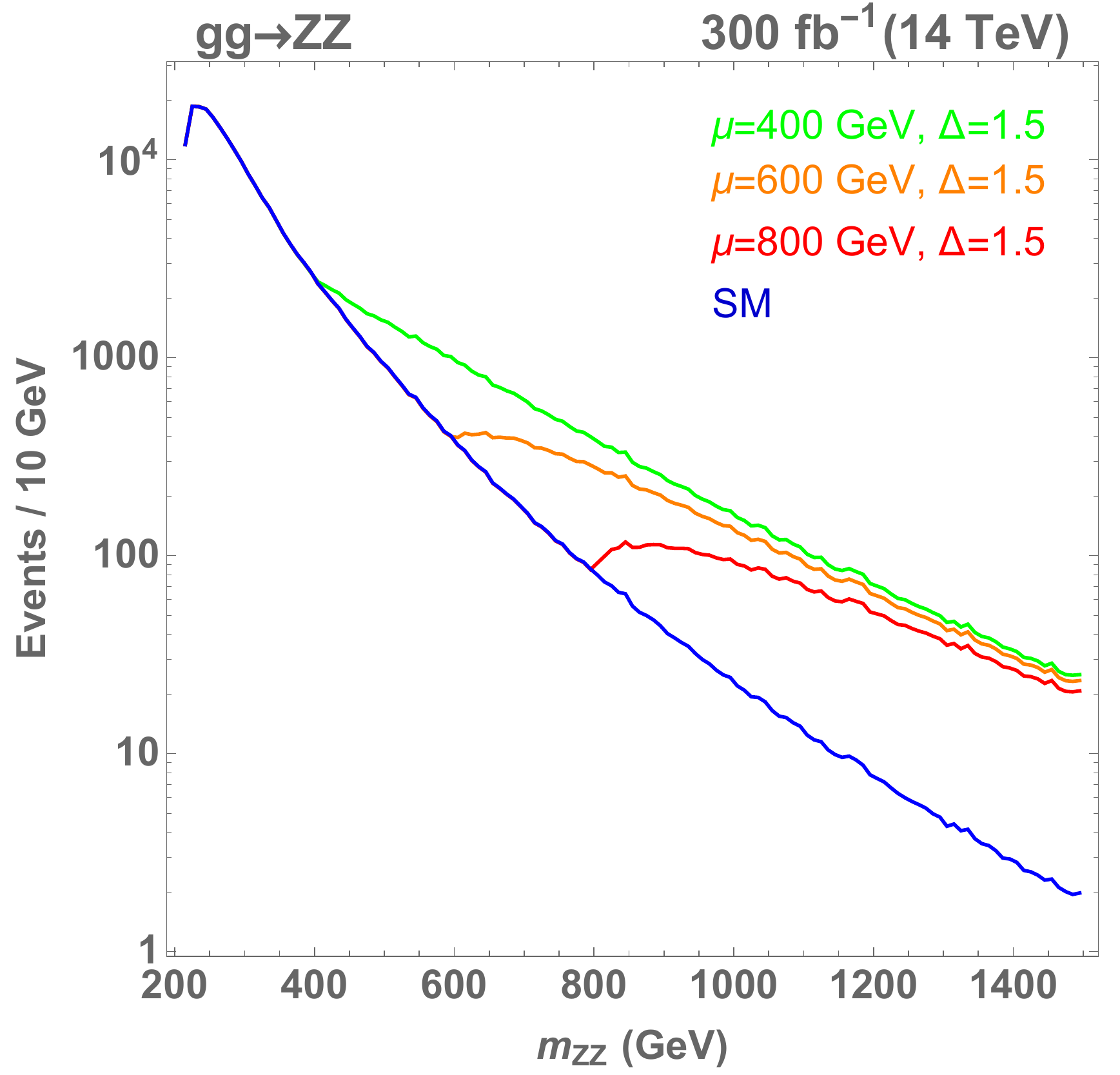}
\end{center}
\caption{The effects of a quantum critical Higgs two point function in the production of on-shell $Z$-boson pairs, with different values of the threshold $\mu$ \cite{Bellazzini:2015cgj}. (Color online.)}
\label{fig:ZZinterference}
\end{figure}

The LHC constraints from the production cross-section on a relatively heavy quantum critical Higgs using the $h\rightarrow ZZ\rightarrow 4 \ell$ channel are straightforward \cite{Englert:2012cb}, but for masses below 200\,GeV (as needed to identify the quantum critical Higgs pole with the Higgs-like boson discovered at the LHC), the analysis must be done by the 
experimental collaborations. Another search strategy is to look at the interference between the off-shell Higgs process $gg\rightarrow h\rightarrow ZZ\rightarrow 4 \ell$
and the QCD process $gg\rightarrow  ZZ\rightarrow 4 \ell$ with just a top loop in the intermediate state. As shown in Fig.~\ref{fig:ZZinterference} this can uncover the continuum threshold, but it also provides access to the scaling dimension $\Delta$ \cite{Bellazzini:2015cgj}. Double Higgs production can also yield complimentary information  \cite{Bellazzini:2015cgj}.

\section{CONCLUSIONS}

All the models we have considered should have some deviations in the Higgs coupling to SM particles, so the model independent 
approach of constraining the couplings of an effective Higgs Lagrangian (section \ref{subsection:effective}) is a robust method for
attacking all models with non-elementary Higgs bosons. A good deal can be learned from the LHC, but some kind of  Higgs factory is
ultimately required. An especially interesting coupling is the Higgs to $WW$ coupling, since deviations in this coupling imply TeV scale new
physics in order to unitarize $WW$ scattering.

In addition different classes of models have additional signatures that are being searched for.
The little  Higgs models (section \ref{subsection:pseudoLHCsearch}) need fermionic top partners in order to cancel the one-loop divergence 
in the Higgs mass.
RS models (section \ref{subsection:RSsearch}) have KK gluons as well as 
KK $W$s, KK $Z$s, (aka $W^\prime$ and $Z^\prime$) and KK gravitons. The most promising LHC search is for the KK gluons, which
can be enhanced using a top-tagger.
Bulk Higgs models (section \ref{subsection:gaugephobic:searches}) share these KK gauge bosons but 
the $W$ and $Z$ resonances need to be much lighter than in RS since they need to contribute to unitarization.
However a thorough $W^\prime$ search using $WZ$ scattering would require around 300\,fb$^{-1}$ of luminosity.
Some bulk Higgs models can share a feature of certain pseudo-Goldstone boson  Higgs  models (section \ref{subsection:pseudoLHCsearch})
which is top-partner fermions with exotic electric charges (like $+5/3$).
The quantum critical Higgs scenario (section \ref{sec:UnhiggsSignals})
is starting to be constrained by precision top decay polarization measurements.
``Fat" SUSY Higgs models (section \ref{subsection:fatLHCsearch}) have the usual superpartners but with an
 unusual mass spectrum.

The particular searches mentioned above are well-known and bounds have already been presented.
In addition to these there are searches that should be done in the future.
The bulk Higgs has a interesting signature in $pp \to W H$ through an intermediate $W'$ which should be looked for.
The  quantum critical Higgs models can be tested using the 
 $h\rightarrow ZZ\rightarrow 4 \ell$ channel. Twin Higgs models (section \ref{sec:twinHiggspheno}) can be probed through invisible decays and displaced vertices.
Finally pseudo-Goldstone boson  Higgs  and dilaton Higgs models can be probed through searching for an enhanced double Higgs production process.

More than ever Higgs physics is a heuristic compass in the quest for new physics beyond the standard model. 
The Run-2  LHC program is set to reveal the first glimpses of this uncharted territory that will be fully explored later at future machines like ILC, CLIC, CepC, SppC and FCC \cite{FabiolaG:2015}.

\section*{Acknowledgments}

We would like to thank H.C.~Cheng, R.~Contino, G.~Giudice, J.~Gunion, M.~Luty, A.~Pomarol, R.~Rattazzi, A.~Weiler for useful discussions. C.C. is supported in part by the NSF grant PHY-0757868. 
C.G. is supported by Spanish Ministry MEC under grants FPA2014-55613-P and FPA2011-25948, by the Generalitat de Catalunya grant 2014-SGR-1450, by the Severo Ochoa excellence program of MINECO (grant SO-2012-0234), 
by the European Commission through the Marie Curie Career Integration Grant 631962
and by the Helmholtz Association.
J.T. is supported in part by the DOE under grant DE-SC-000999.
Part of this work was completed at the Aspen Center for Physics.

\bibliographystyle{apsrmp}
\bibliography{alternatives}

\begin{thebibliography}{382}
\expandafter\ifx\csname natexlab\endcsname\relax\def\natexlab#1{#1}\fi
\expandafter\ifx\csname bibnamefont\endcsname\relax
  \def\bibnamefont#1{#1}\fi
\expandafter\ifx\csname bibfnamefont\endcsname\relax
  \def\bibfnamefont#1{#1}\fi
\expandafter\ifx\csname citenamefont\endcsname\relax
  \def\citenamefont#1{#1}\fi
\expandafter\ifx\csname url\endcsname\relax
  \def\url#1{\texttt{#1}}\fi
\expandafter\ifx\csname urlprefix\endcsname\relax\def\urlprefix{URL }\fi
\providecommand{\bibinfo}[2]{#2}
\providecommand{\eprint}[2][]{\url{#2}}

\bibitem[{Aad \emph{et~al.}(2012{\natexlab{a}})\citenamefont{Aad}
  \emph{et~al.}}]{Aad:2012ky}
\bibinfo{author}{\bibnamefont{Aad}, \bibfnamefont{G.}}, \emph{et~al.}
  (\bibinfo{collaboration}{ATLAS Collaboration}),
  \bibinfo{year}{2012}{\natexlab{a}}, \bibinfo{journal}{JHEP}
  \textbf{\bibinfo{volume}{1206}}, \bibinfo{pages}{088}.

\bibitem[{Aad \emph{et~al.}(2012{\natexlab{b}})\citenamefont{Aad}
  \emph{et~al.}}]{Aad:2012tfa}
\bibinfo{author}{\bibnamefont{Aad}, \bibfnamefont{G.}}, \emph{et~al.}
  (\bibinfo{collaboration}{ATLAS Collaboration}),
  \bibinfo{year}{2012}{\natexlab{b}}, \bibinfo{journal}{Phys.Lett.}
  \textbf{\bibinfo{volume}{B716}}, \bibinfo{pages}{1}.

\bibitem[{Aad \emph{et~al.}(2013)\citenamefont{Aad} \emph{et~al.}}]{Aad:2012cy}
\bibinfo{author}{\bibnamefont{Aad}, \bibfnamefont{G.}}, \emph{et~al.}
  (\bibinfo{collaboration}{ATLAS Collaboration}), \bibinfo{year}{2013},
  \bibinfo{journal}{New J.Phys.} \textbf{\bibinfo{volume}{15}},
  \bibinfo{pages}{043007}.

\bibitem[{Aad \emph{et~al.}(2015{\natexlab{a}})\citenamefont{Aad}
  \emph{et~al.}}]{Aad:2014mfk}
\bibinfo{author}{\bibnamefont{Aad}, \bibfnamefont{G.}}, \emph{et~al.}
  (\bibinfo{collaboration}{ATLAS Collaboration}),
  \bibinfo{year}{2015}{\natexlab{a}}, \bibinfo{journal}{Phys. Rev. Lett.}
  \textbf{\bibinfo{volume}{114}}(\bibinfo{number}{14}),
  \bibinfo{pages}{142001}.

\bibitem[{Aad \emph{et~al.}(2015{\natexlab{b}})\citenamefont{Aad}
  \emph{et~al.}}]{Aad:2015yza}
\bibinfo{author}{\bibnamefont{Aad}, \bibfnamefont{G.}}, \emph{et~al.}
  (\bibinfo{collaboration}{ATLAS Collaboration}),
  \bibinfo{year}{2015}{\natexlab{b}}, \bibinfo{journal}{Eur. Phys. J.}
  \textbf{\bibinfo{volume}{C75}}(\bibinfo{number}{6}), \bibinfo{pages}{263}.

\bibitem[{\citenamefont{Abe} \emph{et~al.}(2012)\citenamefont{Abe, Kitano,
  Konishi, Oda, Sato} \emph{et~al.}}]{Abe:2012eu}
\bibinfo{author}{\bibnamefont{Abe}, \bibfnamefont{T.}},
  \bibinfo{author}{\bibfnamefont{R.}~\bibnamefont{Kitano}},
  \bibinfo{author}{\bibfnamefont{Y.}~\bibnamefont{Konishi}},
  \bibinfo{author}{\bibfnamefont{K.-y.} \bibnamefont{Oda}},
  \bibinfo{author}{\bibfnamefont{J.}~\bibnamefont{Sato}}, \emph{et~al.},
  \bibinfo{year}{2012}, \bibinfo{journal}{Phys.Rev.}
  \textbf{\bibinfo{volume}{D86}}, \bibinfo{pages}{115016}.

\bibitem[{\citenamefont{Agashe} \emph{et~al.}(2008)\citenamefont{Agashe,
  Belyaev, Krupovnickas, Perez, and Virzi}}]{Agashe:2006hk}
\bibinfo{author}{\bibnamefont{Agashe}, \bibfnamefont{K.}},
  \bibinfo{author}{\bibfnamefont{A.}~\bibnamefont{Belyaev}},
  \bibinfo{author}{\bibfnamefont{T.}~\bibnamefont{Krupovnickas}},
  \bibinfo{author}{\bibfnamefont{G.}~\bibnamefont{Perez}}, and
  \bibinfo{author}{\bibfnamefont{J.}~\bibnamefont{Virzi}},
  \bibinfo{year}{2008}, \bibinfo{journal}{Phys.Rev.}
  \textbf{\bibinfo{volume}{D77}}, \bibinfo{pages}{015003}.

\bibitem[{\citenamefont{Agashe and Contino}(2006)}]{Agashe:2005dk}
\bibinfo{author}{\bibnamefont{Agashe}, \bibfnamefont{K.}}, and
  \bibinfo{author}{\bibfnamefont{R.}~\bibnamefont{Contino}},
  \bibinfo{year}{2006}, \bibinfo{journal}{Nucl.Phys.}
  \textbf{\bibinfo{volume}{B742}}, \bibinfo{pages}{59}.

\bibitem[{\citenamefont{Agashe} \emph{et~al.}(2006)\citenamefont{Agashe,
  Contino, Da~Rold, and Pomarol}}]{Agashe:2006at}
\bibinfo{author}{\bibnamefont{Agashe}, \bibfnamefont{K.}},
  \bibinfo{author}{\bibfnamefont{R.}~\bibnamefont{Contino}},
  \bibinfo{author}{\bibfnamefont{L.}~\bibnamefont{Da~Rold}}, and
  \bibinfo{author}{\bibfnamefont{A.}~\bibnamefont{Pomarol}},
  \bibinfo{year}{2006}, \bibinfo{journal}{Phys.Lett.}
  \textbf{\bibinfo{volume}{B641}}, \bibinfo{pages}{62}.

\bibitem[{\citenamefont{Agashe} \emph{et~al.}(2005)\citenamefont{Agashe,
  Contino, and Pomarol}}]{Agashe:2004rs}
\bibinfo{author}{\bibnamefont{Agashe}, \bibfnamefont{K.}},
  \bibinfo{author}{\bibfnamefont{R.}~\bibnamefont{Contino}}, and
  \bibinfo{author}{\bibfnamefont{A.}~\bibnamefont{Pomarol}},
  \bibinfo{year}{2005}, \bibinfo{journal}{Nucl.Phys.}
  \textbf{\bibinfo{volume}{B719}}, \bibinfo{pages}{165}.

\bibitem[{\citenamefont{Agashe} \emph{et~al.}(2007)\citenamefont{Agashe,
  Davoudiasl, Gopalakrishna, Han, Huang} \emph{et~al.}}]{Agashe:2007ki}
\bibinfo{author}{\bibnamefont{Agashe}, \bibfnamefont{K.}},
  \bibinfo{author}{\bibfnamefont{H.}~\bibnamefont{Davoudiasl}},
  \bibinfo{author}{\bibfnamefont{S.}~\bibnamefont{Gopalakrishna}},
  \bibinfo{author}{\bibfnamefont{T.}~\bibnamefont{Han}},
  \bibinfo{author}{\bibfnamefont{G.-Y.} \bibnamefont{Huang}}, \emph{et~al.},
  \bibinfo{year}{2007}, \bibinfo{journal}{Phys.Rev.}
  \textbf{\bibinfo{volume}{D76}}, \bibinfo{pages}{115015}.

\bibitem[{\citenamefont{Agashe} \emph{et~al.}(2003)\citenamefont{Agashe,
  Delgado, May, and Sundrum}}]{Agashe:2003zs}
\bibinfo{author}{\bibnamefont{Agashe}, \bibfnamefont{K.}},
  \bibinfo{author}{\bibfnamefont{A.}~\bibnamefont{Delgado}},
  \bibinfo{author}{\bibfnamefont{M.~J.} \bibnamefont{May}}, and
  \bibinfo{author}{\bibfnamefont{R.}~\bibnamefont{Sundrum}},
  \bibinfo{year}{2003}, \bibinfo{journal}{JHEP}
  \textbf{\bibinfo{volume}{0308}}, \bibinfo{pages}{050}.

\bibitem[{\citenamefont{Agashe} \emph{et~al.}(2009)\citenamefont{Agashe,
  Gopalakrishna, Han, Huang, and Soni}}]{Agashe:2008jb}
\bibinfo{author}{\bibnamefont{Agashe}, \bibfnamefont{K.}},
  \bibinfo{author}{\bibfnamefont{S.}~\bibnamefont{Gopalakrishna}},
  \bibinfo{author}{\bibfnamefont{T.}~\bibnamefont{Han}},
  \bibinfo{author}{\bibfnamefont{G.-Y.} \bibnamefont{Huang}}, and
  \bibinfo{author}{\bibfnamefont{A.}~\bibnamefont{Soni}}, \bibinfo{year}{2009},
  \bibinfo{journal}{Phys.Rev.} \textbf{\bibinfo{volume}{D80}},
  \bibinfo{pages}{075007}.

\bibitem[{Agashe \emph{et~al.}(2013)\citenamefont{Agashe}
  \emph{et~al.}}]{Agashe:2013hma}
\bibinfo{author}{\bibnamefont{Agashe}, \bibfnamefont{K.}}, \emph{et~al.}
  (\bibinfo{collaboration}{Top Quark Working Group}), \bibinfo{year}{2013}, in
  \emph{\bibinfo{booktitle}{{Community Summer Study 2013: Snowmass on the
  Mississippi (CSS2013) Minneapolis, MN, USA, July 29-August 6, 2013}}},
  \eprint{1311.2028}.

\bibitem[{\citenamefont{Aguilar-Saavedra}(2009)}]{AguilarSaavedra:2009es}
\bibinfo{author}{\bibnamefont{Aguilar-Saavedra}, \bibfnamefont{J.}},
  \bibinfo{year}{2009}, \bibinfo{journal}{JHEP}
  \textbf{\bibinfo{volume}{0911}}, \bibinfo{pages}{030}.

\bibitem[{\citenamefont{Alonso} \emph{et~al.}(2014)\citenamefont{Alonso,
  Jenkins, Manohar, and Trott}}]{Alonso:2013hga}
\bibinfo{author}{\bibnamefont{Alonso}, \bibfnamefont{R.}},
  \bibinfo{author}{\bibfnamefont{E.~E.} \bibnamefont{Jenkins}},
  \bibinfo{author}{\bibfnamefont{A.~V.} \bibnamefont{Manohar}}, and
  \bibinfo{author}{\bibfnamefont{M.}~\bibnamefont{Trott}},
  \bibinfo{year}{2014}, \bibinfo{journal}{JHEP} \textbf{\bibinfo{volume}{04}},
  \bibinfo{pages}{159}.

\bibitem[{\citenamefont{Altarelli and Barbieri}(1991)}]{Altarelli:1990zd}
\bibinfo{author}{\bibnamefont{Altarelli}, \bibfnamefont{G.}}, and
  \bibinfo{author}{\bibfnamefont{R.}~\bibnamefont{Barbieri}},
  \bibinfo{year}{1991}, \bibinfo{journal}{Phys. Lett.}
  \textbf{\bibinfo{volume}{B253}}, \bibinfo{pages}{161}.

\bibitem[{\citenamefont{Anderson}(1963)}]{PhysRev.130.439}
\bibinfo{author}{\bibnamefont{Anderson}, \bibfnamefont{P.~W.}},
  \bibinfo{year}{1963}, \bibinfo{journal}{Phys. Rev.}
  \textbf{\bibinfo{volume}{130}}, \bibinfo{pages}{439}.

\bibitem[{\citenamefont{Andreev} \emph{et~al.}(2014)\citenamefont{Andreev,
  Osland, and Pankov}}]{Andreev:2014fwa}
\bibinfo{author}{\bibnamefont{Andreev}, \bibfnamefont{V.}},
  \bibinfo{author}{\bibfnamefont{P.}~\bibnamefont{Osland}}, and
  \bibinfo{author}{\bibfnamefont{A.}~\bibnamefont{Pankov}},
  \bibinfo{year}{2014}, \bibinfo{journal}{Phys.Rev.}
  \textbf{\bibinfo{volume}{D90}}(\bibinfo{number}{5}), \bibinfo{pages}{055025}.

\bibitem[{\citenamefont{Anguelova} \emph{et~al.}(2012)\citenamefont{Anguelova,
  Suranyi, and Wijewardhana}}]{Anguelova:2012ka}
\bibinfo{author}{\bibnamefont{Anguelova}, \bibfnamefont{L.}},
  \bibinfo{author}{\bibfnamefont{P.}~\bibnamefont{Suranyi}}, and
  \bibinfo{author}{\bibfnamefont{L.~R.} \bibnamefont{Wijewardhana}},
  \bibinfo{year}{2012}, \bibinfo{journal}{Nucl.Phys.}
  \textbf{\bibinfo{volume}{B862}}, \bibinfo{pages}{671}.

\bibitem[{\citenamefont{Antoniadis}
  \emph{et~al.}(1998)\citenamefont{Antoniadis, Arkani-Hamed, Dimopoulos, and
  Dvali}}]{Antoniadis:1998ig}
\bibinfo{author}{\bibnamefont{Antoniadis}, \bibfnamefont{I.}},
  \bibinfo{author}{\bibfnamefont{N.}~\bibnamefont{Arkani-Hamed}},
  \bibinfo{author}{\bibfnamefont{S.}~\bibnamefont{Dimopoulos}}, and
  \bibinfo{author}{\bibfnamefont{G.}~\bibnamefont{Dvali}},
  \bibinfo{year}{1998}, \bibinfo{journal}{Phys.Lett.}
  \textbf{\bibinfo{volume}{B436}}, \bibinfo{pages}{257}.

\bibitem[{\citenamefont{Appelquist and Bai}(2010)}]{Appelquist:2010gy}
\bibinfo{author}{\bibnamefont{Appelquist}, \bibfnamefont{T.}}, and
  \bibinfo{author}{\bibfnamefont{Y.}~\bibnamefont{Bai}}, \bibinfo{year}{2010},
  \bibinfo{journal}{Phys.Rev.} \textbf{\bibinfo{volume}{D82}},
  \bibinfo{pages}{071701}.

\bibitem[{\citenamefont{Appelquist}
  \emph{et~al.}(2001)\citenamefont{Appelquist, Cheng, and
  Dobrescu}}]{Appelquist:2000nn}
\bibinfo{author}{\bibnamefont{Appelquist}, \bibfnamefont{T.}},
  \bibinfo{author}{\bibfnamefont{H.-C.} \bibnamefont{Cheng}}, and
  \bibinfo{author}{\bibfnamefont{B.~A.} \bibnamefont{Dobrescu}},
  \bibinfo{year}{2001}, \bibinfo{journal}{Phys.Rev.}
  \textbf{\bibinfo{volume}{D64}}, \bibinfo{pages}{035002}.

\bibitem[{\citenamefont{Appelquist and Wijewardhana}(1987)}]{Appelquist:1987tr}
\bibinfo{author}{\bibnamefont{Appelquist}, \bibfnamefont{T.}}, and
  \bibinfo{author}{\bibfnamefont{L.~C.~R.} \bibnamefont{Wijewardhana}},
  \bibinfo{year}{1987}, \bibinfo{journal}{Phys. Rev.}
  \textbf{\bibinfo{volume}{D35}}, \bibinfo{pages}{774}.

\bibitem[{\citenamefont{Appelquist}
  \emph{et~al.}(1986)\citenamefont{Appelquist, Karabali, and
  Wijewardhana}}]{Appelquist:1986an}
\bibinfo{author}{\bibnamefont{Appelquist}, \bibfnamefont{T.~W.}},
  \bibinfo{author}{\bibfnamefont{D.}~\bibnamefont{Karabali}}, and
  \bibinfo{author}{\bibfnamefont{L.~C.~R.} \bibnamefont{Wijewardhana}},
  \bibinfo{year}{1986}, \bibinfo{journal}{Phys. Rev. Lett.}
  \textbf{\bibinfo{volume}{57}}, \bibinfo{pages}{957}.

\bibitem[{\citenamefont{Archer}(2012)}]{Archer:2012qa}
\bibinfo{author}{\bibnamefont{Archer}, \bibfnamefont{P.~R.}},
  \bibinfo{year}{2012}, \bibinfo{journal}{JHEP}
  \textbf{\bibinfo{volume}{1209}}, \bibinfo{pages}{095}.

\bibitem[{\citenamefont{Arkani-Hamed}
  \emph{et~al.}(2002{\natexlab{a}})\citenamefont{Arkani-Hamed, Cohen, Katz, and
  Nelson}}]{ArkaniHamed:2002qy}
\bibinfo{author}{\bibnamefont{Arkani-Hamed}, \bibfnamefont{N.}},
  \bibinfo{author}{\bibfnamefont{A.}~\bibnamefont{Cohen}},
  \bibinfo{author}{\bibfnamefont{E.}~\bibnamefont{Katz}}, and
  \bibinfo{author}{\bibfnamefont{A.}~\bibnamefont{Nelson}},
  \bibinfo{year}{2002}{\natexlab{a}}, \bibinfo{journal}{JHEP}
  \textbf{\bibinfo{volume}{0207}}, \bibinfo{pages}{034}.

\bibitem[{\citenamefont{Arkani-Hamed}
  \emph{et~al.}(2002{\natexlab{b}})\citenamefont{Arkani-Hamed, Cohen, Katz,
  Nelson, Gregoire} \emph{et~al.}}]{ArkaniHamed:2002qx}
\bibinfo{author}{\bibnamefont{Arkani-Hamed}, \bibfnamefont{N.}},
  \bibinfo{author}{\bibfnamefont{A.}~\bibnamefont{Cohen}},
  \bibinfo{author}{\bibfnamefont{E.}~\bibnamefont{Katz}},
  \bibinfo{author}{\bibfnamefont{A.}~\bibnamefont{Nelson}},
  \bibinfo{author}{\bibfnamefont{T.}~\bibnamefont{Gregoire}}, \emph{et~al.},
  \bibinfo{year}{2002}{\natexlab{b}}, \bibinfo{journal}{JHEP}
  \textbf{\bibinfo{volume}{0208}}, \bibinfo{pages}{021}.

\bibitem[{\citenamefont{Arkani-Hamed}
  \emph{et~al.}(2001{\natexlab{a}})\citenamefont{Arkani-Hamed, Cohen, and
  Georgi}}]{ArkaniHamed:2001ca}
\bibinfo{author}{\bibnamefont{Arkani-Hamed}, \bibfnamefont{N.}},
  \bibinfo{author}{\bibfnamefont{A.~G.} \bibnamefont{Cohen}}, and
  \bibinfo{author}{\bibfnamefont{H.}~\bibnamefont{Georgi}},
  \bibinfo{year}{2001}{\natexlab{a}}, \bibinfo{journal}{Phys. Rev. Lett.}
  \textbf{\bibinfo{volume}{86}}, \bibinfo{pages}{4757}.

\bibitem[{\citenamefont{Arkani-Hamed}
  \emph{et~al.}(2001{\natexlab{b}})\citenamefont{Arkani-Hamed, Cohen, and
  Georgi}}]{ArkaniHamed:2001nc}
\bibinfo{author}{\bibnamefont{Arkani-Hamed}, \bibfnamefont{N.}},
  \bibinfo{author}{\bibfnamefont{A.~G.} \bibnamefont{Cohen}}, and
  \bibinfo{author}{\bibfnamefont{H.}~\bibnamefont{Georgi}},
  \bibinfo{year}{2001}{\natexlab{b}}, \bibinfo{journal}{Phys.Lett.}
  \textbf{\bibinfo{volume}{B513}}, \bibinfo{pages}{232}.

\bibitem[{\citenamefont{Arkani-Hamed}
  \emph{et~al.}(2002{\natexlab{c}})\citenamefont{Arkani-Hamed, Cohen, Gregoire,
  and Wacker}}]{ArkaniHamed:2002pa}
\bibinfo{author}{\bibnamefont{Arkani-Hamed}, \bibfnamefont{N.}},
  \bibinfo{author}{\bibfnamefont{A.~G.} \bibnamefont{Cohen}},
  \bibinfo{author}{\bibfnamefont{T.}~\bibnamefont{Gregoire}}, and
  \bibinfo{author}{\bibfnamefont{J.~G.} \bibnamefont{Wacker}},
  \bibinfo{year}{2002}{\natexlab{c}}, \bibinfo{journal}{JHEP}
  \textbf{\bibinfo{volume}{0208}}, \bibinfo{pages}{020}.

\bibitem[{\citenamefont{Arkani-Hamed}
  \emph{et~al.}(1998)\citenamefont{Arkani-Hamed, Dimopoulos, and
  Dvali}}]{ArkaniHamed:1998rs}
\bibinfo{author}{\bibnamefont{Arkani-Hamed}, \bibfnamefont{N.}},
  \bibinfo{author}{\bibfnamefont{S.}~\bibnamefont{Dimopoulos}}, and
  \bibinfo{author}{\bibfnamefont{G.}~\bibnamefont{Dvali}},
  \bibinfo{year}{1998}, \bibinfo{journal}{Phys.Lett.}
  \textbf{\bibinfo{volume}{B429}}, \bibinfo{pages}{263}.

\bibitem[{\citenamefont{Arkani-Hamed}
  \emph{et~al.}(2001{\natexlab{c}})\citenamefont{Arkani-Hamed, Porrati, and
  Randall}}]{ArkaniHamed:2000ds}
\bibinfo{author}{\bibnamefont{Arkani-Hamed}, \bibfnamefont{N.}},
  \bibinfo{author}{\bibfnamefont{M.}~\bibnamefont{Porrati}}, and
  \bibinfo{author}{\bibfnamefont{L.}~\bibnamefont{Randall}},
  \bibinfo{year}{2001}{\natexlab{c}}, \bibinfo{journal}{JHEP}
  \textbf{\bibinfo{volume}{0108}}, \bibinfo{pages}{017}.

\bibitem[{\citenamefont{Arkani-Hamed and Rattazzi}(1999)}]{ArkaniHamed:1998wc}
\bibinfo{author}{\bibnamefont{Arkani-Hamed}, \bibfnamefont{N.}}, and
  \bibinfo{author}{\bibfnamefont{R.}~\bibnamefont{Rattazzi}},
  \bibinfo{year}{1999}, \bibinfo{journal}{Phys.Lett.}
  \textbf{\bibinfo{volume}{B454}}, \bibinfo{pages}{290}.

\bibitem[{\citenamefont{{ATLAS
  collaboration}}(2013)}]{TheATLAScollaboration:2013jha}
\bibinfo{author}{\bibnamefont{{ATLAS collaboration}}}, \bibinfo{year}{2013},
  \bibinfo{note}{{ATLAS-CONF-2013-051}}.

\bibitem[{\citenamefont{Azatov}
  \emph{et~al.}(2012{\natexlab{a}})\citenamefont{Azatov, Contino, Del~Re,
  Galloway, Grassi} \emph{et~al.}}]{Azatov:2012rd}
\bibinfo{author}{\bibnamefont{Azatov}, \bibfnamefont{A.}},
  \bibinfo{author}{\bibfnamefont{R.}~\bibnamefont{Contino}},
  \bibinfo{author}{\bibfnamefont{D.}~\bibnamefont{Del~Re}},
  \bibinfo{author}{\bibfnamefont{J.}~\bibnamefont{Galloway}},
  \bibinfo{author}{\bibfnamefont{M.}~\bibnamefont{Grassi}}, \emph{et~al.},
  \bibinfo{year}{2012}{\natexlab{a}}, \bibinfo{journal}{JHEP}
  \textbf{\bibinfo{volume}{1206}}, \bibinfo{pages}{134}.

\bibitem[{\citenamefont{Azatov}
  \emph{et~al.}(2012{\natexlab{b}})\citenamefont{Azatov, Contino, and
  Galloway}}]{Azatov:2012bz}
\bibinfo{author}{\bibnamefont{Azatov}, \bibfnamefont{A.}},
  \bibinfo{author}{\bibfnamefont{R.}~\bibnamefont{Contino}}, and
  \bibinfo{author}{\bibfnamefont{J.}~\bibnamefont{Galloway}},
  \bibinfo{year}{2012}{\natexlab{b}}, \bibinfo{journal}{JHEP}
  \textbf{\bibinfo{volume}{1204}}, \bibinfo{pages}{127}.

\bibitem[{\citenamefont{Azatov and Galloway}(2013)}]{Azatov:2012qz}
\bibinfo{author}{\bibnamefont{Azatov}, \bibfnamefont{A.}}, and
  \bibinfo{author}{\bibfnamefont{J.}~\bibnamefont{Galloway}},
  \bibinfo{year}{2013}, \bibinfo{journal}{Int.J.Mod.Phys.}
  \textbf{\bibinfo{volume}{A28}}, \bibinfo{pages}{1330004}.

\bibitem[{\citenamefont{Azatov}
  \emph{et~al.}(2012{\natexlab{c}})\citenamefont{Azatov, Galloway, and
  Luty}}]{Azatov:2011ht}
\bibinfo{author}{\bibnamefont{Azatov}, \bibfnamefont{A.}},
  \bibinfo{author}{\bibfnamefont{J.}~\bibnamefont{Galloway}}, and
  \bibinfo{author}{\bibfnamefont{M.~A.} \bibnamefont{Luty}},
  \bibinfo{year}{2012}{\natexlab{c}}, \bibinfo{journal}{Phys.Rev.Lett.}
  \textbf{\bibinfo{volume}{108}}, \bibinfo{pages}{041802}.

\bibitem[{\citenamefont{Azuelos} \emph{et~al.}(2005)\citenamefont{Azuelos,
  Benslama, Costanzo, Couture, Garcia} \emph{et~al.}}]{Azuelos:2004dm}
\bibinfo{author}{\bibnamefont{Azuelos}, \bibfnamefont{G.}},
  \bibinfo{author}{\bibfnamefont{K.}~\bibnamefont{Benslama}},
  \bibinfo{author}{\bibfnamefont{D.}~\bibnamefont{Costanzo}},
  \bibinfo{author}{\bibfnamefont{G.}~\bibnamefont{Couture}},
  \bibinfo{author}{\bibfnamefont{J.}~\bibnamefont{Garcia}}, \emph{et~al.},
  \bibinfo{year}{2005}, \bibinfo{journal}{Eur.Phys.J.}
  \textbf{\bibinfo{volume}{C39S2}}, \bibinfo{pages}{13}.

\bibitem[{\citenamefont{Bando} \emph{et~al.}(1986)\citenamefont{Bando, iti
  Matumoto, and Yamawaki}}]{Bando:1986bg}
\bibinfo{author}{\bibnamefont{Bando}, \bibfnamefont{M.}},
  \bibinfo{author}{\bibfnamefont{K.}~\bibnamefont{iti Matumoto}}, and
  \bibinfo{author}{\bibfnamefont{K.}~\bibnamefont{Yamawaki}},
  \bibinfo{year}{1986}, \bibinfo{journal}{Phys. Lett.}
  \textbf{\bibinfo{volume}{B178}}, \bibinfo{pages}{308}.

\bibitem[{\citenamefont{Barbieri} \emph{et~al.}(2007)\citenamefont{Barbieri,
  Bellazzini, Rychkov, and Varagnolo}}]{Barbieri:2007bh}
\bibinfo{author}{\bibnamefont{Barbieri}, \bibfnamefont{R.}},
  \bibinfo{author}{\bibfnamefont{B.}~\bibnamefont{Bellazzini}},
  \bibinfo{author}{\bibfnamefont{V.~S.} \bibnamefont{Rychkov}}, and
  \bibinfo{author}{\bibfnamefont{A.}~\bibnamefont{Varagnolo}},
  \bibinfo{year}{2007}, \bibinfo{journal}{Phys.Rev.}
  \textbf{\bibinfo{volume}{D76}}, \bibinfo{pages}{115008}.

\bibitem[{\citenamefont{Barbieri} \emph{et~al.}(2015)\citenamefont{Barbieri,
  Greco, Rattazzi, and Wulzer}}]{Barbieri:2015lqa}
\bibinfo{author}{\bibnamefont{Barbieri}, \bibfnamefont{R.}},
  \bibinfo{author}{\bibfnamefont{D.}~\bibnamefont{Greco}},
  \bibinfo{author}{\bibfnamefont{R.}~\bibnamefont{Rattazzi}}, and
  \bibinfo{author}{\bibfnamefont{A.}~\bibnamefont{Wulzer}},
  \bibinfo{year}{2015}, \bibinfo{journal}{JHEP} \textbf{\bibinfo{volume}{08}},
  \bibinfo{pages}{161}.

\bibitem[{\citenamefont{Bardeen} \emph{et~al.}(1957)\citenamefont{Bardeen,
  Cooper, and Schrieffer}}]{PhysRev.108.1175}
\bibinfo{author}{\bibnamefont{Bardeen}, \bibfnamefont{J.}},
  \bibinfo{author}{\bibfnamefont{L.~N.} \bibnamefont{Cooper}}, and
  \bibinfo{author}{\bibfnamefont{J.~R.} \bibnamefont{Schrieffer}},
  \bibinfo{year}{1957}, \bibinfo{journal}{Phys. Rev.}
  \textbf{\bibinfo{volume}{108}}, \bibinfo{pages}{1175}.

\bibitem[{\citenamefont{Barducci} \emph{et~al.}(2014)\citenamefont{Barducci,
  Belyaev, Buchkremer, Cacciapaglia, Deandrea}
  \emph{et~al.}}]{Barducci:2014ila}
\bibinfo{author}{\bibnamefont{Barducci}, \bibfnamefont{D.}},
  \bibinfo{author}{\bibfnamefont{A.}~\bibnamefont{Belyaev}},
  \bibinfo{author}{\bibfnamefont{M.}~\bibnamefont{Buchkremer}},
  \bibinfo{author}{\bibfnamefont{G.}~\bibnamefont{Cacciapaglia}},
  \bibinfo{author}{\bibfnamefont{A.}~\bibnamefont{Deandrea}}, \emph{et~al.},
  \bibinfo{year}{2014}, \bibinfo{journal}{JHEP}
  \textbf{\bibinfo{volume}{1412}}, \bibinfo{pages}{080}.

\bibitem[{\citenamefont{Barger}
  \emph{et~al.}(2012{\natexlab{a}})\citenamefont{Barger, Ishida, and
  Keung}}]{Barger:2011hu}
\bibinfo{author}{\bibnamefont{Barger}, \bibfnamefont{V.}},
  \bibinfo{author}{\bibfnamefont{M.}~\bibnamefont{Ishida}}, and
  \bibinfo{author}{\bibfnamefont{W.-Y.} \bibnamefont{Keung}},
  \bibinfo{year}{2012}{\natexlab{a}}, \bibinfo{journal}{Phys.Rev.Lett.}
  \textbf{\bibinfo{volume}{108}}, \bibinfo{pages}{101802}.

\bibitem[{\citenamefont{Barger}
  \emph{et~al.}(2012{\natexlab{b}})\citenamefont{Barger, Ishida, and
  Keung}}]{Barger:2011nu}
\bibinfo{author}{\bibnamefont{Barger}, \bibfnamefont{V.}},
  \bibinfo{author}{\bibfnamefont{M.}~\bibnamefont{Ishida}}, and
  \bibinfo{author}{\bibfnamefont{W.-Y.} \bibnamefont{Keung}},
  \bibinfo{year}{2012}{\natexlab{b}}, \bibinfo{journal}{Phys.Rev.}
  \textbf{\bibinfo{volume}{D85}}, \bibinfo{pages}{015024}.

\bibitem[{\citenamefont{Batra and Chacko}(2009)}]{Batra:2008jy}
\bibinfo{author}{\bibnamefont{Batra}, \bibfnamefont{P.}}, and
  \bibinfo{author}{\bibfnamefont{Z.}~\bibnamefont{Chacko}},
  \bibinfo{year}{2009}, \bibinfo{journal}{Phys. Rev.}
  \textbf{\bibinfo{volume}{D79}}, \bibinfo{pages}{095012}.

\bibitem[{\citenamefont{Batra and Kaplan}(2005)}]{Batra:2004ah}
\bibinfo{author}{\bibnamefont{Batra}, \bibfnamefont{P.}}, and
  \bibinfo{author}{\bibfnamefont{D.~E.} \bibnamefont{Kaplan}},
  \bibinfo{year}{2005}, \bibinfo{journal}{JHEP}
  \textbf{\bibinfo{volume}{0503}}, \bibinfo{pages}{028}.

\bibitem[{\citenamefont{Bazzocchi} \emph{et~al.}(2005)\citenamefont{Bazzocchi,
  Fabbrichesi, and Piai}}]{Bazzocchi:2005gs}
\bibinfo{author}{\bibnamefont{Bazzocchi}, \bibfnamefont{F.}},
  \bibinfo{author}{\bibfnamefont{M.}~\bibnamefont{Fabbrichesi}}, and
  \bibinfo{author}{\bibfnamefont{M.}~\bibnamefont{Piai}}, \bibinfo{year}{2005},
  \bibinfo{journal}{Phys.Rev.} \textbf{\bibinfo{volume}{D72}},
  \bibinfo{pages}{095019}.

\bibitem[{\citenamefont{Bellazzini}
  \emph{et~al.}(2015)\citenamefont{Bellazzini, Cs\'aki, Hubisz, Lee, Serra, and
  Terning}}]{Bellazzini:2015cgj}
\bibinfo{author}{\bibnamefont{Bellazzini}, \bibfnamefont{B.}},
  \bibinfo{author}{\bibfnamefont{C.}~\bibnamefont{Cs\'aki}},
  \bibinfo{author}{\bibfnamefont{J.}~\bibnamefont{Hubisz}},
  \bibinfo{author}{\bibfnamefont{S.~J.} \bibnamefont{Lee}},
  \bibinfo{author}{\bibfnamefont{J.}~\bibnamefont{Serra}}, and
  \bibinfo{author}{\bibfnamefont{J.}~\bibnamefont{Terning}},
  \bibinfo{year}{2015}, \eprint{1511.08218}.

\bibitem[{\citenamefont{Bellazzini}
  \emph{et~al.}(2012)\citenamefont{Bellazzini, Cs\'aki, Hubisz, Serra, and
  Terning}}]{Bellazzini:2012tv}
\bibinfo{author}{\bibnamefont{Bellazzini}, \bibfnamefont{B.}},
  \bibinfo{author}{\bibfnamefont{C.}~\bibnamefont{Cs\'aki}},
  \bibinfo{author}{\bibfnamefont{J.}~\bibnamefont{Hubisz}},
  \bibinfo{author}{\bibfnamefont{J.}~\bibnamefont{Serra}}, and
  \bibinfo{author}{\bibfnamefont{J.}~\bibnamefont{Terning}},
  \bibinfo{year}{2012}, \bibinfo{journal}{JHEP}
  \textbf{\bibinfo{volume}{1211}}, \bibinfo{pages}{003}.

\bibitem[{\citenamefont{Bellazzini}
  \emph{et~al.}(2013)\citenamefont{Bellazzini, Cs\'aki, Hubisz, Serra, and
  Terning}}]{Bellazzini:2012vz}
\bibinfo{author}{\bibnamefont{Bellazzini}, \bibfnamefont{B.}},
  \bibinfo{author}{\bibfnamefont{C.}~\bibnamefont{Cs\'aki}},
  \bibinfo{author}{\bibfnamefont{J.}~\bibnamefont{Hubisz}},
  \bibinfo{author}{\bibfnamefont{J.}~\bibnamefont{Serra}}, and
  \bibinfo{author}{\bibfnamefont{J.}~\bibnamefont{Terning}},
  \bibinfo{year}{2013}, \bibinfo{journal}{Eur.Phys.J.}
  \textbf{\bibinfo{volume}{C73}}, \bibinfo{pages}{2333}.

\bibitem[{\citenamefont{Bellazzini}
  \emph{et~al.}(2014)\citenamefont{Bellazzini, Cs\'aki, and
  Serra}}]{Bellazzini:2014yua}
\bibinfo{author}{\bibnamefont{Bellazzini}, \bibfnamefont{B.}},
  \bibinfo{author}{\bibfnamefont{C.}~\bibnamefont{Cs\'aki}}, and
  \bibinfo{author}{\bibfnamefont{J.}~\bibnamefont{Serra}},
  \bibinfo{year}{2014}, \bibinfo{journal}{Eur. Phys. J.}
  \textbf{\bibinfo{volume}{C74}}(\bibinfo{number}{5}), \bibinfo{pages}{2766}.

\bibitem[{\citenamefont{Belyaev} \emph{et~al.}(2006)\citenamefont{Belyaev,
  Chen, Tobe, and Yuan}}]{Belyaev:2006jh}
\bibinfo{author}{\bibnamefont{Belyaev}, \bibfnamefont{A.}},
  \bibinfo{author}{\bibfnamefont{C.-R.} \bibnamefont{Chen}},
  \bibinfo{author}{\bibfnamefont{K.}~\bibnamefont{Tobe}}, and
  \bibinfo{author}{\bibfnamefont{C.-P.} \bibnamefont{Yuan}},
  \bibinfo{year}{2006}, \bibinfo{journal}{Phys.Rev.}
  \textbf{\bibinfo{volume}{D74}}, \bibinfo{pages}{115020}.

\bibitem[{\citenamefont{Beneke} \emph{et~al.}(2011)\citenamefont{Beneke,
  Knechtges, and Muck}}]{Beneke:2011jk}
\bibinfo{author}{\bibnamefont{Beneke}, \bibfnamefont{M.}},
  \bibinfo{author}{\bibfnamefont{P.}~\bibnamefont{Knechtges}}, and
  \bibinfo{author}{\bibfnamefont{A.}~\bibnamefont{Muck}}, \bibinfo{year}{2011},
  \bibinfo{journal}{JHEP} \textbf{\bibinfo{volume}{1110}},
  \bibinfo{pages}{076}.

\bibitem[{\citenamefont{Berger} \emph{et~al.}(2012)\citenamefont{Berger,
  Hubisz, and Perelstein}}]{Berger:2012ec}
\bibinfo{author}{\bibnamefont{Berger}, \bibfnamefont{J.}},
  \bibinfo{author}{\bibfnamefont{J.}~\bibnamefont{Hubisz}}, and
  \bibinfo{author}{\bibfnamefont{M.}~\bibnamefont{Perelstein}},
  \bibinfo{year}{2012}, \bibinfo{journal}{JHEP}
  \textbf{\bibinfo{volume}{1207}}, \bibinfo{pages}{016}.

\bibitem[{\citenamefont{Bhattacharya}
  \emph{et~al.}(2015)\citenamefont{Bhattacharya, Frank, Huitu, Maitra,
  Mukhopadhyaya} \emph{et~al.}}]{Bhattacharya:2014wha}
\bibinfo{author}{\bibnamefont{Bhattacharya}, \bibfnamefont{S.}},
  \bibinfo{author}{\bibfnamefont{M.}~\bibnamefont{Frank}},
  \bibinfo{author}{\bibfnamefont{K.}~\bibnamefont{Huitu}},
  \bibinfo{author}{\bibfnamefont{U.}~\bibnamefont{Maitra}},
  \bibinfo{author}{\bibfnamefont{B.}~\bibnamefont{Mukhopadhyaya}},
  \emph{et~al.}, \bibinfo{year}{2015}, \bibinfo{journal}{Phys.Rev.}
  \textbf{\bibinfo{volume}{D91}}, \bibinfo{pages}{016008}.

\bibitem[{\citenamefont{Bhattacharyya}(2011)}]{Bhattacharyya:2009gw}
\bibinfo{author}{\bibnamefont{Bhattacharyya}, \bibfnamefont{G.}},
  \bibinfo{year}{2011}, \bibinfo{journal}{Rept.Prog.Phys.}
  \textbf{\bibinfo{volume}{74}}, \bibinfo{pages}{026201}.

\bibitem[{\citenamefont{van~der Bij and Dilcher}(2007)}]{unother2}
\bibinfo{author}{\bibnamefont{van~der Bij}, \bibfnamefont{J.}}, and
  \bibinfo{author}{\bibfnamefont{S.}~\bibnamefont{Dilcher}},
  \bibinfo{year}{2007}, \bibinfo{journal}{Phys.Lett.}
  \textbf{\bibinfo{volume}{B655}}, \bibinfo{pages}{183}.

\bibitem[{\citenamefont{Birkedal} \emph{et~al.}(2004)\citenamefont{Birkedal,
  Chacko, and Gaillard}}]{Birkedal:2004xi}
\bibinfo{author}{\bibnamefont{Birkedal}, \bibfnamefont{A.}},
  \bibinfo{author}{\bibfnamefont{Z.}~\bibnamefont{Chacko}}, and
  \bibinfo{author}{\bibfnamefont{M.~K.} \bibnamefont{Gaillard}},
  \bibinfo{year}{2004}, \bibinfo{journal}{JHEP}
  \textbf{\bibinfo{volume}{0410}}, \bibinfo{pages}{036}.

\bibitem[{\citenamefont{Birkedal} \emph{et~al.}(2005)\citenamefont{Birkedal,
  Matchev, and Perelstein}}]{Birkedal:2005yg}
\bibinfo{author}{\bibnamefont{Birkedal}, \bibfnamefont{A.}},
  \bibinfo{author}{\bibfnamefont{K.~T.} \bibnamefont{Matchev}}, and
  \bibinfo{author}{\bibfnamefont{M.}~\bibnamefont{Perelstein}},
  \bibinfo{year}{2005}, \bibinfo{journal}{eConf}
  \textbf{\bibinfo{volume}{C050318}}, \bibinfo{pages}{0314}.

\bibitem[{\citenamefont{Birkedal-Hansen and
  Wacker}(2004)}]{BirkedalHansen:2003mpa}
\bibinfo{author}{\bibnamefont{Birkedal-Hansen}, \bibfnamefont{A.}}, and
  \bibinfo{author}{\bibfnamefont{J.~G.} \bibnamefont{Wacker}},
  \bibinfo{year}{2004}, \bibinfo{journal}{Phys.Rev.}
  \textbf{\bibinfo{volume}{D69}}, \bibinfo{pages}{065022}.

\bibitem[{\citenamefont{Buchkremer}
  \emph{et~al.}(2013)\citenamefont{Buchkremer, Cacciapaglia, Deandrea, and
  Panizzi}}]{Buchkremer:2013bha}
\bibinfo{author}{\bibnamefont{Buchkremer}, \bibfnamefont{M.}},
  \bibinfo{author}{\bibfnamefont{G.}~\bibnamefont{Cacciapaglia}},
  \bibinfo{author}{\bibfnamefont{A.}~\bibnamefont{Deandrea}}, and
  \bibinfo{author}{\bibfnamefont{L.}~\bibnamefont{Panizzi}},
  \bibinfo{year}{2013}, \bibinfo{journal}{Nucl. Phys.}
  \textbf{\bibinfo{volume}{B876}}, \bibinfo{pages}{376}.

\bibitem[{\citenamefont{Buras} \emph{et~al.}(2005)\citenamefont{Buras,
  Poschenrieder, and Uhlig}}]{Buras:2004kq}
\bibinfo{author}{\bibnamefont{Buras}, \bibfnamefont{A.~J.}},
  \bibinfo{author}{\bibfnamefont{A.}~\bibnamefont{Poschenrieder}}, and
  \bibinfo{author}{\bibfnamefont{S.}~\bibnamefont{Uhlig}},
  \bibinfo{year}{2005}, \bibinfo{journal}{Nucl.Phys.}
  \textbf{\bibinfo{volume}{B716}}, \bibinfo{pages}{173}.

\bibitem[{\citenamefont{Burdman} \emph{et~al.}(2007)\citenamefont{Burdman,
  Chacko, Goh, and Harnik}}]{Burdman:2006tz}
\bibinfo{author}{\bibnamefont{Burdman}, \bibfnamefont{G.}},
  \bibinfo{author}{\bibfnamefont{Z.}~\bibnamefont{Chacko}},
  \bibinfo{author}{\bibfnamefont{H.-S.} \bibnamefont{Goh}}, and
  \bibinfo{author}{\bibfnamefont{R.}~\bibnamefont{Harnik}},
  \bibinfo{year}{2007}, \bibinfo{journal}{JHEP} \textbf{\bibinfo{volume}{02}},
  \bibinfo{pages}{009}.

\bibitem[{\citenamefont{Burdman} \emph{et~al.}(2015)\citenamefont{Burdman,
  Chacko, Harnik, de~Lima, and Verhaaren}}]{Burdman:2014zta}
\bibinfo{author}{\bibnamefont{Burdman}, \bibfnamefont{G.}},
  \bibinfo{author}{\bibfnamefont{Z.}~\bibnamefont{Chacko}},
  \bibinfo{author}{\bibfnamefont{R.}~\bibnamefont{Harnik}},
  \bibinfo{author}{\bibfnamefont{L.}~\bibnamefont{de~Lima}}, and
  \bibinfo{author}{\bibfnamefont{C.~B.} \bibnamefont{Verhaaren}},
  \bibinfo{year}{2015}, \bibinfo{journal}{Phys. Rev.}
  \textbf{\bibinfo{volume}{D91}}(\bibinfo{number}{5}), \bibinfo{pages}{055007}.

\bibitem[{\citenamefont{Burdman} \emph{et~al.}(2003)\citenamefont{Burdman,
  Perelstein, and Pierce}}]{Burdman:2002ns}
\bibinfo{author}{\bibnamefont{Burdman}, \bibfnamefont{G.}},
  \bibinfo{author}{\bibfnamefont{M.}~\bibnamefont{Perelstein}}, and
  \bibinfo{author}{\bibfnamefont{A.}~\bibnamefont{Pierce}},
  \bibinfo{year}{2003}, \bibinfo{journal}{Phys.Rev.Lett.}
  \textbf{\bibinfo{volume}{90}}, \bibinfo{pages}{241802}.

\bibitem[{\citenamefont{Cabrer} \emph{et~al.}(2011)\citenamefont{Cabrer, von
  Gersdorff, and Quiros}}]{Cabrer:2011vu}
\bibinfo{author}{\bibnamefont{Cabrer}, \bibfnamefont{J.~A.}},
  \bibinfo{author}{\bibfnamefont{G.}~\bibnamefont{von Gersdorff}}, and
  \bibinfo{author}{\bibfnamefont{M.}~\bibnamefont{Quiros}},
  \bibinfo{year}{2011}, \bibinfo{journal}{Phys.Rev.}
  \textbf{\bibinfo{volume}{D84}}, \bibinfo{pages}{035024}.

\bibitem[{\citenamefont{Cacciapaglia}
  \emph{et~al.}(2007)\citenamefont{Cacciapaglia, Cs\'aki, Marandella, and
  Terning}}]{gaugephobic}
\bibinfo{author}{\bibnamefont{Cacciapaglia}, \bibfnamefont{G.}},
  \bibinfo{author}{\bibfnamefont{C.}~\bibnamefont{Cs\'aki}},
  \bibinfo{author}{\bibfnamefont{G.}~\bibnamefont{Marandella}}, and
  \bibinfo{author}{\bibfnamefont{J.}~\bibnamefont{Terning}},
  \bibinfo{year}{2007}, \bibinfo{journal}{JHEP}
  \textbf{\bibinfo{volume}{0702}}, \bibinfo{pages}{036}.

\bibitem[{\citenamefont{Cacciapaglia}
  \emph{et~al.}(2013{\natexlab{a}})\citenamefont{Cacciapaglia, Deandrea, Ellis,
  Marrouche, and Panizzi}}]{Cacciapaglia:2013wha}
\bibinfo{author}{\bibnamefont{Cacciapaglia}, \bibfnamefont{G.}},
  \bibinfo{author}{\bibfnamefont{A.}~\bibnamefont{Deandrea}},
  \bibinfo{author}{\bibfnamefont{J.}~\bibnamefont{Ellis}},
  \bibinfo{author}{\bibfnamefont{J.}~\bibnamefont{Marrouche}}, and
  \bibinfo{author}{\bibfnamefont{L.}~\bibnamefont{Panizzi}},
  \bibinfo{year}{2013}{\natexlab{a}}, \bibinfo{journal}{Phys.Rev.}
  \textbf{\bibinfo{volume}{D87}}(\bibinfo{number}{7}), \bibinfo{pages}{075006}.

\bibitem[{\citenamefont{Cacciapaglia}
  \emph{et~al.}(2013{\natexlab{b}})\citenamefont{Cacciapaglia, Deandrea,
  La~Rochelle, and Flament}}]{Cacciapaglia:2012wb}
\bibinfo{author}{\bibnamefont{Cacciapaglia}, \bibfnamefont{G.}},
  \bibinfo{author}{\bibfnamefont{A.}~\bibnamefont{Deandrea}},
  \bibinfo{author}{\bibfnamefont{G.~D.} \bibnamefont{La~Rochelle}}, and
  \bibinfo{author}{\bibfnamefont{J.-B.} \bibnamefont{Flament}},
  \bibinfo{year}{2013}{\natexlab{b}}, \bibinfo{journal}{JHEP}
  \textbf{\bibinfo{volume}{1303}}, \bibinfo{pages}{029}.

\bibitem[{\citenamefont{Cacciapaglia}
  \emph{et~al.}(2008)\citenamefont{Cacciapaglia, Marandella, and
  Terning}}]{coloredunparticles}
\bibinfo{author}{\bibnamefont{Cacciapaglia}, \bibfnamefont{G.}},
  \bibinfo{author}{\bibfnamefont{G.}~\bibnamefont{Marandella}}, and
  \bibinfo{author}{\bibfnamefont{J.}~\bibnamefont{Terning}},
  \bibinfo{year}{2008}, \bibinfo{journal}{JHEP}
  \textbf{\bibinfo{volume}{0801}}, \bibinfo{pages}{070}.

\bibitem[{\citenamefont{Cacciapaglia}
  \emph{et~al.}(2009)\citenamefont{Cacciapaglia, Marandella, and
  Terning}}]{Cacciapaglia:2008ns}
\bibinfo{author}{\bibnamefont{Cacciapaglia}, \bibfnamefont{G.}},
  \bibinfo{author}{\bibfnamefont{G.}~\bibnamefont{Marandella}}, and
  \bibinfo{author}{\bibfnamefont{J.}~\bibnamefont{Terning}},
  \bibinfo{year}{2009}, \bibinfo{journal}{JHEP}
  \textbf{\bibinfo{volume}{0902}}, \bibinfo{pages}{049}.

\bibitem[{\citenamefont{Callan} \emph{et~al.}(1969)\citenamefont{Callan,
  Coleman, Wess, and Zumino}}]{Callan:1969sn}
\bibinfo{author}{\bibnamefont{Callan}, \bibfnamefont{J., Curtis~G.}},
  \bibinfo{author}{\bibfnamefont{S.~R.} \bibnamefont{Coleman}},
  \bibinfo{author}{\bibfnamefont{J.}~\bibnamefont{Wess}}, and
  \bibinfo{author}{\bibfnamefont{B.}~\bibnamefont{Zumino}},
  \bibinfo{year}{1969}, \bibinfo{journal}{Phys.Rev.}
  \textbf{\bibinfo{volume}{177}}, \bibinfo{pages}{2247}.

\bibitem[{\citenamefont{Campbell} \emph{et~al.}(2012)\citenamefont{Campbell,
  Ellis, and Olive}}]{Campbell:2011iw}
\bibinfo{author}{\bibnamefont{Campbell}, \bibfnamefont{B.~A.}},
  \bibinfo{author}{\bibfnamefont{J.}~\bibnamefont{Ellis}}, and
  \bibinfo{author}{\bibfnamefont{K.~A.} \bibnamefont{Olive}},
  \bibinfo{year}{2012}, \bibinfo{journal}{JHEP}
  \textbf{\bibinfo{volume}{1203}}, \bibinfo{pages}{026}.

\bibitem[{\citenamefont{Carena} \emph{et~al.}(2003)\citenamefont{Carena,
  Delgado, Ponton, Tait, and Wagner}}]{Carena:2003fx}
\bibinfo{author}{\bibnamefont{Carena}, \bibfnamefont{M.~S.}},
  \bibinfo{author}{\bibfnamefont{A.}~\bibnamefont{Delgado}},
  \bibinfo{author}{\bibfnamefont{E.}~\bibnamefont{Ponton}},
  \bibinfo{author}{\bibfnamefont{T.~M.} \bibnamefont{Tait}}, and
  \bibinfo{author}{\bibfnamefont{C.}~\bibnamefont{Wagner}},
  \bibinfo{year}{2003}, \bibinfo{journal}{Phys.Rev.}
  \textbf{\bibinfo{volume}{D68}}, \bibinfo{pages}{035010}.

\bibitem[{\citenamefont{Carena} \emph{et~al.}(2007)\citenamefont{Carena,
  Ponton, Santiago, and Wagner}}]{Carena:2007ua}
\bibinfo{author}{\bibnamefont{Carena}, \bibfnamefont{M.~S.}},
  \bibinfo{author}{\bibfnamefont{E.}~\bibnamefont{Ponton}},
  \bibinfo{author}{\bibfnamefont{J.}~\bibnamefont{Santiago}}, and
  \bibinfo{author}{\bibfnamefont{C.}~\bibnamefont{Wagner}},
  \bibinfo{year}{2007}, \bibinfo{journal}{Phys.Rev.}
  \textbf{\bibinfo{volume}{D76}}, \bibinfo{pages}{035006}.

\bibitem[{\citenamefont{Carena} \emph{et~al.}(2006)\citenamefont{Carena,
  Ponton, Santiago, and Wagner}}]{Carena:2006bn}
\bibinfo{author}{\bibnamefont{Carena}, \bibfnamefont{M.~S.}},
  \bibinfo{author}{\bibfnamefont{E.}~\bibnamefont{Ponton}},
  \bibinfo{author}{\bibfnamefont{J.}~\bibnamefont{Santiago}}, and
  \bibinfo{author}{\bibfnamefont{C.~E.} \bibnamefont{Wagner}},
  \bibinfo{year}{2006}, \bibinfo{journal}{Nucl.Phys.}
  \textbf{\bibinfo{volume}{B759}}, \bibinfo{pages}{202}.

\bibitem[{\citenamefont{Casalbuoni}
  \emph{et~al.}(2004)\citenamefont{Casalbuoni, Deandrea, and
  Oertel}}]{Casalbuoni:2003ft}
\bibinfo{author}{\bibnamefont{Casalbuoni}, \bibfnamefont{R.}},
  \bibinfo{author}{\bibfnamefont{A.}~\bibnamefont{Deandrea}}, and
  \bibinfo{author}{\bibfnamefont{M.}~\bibnamefont{Oertel}},
  \bibinfo{year}{2004}, \bibinfo{journal}{JHEP}
  \textbf{\bibinfo{volume}{0402}}, \bibinfo{pages}{032}.

\bibitem[{\citenamefont{Casas} \emph{et~al.}(2005)\citenamefont{Casas,
  Espinosa, and Hidalgo}}]{Casas:2005ev}
\bibinfo{author}{\bibnamefont{Casas}, \bibfnamefont{J.~A.}},
  \bibinfo{author}{\bibfnamefont{J.~R.} \bibnamefont{Espinosa}}, and
  \bibinfo{author}{\bibfnamefont{I.}~\bibnamefont{Hidalgo}},
  \bibinfo{year}{2005}, \bibinfo{journal}{JHEP}
  \textbf{\bibinfo{volume}{0503}}, \bibinfo{pages}{038}.

\bibitem[{\citenamefont{Chacko}
  \emph{et~al.}(2013{\natexlab{a}})\citenamefont{Chacko, Franceschini, and
  Mishra}}]{Chacko:2012vm}
\bibinfo{author}{\bibnamefont{Chacko}, \bibfnamefont{Z.}},
  \bibinfo{author}{\bibfnamefont{R.}~\bibnamefont{Franceschini}}, and
  \bibinfo{author}{\bibfnamefont{R.~K.} \bibnamefont{Mishra}},
  \bibinfo{year}{2013}{\natexlab{a}}, \bibinfo{journal}{JHEP}
  \textbf{\bibinfo{volume}{1304}}, \bibinfo{pages}{015}.

\bibitem[{\citenamefont{Chacko} \emph{et~al.}(2006)\citenamefont{Chacko, Goh,
  and Harnik}}]{Chacko:2005pe}
\bibinfo{author}{\bibnamefont{Chacko}, \bibfnamefont{Z.}},
  \bibinfo{author}{\bibfnamefont{H.-S.} \bibnamefont{Goh}}, and
  \bibinfo{author}{\bibfnamefont{R.}~\bibnamefont{Harnik}},
  \bibinfo{year}{2006}, \bibinfo{journal}{Phys. Rev. Lett.}
  \textbf{\bibinfo{volume}{96}}, \bibinfo{pages}{231802}.

\bibitem[{\citenamefont{Chacko and Mishra}(2013)}]{Chacko:2012sy}
\bibinfo{author}{\bibnamefont{Chacko}, \bibfnamefont{Z.}}, and
  \bibinfo{author}{\bibfnamefont{R.~K.} \bibnamefont{Mishra}},
  \bibinfo{year}{2013}, \bibinfo{journal}{Phys.Rev.}
  \textbf{\bibinfo{volume}{D87}}, \bibinfo{pages}{115006}.

\bibitem[{\citenamefont{Chacko}
  \emph{et~al.}(2013{\natexlab{b}})\citenamefont{Chacko, Mishra, and
  Stolarski}}]{Chacko:2013dra}
\bibinfo{author}{\bibnamefont{Chacko}, \bibfnamefont{Z.}},
  \bibinfo{author}{\bibfnamefont{R.~K.} \bibnamefont{Mishra}}, and
  \bibinfo{author}{\bibfnamefont{D.}~\bibnamefont{Stolarski}},
  \bibinfo{year}{2013}{\natexlab{b}}, \bibinfo{journal}{JHEP}
  \textbf{\bibinfo{volume}{09}}, \bibinfo{pages}{121}.

\bibitem[{\citenamefont{Chang and Ng}(2000)}]{Chang:2000zh}
\bibinfo{author}{\bibnamefont{Chang}, \bibfnamefont{C.-H.~V.}}, and
  \bibinfo{author}{\bibfnamefont{J.}~\bibnamefont{Ng}}, \bibinfo{year}{2000},
  \bibinfo{journal}{Phys.Lett.} \textbf{\bibinfo{volume}{B488}},
  \bibinfo{pages}{390}.

\bibitem[{\citenamefont{Chang} \emph{et~al.}(2012)\citenamefont{Chang, Cheung,
  Tseng, and Yuan}}]{Chang:2012tb}
\bibinfo{author}{\bibnamefont{Chang}, \bibfnamefont{J.}},
  \bibinfo{author}{\bibfnamefont{K.}~\bibnamefont{Cheung}},
  \bibinfo{author}{\bibfnamefont{P.-Y.} \bibnamefont{Tseng}}, and
  \bibinfo{author}{\bibfnamefont{T.-C.} \bibnamefont{Yuan}},
  \bibinfo{year}{2012}, \bibinfo{journal}{JHEP}
  \textbf{\bibinfo{volume}{1212}}, \bibinfo{pages}{058}.

\bibitem[{\citenamefont{Chang}(2003)}]{Chang:2003zn}
\bibinfo{author}{\bibnamefont{Chang}, \bibfnamefont{S.}}, \bibinfo{year}{2003},
  \bibinfo{journal}{JHEP} \textbf{\bibinfo{volume}{0312}},
  \bibinfo{pages}{057}.

\bibitem[{\citenamefont{Chang} \emph{et~al.}(2015)\citenamefont{Chang,
  Galloway, Luty, Salvioni, and Tsai}}]{Chang:2014ida}
\bibinfo{author}{\bibnamefont{Chang}, \bibfnamefont{S.}},
  \bibinfo{author}{\bibfnamefont{J.}~\bibnamefont{Galloway}},
  \bibinfo{author}{\bibfnamefont{M.}~\bibnamefont{Luty}},
  \bibinfo{author}{\bibfnamefont{E.}~\bibnamefont{Salvioni}}, and
  \bibinfo{author}{\bibfnamefont{Y.}~\bibnamefont{Tsai}}, \bibinfo{year}{2015},
  \bibinfo{journal}{JHEP} \textbf{\bibinfo{volume}{03}}, \bibinfo{pages}{017}.

\bibitem[{\citenamefont{Chang} \emph{et~al.}(2007)\citenamefont{Chang, Hall,
  and Weiner}}]{Chang:2006ra}
\bibinfo{author}{\bibnamefont{Chang}, \bibfnamefont{S.}},
  \bibinfo{author}{\bibfnamefont{L.~J.} \bibnamefont{Hall}}, and
  \bibinfo{author}{\bibfnamefont{N.}~\bibnamefont{Weiner}},
  \bibinfo{year}{2007}, \bibinfo{journal}{Phys. Rev.}
  \textbf{\bibinfo{volume}{D75}}, \bibinfo{pages}{035009}.

\bibitem[{\citenamefont{Chang and Wacker}(2004)}]{Chang:2003un}
\bibinfo{author}{\bibnamefont{Chang}, \bibfnamefont{S.}}, and
  \bibinfo{author}{\bibfnamefont{J.~G.} \bibnamefont{Wacker}},
  \bibinfo{year}{2004}, \bibinfo{journal}{Phys.Rev.}
  \textbf{\bibinfo{volume}{D69}}, \bibinfo{pages}{035002}.

\bibitem[{Chatrchyan \emph{et~al.}(2012{\natexlab{a}})\citenamefont{Chatrchyan}
  \emph{et~al.}}]{Chatrchyan:2012ufa}
\bibinfo{author}{\bibnamefont{Chatrchyan}, \bibfnamefont{S.}}, \emph{et~al.}
  (\bibinfo{collaboration}{CMS Collaboration}),
  \bibinfo{year}{2012}{\natexlab{a}}, \bibinfo{journal}{Phys.Lett.}
  \textbf{\bibinfo{volume}{B716}}, \bibinfo{pages}{30},
  \bibinfo{note}{{CMS-HIG-12-028}}.

\bibitem[{Chatrchyan \emph{et~al.}(2012{\natexlab{b}})\citenamefont{Chatrchyan}
  \emph{et~al.}}]{Chatrchyan:2012ku}
\bibinfo{author}{\bibnamefont{Chatrchyan}, \bibfnamefont{S.}}, \emph{et~al.}
  (\bibinfo{collaboration}{CMS Collaboration}),
  \bibinfo{year}{2012}{\natexlab{b}}, \bibinfo{journal}{JHEP}
  \textbf{\bibinfo{volume}{1209}}, \bibinfo{pages}{029}.

\bibitem[{Chatrchyan \emph{et~al.}(2014{\natexlab{a}})\citenamefont{Chatrchyan}
  \emph{et~al.}}]{Chatrchyan:2013uxa}
\bibinfo{author}{\bibnamefont{Chatrchyan}, \bibfnamefont{S.}}, \emph{et~al.}
  (\bibinfo{collaboration}{CMS Collaboration}),
  \bibinfo{year}{2014}{\natexlab{a}}, \bibinfo{journal}{Phys. Lett.}
  \textbf{\bibinfo{volume}{B729}}, \bibinfo{pages}{149}.

\bibitem[{Chatrchyan \emph{et~al.}(2014{\natexlab{b}})\citenamefont{Chatrchyan}
  \emph{et~al.}}]{CMSprelimTpartner}
\bibinfo{author}{\bibnamefont{Chatrchyan}, \bibfnamefont{S.}}, \emph{et~al.}
  (\bibinfo{collaboration}{CMS Collaboration}),
  \bibinfo{year}{2014}{\natexlab{b}}, \bibinfo{journal}{Phys. Lett.}
  \textbf{\bibinfo{volume}{B729}}, \bibinfo{pages}{149}.

\bibitem[{Chatrchyan \emph{et~al.}(2014{\natexlab{c}})\citenamefont{Chatrchyan}
  \emph{et~al.}}]{Chatrchyan:2013wfa}
\bibinfo{author}{\bibnamefont{Chatrchyan}, \bibfnamefont{S.}}, \emph{et~al.}
  (\bibinfo{collaboration}{CMS Collaboration}),
  \bibinfo{year}{2014}{\natexlab{c}}, \bibinfo{journal}{Phys. Rev. Lett.}
  \textbf{\bibinfo{volume}{112}}(\bibinfo{number}{17}),
  \bibinfo{pages}{171801}.

\bibitem[{\citenamefont{Chen}(2006)}]{Chen:2006dy}
\bibinfo{author}{\bibnamefont{Chen}, \bibfnamefont{M.-C.}},
  \bibinfo{year}{2006}, \bibinfo{journal}{Mod.Phys.Lett.}
  \textbf{\bibinfo{volume}{A21}}, \bibinfo{pages}{621}.

\bibitem[{\citenamefont{Cheng}(2007)}]{Cheng:2007bu}
\bibinfo{author}{\bibnamefont{Cheng}, \bibfnamefont{H.-C.}},
  \bibinfo{year}{2007}, \eprint{0710.3407}.

\bibitem[{\citenamefont{Cheng}(2010)}]{Cheng:2010pt}
\bibinfo{author}{\bibnamefont{Cheng}, \bibfnamefont{H.-C.}},
  \bibinfo{year}{2010}, \eprint{1003.1162}.

\bibitem[{\citenamefont{Cheng and Low}(2003)}]{Cheng:2003ju}
\bibinfo{author}{\bibnamefont{Cheng}, \bibfnamefont{H.-C.}}, and
  \bibinfo{author}{\bibfnamefont{I.}~\bibnamefont{Low}}, \bibinfo{year}{2003},
  \bibinfo{journal}{JHEP} \textbf{\bibinfo{volume}{0309}},
  \bibinfo{pages}{051}.

\bibitem[{\citenamefont{Cheng and Low}(2004)}]{Cheng:2004yc}
\bibinfo{author}{\bibnamefont{Cheng}, \bibfnamefont{H.-C.}}, and
  \bibinfo{author}{\bibfnamefont{I.}~\bibnamefont{Low}}, \bibinfo{year}{2004},
  \bibinfo{journal}{JHEP} \textbf{\bibinfo{volume}{0408}},
  \bibinfo{pages}{061}.

\bibitem[{\citenamefont{Cheng} \emph{et~al.}(2006)\citenamefont{Cheng, Low, and
  Wang}}]{Cheng:2005as}
\bibinfo{author}{\bibnamefont{Cheng}, \bibfnamefont{H.-C.}},
  \bibinfo{author}{\bibfnamefont{I.}~\bibnamefont{Low}}, and
  \bibinfo{author}{\bibfnamefont{L.-T.} \bibnamefont{Wang}},
  \bibinfo{year}{2006}, \bibinfo{journal}{Phys.Rev.}
  \textbf{\bibinfo{volume}{D74}}, \bibinfo{pages}{055001}.

\bibitem[{\citenamefont{Cheng and Shadmi}(1998)}]{Cheng:1998xg}
\bibinfo{author}{\bibnamefont{Cheng}, \bibfnamefont{H.-C.}}, and
  \bibinfo{author}{\bibfnamefont{Y.}~\bibnamefont{Shadmi}},
  \bibinfo{year}{1998}, \bibinfo{journal}{Nucl.Phys.}
  \textbf{\bibinfo{volume}{B531}}, \bibinfo{pages}{125}.

\bibitem[{\citenamefont{Cheung and Shen}(2015)}]{Cheung:2015aba}
\bibinfo{author}{\bibnamefont{Cheung}, \bibfnamefont{C.}}, and
  \bibinfo{author}{\bibfnamefont{C.-H.} \bibnamefont{Shen}},
  \bibinfo{year}{2015}, \bibinfo{journal}{Phys. Rev. Lett.}
  \textbf{\bibinfo{volume}{115}}(\bibinfo{number}{7}), \bibinfo{pages}{071601}.

\bibitem[{\citenamefont{Chivukula}
  \emph{et~al.}(2002{\natexlab{a}})\citenamefont{Chivukula, Dicus, and
  He}}]{SekharChivukula:2001hz}
\bibinfo{author}{\bibnamefont{Chivukula}, \bibfnamefont{R.~S.}},
  \bibinfo{author}{\bibfnamefont{D.~A.} \bibnamefont{Dicus}}, and
  \bibinfo{author}{\bibfnamefont{H.-J.} \bibnamefont{He}},
  \bibinfo{year}{2002}{\natexlab{a}}, \bibinfo{journal}{Phys. Lett.}
  \textbf{\bibinfo{volume}{B525}}, \bibinfo{pages}{175}.

\bibitem[{\citenamefont{Chivukula}
  \emph{et~al.}(2002{\natexlab{b}})\citenamefont{Chivukula, Evans, and
  Simmons}}]{Chivukula:2002ww}
\bibinfo{author}{\bibnamefont{Chivukula}, \bibfnamefont{R.~S.}},
  \bibinfo{author}{\bibfnamefont{N.~J.} \bibnamefont{Evans}}, and
  \bibinfo{author}{\bibfnamefont{E.~H.} \bibnamefont{Simmons}},
  \bibinfo{year}{2002}{\natexlab{b}}, \bibinfo{journal}{Phys.Rev.}
  \textbf{\bibinfo{volume}{D66}}, \bibinfo{pages}{035008}.

\bibitem[{\citenamefont{Chivukula and He}(2002)}]{Chivukula:2002ej}
\bibinfo{author}{\bibnamefont{Chivukula}, \bibfnamefont{R.~S.}}, and
  \bibinfo{author}{\bibfnamefont{H.-J.} \bibnamefont{He}},
  \bibinfo{year}{2002}, \bibinfo{journal}{Phys. Lett.}
  \textbf{\bibinfo{volume}{B532}}, \bibinfo{pages}{121}.

\bibitem[{\citenamefont{Cho} \emph{et~al.}(2013)\citenamefont{Cho, Nomura, and
  Ohno}}]{Cho:2013mva}
\bibinfo{author}{\bibnamefont{Cho}, \bibfnamefont{G.-C.}},
  \bibinfo{author}{\bibfnamefont{D.}~\bibnamefont{Nomura}}, and
  \bibinfo{author}{\bibfnamefont{Y.}~\bibnamefont{Ohno}}, \bibinfo{year}{2013},
  \bibinfo{journal}{Mod.Phys.Lett.} \textbf{\bibinfo{volume}{A28}},
  \bibinfo{pages}{1350148}.

\bibitem[{\citenamefont{Choudhury} \emph{et~al.}(2004)\citenamefont{Choudhury,
  Gaur, Goyal, and Mahajan}}]{Choudhury:2004bh}
\bibinfo{author}{\bibnamefont{Choudhury}, \bibfnamefont{S.~R.}},
  \bibinfo{author}{\bibfnamefont{N.}~\bibnamefont{Gaur}},
  \bibinfo{author}{\bibfnamefont{A.}~\bibnamefont{Goyal}}, and
  \bibinfo{author}{\bibfnamefont{N.}~\bibnamefont{Mahajan}},
  \bibinfo{year}{2004}, \bibinfo{journal}{Phys.Lett.}
  \textbf{\bibinfo{volume}{B601}}, \bibinfo{pages}{164}.

\bibitem[{\citenamefont{Ciuchini} \emph{et~al.}(2013)\citenamefont{Ciuchini,
  Franco, Mishima, and Silvestrini}}]{Ciuchini:2013pca}
\bibinfo{author}{\bibnamefont{Ciuchini}, \bibfnamefont{M.}},
  \bibinfo{author}{\bibfnamefont{E.}~\bibnamefont{Franco}},
  \bibinfo{author}{\bibfnamefont{S.}~\bibnamefont{Mishima}}, and
  \bibinfo{author}{\bibfnamefont{L.}~\bibnamefont{Silvestrini}},
  \bibinfo{year}{2013}, \bibinfo{journal}{JHEP} \textbf{\bibinfo{volume}{08}},
  \bibinfo{pages}{106}.

\bibitem[{\citenamefont{Cleary and Terning}(2015)}]{Cleary:2015koc}
\bibinfo{author}{\bibnamefont{Cleary}, \bibfnamefont{K.~F.}}, and
  \bibinfo{author}{\bibfnamefont{J.}~\bibnamefont{Terning}},
  \bibinfo{year}{2015}, \eprint{1511.08216}.

\bibitem[{\citenamefont{{CMS Collaboration}}(2015)}]{CMS:2015nza}
\bibinfo{author}{\bibnamefont{{CMS Collaboration}}}, \bibinfo{year}{2015},
  \bibinfo{note}{{CMS-PAS-B2G-13-008}}.

\bibitem[{\citenamefont{Cohen} \emph{et~al.}(2015)\citenamefont{Cohen, Craig,
  Lou, and Pinner}}]{Cohen:2015gaa}
\bibinfo{author}{\bibnamefont{Cohen}, \bibfnamefont{T.}},
  \bibinfo{author}{\bibfnamefont{N.}~\bibnamefont{Craig}},
  \bibinfo{author}{\bibfnamefont{H.~K.} \bibnamefont{Lou}}, and
  \bibinfo{author}{\bibfnamefont{D.}~\bibnamefont{Pinner}},
  \bibinfo{year}{2015}, \eprint{1508.05396}.

\bibitem[{\citenamefont{Coleman}(1985)}]{Coleman}
\bibinfo{author}{\bibnamefont{Coleman}, \bibfnamefont{S.}},
  \bibinfo{year}{1985}, \bibinfo{note}{{Dilatations in ``Aspects of Symmetry"
  (Cambridge University Press)}}.

\bibitem[{\citenamefont{Coleman and Weinberg}(1973)}]{Coleman:1973jx}
\bibinfo{author}{\bibnamefont{Coleman}, \bibfnamefont{S.~R.}}, and
  \bibinfo{author}{\bibfnamefont{E.~J.} \bibnamefont{Weinberg}},
  \bibinfo{year}{1973}, \bibinfo{journal}{Phys.Rev.}
  \textbf{\bibinfo{volume}{D7}}, \bibinfo{pages}{1888}.

\bibitem[{\citenamefont{Coleman} \emph{et~al.}(1969)\citenamefont{Coleman,
  Wess, and Zumino}}]{Coleman:1969sm}
\bibinfo{author}{\bibnamefont{Coleman}, \bibfnamefont{S.~R.}},
  \bibinfo{author}{\bibfnamefont{J.}~\bibnamefont{Wess}}, and
  \bibinfo{author}{\bibfnamefont{B.}~\bibnamefont{Zumino}},
  \bibinfo{year}{1969}, \bibinfo{journal}{Phys.Rev.}
  \textbf{\bibinfo{volume}{177}}, \bibinfo{pages}{2239}.

\bibitem[{\citenamefont{Coleppa} \emph{et~al.}(2012)\citenamefont{Coleppa,
  Gregoire, and Logan}}]{Coleppa:2011zx}
\bibinfo{author}{\bibnamefont{Coleppa}, \bibfnamefont{B.}},
  \bibinfo{author}{\bibfnamefont{T.}~\bibnamefont{Gregoire}}, and
  \bibinfo{author}{\bibfnamefont{H.~E.} \bibnamefont{Logan}},
  \bibinfo{year}{2012}, \bibinfo{journal}{Phys.Rev.}
  \textbf{\bibinfo{volume}{D85}}, \bibinfo{pages}{055001}.

\bibitem[{\citenamefont{Contino}(2011)}]{Contino:2010rs}
\bibinfo{author}{\bibnamefont{Contino}, \bibfnamefont{R.}},
  \bibinfo{year}{2011}, in \emph{\bibinfo{booktitle}{{Physics of the large and
  the small, TASI 09, proceedings of the Theoretical Advanced Study Institute
  in Elementary Particle Physics, Boulder, Colorado, USA, 1-26 June 2009}}},
  pp. \bibinfo{pages}{235--306}, \eprint{1005.4269}.

\bibitem[{\citenamefont{Contino} \emph{et~al.}(2007)\citenamefont{Contino,
  Da~Rold, and Pomarol}}]{Contino:2006qr}
\bibinfo{author}{\bibnamefont{Contino}, \bibfnamefont{R.}},
  \bibinfo{author}{\bibfnamefont{L.}~\bibnamefont{Da~Rold}}, and
  \bibinfo{author}{\bibfnamefont{A.}~\bibnamefont{Pomarol}},
  \bibinfo{year}{2007}, \bibinfo{journal}{Phys.Rev.}
  \textbf{\bibinfo{volume}{D75}}, \bibinfo{pages}{055014}.

\bibitem[{\citenamefont{Contino} \emph{et~al.}(2013)\citenamefont{Contino,
  Ghezzi, Grojean, Muhlleitner, and Spira}}]{Contino:2013kra}
\bibinfo{author}{\bibnamefont{Contino}, \bibfnamefont{R.}},
  \bibinfo{author}{\bibfnamefont{M.}~\bibnamefont{Ghezzi}},
  \bibinfo{author}{\bibfnamefont{C.}~\bibnamefont{Grojean}},
  \bibinfo{author}{\bibfnamefont{M.}~\bibnamefont{Muhlleitner}}, and
  \bibinfo{author}{\bibfnamefont{M.}~\bibnamefont{Spira}},
  \bibinfo{year}{2013}, \eprint{1303.3876}.

\bibitem[{\citenamefont{Contino} \emph{et~al.}(2010)\citenamefont{Contino,
  Grojean, Moretti, Piccinini, and Rattazzi}}]{Contino:2010mh}
\bibinfo{author}{\bibnamefont{Contino}, \bibfnamefont{R.}},
  \bibinfo{author}{\bibfnamefont{C.}~\bibnamefont{Grojean}},
  \bibinfo{author}{\bibfnamefont{M.}~\bibnamefont{Moretti}},
  \bibinfo{author}{\bibfnamefont{F.}~\bibnamefont{Piccinini}}, and
  \bibinfo{author}{\bibfnamefont{R.}~\bibnamefont{Rattazzi}},
  \bibinfo{year}{2010}, \bibinfo{journal}{JHEP}
  \textbf{\bibinfo{volume}{1005}}, \bibinfo{pages}{089}.

\bibitem[{\citenamefont{Contino} \emph{et~al.}(2003)\citenamefont{Contino,
  Nomura, and Pomarol}}]{Contino:2003ve}
\bibinfo{author}{\bibnamefont{Contino}, \bibfnamefont{R.}},
  \bibinfo{author}{\bibfnamefont{Y.}~\bibnamefont{Nomura}}, and
  \bibinfo{author}{\bibfnamefont{A.}~\bibnamefont{Pomarol}},
  \bibinfo{year}{2003}, \bibinfo{journal}{Nucl.Phys.}
  \textbf{\bibinfo{volume}{B671}}, \bibinfo{pages}{148}.

\bibitem[{\citenamefont{Contino and Servant}(2008)}]{Contino:2008hi}
\bibinfo{author}{\bibnamefont{Contino}, \bibfnamefont{R.}}, and
  \bibinfo{author}{\bibfnamefont{G.}~\bibnamefont{Servant}},
  \bibinfo{year}{2008}, \bibinfo{journal}{JHEP}
  \textbf{\bibinfo{volume}{0806}}, \bibinfo{pages}{026}.

\bibitem[{\citenamefont{Coriano} \emph{et~al.}(2012)\citenamefont{Coriano,
  Delle~Rose, Marzo, and Serino}}]{Coriano:2012dg}
\bibinfo{author}{\bibnamefont{Coriano}, \bibfnamefont{C.}},
  \bibinfo{author}{\bibfnamefont{L.}~\bibnamefont{Delle~Rose}},
  \bibinfo{author}{\bibfnamefont{C.}~\bibnamefont{Marzo}}, and
  \bibinfo{author}{\bibfnamefont{M.}~\bibnamefont{Serino}},
  \bibinfo{year}{2012}, \bibinfo{journal}{Phys.Lett.}
  \textbf{\bibinfo{volume}{B717}}, \bibinfo{pages}{182}.

\bibitem[{\citenamefont{Coriano} \emph{et~al.}(2013)\citenamefont{Coriano,
  Delle~Rose, Quintavalle, and Serino}}]{Coriano:2012nm}
\bibinfo{author}{\bibnamefont{Coriano}, \bibfnamefont{C.}},
  \bibinfo{author}{\bibfnamefont{L.}~\bibnamefont{Delle~Rose}},
  \bibinfo{author}{\bibfnamefont{A.}~\bibnamefont{Quintavalle}}, and
  \bibinfo{author}{\bibfnamefont{M.}~\bibnamefont{Serino}},
  \bibinfo{year}{2013}, \bibinfo{journal}{JHEP} \textbf{\bibinfo{volume}{06}},
  \bibinfo{pages}{077}.

\bibitem[{\citenamefont{Craig and Howe}(2014)}]{Craig:2013fga}
\bibinfo{author}{\bibnamefont{Craig}, \bibfnamefont{N.}}, and
  \bibinfo{author}{\bibfnamefont{K.}~\bibnamefont{Howe}}, \bibinfo{year}{2014},
  \bibinfo{journal}{JHEP} \textbf{\bibinfo{volume}{03}}, \bibinfo{pages}{140}.

\bibitem[{\citenamefont{Craig}
  \emph{et~al.}(2015{\natexlab{a}})\citenamefont{Craig, Katz, Strassler, and
  Sundrum}}]{Craig:2015pha}
\bibinfo{author}{\bibnamefont{Craig}, \bibfnamefont{N.}},
  \bibinfo{author}{\bibfnamefont{A.}~\bibnamefont{Katz}},
  \bibinfo{author}{\bibfnamefont{M.}~\bibnamefont{Strassler}}, and
  \bibinfo{author}{\bibfnamefont{R.}~\bibnamefont{Sundrum}},
  \bibinfo{year}{2015}{\natexlab{a}}, \bibinfo{journal}{JHEP}
  \textbf{\bibinfo{volume}{07}}, \bibinfo{pages}{105}.

\bibitem[{\citenamefont{Craig}
  \emph{et~al.}(2015{\natexlab{b}})\citenamefont{Craig, Knapen, and
  Longhi}}]{Craig:2014aea}
\bibinfo{author}{\bibnamefont{Craig}, \bibfnamefont{N.}},
  \bibinfo{author}{\bibfnamefont{S.}~\bibnamefont{Knapen}}, and
  \bibinfo{author}{\bibfnamefont{P.}~\bibnamefont{Longhi}},
  \bibinfo{year}{2015}{\natexlab{b}}, \bibinfo{journal}{Phys. Rev. Lett.}
  \textbf{\bibinfo{volume}{114}}(\bibinfo{number}{6}), \bibinfo{pages}{061803}.

\bibitem[{\citenamefont{Craig}
  \emph{et~al.}(2015{\natexlab{c}})\citenamefont{Craig, Knapen, and
  Longhi}}]{Craig:2014roa}
\bibinfo{author}{\bibnamefont{Craig}, \bibfnamefont{N.}},
  \bibinfo{author}{\bibfnamefont{S.}~\bibnamefont{Knapen}}, and
  \bibinfo{author}{\bibfnamefont{P.}~\bibnamefont{Longhi}},
  \bibinfo{year}{2015}{\natexlab{c}}, \bibinfo{journal}{JHEP}
  \textbf{\bibinfo{volume}{03}}, \bibinfo{pages}{106}.

\bibitem[{\citenamefont{Cs\'aki} \emph{et~al.}(2002)\citenamefont{Cs\'aki,
  Erlich, and Terning}}]{Csaki:2002gy}
\bibinfo{author}{\bibnamefont{Cs\'aki}, \bibfnamefont{C.}},
  \bibinfo{author}{\bibfnamefont{J.}~\bibnamefont{Erlich}}, and
  \bibinfo{author}{\bibfnamefont{J.}~\bibnamefont{Terning}},
  \bibinfo{year}{2002}, \bibinfo{journal}{Phys. Rev.}
  \textbf{\bibinfo{volume}{D66}}, \bibinfo{pages}{064021}.

\bibitem[{\citenamefont{Cs\'aki} \emph{et~al.}(2008)\citenamefont{Cs\'aki,
  Falkowski, and Weiler}}]{Csaki:2008zd}
\bibinfo{author}{\bibnamefont{Cs\'aki}, \bibfnamefont{C.}},
  \bibinfo{author}{\bibfnamefont{A.}~\bibnamefont{Falkowski}}, and
  \bibinfo{author}{\bibfnamefont{A.}~\bibnamefont{Weiler}},
  \bibinfo{year}{2008}, \bibinfo{journal}{JHEP}
  \textbf{\bibinfo{volume}{0809}}, \bibinfo{pages}{008}.

\bibitem[{\citenamefont{Cs\'aki} \emph{et~al.}(1999)\citenamefont{Cs\'aki,
  Graesser, Kolda, and Terning}}]{Csaki:1999jh}
\bibinfo{author}{\bibnamefont{Cs\'aki}, \bibfnamefont{C.}},
  \bibinfo{author}{\bibfnamefont{M.}~\bibnamefont{Graesser}},
  \bibinfo{author}{\bibfnamefont{C.~F.} \bibnamefont{Kolda}}, and
  \bibinfo{author}{\bibfnamefont{J.}~\bibnamefont{Terning}},
  \bibinfo{year}{1999}, \bibinfo{journal}{Phys.Lett.}
  \textbf{\bibinfo{volume}{B462}}, \bibinfo{pages}{34}.

\bibitem[{\citenamefont{Cs\'aki} \emph{et~al.}(2000)\citenamefont{Cs\'aki,
  Graesser, Randall, and Terning}}]{Csaki:1999mp}
\bibinfo{author}{\bibnamefont{Cs\'aki}, \bibfnamefont{C.}},
  \bibinfo{author}{\bibfnamefont{M.}~\bibnamefont{Graesser}},
  \bibinfo{author}{\bibfnamefont{L.}~\bibnamefont{Randall}}, and
  \bibinfo{author}{\bibfnamefont{J.}~\bibnamefont{Terning}},
  \bibinfo{year}{2000}, \bibinfo{journal}{Phys.Rev.}
  \textbf{\bibinfo{volume}{D62}}, \bibinfo{pages}{045015}.

\bibitem[{\citenamefont{Cs\'aki} \emph{et~al.}(2001)\citenamefont{Cs\'aki,
  Graesser, and Kribs}}]{Csaki:2000zn}
\bibinfo{author}{\bibnamefont{Cs\'aki}, \bibfnamefont{C.}},
  \bibinfo{author}{\bibfnamefont{M.~L.} \bibnamefont{Graesser}}, and
  \bibinfo{author}{\bibfnamefont{G.~D.} \bibnamefont{Kribs}},
  \bibinfo{year}{2001}, \bibinfo{journal}{Phys.Rev.}
  \textbf{\bibinfo{volume}{D63}}, \bibinfo{pages}{065002}.

\bibitem[{\citenamefont{Cs\'aki}
  \emph{et~al.}(2003{\natexlab{a}})\citenamefont{Cs\'aki, Grojean, and
  Murayama}}]{Csaki:2002ur}
\bibinfo{author}{\bibnamefont{Cs\'aki}, \bibfnamefont{C.}},
  \bibinfo{author}{\bibfnamefont{C.}~\bibnamefont{Grojean}}, and
  \bibinfo{author}{\bibfnamefont{H.}~\bibnamefont{Murayama}},
  \bibinfo{year}{2003}{\natexlab{a}}, \bibinfo{journal}{Phys.Rev.}
  \textbf{\bibinfo{volume}{D67}}, \bibinfo{pages}{085012}.

\bibitem[{\citenamefont{Cs\'aki}
  \emph{et~al.}(2004{\natexlab{a}})\citenamefont{Cs\'aki, Grojean, Murayama,
  Pilo, and Terning}}]{Csaki:2003dt}
\bibinfo{author}{\bibnamefont{Cs\'aki}, \bibfnamefont{C.}},
  \bibinfo{author}{\bibfnamefont{C.}~\bibnamefont{Grojean}},
  \bibinfo{author}{\bibfnamefont{H.}~\bibnamefont{Murayama}},
  \bibinfo{author}{\bibfnamefont{L.}~\bibnamefont{Pilo}}, and
  \bibinfo{author}{\bibfnamefont{J.}~\bibnamefont{Terning}},
  \bibinfo{year}{2004}{\natexlab{a}}, \bibinfo{journal}{Phys. Rev.}
  \textbf{\bibinfo{volume}{D69}}, \bibinfo{pages}{055006}.

\bibitem[{\citenamefont{Cs\'aki}
  \emph{et~al.}(2004{\natexlab{b}})\citenamefont{Cs\'aki, Grojean, Pilo, and
  Terning}}]{CGPT}
\bibinfo{author}{\bibnamefont{Cs\'aki}, \bibfnamefont{C.}},
  \bibinfo{author}{\bibfnamefont{C.}~\bibnamefont{Grojean}},
  \bibinfo{author}{\bibfnamefont{L.}~\bibnamefont{Pilo}}, and
  \bibinfo{author}{\bibfnamefont{J.}~\bibnamefont{Terning}},
  \bibinfo{year}{2004}{\natexlab{b}}, \bibinfo{journal}{Phys.Rev.Lett.}
  \textbf{\bibinfo{volume}{92}}, \bibinfo{pages}{101802}.

\bibitem[{\citenamefont{Cs\'aki}
  \emph{et~al.}(2003{\natexlab{b}})\citenamefont{Cs\'aki, Hubisz, Kribs, Meade,
  and Terning}}]{Csaki:2002qg}
\bibinfo{author}{\bibnamefont{Cs\'aki}, \bibfnamefont{C.}},
  \bibinfo{author}{\bibfnamefont{J.}~\bibnamefont{Hubisz}},
  \bibinfo{author}{\bibfnamefont{G.~D.} \bibnamefont{Kribs}},
  \bibinfo{author}{\bibfnamefont{P.}~\bibnamefont{Meade}}, and
  \bibinfo{author}{\bibfnamefont{J.}~\bibnamefont{Terning}},
  \bibinfo{year}{2003}{\natexlab{b}}, \bibinfo{journal}{Phys.Rev.}
  \textbf{\bibinfo{volume}{D67}}, \bibinfo{pages}{115002}.

\bibitem[{\citenamefont{Cs\'aki}
  \emph{et~al.}(2003{\natexlab{c}})\citenamefont{Cs\'aki, Hubisz, Kribs, Meade,
  and Terning}}]{Csaki:2003si}
\bibinfo{author}{\bibnamefont{Cs\'aki}, \bibfnamefont{C.}},
  \bibinfo{author}{\bibfnamefont{J.}~\bibnamefont{Hubisz}},
  \bibinfo{author}{\bibfnamefont{G.~D.} \bibnamefont{Kribs}},
  \bibinfo{author}{\bibfnamefont{P.}~\bibnamefont{Meade}}, and
  \bibinfo{author}{\bibfnamefont{J.}~\bibnamefont{Terning}},
  \bibinfo{year}{2003}{\natexlab{c}}, \bibinfo{journal}{Phys.Rev.}
  \textbf{\bibinfo{volume}{D68}}, \bibinfo{pages}{035009}.

\bibitem[{\citenamefont{Cs\'aki} \emph{et~al.}(2007)\citenamefont{Cs\'aki,
  Hubisz, and Lee}}]{Csaki:2007ns}
\bibinfo{author}{\bibnamefont{Cs\'aki}, \bibfnamefont{C.}},
  \bibinfo{author}{\bibfnamefont{J.}~\bibnamefont{Hubisz}}, and
  \bibinfo{author}{\bibfnamefont{S.~J.} \bibnamefont{Lee}},
  \bibinfo{year}{2007}, \bibinfo{journal}{Phys.Rev.}
  \textbf{\bibinfo{volume}{D76}}, \bibinfo{pages}{125015}.

\bibitem[{\citenamefont{Cs\'aki} \emph{et~al.}(2005)\citenamefont{Cs\'aki,
  Hubisz, and Meade}}]{Csaki:2005vy}
\bibinfo{author}{\bibnamefont{Cs\'aki}, \bibfnamefont{C.}},
  \bibinfo{author}{\bibfnamefont{J.}~\bibnamefont{Hubisz}}, and
  \bibinfo{author}{\bibfnamefont{P.}~\bibnamefont{Meade}},
  \bibinfo{year}{2005}, in \emph{\bibinfo{booktitle}{{Physics in D $\geq$ 4.
  Proceedings, Theoretical Advanced Study Institute in elementary particle
  physics, TASI 2004, Boulder, USA, June 6-July 2, 2004}}}, pp.
  \bibinfo{pages}{703--776}, \eprint{hep-ph/0510275}.

\bibitem[{\citenamefont{Cs\'aki} \emph{et~al.}(2015)\citenamefont{Cs\'aki,
  Kuflik, Lombardo, and Slone}}]{Csaki:2015fba}
\bibinfo{author}{\bibnamefont{Cs\'aki}, \bibfnamefont{C.}},
  \bibinfo{author}{\bibfnamefont{E.}~\bibnamefont{Kuflik}},
  \bibinfo{author}{\bibfnamefont{S.}~\bibnamefont{Lombardo}}, and
  \bibinfo{author}{\bibfnamefont{O.}~\bibnamefont{Slone}},
  \bibinfo{year}{2015}, \bibinfo{journal}{Phys. Rev.}
  \textbf{\bibinfo{volume}{D92}}(\bibinfo{number}{7}), \bibinfo{pages}{073008}.

\bibitem[{\citenamefont{Cs\'aki} \emph{et~al.}(2006)\citenamefont{Cs\'aki,
  Marandella, Shirman, and Strumia}}]{Csaki:2005fc}
\bibinfo{author}{\bibnamefont{Cs\'aki}, \bibfnamefont{C.}},
  \bibinfo{author}{\bibfnamefont{G.}~\bibnamefont{Marandella}},
  \bibinfo{author}{\bibfnamefont{Y.}~\bibnamefont{Shirman}}, and
  \bibinfo{author}{\bibfnamefont{A.}~\bibnamefont{Strumia}},
  \bibinfo{year}{2006}, \bibinfo{journal}{Phys.Rev.}
  \textbf{\bibinfo{volume}{D73}}, \bibinfo{pages}{035006}.

\bibitem[{\citenamefont{Cs\'aki} \emph{et~al.}(2012)\citenamefont{Cs\'aki,
  Randall, and Terning}}]{Csaki:2012fh}
\bibinfo{author}{\bibnamefont{Cs\'aki}, \bibfnamefont{C.}},
  \bibinfo{author}{\bibfnamefont{L.}~\bibnamefont{Randall}}, and
  \bibinfo{author}{\bibfnamefont{J.}~\bibnamefont{Terning}},
  \bibinfo{year}{2012}, \bibinfo{journal}{Phys. Rev.}
  \textbf{\bibinfo{volume}{D86}}, \bibinfo{pages}{075009}.

\bibitem[{\citenamefont{Cs\'aki} \emph{et~al.}(2011)\citenamefont{Cs\'aki,
  Shirman, and Terning}}]{Csaki:2011xn}
\bibinfo{author}{\bibnamefont{Cs\'aki}, \bibfnamefont{C.}},
  \bibinfo{author}{\bibfnamefont{Y.}~\bibnamefont{Shirman}}, and
  \bibinfo{author}{\bibfnamefont{J.}~\bibnamefont{Terning}},
  \bibinfo{year}{2011}, \bibinfo{journal}{Phys.Rev.}
  \textbf{\bibinfo{volume}{D84}}, \bibinfo{pages}{095011}.

\bibitem[{\citenamefont{Curtin and Verhaaren}(2015)}]{Curtin:2015fna}
\bibinfo{author}{\bibnamefont{Curtin}, \bibfnamefont{D.}}, and
  \bibinfo{author}{\bibfnamefont{C.~B.} \bibnamefont{Verhaaren}},
  \bibinfo{year}{2015}, \eprint{1506.06141}.

\bibitem[{\citenamefont{Das} \emph{et~al.}(2011)\citenamefont{Das, Hundi, and
  SenGupta}}]{Das:2011fb}
\bibinfo{author}{\bibnamefont{Das}, \bibfnamefont{A.}},
  \bibinfo{author}{\bibfnamefont{R.}~\bibnamefont{Hundi}}, and
  \bibinfo{author}{\bibfnamefont{S.}~\bibnamefont{SenGupta}},
  \bibinfo{year}{2011}, \bibinfo{journal}{Phys.Rev.}
  \textbf{\bibinfo{volume}{D83}}, \bibinfo{pages}{116003}.

\bibitem[{\citenamefont{Davoudiasl}
  \emph{et~al.}(2000{\natexlab{a}})\citenamefont{Davoudiasl, Hewett, and
  Rizzo}}]{Davoudiasl:1999tf}
\bibinfo{author}{\bibnamefont{Davoudiasl}, \bibfnamefont{H.}},
  \bibinfo{author}{\bibfnamefont{J.}~\bibnamefont{Hewett}}, and
  \bibinfo{author}{\bibfnamefont{T.}~\bibnamefont{Rizzo}},
  \bibinfo{year}{2000}{\natexlab{a}}, \bibinfo{journal}{Phys.Lett.}
  \textbf{\bibinfo{volume}{B473}}, \bibinfo{pages}{43}.

\bibitem[{\citenamefont{Davoudiasl}
  \emph{et~al.}(2000{\natexlab{b}})\citenamefont{Davoudiasl, Hewett, and
  Rizzo}}]{Davoudiasl:1999jd}
\bibinfo{author}{\bibnamefont{Davoudiasl}, \bibfnamefont{H.}},
  \bibinfo{author}{\bibfnamefont{J.}~\bibnamefont{Hewett}}, and
  \bibinfo{author}{\bibfnamefont{T.}~\bibnamefont{Rizzo}},
  \bibinfo{year}{2000}{\natexlab{b}}, \bibinfo{journal}{Phys.Rev.Lett.}
  \textbf{\bibinfo{volume}{84}}, \bibinfo{pages}{2080}.

\bibitem[{\citenamefont{Davoudiasl}
  \emph{et~al.}(2001)\citenamefont{Davoudiasl, Hewett, and
  Rizzo}}]{Davoudiasl:2000wi}
\bibinfo{author}{\bibnamefont{Davoudiasl}, \bibfnamefont{H.}},
  \bibinfo{author}{\bibfnamefont{J.}~\bibnamefont{Hewett}}, and
  \bibinfo{author}{\bibfnamefont{T.}~\bibnamefont{Rizzo}},
  \bibinfo{year}{2001}, \bibinfo{journal}{Phys.Rev.}
  \textbf{\bibinfo{volume}{D63}}, \bibinfo{pages}{075004}.

\bibitem[{\citenamefont{Davoudiasl}
  \emph{et~al.}(2006)\citenamefont{Davoudiasl, Lillie, and
  Rizzo}}]{Davoudiasl:2005uu}
\bibinfo{author}{\bibnamefont{Davoudiasl}, \bibfnamefont{H.}},
  \bibinfo{author}{\bibfnamefont{B.}~\bibnamefont{Lillie}}, and
  \bibinfo{author}{\bibfnamefont{T.~G.} \bibnamefont{Rizzo}},
  \bibinfo{year}{2006}, \bibinfo{journal}{JHEP}
  \textbf{\bibinfo{volume}{0608}}, \bibinfo{pages}{042}.

\bibitem[{\citenamefont{De~Simone} \emph{et~al.}(2013)\citenamefont{De~Simone,
  Matsedonskyi, Rattazzi, and Wulzer}}]{DeSimone:2012fs}
\bibinfo{author}{\bibnamefont{De~Simone}, \bibfnamefont{A.}},
  \bibinfo{author}{\bibfnamefont{O.}~\bibnamefont{Matsedonskyi}},
  \bibinfo{author}{\bibfnamefont{R.}~\bibnamefont{Rattazzi}}, and
  \bibinfo{author}{\bibfnamefont{A.}~\bibnamefont{Wulzer}},
  \bibinfo{year}{2013}, \bibinfo{journal}{JHEP} \textbf{\bibinfo{volume}{04}},
  \bibinfo{pages}{004}.

\bibitem[{\citenamefont{Delgado}
  \emph{et~al.}(2008{\natexlab{a}})\citenamefont{Delgado, Espinosa, No, and
  Quiros}}]{unother6}
\bibinfo{author}{\bibnamefont{Delgado}, \bibfnamefont{A.}},
  \bibinfo{author}{\bibfnamefont{J.}~\bibnamefont{Espinosa}},
  \bibinfo{author}{\bibfnamefont{J.}~\bibnamefont{No}}, and
  \bibinfo{author}{\bibfnamefont{M.}~\bibnamefont{Quiros}},
  \bibinfo{year}{2008}{\natexlab{a}}, \bibinfo{journal}{JHEP}
  \textbf{\bibinfo{volume}{0811}}, \bibinfo{pages}{071}.

\bibitem[{\citenamefont{Delgado}
  \emph{et~al.}(2008{\natexlab{b}})\citenamefont{Delgado, Espinosa, No, and
  Quiros}}]{unother5}
\bibinfo{author}{\bibnamefont{Delgado}, \bibfnamefont{A.}},
  \bibinfo{author}{\bibfnamefont{J.}~\bibnamefont{Espinosa}},
  \bibinfo{author}{\bibfnamefont{J.}~\bibnamefont{No}}, and
  \bibinfo{author}{\bibfnamefont{M.}~\bibnamefont{Quiros}},
  \bibinfo{year}{2008}{\natexlab{b}}, \bibinfo{journal}{JHEP}
  \textbf{\bibinfo{volume}{0804}}, \bibinfo{pages}{028}.

\bibitem[{\citenamefont{Delgado} \emph{et~al.}(2007)\citenamefont{Delgado,
  Espinosa, and Quiros}}]{unother4}
\bibinfo{author}{\bibnamefont{Delgado}, \bibfnamefont{A.}},
  \bibinfo{author}{\bibfnamefont{J.~R.} \bibnamefont{Espinosa}}, and
  \bibinfo{author}{\bibfnamefont{M.}~\bibnamefont{Quiros}},
  \bibinfo{year}{2007}, \bibinfo{journal}{JHEP}
  \textbf{\bibinfo{volume}{0710}}, \bibinfo{pages}{094}.

\bibitem[{\citenamefont{Delgado and Tait}(2005)}]{Delgado:2005fq}
\bibinfo{author}{\bibnamefont{Delgado}, \bibfnamefont{A.}}, and
  \bibinfo{author}{\bibfnamefont{T.~M.} \bibnamefont{Tait}},
  \bibinfo{year}{2005}, \bibinfo{journal}{JHEP}
  \textbf{\bibinfo{volume}{0507}}, \bibinfo{pages}{023}.

\bibitem[{\citenamefont{Desai} \emph{et~al.}(2013)\citenamefont{Desai, Maitra,
  and Mukhopadhyaya}}]{Desai:2013pga}
\bibinfo{author}{\bibnamefont{Desai}, \bibfnamefont{N.}},
  \bibinfo{author}{\bibfnamefont{U.}~\bibnamefont{Maitra}}, and
  \bibinfo{author}{\bibfnamefont{B.}~\bibnamefont{Mukhopadhyaya}},
  \bibinfo{year}{2013}, \bibinfo{journal}{JHEP}
  \textbf{\bibinfo{volume}{1310}}, \bibinfo{pages}{093}.

\bibitem[{\citenamefont{Dey} \emph{et~al.}(2010)\citenamefont{Dey,
  Mukhopadhyaya, and SenGupta}}]{Dey:2009gf}
\bibinfo{author}{\bibnamefont{Dey}, \bibfnamefont{P.}},
  \bibinfo{author}{\bibfnamefont{B.}~\bibnamefont{Mukhopadhyaya}}, and
  \bibinfo{author}{\bibfnamefont{S.}~\bibnamefont{SenGupta}},
  \bibinfo{year}{2010}, \bibinfo{journal}{Phys.Rev.}
  \textbf{\bibinfo{volume}{D81}}, \bibinfo{pages}{036011}.

\bibitem[{\citenamefont{Dib} \emph{et~al.}(2003)\citenamefont{Dib, Rosenfeld,
  and Zerwekh}}]{Dib:2003zj}
\bibinfo{author}{\bibnamefont{Dib}, \bibfnamefont{C.}},
  \bibinfo{author}{\bibfnamefont{R.}~\bibnamefont{Rosenfeld}}, and
  \bibinfo{author}{\bibfnamefont{A.}~\bibnamefont{Zerwekh}},
  \bibinfo{year}{2003}, \eprint{hep-ph/0302068}.

\bibitem[{\citenamefont{Dietrich} \emph{et~al.}(2005)\citenamefont{Dietrich,
  Sannino, and Tuominen}}]{Dietrich:2005jn}
\bibinfo{author}{\bibnamefont{Dietrich}, \bibfnamefont{D.~D.}},
  \bibinfo{author}{\bibfnamefont{F.}~\bibnamefont{Sannino}}, and
  \bibinfo{author}{\bibfnamefont{K.}~\bibnamefont{Tuominen}},
  \bibinfo{year}{2005}, \bibinfo{journal}{Phys.Rev.}
  \textbf{\bibinfo{volume}{D72}}, \bibinfo{pages}{055001}.

\bibitem[{\citenamefont{Dimopoulos and Susskind}(1979)}]{Dimopoulos:1979es}
\bibinfo{author}{\bibnamefont{Dimopoulos}, \bibfnamefont{S.}}, and
  \bibinfo{author}{\bibfnamefont{L.}~\bibnamefont{Susskind}},
  \bibinfo{year}{1979}, \bibinfo{journal}{Nucl.Phys.}
  \textbf{\bibinfo{volume}{B155}}, \bibinfo{pages}{237}.

\bibitem[{\citenamefont{Dissertori}
  \emph{et~al.}(2010)\citenamefont{Dissertori, Furlan, Moortgat, and
  Nef}}]{Dissertori:2010ug}
\bibinfo{author}{\bibnamefont{Dissertori}, \bibfnamefont{G.}},
  \bibinfo{author}{\bibfnamefont{E.}~\bibnamefont{Furlan}},
  \bibinfo{author}{\bibfnamefont{F.}~\bibnamefont{Moortgat}}, and
  \bibinfo{author}{\bibfnamefont{P.}~\bibnamefont{Nef}}, \bibinfo{year}{2010},
  \bibinfo{journal}{JHEP} \textbf{\bibinfo{volume}{1009}},
  \bibinfo{pages}{019}.

\bibitem[{\citenamefont{Dolan} \emph{et~al.}(2013)\citenamefont{Dolan, Englert,
  and Spannowsky}}]{Dolan:2012ac}
\bibinfo{author}{\bibnamefont{Dolan}, \bibfnamefont{M.~J.}},
  \bibinfo{author}{\bibfnamefont{C.}~\bibnamefont{Englert}}, and
  \bibinfo{author}{\bibfnamefont{M.}~\bibnamefont{Spannowsky}},
  \bibinfo{year}{2013}, \bibinfo{journal}{Phys.Rev.}
  \textbf{\bibinfo{volume}{D87}}, \bibinfo{pages}{055002}.

\bibitem[{\citenamefont{Dugan} \emph{et~al.}(1985)\citenamefont{Dugan, Georgi,
  and Kaplan}}]{Dugan:1984hq}
\bibinfo{author}{\bibnamefont{Dugan}, \bibfnamefont{M.~J.}},
  \bibinfo{author}{\bibfnamefont{H.}~\bibnamefont{Georgi}}, and
  \bibinfo{author}{\bibfnamefont{D.~B.} \bibnamefont{Kaplan}},
  \bibinfo{year}{1985}, \bibinfo{journal}{Nucl.Phys.}
  \textbf{\bibinfo{volume}{B254}}, \bibinfo{pages}{299}.

\bibitem[{\citenamefont{Duhrssen} \emph{et~al.}(2004)\citenamefont{Duhrssen,
  Heinemeyer, Logan, Rainwater, Weiglein} \emph{et~al.}}]{Duhrssen:2004cv}
\bibinfo{author}{\bibnamefont{Duhrssen}, \bibfnamefont{M.}},
  \bibinfo{author}{\bibfnamefont{S.}~\bibnamefont{Heinemeyer}},
  \bibinfo{author}{\bibfnamefont{H.}~\bibnamefont{Logan}},
  \bibinfo{author}{\bibfnamefont{D.}~\bibnamefont{Rainwater}},
  \bibinfo{author}{\bibfnamefont{G.}~\bibnamefont{Weiglein}}, \emph{et~al.},
  \bibinfo{year}{2004}, \bibinfo{journal}{Phys.Rev.}
  \textbf{\bibinfo{volume}{D70}}, \bibinfo{pages}{113009}.

\bibitem[{\citenamefont{Eboli} \emph{et~al.}(2012)\citenamefont{Eboli,
  Gonzalez-Fraile, and Gonzalez-Garcia}}]{Eboli:2011ye}
\bibinfo{author}{\bibnamefont{Eboli}, \bibfnamefont{O.~J.~P.}},
  \bibinfo{author}{\bibfnamefont{J.}~\bibnamefont{Gonzalez-Fraile}}, and
  \bibinfo{author}{\bibfnamefont{M.~C.} \bibnamefont{Gonzalez-Garcia}},
  \bibinfo{year}{2012}, \bibinfo{journal}{Phys. Rev.}
  \textbf{\bibinfo{volume}{D85}}, \bibinfo{pages}{055019}.

\bibitem[{\citenamefont{Eichten and Lane}(1980)}]{Eichten:1979ah}
\bibinfo{author}{\bibnamefont{Eichten}, \bibfnamefont{E.}}, and
  \bibinfo{author}{\bibfnamefont{K.}~\bibnamefont{Lane}}, \bibinfo{year}{1980},
  \bibinfo{journal}{Phys. Lett.} \textbf{\bibinfo{volume}{B90}},
  \bibinfo{pages}{125}.

\bibitem[{\citenamefont{Elander and Piai}(2011)}]{Elander:2010wd}
\bibinfo{author}{\bibnamefont{Elander}, \bibfnamefont{D.}}, and
  \bibinfo{author}{\bibfnamefont{M.}~\bibnamefont{Piai}}, \bibinfo{year}{2011},
  \bibinfo{journal}{JHEP} \textbf{\bibinfo{volume}{1101}},
  \bibinfo{pages}{026}.

\bibitem[{\citenamefont{Elander and Piai}(2012)}]{Elander:2011aa}
\bibinfo{author}{\bibnamefont{Elander}, \bibfnamefont{D.}}, and
  \bibinfo{author}{\bibfnamefont{M.}~\bibnamefont{Piai}}, \bibinfo{year}{2012},
  \bibinfo{journal}{Nucl.Phys.} \textbf{\bibinfo{volume}{B864}},
  \bibinfo{pages}{241}.

\bibitem[{\citenamefont{Elander and Piai}(2013)}]{Elander:2012fk}
\bibinfo{author}{\bibnamefont{Elander}, \bibfnamefont{D.}}, and
  \bibinfo{author}{\bibfnamefont{M.}~\bibnamefont{Piai}}, \bibinfo{year}{2013},
  \bibinfo{journal}{Nucl.Phys.} \textbf{\bibinfo{volume}{B867}},
  \bibinfo{pages}{779Ð809}.

\bibitem[{\citenamefont{Elias-Mir\'o}
  \emph{et~al.}(2013)\citenamefont{Elias-Mir\'o, Espinosa, Masso, and
  Pomarol}}]{Elias-Miro:2013gya}
\bibinfo{author}{\bibnamefont{Elias-Mir\'o}, \bibfnamefont{J.}},
  \bibinfo{author}{\bibfnamefont{J.~R.} \bibnamefont{Espinosa}},
  \bibinfo{author}{\bibfnamefont{E.}~\bibnamefont{Masso}}, and
  \bibinfo{author}{\bibfnamefont{A.}~\bibnamefont{Pomarol}},
  \bibinfo{year}{2013}, \bibinfo{journal}{JHEP} \textbf{\bibinfo{volume}{08}},
  \bibinfo{pages}{033}.

\bibitem[{\citenamefont{Elias-Mir\'o}
  \emph{et~al.}(2015)\citenamefont{Elias-Mir\'o, Espinosa, and
  Pomarol}}]{Elias-Miro:2014eia}
\bibinfo{author}{\bibnamefont{Elias-Mir\'o}, \bibfnamefont{J.}},
  \bibinfo{author}{\bibfnamefont{J.~R.} \bibnamefont{Espinosa}}, and
  \bibinfo{author}{\bibfnamefont{A.}~\bibnamefont{Pomarol}},
  \bibinfo{year}{2015}, \bibinfo{journal}{Phys. Lett.}
  \textbf{\bibinfo{volume}{B747}}, \bibinfo{pages}{272}.

\bibitem[{\citenamefont{Elias-Mir\'o}
  \emph{et~al.}(2014)\citenamefont{Elias-Mir\'o, Grojean, Gupta, and
  Marzocca}}]{Elias-Miro:2013eta}
\bibinfo{author}{\bibnamefont{Elias-Mir\'o}, \bibfnamefont{J.}},
  \bibinfo{author}{\bibfnamefont{C.}~\bibnamefont{Grojean}},
  \bibinfo{author}{\bibfnamefont{R.~S.} \bibnamefont{Gupta}}, and
  \bibinfo{author}{\bibfnamefont{D.}~\bibnamefont{Marzocca}},
  \bibinfo{year}{2014}, \bibinfo{journal}{JHEP} \textbf{\bibinfo{volume}{05}},
  \bibinfo{pages}{019}.

\bibitem[{\citenamefont{Ellis} \emph{et~al.}(1976)\citenamefont{Ellis,
  Gaillard, and Nanopoulos}}]{Ellis:1975ap}
\bibinfo{author}{\bibnamefont{Ellis}, \bibfnamefont{J.~R.}},
  \bibinfo{author}{\bibfnamefont{M.~K.} \bibnamefont{Gaillard}}, and
  \bibinfo{author}{\bibfnamefont{D.~V.} \bibnamefont{Nanopoulos}},
  \bibinfo{year}{1976}, \bibinfo{journal}{Nucl.Phys.}
  \textbf{\bibinfo{volume}{B106}}, \bibinfo{pages}{292}.

\bibitem[{\citenamefont{Ellis} \emph{et~al.}(2012)\citenamefont{Ellis, Hornig,
  Roy, Krohn, and Schwartz}}]{Ellis:2012sn}
\bibinfo{author}{\bibnamefont{Ellis}, \bibfnamefont{S.~D.}},
  \bibinfo{author}{\bibfnamefont{A.}~\bibnamefont{Hornig}},
  \bibinfo{author}{\bibfnamefont{T.~S.} \bibnamefont{Roy}},
  \bibinfo{author}{\bibfnamefont{D.}~\bibnamefont{Krohn}}, and
  \bibinfo{author}{\bibfnamefont{M.~D.} \bibnamefont{Schwartz}},
  \bibinfo{year}{2012}, \bibinfo{journal}{Phys.Rev.Lett.}
  \textbf{\bibinfo{volume}{108}}, \bibinfo{pages}{182003}.

\bibitem[{\citenamefont{Englert}
  \emph{et~al.}(2012{\natexlab{a}})\citenamefont{Englert, Jaeckel, Re, and
  Spannowsky}}]{Englert:2011us}
\bibinfo{author}{\bibnamefont{Englert}, \bibfnamefont{C.}},
  \bibinfo{author}{\bibfnamefont{J.}~\bibnamefont{Jaeckel}},
  \bibinfo{author}{\bibfnamefont{E.}~\bibnamefont{Re}}, and
  \bibinfo{author}{\bibfnamefont{M.}~\bibnamefont{Spannowsky}},
  \bibinfo{year}{2012}{\natexlab{a}}, \bibinfo{journal}{Phys.Rev.}
  \textbf{\bibinfo{volume}{D85}}, \bibinfo{pages}{035008}.

\bibitem[{\citenamefont{Englert}
  \emph{et~al.}(2012{\natexlab{b}})\citenamefont{Englert, Netto, Spannowsky,
  and Terning}}]{Englert:2012cb}
\bibinfo{author}{\bibnamefont{Englert}, \bibfnamefont{C.}},
  \bibinfo{author}{\bibfnamefont{D.~G.} \bibnamefont{Netto}},
  \bibinfo{author}{\bibfnamefont{M.}~\bibnamefont{Spannowsky}}, and
  \bibinfo{author}{\bibfnamefont{J.}~\bibnamefont{Terning}},
  \bibinfo{year}{2012}{\natexlab{b}}, \bibinfo{journal}{Phys.Rev.}
  \textbf{\bibinfo{volume}{D86}}, \bibinfo{pages}{035010}.

\bibitem[{\citenamefont{Englert}
  \emph{et~al.}(2012{\natexlab{c}})\citenamefont{Englert, Spannowsky, Stancato,
  and Terning}}]{Englert:2012dq}
\bibinfo{author}{\bibnamefont{Englert}, \bibfnamefont{C.}},
  \bibinfo{author}{\bibfnamefont{M.}~\bibnamefont{Spannowsky}},
  \bibinfo{author}{\bibfnamefont{D.}~\bibnamefont{Stancato}}, and
  \bibinfo{author}{\bibfnamefont{J.}~\bibnamefont{Terning}},
  \bibinfo{year}{2012}{\natexlab{c}}, \bibinfo{journal}{Phys.Rev.}
  \textbf{\bibinfo{volume}{D85}}, \bibinfo{pages}{095003}.

\bibitem[{\citenamefont{Englert and Brout}(1964)}]{Englert:1964et}
\bibinfo{author}{\bibnamefont{Englert}, \bibfnamefont{F.}}, and
  \bibinfo{author}{\bibfnamefont{R.}~\bibnamefont{Brout}},
  \bibinfo{year}{1964}, \bibinfo{journal}{Phys.Rev.Lett.}
  \textbf{\bibinfo{volume}{13}}, \bibinfo{pages}{321}.

\bibitem[{\citenamefont{Espinosa} \emph{et~al.}(2010)\citenamefont{Espinosa,
  Grojean, and Muhlleitner}}]{Espinosa:2010vn}
\bibinfo{author}{\bibnamefont{Espinosa}, \bibfnamefont{J.}},
  \bibinfo{author}{\bibfnamefont{C.}~\bibnamefont{Grojean}}, and
  \bibinfo{author}{\bibfnamefont{M.}~\bibnamefont{Muhlleitner}},
  \bibinfo{year}{2010}, \bibinfo{journal}{JHEP}
  \textbf{\bibinfo{volume}{1005}}, \bibinfo{pages}{065}.

\bibitem[{\citenamefont{Espinosa}
  \emph{et~al.}(2012{\natexlab{a}})\citenamefont{Espinosa, Grojean, and
  Muhlleitner}}]{Espinosa:2012qj}
\bibinfo{author}{\bibnamefont{Espinosa}, \bibfnamefont{J.}},
  \bibinfo{author}{\bibfnamefont{C.}~\bibnamefont{Grojean}}, and
  \bibinfo{author}{\bibfnamefont{M.}~\bibnamefont{Muhlleitner}},
  \bibinfo{year}{2012}{\natexlab{a}}, \bibinfo{journal}{EPJ Web Conf.}
  \textbf{\bibinfo{volume}{28}}, \bibinfo{pages}{08004}.

\bibitem[{\citenamefont{Espinosa}
  \emph{et~al.}(2012{\natexlab{b}})\citenamefont{Espinosa, Grojean,
  Muhlleitner, and Trott}}]{Espinosa:2012ir}
\bibinfo{author}{\bibnamefont{Espinosa}, \bibfnamefont{J.}},
  \bibinfo{author}{\bibfnamefont{C.}~\bibnamefont{Grojean}},
  \bibinfo{author}{\bibfnamefont{M.}~\bibnamefont{Muhlleitner}}, and
  \bibinfo{author}{\bibfnamefont{M.}~\bibnamefont{Trott}},
  \bibinfo{year}{2012}{\natexlab{b}}, \bibinfo{journal}{JHEP}
  \textbf{\bibinfo{volume}{1205}}, \bibinfo{pages}{097}.

\bibitem[{\citenamefont{Espinosa} \emph{et~al.}(2015)\citenamefont{Espinosa,
  Grojean, Panico, Pomarol, Pujol{\`a}s, and Servant}}]{Espinosa:2015eda}
\bibinfo{author}{\bibnamefont{Espinosa}, \bibfnamefont{J.~R.}},
  \bibinfo{author}{\bibfnamefont{C.}~\bibnamefont{Grojean}},
  \bibinfo{author}{\bibfnamefont{G.}~\bibnamefont{Panico}},
  \bibinfo{author}{\bibfnamefont{A.}~\bibnamefont{Pomarol}},
  \bibinfo{author}{\bibfnamefont{O.}~\bibnamefont{Pujol{\`a}s}}, and
  \bibinfo{author}{\bibfnamefont{G.}~\bibnamefont{Servant}},
  \bibinfo{year}{2015}, \eprint{1506.09217}.

\bibitem[{\citenamefont{Espinosa and Gunion}(1999)}]{unother1}
\bibinfo{author}{\bibnamefont{Espinosa}, \bibfnamefont{J.~R.}}, and
  \bibinfo{author}{\bibfnamefont{J.~F.} \bibnamefont{Gunion}},
  \bibinfo{year}{1999}, \bibinfo{journal}{Phys.Rev.Lett.}
  \textbf{\bibinfo{volume}{82}}, \bibinfo{pages}{1084}.

\bibitem[{\citenamefont{Evans} \emph{et~al.}(2011)\citenamefont{Evans,
  Galloway, Luty, and Tacchi}}]{Evans:2010ed}
\bibinfo{author}{\bibnamefont{Evans}, \bibfnamefont{J.~A.}},
  \bibinfo{author}{\bibfnamefont{J.}~\bibnamefont{Galloway}},
  \bibinfo{author}{\bibfnamefont{M.~A.} \bibnamefont{Luty}}, and
  \bibinfo{author}{\bibfnamefont{R.~A.} \bibnamefont{Tacchi}},
  \bibinfo{year}{2011}, \bibinfo{journal}{JHEP}
  \textbf{\bibinfo{volume}{1104}}, \bibinfo{pages}{003}.

\bibitem[{\citenamefont{Evans and Tuominen}(2013)}]{Evans:2013vca}
\bibinfo{author}{\bibnamefont{Evans}, \bibfnamefont{N.}}, and
  \bibinfo{author}{\bibfnamefont{K.}~\bibnamefont{Tuominen}},
  \bibinfo{year}{2013}, \bibinfo{journal}{Phys.Rev.}
  \textbf{\bibinfo{volume}{D87}}, \bibinfo{pages}{086003}.

\bibitem[{\citenamefont{Falkowski} \emph{et~al.}(2011)\citenamefont{Falkowski,
  Grojean, Kaminska, Pokorski, and Weiler}}]{Falkowski:2011ua}
\bibinfo{author}{\bibnamefont{Falkowski}, \bibfnamefont{A.}},
  \bibinfo{author}{\bibfnamefont{C.}~\bibnamefont{Grojean}},
  \bibinfo{author}{\bibfnamefont{A.}~\bibnamefont{Kaminska}},
  \bibinfo{author}{\bibfnamefont{S.}~\bibnamefont{Pokorski}}, and
  \bibinfo{author}{\bibfnamefont{A.}~\bibnamefont{Weiler}},
  \bibinfo{year}{2011}, \bibinfo{journal}{JHEP} \textbf{\bibinfo{volume}{11}},
  \bibinfo{pages}{028}.

\bibitem[{\citenamefont{Falkowski and
  Perez-Victoria}(2009{\natexlab{a}})}]{Falkowski:2009uy}
\bibinfo{author}{\bibnamefont{Falkowski}, \bibfnamefont{A.}}, and
  \bibinfo{author}{\bibfnamefont{M.}~\bibnamefont{Perez-Victoria}},
  \bibinfo{year}{2009}{\natexlab{a}}, \bibinfo{journal}{JHEP}
  \textbf{\bibinfo{volume}{0912}}, \bibinfo{pages}{061}.

\bibitem[{\citenamefont{Falkowski and
  Perez-Victoria}(2009{\natexlab{b}})}]{Falkowski:2008yr}
\bibinfo{author}{\bibnamefont{Falkowski}, \bibfnamefont{A.}}, and
  \bibinfo{author}{\bibfnamefont{M.}~\bibnamefont{Perez-Victoria}},
  \bibinfo{year}{2009}{\natexlab{b}}, \bibinfo{journal}{Phys.Rev.}
  \textbf{\bibinfo{volume}{D79}}, \bibinfo{pages}{035005}.

\bibitem[{\citenamefont{Falkowski} \emph{et~al.}(2012)\citenamefont{Falkowski,
  Rychkov, and Urbano}}]{Falkowski:2012vh}
\bibinfo{author}{\bibnamefont{Falkowski}, \bibfnamefont{A.}},
  \bibinfo{author}{\bibfnamefont{S.}~\bibnamefont{Rychkov}}, and
  \bibinfo{author}{\bibfnamefont{A.}~\bibnamefont{Urbano}},
  \bibinfo{year}{2012}, \bibinfo{journal}{JHEP}
  \textbf{\bibinfo{volume}{1204}}, \bibinfo{pages}{073}.

\bibitem[{\citenamefont{Fan} \emph{et~al.}(2011)\citenamefont{Fan, Reece, and
  Ruderman}}]{Fan:2011yu}
\bibinfo{author}{\bibnamefont{Fan}, \bibfnamefont{J.}},
  \bibinfo{author}{\bibfnamefont{M.}~\bibnamefont{Reece}}, and
  \bibinfo{author}{\bibfnamefont{J.~T.} \bibnamefont{Ruderman}},
  \bibinfo{year}{2011}, \bibinfo{journal}{JHEP}
  \textbf{\bibinfo{volume}{1111}}, \bibinfo{pages}{012}.

\bibitem[{\citenamefont{Fitzpatrick and Shih}(2011)}]{Fitzpatrick:2011hh}
\bibinfo{author}{\bibnamefont{Fitzpatrick}, \bibfnamefont{A.~L.}}, and
  \bibinfo{author}{\bibfnamefont{D.}~\bibnamefont{Shih}}, \bibinfo{year}{2011},
  \bibinfo{journal}{JHEP} \textbf{\bibinfo{volume}{1110}},
  \bibinfo{pages}{113}.

\bibitem[{\citenamefont{Fodor} \emph{et~al.}(2012)\citenamefont{Fodor, Holland,
  Kuti, Nogradi, Schroeder} \emph{et~al.}}]{Fodor:2012ty}
\bibinfo{author}{\bibnamefont{Fodor}, \bibfnamefont{Z.}},
  \bibinfo{author}{\bibfnamefont{K.}~\bibnamefont{Holland}},
  \bibinfo{author}{\bibfnamefont{J.}~\bibnamefont{Kuti}},
  \bibinfo{author}{\bibfnamefont{D.}~\bibnamefont{Nogradi}},
  \bibinfo{author}{\bibfnamefont{C.}~\bibnamefont{Schroeder}}, \emph{et~al.},
  \bibinfo{year}{2012}, \bibinfo{journal}{Phys.Lett.}
  \textbf{\bibinfo{volume}{B718}}, \bibinfo{pages}{657}.

\bibitem[{\citenamefont{Foot} \emph{et~al.}(2010)\citenamefont{Foot,
  Kobakhidze, and McDonald}}]{Foot:2008tz}
\bibinfo{author}{\bibnamefont{Foot}, \bibfnamefont{R.}},
  \bibinfo{author}{\bibfnamefont{A.}~\bibnamefont{Kobakhidze}}, and
  \bibinfo{author}{\bibfnamefont{K.~L.} \bibnamefont{McDonald}},
  \bibinfo{year}{2010}, \bibinfo{journal}{Eur.Phys.J.}
  \textbf{\bibinfo{volume}{C68}}, \bibinfo{pages}{421}.

\bibitem[{\citenamefont{Fox} \emph{et~al.}(2007)\citenamefont{Fox, Rajaraman,
  and Shirman}}]{Fox:2007sy}
\bibinfo{author}{\bibnamefont{Fox}, \bibfnamefont{P.~J.}},
  \bibinfo{author}{\bibfnamefont{A.}~\bibnamefont{Rajaraman}}, and
  \bibinfo{author}{\bibfnamefont{Y.}~\bibnamefont{Shirman}},
  \bibinfo{year}{2007}, \bibinfo{journal}{Phys.Rev.}
  \textbf{\bibinfo{volume}{D76}}, \bibinfo{pages}{075004}.

\bibitem[{\citenamefont{Fox} \emph{et~al.}(2011)\citenamefont{Fox,
  Tucker-Smith, and Weiner}}]{Fox:2011qc}
\bibinfo{author}{\bibnamefont{Fox}, \bibfnamefont{P.~J.}},
  \bibinfo{author}{\bibfnamefont{D.}~\bibnamefont{Tucker-Smith}}, and
  \bibinfo{author}{\bibfnamefont{N.}~\bibnamefont{Weiner}},
  \bibinfo{year}{2011}, \bibinfo{journal}{JHEP}
  \textbf{\bibinfo{volume}{1106}}, \bibinfo{pages}{127}.

\bibitem[{\citenamefont{Frank} \emph{et~al.}(2013)\citenamefont{Frank,
  Pourtolami, and Toharia}}]{Frank:2013un}
\bibinfo{author}{\bibnamefont{Frank}, \bibfnamefont{M.}},
  \bibinfo{author}{\bibfnamefont{N.}~\bibnamefont{Pourtolami}}, and
  \bibinfo{author}{\bibfnamefont{M.}~\bibnamefont{Toharia}},
  \bibinfo{year}{2013}, \eprint{1301.7692}.

\bibitem[{Fujii \emph{et~al.}(2015)\citenamefont{Fujii}
  \emph{et~al.}}]{Fujii:2015jha}
\bibinfo{author}{\bibnamefont{Fujii}, \bibfnamefont{K.}}, \emph{et~al.},
  \bibinfo{year}{2015}, \eprint{1506.05992}.

\bibitem[{\citenamefont{Fukushima} \emph{et~al.}(2011)\citenamefont{Fukushima,
  Kitano, and Yamaguchi}}]{Fukushima:2010pm}
\bibinfo{author}{\bibnamefont{Fukushima}, \bibfnamefont{H.}},
  \bibinfo{author}{\bibfnamefont{R.}~\bibnamefont{Kitano}}, and
  \bibinfo{author}{\bibfnamefont{M.}~\bibnamefont{Yamaguchi}},
  \bibinfo{year}{2011}, \bibinfo{journal}{JHEP}
  \textbf{\bibinfo{volume}{1101}}, \bibinfo{pages}{111}.

\bibitem[{\citenamefont{Galloway} \emph{et~al.}(2010)\citenamefont{Galloway,
  Evans, Luty, and Tacchi}}]{Galloway:2010bp}
\bibinfo{author}{\bibnamefont{Galloway}, \bibfnamefont{J.}},
  \bibinfo{author}{\bibfnamefont{J.~A.} \bibnamefont{Evans}},
  \bibinfo{author}{\bibfnamefont{M.~A.} \bibnamefont{Luty}}, and
  \bibinfo{author}{\bibfnamefont{R.~A.} \bibnamefont{Tacchi}},
  \bibinfo{year}{2010}, \bibinfo{journal}{JHEP}
  \textbf{\bibinfo{volume}{1010}}, \bibinfo{pages}{086}.

\bibitem[{\citenamefont{Galloway}
  \emph{et~al.}(2009{\natexlab{a}})\citenamefont{Galloway, McElrath, McRaven,
  and Terning}}]{Galloway:2009xn}
\bibinfo{author}{\bibnamefont{Galloway}, \bibfnamefont{J.}},
  \bibinfo{author}{\bibfnamefont{B.}~\bibnamefont{McElrath}},
  \bibinfo{author}{\bibfnamefont{J.}~\bibnamefont{McRaven}}, and
  \bibinfo{author}{\bibfnamefont{J.}~\bibnamefont{Terning}},
  \bibinfo{year}{2009}{\natexlab{a}}, \bibinfo{journal}{JHEP}
  \textbf{\bibinfo{volume}{0911}}, \bibinfo{pages}{031}.

\bibitem[{\citenamefont{Galloway}
  \emph{et~al.}(2009{\natexlab{b}})\citenamefont{Galloway, McRaven, and
  Terning}}]{Galloway:2008sq}
\bibinfo{author}{\bibnamefont{Galloway}, \bibfnamefont{J.}},
  \bibinfo{author}{\bibfnamefont{J.}~\bibnamefont{McRaven}}, and
  \bibinfo{author}{\bibfnamefont{J.}~\bibnamefont{Terning}},
  \bibinfo{year}{2009}{\natexlab{b}}, \bibinfo{journal}{Phys.Rev.}
  \textbf{\bibinfo{volume}{D80}}, \bibinfo{pages}{105017}.

\bibitem[{\citenamefont{Geller and Telem}(2015)}]{Geller:2014kta}
\bibinfo{author}{\bibnamefont{Geller}, \bibfnamefont{M.}}, and
  \bibinfo{author}{\bibfnamefont{O.}~\bibnamefont{Telem}},
  \bibinfo{year}{2015}, \bibinfo{journal}{Phys. Rev. Lett.}
  \textbf{\bibinfo{volume}{114}}, \bibinfo{pages}{191801}.

\bibitem[{\citenamefont{Georgi}(2007{\natexlab{a}})}]{Georgi:2007si}
\bibinfo{author}{\bibnamefont{Georgi}, \bibfnamefont{H.}},
  \bibinfo{year}{2007}{\natexlab{a}}, \bibinfo{journal}{Phys.Lett.}
  \textbf{\bibinfo{volume}{B650}}, \bibinfo{pages}{275}.

\bibitem[{\citenamefont{Georgi}(2007{\natexlab{b}})}]{Georgi:2007ek}
\bibinfo{author}{\bibnamefont{Georgi}, \bibfnamefont{H.}},
  \bibinfo{year}{2007}{\natexlab{b}}, \bibinfo{journal}{Phys.Rev.Lett.}
  \textbf{\bibinfo{volume}{98}}, \bibinfo{pages}{221601}.

\bibitem[{\citenamefont{Georgi and Kaplan}(1984)}]{Georgi:1984af}
\bibinfo{author}{\bibnamefont{Georgi}, \bibfnamefont{H.}}, and
  \bibinfo{author}{\bibfnamefont{D.~B.} \bibnamefont{Kaplan}},
  \bibinfo{year}{1984}, \bibinfo{journal}{Phys.Lett.}
  \textbf{\bibinfo{volume}{B145}}, \bibinfo{pages}{216}.

\bibitem[{\citenamefont{Georgi} \emph{et~al.}(1984)\citenamefont{Georgi,
  Kaplan, and Galison}}]{Georgi:1984ef}
\bibinfo{author}{\bibnamefont{Georgi}, \bibfnamefont{H.}},
  \bibinfo{author}{\bibfnamefont{D.~B.} \bibnamefont{Kaplan}}, and
  \bibinfo{author}{\bibfnamefont{P.}~\bibnamefont{Galison}},
  \bibinfo{year}{1984}, \bibinfo{journal}{Phys.Lett.}
  \textbf{\bibinfo{volume}{B143}}, \bibinfo{pages}{152}.

\bibitem[{\citenamefont{Georgi and Kats}(2010)}]{Georgi:2009xq}
\bibinfo{author}{\bibnamefont{Georgi}, \bibfnamefont{H.}}, and
  \bibinfo{author}{\bibfnamefont{Y.}~\bibnamefont{Kats}}, \bibinfo{year}{2010},
  \bibinfo{journal}{JHEP} \textbf{\bibinfo{volume}{1002}},
  \bibinfo{pages}{065}.

\bibitem[{\citenamefont{Georgi and Pais}(1974)}]{Georgi:1974yw}
\bibinfo{author}{\bibnamefont{Georgi}, \bibfnamefont{H.}}, and
  \bibinfo{author}{\bibfnamefont{A.}~\bibnamefont{Pais}}, \bibinfo{year}{1974},
  \bibinfo{journal}{Phys.Rev.} \textbf{\bibinfo{volume}{D10}},
  \bibinfo{pages}{539}.

\bibitem[{\citenamefont{Georgi and Pais}(1975)}]{Georgi:1975tz}
\bibinfo{author}{\bibnamefont{Georgi}, \bibfnamefont{H.}}, and
  \bibinfo{author}{\bibfnamefont{A.}~\bibnamefont{Pais}}, \bibinfo{year}{1975},
  \bibinfo{journal}{Phys.Rev.} \textbf{\bibinfo{volume}{D12}},
  \bibinfo{pages}{508}.

\bibitem[{\citenamefont{Gherghetta}(2006)}]{Gherghetta:2006ha}
\bibinfo{author}{\bibnamefont{Gherghetta}, \bibfnamefont{T.}},
  \bibinfo{year}{2006}, \bibinfo{pages}{263}, \eprint{hep-ph/0601213}.

\bibitem[{\citenamefont{Gherghetta}(2011)}]{Gherghetta:2010cj}
\bibinfo{author}{\bibnamefont{Gherghetta}, \bibfnamefont{T.}},
  \bibinfo{year}{2011}, in \emph{\bibinfo{booktitle}{{Physics of the large and
  the small, TASI 09, proceedings of the Theoretical Advanced Study Institute
  in Elementary Particle Physics, Boulder, Colorado, USA, 1-26 June 2009}}},
  pp. \bibinfo{pages}{165--232}, \eprint{1008.2570}.

\bibitem[{\citenamefont{Gherghetta and Pomarol}(2000)}]{Gherghetta:2000qt}
\bibinfo{author}{\bibnamefont{Gherghetta}, \bibfnamefont{T.}}, and
  \bibinfo{author}{\bibfnamefont{A.}~\bibnamefont{Pomarol}},
  \bibinfo{year}{2000}, \bibinfo{journal}{Nucl.Phys.}
  \textbf{\bibinfo{volume}{B586}}, \bibinfo{pages}{141}.

\bibitem[{\citenamefont{Gherghetta and Pomarol}(2011)}]{Gherghetta:2011na}
\bibinfo{author}{\bibnamefont{Gherghetta}, \bibfnamefont{T.}}, and
  \bibinfo{author}{\bibfnamefont{A.}~\bibnamefont{Pomarol}},
  \bibinfo{year}{2011}, \bibinfo{journal}{JHEP}
  \textbf{\bibinfo{volume}{1112}}, \bibinfo{pages}{069}.

\bibitem[{\citenamefont{Gianoti}(2015)}]{FabiolaG:2015}
\bibinfo{author}{\bibnamefont{Gianoti}, \bibfnamefont{F.}},
  \bibinfo{year}{2015}, \bibinfo{title}{{Talk at the EPS-HEP conference,
  Vienna, ``Outlook: physics prospects at high-energy colliders"}}.

\bibitem[{\citenamefont{Giudice} \emph{et~al.}(2007)\citenamefont{Giudice,
  Grojean, Pomarol, and Rattazzi}}]{SILH}
\bibinfo{author}{\bibnamefont{Giudice}, \bibfnamefont{G.}},
  \bibinfo{author}{\bibfnamefont{C.}~\bibnamefont{Grojean}},
  \bibinfo{author}{\bibfnamefont{A.}~\bibnamefont{Pomarol}}, and
  \bibinfo{author}{\bibfnamefont{R.}~\bibnamefont{Rattazzi}},
  \bibinfo{year}{2007}, \bibinfo{journal}{JHEP}
  \textbf{\bibinfo{volume}{0706}}, \bibinfo{pages}{045}.

\bibitem[{\citenamefont{Giudice} \emph{et~al.}(2001)\citenamefont{Giudice,
  Rattazzi, and Wells}}]{Giudice:2000av}
\bibinfo{author}{\bibnamefont{Giudice}, \bibfnamefont{G.~F.}},
  \bibinfo{author}{\bibfnamefont{R.}~\bibnamefont{Rattazzi}}, and
  \bibinfo{author}{\bibfnamefont{J.~D.} \bibnamefont{Wells}},
  \bibinfo{year}{2001}, \bibinfo{journal}{Nucl.Phys.}
  \textbf{\bibinfo{volume}{B595}}, \bibinfo{pages}{250}.

\bibitem[{\citenamefont{Godfrey} \emph{et~al.}(2012)\citenamefont{Godfrey,
  Gregoire, Kalyniak, Martin, and Moats}}]{Godfrey:2012tf}
\bibinfo{author}{\bibnamefont{Godfrey}, \bibfnamefont{S.}},
  \bibinfo{author}{\bibfnamefont{T.}~\bibnamefont{Gregoire}},
  \bibinfo{author}{\bibfnamefont{P.}~\bibnamefont{Kalyniak}},
  \bibinfo{author}{\bibfnamefont{T.~A.} \bibnamefont{Martin}}, and
  \bibinfo{author}{\bibfnamefont{K.}~\bibnamefont{Moats}},
  \bibinfo{year}{2012}, \bibinfo{journal}{JHEP}
  \textbf{\bibinfo{volume}{1204}}, \bibinfo{pages}{032}.

\bibitem[{\citenamefont{Goldberger}
  \emph{et~al.}(2008)\citenamefont{Goldberger, Grinstein, and
  Skiba}}]{Goldberger:2007zk}
\bibinfo{author}{\bibnamefont{Goldberger}, \bibfnamefont{W.~D.}},
  \bibinfo{author}{\bibfnamefont{B.}~\bibnamefont{Grinstein}}, and
  \bibinfo{author}{\bibfnamefont{W.}~\bibnamefont{Skiba}},
  \bibinfo{year}{2008}, \bibinfo{journal}{Phys.Rev.Lett.}
  \textbf{\bibinfo{volume}{100}}, \bibinfo{pages}{111802}.

\bibitem[{\citenamefont{Goldberger and Wise}(1999)}]{Goldberger:1999uk}
\bibinfo{author}{\bibnamefont{Goldberger}, \bibfnamefont{W.~D.}}, and
  \bibinfo{author}{\bibfnamefont{M.~B.} \bibnamefont{Wise}},
  \bibinfo{year}{1999}, \bibinfo{journal}{Phys.Rev.Lett.}
  \textbf{\bibinfo{volume}{83}}, \bibinfo{pages}{4922}.

\bibitem[{\citenamefont{Goldberger and Wise}(2000)}]{Goldberger:1999un}
\bibinfo{author}{\bibnamefont{Goldberger}, \bibfnamefont{W.~D.}}, and
  \bibinfo{author}{\bibfnamefont{M.~B.} \bibnamefont{Wise}},
  \bibinfo{year}{2000}, \bibinfo{journal}{Phys.Lett.}
  \textbf{\bibinfo{volume}{B475}}, \bibinfo{pages}{275}.

\bibitem[{\citenamefont{Golden and Randall}(1991)}]{Golden:1990ig}
\bibinfo{author}{\bibnamefont{Golden}, \bibfnamefont{M.}}, and
  \bibinfo{author}{\bibfnamefont{L.}~\bibnamefont{Randall}},
  \bibinfo{year}{1991}, \bibinfo{journal}{Nucl. Phys.}
  \textbf{\bibinfo{volume}{B361}}, \bibinfo{pages}{3}.

\bibitem[{\citenamefont{Goldstone}(1961)}]{Goldstone:1961eq}
\bibinfo{author}{\bibnamefont{Goldstone}, \bibfnamefont{J.}},
  \bibinfo{year}{1961}, \bibinfo{journal}{Nuovo Cim.}
  \textbf{\bibinfo{volume}{19}}, \bibinfo{pages}{154}.

\bibitem[{\citenamefont{Gonzalez-Fraile}(2012)}]{GonzalezFraile:2012fq}
\bibinfo{author}{\bibnamefont{Gonzalez-Fraile}, \bibfnamefont{J.}},
  \bibinfo{year}{2012}, \eprint{1205.5802}.

\bibitem[{\citenamefont{Gonzalez-Sprinberg}
  \emph{et~al.}(2005)\citenamefont{Gonzalez-Sprinberg, Martinez, and
  Rodriguez}}]{GonzalezSprinberg:2004bb}
\bibinfo{author}{\bibnamefont{Gonzalez-Sprinberg}, \bibfnamefont{G.}},
  \bibinfo{author}{\bibfnamefont{R.}~\bibnamefont{Martinez}}, and
  \bibinfo{author}{\bibfnamefont{J.~A.} \bibnamefont{Rodriguez}},
  \bibinfo{year}{2005}, \bibinfo{journal}{Phys.Rev.}
  \textbf{\bibinfo{volume}{D71}}, \bibinfo{pages}{035003}.

\bibitem[{\citenamefont{Gouzevitch}
  \emph{et~al.}(2013)\citenamefont{Gouzevitch, Oliveira, Rojo, Rosenfeld,
  Salam} \emph{et~al.}}]{Gouzevitch:2013qca}
\bibinfo{author}{\bibnamefont{Gouzevitch}, \bibfnamefont{M.}},
  \bibinfo{author}{\bibfnamefont{A.}~\bibnamefont{Oliveira}},
  \bibinfo{author}{\bibfnamefont{J.}~\bibnamefont{Rojo}},
  \bibinfo{author}{\bibfnamefont{R.}~\bibnamefont{Rosenfeld}},
  \bibinfo{author}{\bibfnamefont{G.~P.} \bibnamefont{Salam}}, \emph{et~al.},
  \bibinfo{year}{2013}, \bibinfo{journal}{JHEP}
  \textbf{\bibinfo{volume}{1307}}, \bibinfo{pages}{148}.

\bibitem[{\citenamefont{Graham} \emph{et~al.}(2015)\citenamefont{Graham,
  Kaplan, and Rajendran}}]{Graham:2015cka}
\bibinfo{author}{\bibnamefont{Graham}, \bibfnamefont{P.~W.}},
  \bibinfo{author}{\bibfnamefont{D.~E.} \bibnamefont{Kaplan}}, and
  \bibinfo{author}{\bibfnamefont{S.}~\bibnamefont{Rajendran}},
  \bibinfo{year}{2015}, \bibinfo{journal}{Phys. Rev. Lett.}
  \textbf{\bibinfo{volume}{115}}(\bibinfo{number}{22}),
  \bibinfo{pages}{221801}.

\bibitem[{\citenamefont{Green and Shih}(2012)}]{Green:2012nq}
\bibinfo{author}{\bibnamefont{Green}, \bibfnamefont{D.}}, and
  \bibinfo{author}{\bibfnamefont{D.}~\bibnamefont{Shih}}, \bibinfo{year}{2012},
  \bibinfo{journal}{JHEP} \textbf{\bibinfo{volume}{1209}},
  \bibinfo{pages}{026}.

\bibitem[{\citenamefont{Gregoire} \emph{et~al.}(2004)\citenamefont{Gregoire,
  Tucker-Smith, and Wacker}}]{Gregoire:2003kr}
\bibinfo{author}{\bibnamefont{Gregoire}, \bibfnamefont{T.}},
  \bibinfo{author}{\bibfnamefont{D.}~\bibnamefont{Tucker-Smith}}, and
  \bibinfo{author}{\bibfnamefont{J.~G.} \bibnamefont{Wacker}},
  \bibinfo{year}{2004}, \bibinfo{journal}{Phys.Rev.}
  \textbf{\bibinfo{volume}{D69}}, \bibinfo{pages}{115008}.

\bibitem[{\citenamefont{Gregoire and Wacker}(2002)}]{Gregoire:2002ra}
\bibinfo{author}{\bibnamefont{Gregoire}, \bibfnamefont{T.}}, and
  \bibinfo{author}{\bibfnamefont{J.~G.} \bibnamefont{Wacker}},
  \bibinfo{year}{2002}, \bibinfo{journal}{JHEP}
  \textbf{\bibinfo{volume}{0208}}, \bibinfo{pages}{019}.

\bibitem[{\citenamefont{Greiner}(2007)}]{UnWWscattering}
\bibinfo{author}{\bibnamefont{Greiner}, \bibfnamefont{N.}},
  \bibinfo{year}{2007}, \bibinfo{journal}{Phys.Lett.}
  \textbf{\bibinfo{volume}{B653}}, \bibinfo{pages}{75}.

\bibitem[{\citenamefont{Grinstein and Trott}(2008)}]{Grinstein:2008kt}
\bibinfo{author}{\bibnamefont{Grinstein}, \bibfnamefont{B.}}, and
  \bibinfo{author}{\bibfnamefont{M.}~\bibnamefont{Trott}},
  \bibinfo{year}{2008}, \bibinfo{journal}{JHEP}
  \textbf{\bibinfo{volume}{0811}}, \bibinfo{pages}{064}.

\bibitem[{\citenamefont{Grinstein and Uttayarat}(2011)}]{Grinstein:2011dq}
\bibinfo{author}{\bibnamefont{Grinstein}, \bibfnamefont{B.}}, and
  \bibinfo{author}{\bibfnamefont{P.}~\bibnamefont{Uttayarat}},
  \bibinfo{year}{2011}, \bibinfo{journal}{JHEP}
  \textbf{\bibinfo{volume}{1107}}, \bibinfo{pages}{038}.

\bibitem[{\citenamefont{Gripaios} \emph{et~al.}(2009)\citenamefont{Gripaios,
  Pomarol, Riva, and Serra}}]{Gripaios:2009pe}
\bibinfo{author}{\bibnamefont{Gripaios}, \bibfnamefont{B.}},
  \bibinfo{author}{\bibfnamefont{A.}~\bibnamefont{Pomarol}},
  \bibinfo{author}{\bibfnamefont{F.}~\bibnamefont{Riva}}, and
  \bibinfo{author}{\bibfnamefont{J.}~\bibnamefont{Serra}},
  \bibinfo{year}{2009}, \bibinfo{journal}{JHEP}
  \textbf{\bibinfo{volume}{0904}}, \bibinfo{pages}{070}.

\bibitem[{\citenamefont{Gripaios and West}(2006)}]{Gripaios:2006nn}
\bibinfo{author}{\bibnamefont{Gripaios}, \bibfnamefont{B.}}, and
  \bibinfo{author}{\bibfnamefont{S.~M.} \bibnamefont{West}},
  \bibinfo{year}{2006}, \bibinfo{journal}{Phys.Rev.}
  \textbf{\bibinfo{volume}{D74}}, \bibinfo{pages}{075002}.

\bibitem[{\citenamefont{Grojean}(2009)}]{Grojean:2009fd}
\bibinfo{author}{\bibnamefont{Grojean}, \bibfnamefont{C.}},
  \bibinfo{year}{2009}, \bibinfo{journal}{PoS}
  \textbf{\bibinfo{volume}{EPS-HEP2009}}, \bibinfo{pages}{008}.

\bibitem[{\citenamefont{Grojean}
  \emph{et~al.}(2013{\natexlab{a}})\citenamefont{Grojean, Jenkins, Manohar, and
  Trott}}]{Grojean:2013kd}
\bibinfo{author}{\bibnamefont{Grojean}, \bibfnamefont{C.}},
  \bibinfo{author}{\bibfnamefont{E.~E.} \bibnamefont{Jenkins}},
  \bibinfo{author}{\bibfnamefont{A.~V.} \bibnamefont{Manohar}}, and
  \bibinfo{author}{\bibfnamefont{M.}~\bibnamefont{Trott}},
  \bibinfo{year}{2013}{\natexlab{a}}, \bibinfo{journal}{JHEP}
  \textbf{\bibinfo{volume}{1304}}, \bibinfo{pages}{016}.

\bibitem[{\citenamefont{Grojean}
  \emph{et~al.}(2013{\natexlab{b}})\citenamefont{Grojean, Matsedonskyi, and
  Panico}}]{Grojean:2013qca}
\bibinfo{author}{\bibnamefont{Grojean}, \bibfnamefont{C.}},
  \bibinfo{author}{\bibfnamefont{O.}~\bibnamefont{Matsedonskyi}}, and
  \bibinfo{author}{\bibfnamefont{G.}~\bibnamefont{Panico}},
  \bibinfo{year}{2013}{\natexlab{b}}, \bibinfo{journal}{JHEP}
  \textbf{\bibinfo{volume}{10}}, \bibinfo{pages}{160}.

\bibitem[{\citenamefont{Grossman and Neubert}(2000)}]{Grossman:1999ra}
\bibinfo{author}{\bibnamefont{Grossman}, \bibfnamefont{Y.}}, and
  \bibinfo{author}{\bibfnamefont{M.}~\bibnamefont{Neubert}},
  \bibinfo{year}{2000}, \bibinfo{journal}{Phys.Lett.}
  \textbf{\bibinfo{volume}{B474}}, \bibinfo{pages}{361}.

\bibitem[{\citenamefont{Grzadkowski}
  \emph{et~al.}(2012)\citenamefont{Grzadkowski, Gunion, and
  Toharia}}]{Grzadkowski:2012ng}
\bibinfo{author}{\bibnamefont{Grzadkowski}, \bibfnamefont{B.}},
  \bibinfo{author}{\bibfnamefont{J.~F.} \bibnamefont{Gunion}}, and
  \bibinfo{author}{\bibfnamefont{M.}~\bibnamefont{Toharia}},
  \bibinfo{year}{2012}, \bibinfo{journal}{Phys.Lett.}
  \textbf{\bibinfo{volume}{B712}}, \bibinfo{pages}{70}.

\bibitem[{\citenamefont{Guralnik} \emph{et~al.}(1964)\citenamefont{Guralnik,
  Hagen, and Kibble}}]{Guralnik:1964eu}
\bibinfo{author}{\bibnamefont{Guralnik}, \bibfnamefont{G.}},
  \bibinfo{author}{\bibfnamefont{C.}~\bibnamefont{Hagen}}, and
  \bibinfo{author}{\bibfnamefont{T.}~\bibnamefont{Kibble}},
  \bibinfo{year}{1964}, \bibinfo{journal}{Phys.Rev.Lett.}
  \textbf{\bibinfo{volume}{13}}, \bibinfo{pages}{585}.

\bibitem[{\citenamefont{Han}
  \emph{et~al.}(2003{\natexlab{a}})\citenamefont{Han, Logan, McElrath, and
  Wang}}]{Han:2003gf}
\bibinfo{author}{\bibnamefont{Han}, \bibfnamefont{T.}},
  \bibinfo{author}{\bibfnamefont{H.~E.} \bibnamefont{Logan}},
  \bibinfo{author}{\bibfnamefont{B.}~\bibnamefont{McElrath}}, and
  \bibinfo{author}{\bibfnamefont{L.-T.} \bibnamefont{Wang}},
  \bibinfo{year}{2003}{\natexlab{a}}, \bibinfo{journal}{Phys.Lett.}
  \textbf{\bibinfo{volume}{B563}}, \bibinfo{pages}{191}.

\bibitem[{\citenamefont{Han}
  \emph{et~al.}(2003{\natexlab{b}})\citenamefont{Han, Logan, McElrath, and
  Wang}}]{Han:2003wu}
\bibinfo{author}{\bibnamefont{Han}, \bibfnamefont{T.}},
  \bibinfo{author}{\bibfnamefont{H.~E.} \bibnamefont{Logan}},
  \bibinfo{author}{\bibfnamefont{B.}~\bibnamefont{McElrath}}, and
  \bibinfo{author}{\bibfnamefont{L.-T.} \bibnamefont{Wang}},
  \bibinfo{year}{2003}{\natexlab{b}}, \bibinfo{journal}{Phys.Rev.}
  \textbf{\bibinfo{volume}{D67}}, \bibinfo{pages}{095004}.

\bibitem[{\citenamefont{Han} \emph{et~al.}(2006)\citenamefont{Han, Logan, and
  Wang}}]{Han:2005ru}
\bibinfo{author}{\bibnamefont{Han}, \bibfnamefont{T.}},
  \bibinfo{author}{\bibfnamefont{H.~E.} \bibnamefont{Logan}}, and
  \bibinfo{author}{\bibfnamefont{L.-T.} \bibnamefont{Wang}},
  \bibinfo{year}{2006}, \bibinfo{journal}{JHEP}
  \textbf{\bibinfo{volume}{0601}}, \bibinfo{pages}{099}.

\bibitem[{\citenamefont{Han and Skiba}(2005)}]{Han:2005dz}
\bibinfo{author}{\bibnamefont{Han}, \bibfnamefont{Z.}}, and
  \bibinfo{author}{\bibfnamefont{W.}~\bibnamefont{Skiba}},
  \bibinfo{year}{2005}, \bibinfo{journal}{Phys.Rev.}
  \textbf{\bibinfo{volume}{D72}}, \bibinfo{pages}{035005}.

\bibitem[{\citenamefont{Harnik} \emph{et~al.}(2004)\citenamefont{Harnik, Kribs,
  Larson, and Murayama}}]{Harnik:2003rs}
\bibinfo{author}{\bibnamefont{Harnik}, \bibfnamefont{R.}},
  \bibinfo{author}{\bibfnamefont{G.~D.} \bibnamefont{Kribs}},
  \bibinfo{author}{\bibfnamefont{D.~T.} \bibnamefont{Larson}}, and
  \bibinfo{author}{\bibfnamefont{H.}~\bibnamefont{Murayama}},
  \bibinfo{year}{2004}, \bibinfo{journal}{Phys.Rev.}
  \textbf{\bibinfo{volume}{D70}}, \bibinfo{pages}{015002}.

\bibitem[{\citenamefont{Hashimoto}(2011)}]{Hashimoto:2011cw}
\bibinfo{author}{\bibnamefont{Hashimoto}, \bibfnamefont{M.}},
  \bibinfo{year}{2011}, \bibinfo{journal}{Phys.Rev.}
  \textbf{\bibinfo{volume}{D84}}, \bibinfo{pages}{111901}.

\bibitem[{\citenamefont{He and Wen}(2008)}]{UnHHscattering}
\bibinfo{author}{\bibnamefont{He}, \bibfnamefont{X.-G.}}, and
  \bibinfo{author}{\bibfnamefont{C.-C.} \bibnamefont{Wen}},
  \bibinfo{year}{2008}, \bibinfo{journal}{Phys.Rev.}
  \textbf{\bibinfo{volume}{D78}}, \bibinfo{pages}{017301}.

\bibitem[{\citenamefont{Heckman} \emph{et~al.}(2012)\citenamefont{Heckman,
  Kumar, and Wecht}}]{Heckman:2012nt}
\bibinfo{author}{\bibnamefont{Heckman}, \bibfnamefont{J.~J.}},
  \bibinfo{author}{\bibfnamefont{P.}~\bibnamefont{Kumar}}, and
  \bibinfo{author}{\bibfnamefont{B.}~\bibnamefont{Wecht}},
  \bibinfo{year}{2012}, \bibinfo{journal}{JHEP}
  \textbf{\bibinfo{volume}{1207}}, \bibinfo{pages}{118}.

\bibitem[{\citenamefont{Hektor} \emph{et~al.}(2007)\citenamefont{Hektor,
  Kadastik, Muntel, Raidal, and Rebane}}]{Hektor:2007uu}
\bibinfo{author}{\bibnamefont{Hektor}, \bibfnamefont{A.}},
  \bibinfo{author}{\bibfnamefont{M.}~\bibnamefont{Kadastik}},
  \bibinfo{author}{\bibfnamefont{M.}~\bibnamefont{Muntel}},
  \bibinfo{author}{\bibfnamefont{M.}~\bibnamefont{Raidal}}, and
  \bibinfo{author}{\bibfnamefont{L.}~\bibnamefont{Rebane}},
  \bibinfo{year}{2007}, \bibinfo{journal}{Nucl.Phys.}
  \textbf{\bibinfo{volume}{B787}}, \bibinfo{pages}{198}.

\bibitem[{\citenamefont{Hewett} \emph{et~al.}(2004)\citenamefont{Hewett,
  Lillie, and Rizzo}}]{Hewett:2004dv}
\bibinfo{author}{\bibnamefont{Hewett}, \bibfnamefont{J.}},
  \bibinfo{author}{\bibfnamefont{B.}~\bibnamefont{Lillie}}, and
  \bibinfo{author}{\bibfnamefont{T.}~\bibnamefont{Rizzo}},
  \bibinfo{year}{2004}, \bibinfo{journal}{JHEP}
  \textbf{\bibinfo{volume}{0410}}, \bibinfo{pages}{014}.

\bibitem[{\citenamefont{Hewett} \emph{et~al.}(2003)\citenamefont{Hewett,
  Petriello, and Rizzo}}]{Hewett:2002px}
\bibinfo{author}{\bibnamefont{Hewett}, \bibfnamefont{J.~L.}},
  \bibinfo{author}{\bibfnamefont{F.~J.} \bibnamefont{Petriello}}, and
  \bibinfo{author}{\bibfnamefont{T.~G.} \bibnamefont{Rizzo}},
  \bibinfo{year}{2003}, \bibinfo{journal}{JHEP}
  \textbf{\bibinfo{volume}{0310}}, \bibinfo{pages}{062}.

\bibitem[{\citenamefont{Higgs}(1964{\natexlab{a}})}]{Higgs:1964pj}
\bibinfo{author}{\bibnamefont{Higgs}, \bibfnamefont{P.~W.}},
  \bibinfo{year}{1964}{\natexlab{a}}, \bibinfo{journal}{Phys.Rev.Lett.}
  \textbf{\bibinfo{volume}{13}}, \bibinfo{pages}{508}.

\bibitem[{\citenamefont{Higgs}(1964{\natexlab{b}})}]{Higgs:1964ia}
\bibinfo{author}{\bibnamefont{Higgs}, \bibfnamefont{P.~W.}},
  \bibinfo{year}{1964}{\natexlab{b}}, \bibinfo{journal}{Phys.Lett.}
  \textbf{\bibinfo{volume}{12}}, \bibinfo{pages}{132}.

\bibitem[{\citenamefont{Hill and Hill}(2007)}]{Hill:2007zv}
\bibinfo{author}{\bibnamefont{Hill}, \bibfnamefont{C.~T.}}, and
  \bibinfo{author}{\bibfnamefont{R.~J.} \bibnamefont{Hill}},
  \bibinfo{year}{2007}, \bibinfo{journal}{Phys.Rev.}
  \textbf{\bibinfo{volume}{D76}}, \bibinfo{pages}{115014}.

\bibitem[{\citenamefont{Hill} \emph{et~al.}(2001)\citenamefont{Hill, Pokorski,
  and Wang}}]{Hill:2000mu}
\bibinfo{author}{\bibnamefont{Hill}, \bibfnamefont{C.~T.}},
  \bibinfo{author}{\bibfnamefont{S.}~\bibnamefont{Pokorski}}, and
  \bibinfo{author}{\bibfnamefont{J.}~\bibnamefont{Wang}}, \bibinfo{year}{2001},
  \bibinfo{journal}{Phys. Rev.} \textbf{\bibinfo{volume}{D64}},
  \bibinfo{pages}{105005}.

\bibitem[{\citenamefont{Hill and Simmons}(2003)}]{Hill2003235}
\bibinfo{author}{\bibnamefont{Hill}, \bibfnamefont{C.~T.}}, and
  \bibinfo{author}{\bibfnamefont{E.~H.} \bibnamefont{Simmons}},
  \bibinfo{year}{2003}, \bibinfo{journal}{Physics Reports}
  \textbf{\bibinfo{volume}{381}}(\bibinfo{number}{4‰ÛÒ6}),
  \bibinfo{pages}{235}.

\bibitem[{\citenamefont{Holdom}(1981{\natexlab{a}})}]{PhysRevD.24.1441}
\bibinfo{author}{\bibnamefont{Holdom}, \bibfnamefont{B.}},
  \bibinfo{year}{1981}{\natexlab{a}}, \bibinfo{journal}{Phys. Rev. D}
  \textbf{\bibinfo{volume}{24}}, \bibinfo{pages}{1441}.

\bibitem[{\citenamefont{Holdom}(1981{\natexlab{b}})}]{Holdom:1981rm}
\bibinfo{author}{\bibnamefont{Holdom}, \bibfnamefont{B.}},
  \bibinfo{year}{1981}{\natexlab{b}}, \bibinfo{journal}{Phys. Rev.}
  \textbf{\bibinfo{volume}{D24}}, \bibinfo{pages}{1441}.

\bibitem[{\citenamefont{Holdom}(1985)}]{Holdom1985301}
\bibinfo{author}{\bibnamefont{Holdom}, \bibfnamefont{B.}},
  \bibinfo{year}{1985}, \bibinfo{journal}{Physics Letters B}
  \textbf{\bibinfo{volume}{150}}(\bibinfo{number}{4}), \bibinfo{pages}{301 }.

\bibitem[{\citenamefont{Holdom and Terning}(1988)}]{Holdom:1987yu}
\bibinfo{author}{\bibnamefont{Holdom}, \bibfnamefont{B.}}, and
  \bibinfo{author}{\bibfnamefont{J.}~\bibnamefont{Terning}},
  \bibinfo{year}{1988}, \bibinfo{journal}{Phys.Lett.}
  \textbf{\bibinfo{volume}{B200}}, \bibinfo{pages}{338}.

\bibitem[{\citenamefont{Holdom and Terning}(1990)}]{Holdom:1990tc}
\bibinfo{author}{\bibnamefont{Holdom}, \bibfnamefont{B.}}, and
  \bibinfo{author}{\bibfnamefont{J.}~\bibnamefont{Terning}},
  \bibinfo{year}{1990}, \bibinfo{journal}{Phys. Lett.}
  \textbf{\bibinfo{volume}{B247}}, \bibinfo{pages}{88}.

\bibitem[{\citenamefont{Huber and Shafi}(2001)}]{Huber:2000fh}
\bibinfo{author}{\bibnamefont{Huber}, \bibfnamefont{S.~J.}}, and
  \bibinfo{author}{\bibfnamefont{Q.}~\bibnamefont{Shafi}},
  \bibinfo{year}{2001}, \bibinfo{journal}{Phys.Rev.}
  \textbf{\bibinfo{volume}{D63}}, \bibinfo{pages}{045010}.

\bibitem[{\citenamefont{Hubisz and Meade}(2005)}]{Hubisz:2004ft}
\bibinfo{author}{\bibnamefont{Hubisz}, \bibfnamefont{J.}}, and
  \bibinfo{author}{\bibfnamefont{P.}~\bibnamefont{Meade}},
  \bibinfo{year}{2005}, \bibinfo{journal}{Phys.Rev.}
  \textbf{\bibinfo{volume}{D71}}, \bibinfo{pages}{035016}.

\bibitem[{\citenamefont{Hubisz} \emph{et~al.}(2006)\citenamefont{Hubisz, Meade,
  Noble, and Perelstein}}]{Hubisz:2005tx}
\bibinfo{author}{\bibnamefont{Hubisz}, \bibfnamefont{J.}},
  \bibinfo{author}{\bibfnamefont{P.}~\bibnamefont{Meade}},
  \bibinfo{author}{\bibfnamefont{A.}~\bibnamefont{Noble}}, and
  \bibinfo{author}{\bibfnamefont{M.}~\bibnamefont{Perelstein}},
  \bibinfo{year}{2006}, \bibinfo{journal}{JHEP}
  \textbf{\bibinfo{volume}{0601}}, \bibinfo{pages}{135}.

\bibitem[{\citenamefont{Huo and Zhu}(2003)}]{Huo:2003vd}
\bibinfo{author}{\bibnamefont{Huo}, \bibfnamefont{W.-j.}}, and
  \bibinfo{author}{\bibfnamefont{S.-h.} \bibnamefont{Zhu}},
  \bibinfo{year}{2003}, \bibinfo{journal}{Phys.Rev.}
  \textbf{\bibinfo{volume}{D68}}, \bibinfo{pages}{097301}.

\bibitem[{\citenamefont{Jora}(2009)}]{Jora:2009dh}
\bibinfo{author}{\bibnamefont{Jora}, \bibfnamefont{R.}}, \bibinfo{year}{2009},
  \eprint{0905.0206}.

\bibitem[{\citenamefont{Kaplan and Georgi}(1984)}]{Kaplan:1983fs}
\bibinfo{author}{\bibnamefont{Kaplan}, \bibfnamefont{D.~B.}}, and
  \bibinfo{author}{\bibfnamefont{H.}~\bibnamefont{Georgi}},
  \bibinfo{year}{1984}, \bibinfo{journal}{Phys.Lett.}
  \textbf{\bibinfo{volume}{B136}}, \bibinfo{pages}{183}.

\bibitem[{\citenamefont{Kaplan} \emph{et~al.}(1984)\citenamefont{Kaplan,
  Georgi, and Dimopoulos}}]{Kaplan:1983sm}
\bibinfo{author}{\bibnamefont{Kaplan}, \bibfnamefont{D.~B.}},
  \bibinfo{author}{\bibfnamefont{H.}~\bibnamefont{Georgi}}, and
  \bibinfo{author}{\bibfnamefont{S.}~\bibnamefont{Dimopoulos}},
  \bibinfo{year}{1984}, \bibinfo{journal}{Phys.Lett.}
  \textbf{\bibinfo{volume}{B136}}, \bibinfo{pages}{187}.

\bibitem[{\citenamefont{Kaplan} \emph{et~al.}(2008)\citenamefont{Kaplan,
  Rehermann, Schwartz, and Tweedie}}]{Kaplan:2008ie}
\bibinfo{author}{\bibnamefont{Kaplan}, \bibfnamefont{D.~E.}},
  \bibinfo{author}{\bibfnamefont{K.}~\bibnamefont{Rehermann}},
  \bibinfo{author}{\bibfnamefont{M.~D.} \bibnamefont{Schwartz}}, and
  \bibinfo{author}{\bibfnamefont{B.}~\bibnamefont{Tweedie}},
  \bibinfo{year}{2008}, \bibinfo{journal}{Phys.Rev.Lett.}
  \textbf{\bibinfo{volume}{101}}, \bibinfo{pages}{142001}.

\bibitem[{\citenamefont{Kaplan and Schmaltz}(2003)}]{Kaplan:2003uc}
\bibinfo{author}{\bibnamefont{Kaplan}, \bibfnamefont{D.~E.}}, and
  \bibinfo{author}{\bibfnamefont{M.}~\bibnamefont{Schmaltz}},
  \bibinfo{year}{2003}, \bibinfo{journal}{JHEP}
  \textbf{\bibinfo{volume}{0310}}, \bibinfo{pages}{039}.

\bibitem[{\citenamefont{Kaplan} \emph{et~al.}(2004)\citenamefont{Kaplan,
  Schmaltz, and Skiba}}]{Kaplan:2004cr}
\bibinfo{author}{\bibnamefont{Kaplan}, \bibfnamefont{D.~E.}},
  \bibinfo{author}{\bibfnamefont{M.}~\bibnamefont{Schmaltz}}, and
  \bibinfo{author}{\bibfnamefont{W.}~\bibnamefont{Skiba}},
  \bibinfo{year}{2004}, \bibinfo{journal}{Phys.Rev.}
  \textbf{\bibinfo{volume}{D70}}, \bibinfo{pages}{075009}.

\bibitem[{\citenamefont{Katz} \emph{et~al.}(2005)\citenamefont{Katz, Lee,
  Nelson, and Walker}}]{Katz:2003sn}
\bibinfo{author}{\bibnamefont{Katz}, \bibfnamefont{E.}},
  \bibinfo{author}{\bibfnamefont{J.-y.} \bibnamefont{Lee}},
  \bibinfo{author}{\bibfnamefont{A.~E.} \bibnamefont{Nelson}}, and
  \bibinfo{author}{\bibfnamefont{D.~G.} \bibnamefont{Walker}},
  \bibinfo{year}{2005}, \bibinfo{journal}{JHEP}
  \textbf{\bibinfo{volume}{0510}}, \bibinfo{pages}{088}.

\bibitem[{\citenamefont{Kearney} \emph{et~al.}(2013)\citenamefont{Kearney,
  Pierce, and Thaler}}]{Kearney:2013cca}
\bibinfo{author}{\bibnamefont{Kearney}, \bibfnamefont{J.}},
  \bibinfo{author}{\bibfnamefont{A.}~\bibnamefont{Pierce}}, and
  \bibinfo{author}{\bibfnamefont{J.}~\bibnamefont{Thaler}},
  \bibinfo{year}{2013}, \bibinfo{journal}{JHEP} \textbf{\bibinfo{volume}{10}},
  \bibinfo{pages}{230}.

\bibitem[{Khachatryan \emph{et~al.}(2011)\citenamefont{Khachatryan}
  \emph{et~al.}}]{Khachatryan:2011tk}
\bibinfo{author}{\bibnamefont{Khachatryan}, \bibfnamefont{V.}}, \emph{et~al.}
  (\bibinfo{collaboration}{CMS Collaboration}), \bibinfo{year}{2011},
  \bibinfo{journal}{Phys. Lett.} \textbf{\bibinfo{volume}{B698}},
  \bibinfo{pages}{196}.

\bibitem[{Khachatryan
  \emph{et~al.}(2015{\natexlab{a}})\citenamefont{Khachatryan}
  \emph{et~al.}}]{Khachatryan:2014vma}
\bibinfo{author}{\bibnamefont{Khachatryan}, \bibfnamefont{V.}}, \emph{et~al.}
  (\bibinfo{collaboration}{CMS Collaboration}),
  \bibinfo{year}{2015}{\natexlab{a}}, \bibinfo{journal}{JHEP}
  \textbf{\bibinfo{volume}{01}}, \bibinfo{pages}{053}.

\bibitem[{Khachatryan
  \emph{et~al.}(2015{\natexlab{b}})\citenamefont{Khachatryan}
  \emph{et~al.}}]{Khachatryan:2015bma}
\bibinfo{author}{\bibnamefont{Khachatryan}, \bibfnamefont{V.}}, \emph{et~al.}
  (\bibinfo{collaboration}{CMS Collaboration}),
  \bibinfo{year}{2015}{\natexlab{b}}, \eprint{1506.01443}.

\bibitem[{Khachatryan
  \emph{et~al.}(2015{\natexlab{c}})\citenamefont{Khachatryan}
  \emph{et~al.}}]{Khachatryan:2014xja}
\bibinfo{author}{\bibnamefont{Khachatryan}, \bibfnamefont{V.}}, \emph{et~al.}
  (\bibinfo{collaboration}{CMS Collaboration}),
  \bibinfo{year}{2015}{\natexlab{c}}, \bibinfo{journal}{Phys.Lett.}
  \textbf{\bibinfo{volume}{B740}}, \bibinfo{pages}{83}.

\bibitem[{Khachatryan
  \emph{et~al.}(2015{\natexlab{d}})\citenamefont{Khachatryan}
  \emph{et~al.}}]{Khachatryan:2014fba}
\bibinfo{author}{\bibnamefont{Khachatryan}, \bibfnamefont{V.}}, \emph{et~al.}
  (\bibinfo{collaboration}{CMS Collaboration}),
  \bibinfo{year}{2015}{\natexlab{d}}, \bibinfo{journal}{JHEP}
  \textbf{\bibinfo{volume}{04}}, \bibinfo{pages}{025}.

\bibitem[{Khachatryan
  \emph{et~al.}(2015{\natexlab{e}})\citenamefont{Khachatryan}
  \emph{et~al.}}]{Khachatryan:2015sja}
\bibinfo{author}{\bibnamefont{Khachatryan}, \bibfnamefont{V.}}, \emph{et~al.}
  (\bibinfo{collaboration}{CMS Collaboration}),
  \bibinfo{year}{2015}{\natexlab{e}}, \bibinfo{journal}{Phys.Rev.}
  \textbf{\bibinfo{volume}{D91}}(\bibinfo{number}{5}), \bibinfo{pages}{052009}.

\bibitem[{\citenamefont{Kikuchi and Okada}(2008)}]{unother3}
\bibinfo{author}{\bibnamefont{Kikuchi}, \bibfnamefont{T.}}, and
  \bibinfo{author}{\bibfnamefont{N.}~\bibnamefont{Okada}},
  \bibinfo{year}{2008}, \bibinfo{journal}{Phys.Lett.}
  \textbf{\bibinfo{volume}{B661}}, \bibinfo{pages}{360}.

\bibitem[{\citenamefont{Kilian} \emph{et~al.}(2005)\citenamefont{Kilian,
  Rainwater, and Reuter}}]{Kilian:2004pp}
\bibinfo{author}{\bibnamefont{Kilian}, \bibfnamefont{W.}},
  \bibinfo{author}{\bibfnamefont{D.}~\bibnamefont{Rainwater}}, and
  \bibinfo{author}{\bibfnamefont{J.}~\bibnamefont{Reuter}},
  \bibinfo{year}{2005}, \bibinfo{journal}{Phys.Rev.}
  \textbf{\bibinfo{volume}{D71}}, \bibinfo{pages}{015008}.

\bibitem[{\citenamefont{Kilian and Reuter}(2004)}]{Kilian:2003xt}
\bibinfo{author}{\bibnamefont{Kilian}, \bibfnamefont{W.}}, and
  \bibinfo{author}{\bibfnamefont{J.}~\bibnamefont{Reuter}},
  \bibinfo{year}{2004}, \bibinfo{journal}{Phys.Rev.}
  \textbf{\bibinfo{volume}{D70}}, \bibinfo{pages}{015004}.

\bibitem[{\citenamefont{Kilic and Mahbubani}(2004)}]{Kilic:2003mq}
\bibinfo{author}{\bibnamefont{Kilic}, \bibfnamefont{C.}}, and
  \bibinfo{author}{\bibfnamefont{R.}~\bibnamefont{Mahbubani}},
  \bibinfo{year}{2004}, \bibinfo{journal}{JHEP}
  \textbf{\bibinfo{volume}{0407}}, \bibinfo{pages}{013}.

\bibitem[{\citenamefont{Klebanov and Witten}(1999)}]{Klebanov:1999tb}
\bibinfo{author}{\bibnamefont{Klebanov}, \bibfnamefont{I.~R.}}, and
  \bibinfo{author}{\bibfnamefont{E.}~\bibnamefont{Witten}},
  \bibinfo{year}{1999}, \bibinfo{journal}{Nucl.Phys.}
  \textbf{\bibinfo{volume}{B556}}, \bibinfo{pages}{89}.

\bibitem[{\citenamefont{Kozlov}(2012)}]{Kozlov:2012cr}
\bibinfo{author}{\bibnamefont{Kozlov}, \bibfnamefont{G.}},
  \bibinfo{year}{2012}, \bibinfo{note}{[PoSConfinementX,086(2012)]},
  \eprint{1211.1186}.

\bibitem[{\citenamefont{Kribs}(2006)}]{Kribs:2006mq}
\bibinfo{author}{\bibnamefont{Kribs}, \bibfnamefont{G.~D.}},
  \bibinfo{year}{2006}, in \emph{\bibinfo{booktitle}{{Physics in D $\geq$ 4.
  Proceedings, Theoretical Advanced Study Institute in elementary particle
  physics, TASI 2004, Boulder, USA, June 6-July 2, 2004}}}, pp.
  \bibinfo{pages}{633--699}, \eprint{hep-ph/0605325}.

\bibitem[{\citenamefont{Langacker}(2009)}]{Langacker:2008yv}
\bibinfo{author}{\bibnamefont{Langacker}, \bibfnamefont{P.}},
  \bibinfo{year}{2009}, \bibinfo{journal}{Rev.Mod.Phys.}
  \textbf{\bibinfo{volume}{81}}, \bibinfo{pages}{1199}.

\bibitem[{\citenamefont{Larkoski} \emph{et~al.}(2013)\citenamefont{Larkoski,
  Salam, and Thaler}}]{Larkoski:2013eya}
\bibinfo{author}{\bibnamefont{Larkoski}, \bibfnamefont{A.~J.}},
  \bibinfo{author}{\bibfnamefont{G.~P.} \bibnamefont{Salam}}, and
  \bibinfo{author}{\bibfnamefont{J.}~\bibnamefont{Thaler}},
  \bibinfo{year}{2013}, \bibinfo{journal}{JHEP}
  \textbf{\bibinfo{volume}{1306}}, \bibinfo{pages}{108}.

\bibitem[{\citenamefont{Lawrance and Piai}(2013)}]{Lawrance:2012cg}
\bibinfo{author}{\bibnamefont{Lawrance}, \bibfnamefont{R.}}, and
  \bibinfo{author}{\bibfnamefont{M.}~\bibnamefont{Piai}}, \bibinfo{year}{2013},
  \bibinfo{journal}{Int.J.Mod.Phys.} \textbf{\bibinfo{volume}{A28}},
  \bibinfo{pages}{1350081}.

\bibitem[{\citenamefont{Lee}(2008)}]{Lee:2008ph}
\bibinfo{author}{\bibnamefont{Lee}, \bibfnamefont{J.-P.}},
  \bibinfo{year}{2008}, \eprint{0803.0833}.

\bibitem[{\citenamefont{Lee}(2009)}]{Lee:2008vp}
\bibinfo{author}{\bibnamefont{Lee}, \bibfnamefont{J.-P.}},
  \bibinfo{year}{2009}, \bibinfo{journal}{AIP Conf.Proc.}
  \textbf{\bibinfo{volume}{1078}}, \bibinfo{pages}{626}.

\bibitem[{\citenamefont{Lee}(2004)}]{Lee:2004me}
\bibinfo{author}{\bibnamefont{Lee}, \bibfnamefont{J.~Y.}},
  \bibinfo{year}{2004}, \bibinfo{journal}{JHEP}
  \textbf{\bibinfo{volume}{0412}}, \bibinfo{pages}{065}.

\bibitem[{\citenamefont{Lillie}
  \emph{et~al.}(2007{\natexlab{a}})\citenamefont{Lillie, Randall, and
  Wang}}]{Lillie:2007yh}
\bibinfo{author}{\bibnamefont{Lillie}, \bibfnamefont{B.}},
  \bibinfo{author}{\bibfnamefont{L.}~\bibnamefont{Randall}}, and
  \bibinfo{author}{\bibfnamefont{L.-T.} \bibnamefont{Wang}},
  \bibinfo{year}{2007}{\natexlab{a}}, \bibinfo{journal}{JHEP}
  \textbf{\bibinfo{volume}{0709}}, \bibinfo{pages}{074}.

\bibitem[{\citenamefont{Lillie}
  \emph{et~al.}(2007{\natexlab{b}})\citenamefont{Lillie, Shu, and
  Tait}}]{Lillie:2007ve}
\bibinfo{author}{\bibnamefont{Lillie}, \bibfnamefont{B.}},
  \bibinfo{author}{\bibfnamefont{J.}~\bibnamefont{Shu}}, and
  \bibinfo{author}{\bibfnamefont{T.~M.} \bibnamefont{Tait}},
  \bibinfo{year}{2007}{\natexlab{b}}, \bibinfo{journal}{Phys.Rev.}
  \textbf{\bibinfo{volume}{D76}}, \bibinfo{pages}{115016}.

\bibitem[{\citenamefont{Low}(2004)}]{Low:2004xc}
\bibinfo{author}{\bibnamefont{Low}, \bibfnamefont{I.}}, \bibinfo{year}{2004},
  \bibinfo{journal}{JHEP} \textbf{\bibinfo{volume}{0410}},
  \bibinfo{pages}{067}.

\bibitem[{\citenamefont{Low} \emph{et~al.}(2012)\citenamefont{Low, Lykken, and
  Shaughnessy}}]{Low:2012rj}
\bibinfo{author}{\bibnamefont{Low}, \bibfnamefont{I.}},
  \bibinfo{author}{\bibfnamefont{J.}~\bibnamefont{Lykken}}, and
  \bibinfo{author}{\bibfnamefont{G.}~\bibnamefont{Shaughnessy}},
  \bibinfo{year}{2012}, \bibinfo{journal}{Phys.Rev.}
  \textbf{\bibinfo{volume}{D86}}, \bibinfo{pages}{093012}.

\bibitem[{\citenamefont{Low} \emph{et~al.}(2002)\citenamefont{Low, Skiba, and
  Tucker-Smith}}]{Low:2002ws}
\bibinfo{author}{\bibnamefont{Low}, \bibfnamefont{I.}},
  \bibinfo{author}{\bibfnamefont{W.}~\bibnamefont{Skiba}}, and
  \bibinfo{author}{\bibfnamefont{D.}~\bibnamefont{Tucker-Smith}},
  \bibinfo{year}{2002}, \bibinfo{journal}{Phys.Rev.}
  \textbf{\bibinfo{volume}{D66}}, \bibinfo{pages}{072001}.

\bibitem[{\citenamefont{Low} \emph{et~al.}(2015)\citenamefont{Low, Tesi, and
  Wang}}]{Low:2015nqa}
\bibinfo{author}{\bibnamefont{Low}, \bibfnamefont{M.}},
  \bibinfo{author}{\bibfnamefont{A.}~\bibnamefont{Tesi}}, and
  \bibinfo{author}{\bibfnamefont{L.-T.} \bibnamefont{Wang}},
  \bibinfo{year}{2015}, \bibinfo{journal}{Phys. Rev.}
  \textbf{\bibinfo{volume}{D91}}, \bibinfo{pages}{095012}.

\bibitem[{\citenamefont{Luty}(2009)}]{Luty:2008vs}
\bibinfo{author}{\bibnamefont{Luty}, \bibfnamefont{M.~A.}},
  \bibinfo{year}{2009}, \bibinfo{journal}{JHEP}
  \textbf{\bibinfo{volume}{0904}}, \bibinfo{pages}{050}.

\bibitem[{\citenamefont{Luty and Okui}(2006)}]{Luty:2004ye}
\bibinfo{author}{\bibnamefont{Luty}, \bibfnamefont{M.~A.}}, and
  \bibinfo{author}{\bibfnamefont{T.}~\bibnamefont{Okui}}, \bibinfo{year}{2006},
  \bibinfo{journal}{JHEP} \textbf{\bibinfo{volume}{0609}},
  \bibinfo{pages}{070}.

\bibitem[{\citenamefont{Luty and Rattazzi}(1999)}]{Luty:1999qc}
\bibinfo{author}{\bibnamefont{Luty}, \bibfnamefont{M.~A.}}, and
  \bibinfo{author}{\bibfnamefont{R.}~\bibnamefont{Rattazzi}},
  \bibinfo{year}{1999}, \bibinfo{journal}{JHEP}
  \textbf{\bibinfo{volume}{9911}}, \bibinfo{pages}{001}.

\bibitem[{\citenamefont{Luty} \emph{et~al.}(2001)\citenamefont{Luty, Terning,
  and Grant}}]{Luty:2000fj}
\bibinfo{author}{\bibnamefont{Luty}, \bibfnamefont{M.~A.}},
  \bibinfo{author}{\bibfnamefont{J.}~\bibnamefont{Terning}}, and
  \bibinfo{author}{\bibfnamefont{A.~K.} \bibnamefont{Grant}},
  \bibinfo{year}{2001}, \bibinfo{journal}{Phys.Rev.}
  \textbf{\bibinfo{volume}{D63}}, \bibinfo{pages}{075001}.

\bibitem[{\citenamefont{Maldacena}(1998)}]{Maldacena:1997re}
\bibinfo{author}{\bibnamefont{Maldacena}, \bibfnamefont{J.~M.}},
  \bibinfo{year}{1998}, \bibinfo{journal}{Adv.Theor.Math.Phys.}
  \textbf{\bibinfo{volume}{2}}, \bibinfo{pages}{231}.

\bibitem[{\citenamefont{Mandelstam}(1962)}]{Mandelstam:1962mi}
\bibinfo{author}{\bibnamefont{Mandelstam}, \bibfnamefont{S.}},
  \bibinfo{year}{1962}, \bibinfo{journal}{Annals Phys.}
  \textbf{\bibinfo{volume}{19}}, \bibinfo{pages}{1}.

\bibitem[{\citenamefont{Marandella}
  \emph{et~al.}(2005)\citenamefont{Marandella, Schappacher, and
  Strumia}}]{Marandella:2005wd}
\bibinfo{author}{\bibnamefont{Marandella}, \bibfnamefont{G.}},
  \bibinfo{author}{\bibfnamefont{C.}~\bibnamefont{Schappacher}}, and
  \bibinfo{author}{\bibfnamefont{A.}~\bibnamefont{Strumia}},
  \bibinfo{year}{2005}, \bibinfo{journal}{Phys.Rev.}
  \textbf{\bibinfo{volume}{D72}}, \bibinfo{pages}{035014}.

\bibitem[{\citenamefont{Marzocca} \emph{et~al.}(2012)\citenamefont{Marzocca,
  Serone, and Shu}}]{Marzocca:2012zn}
\bibinfo{author}{\bibnamefont{Marzocca}, \bibfnamefont{D.}},
  \bibinfo{author}{\bibfnamefont{M.}~\bibnamefont{Serone}}, and
  \bibinfo{author}{\bibfnamefont{J.}~\bibnamefont{Shu}}, \bibinfo{year}{2012},
  \bibinfo{journal}{JHEP} \textbf{\bibinfo{volume}{1208}},
  \bibinfo{pages}{013}.

\bibitem[{\citenamefont{Matsumoto} \emph{et~al.}(2007)\citenamefont{Matsumoto,
  Nojiri, and Nomura}}]{Matsumoto:2006ws}
\bibinfo{author}{\bibnamefont{Matsumoto}, \bibfnamefont{S.}},
  \bibinfo{author}{\bibfnamefont{M.~M.} \bibnamefont{Nojiri}}, and
  \bibinfo{author}{\bibfnamefont{D.}~\bibnamefont{Nomura}},
  \bibinfo{year}{2007}, \bibinfo{journal}{Phys.Rev.}
  \textbf{\bibinfo{volume}{D75}}, \bibinfo{pages}{055006}.

\bibitem[{\citenamefont{Matsuzaki and
  Yamawaki}(2012{\natexlab{a}})}]{Matsuzaki:2012xx}
\bibinfo{author}{\bibnamefont{Matsuzaki}, \bibfnamefont{S.}}, and
  \bibinfo{author}{\bibfnamefont{K.}~\bibnamefont{Yamawaki}},
  \bibinfo{year}{2012}{\natexlab{a}}, \bibinfo{journal}{Phys.Rev.}
  \textbf{\bibinfo{volume}{D86}}, \bibinfo{pages}{115004}.

\bibitem[{\citenamefont{Matsuzaki and
  Yamawaki}(2012{\natexlab{b}})}]{Matsuzaki:2012gd}
\bibinfo{author}{\bibnamefont{Matsuzaki}, \bibfnamefont{S.}}, and
  \bibinfo{author}{\bibfnamefont{K.}~\bibnamefont{Yamawaki}},
  \bibinfo{year}{2012}{\natexlab{b}}, \bibinfo{journal}{Phys.Rev.}
  \textbf{\bibinfo{volume}{D85}}, \bibinfo{pages}{095020}.

\bibitem[{\citenamefont{Matsuzaki and
  Yamawaki}(2012{\natexlab{c}})}]{Matsuzaki:2011ie}
\bibinfo{author}{\bibnamefont{Matsuzaki}, \bibfnamefont{S.}}, and
  \bibinfo{author}{\bibfnamefont{K.}~\bibnamefont{Yamawaki}},
  \bibinfo{year}{2012}{\natexlab{c}}, \bibinfo{journal}{Prog.Theor.Phys.}
  \textbf{\bibinfo{volume}{127}}, \bibinfo{pages}{209}.

\bibitem[{\citenamefont{Meade and Reece}(2006)}]{Meade:2006dw}
\bibinfo{author}{\bibnamefont{Meade}, \bibfnamefont{P.}}, and
  \bibinfo{author}{\bibfnamefont{M.}~\bibnamefont{Reece}},
  \bibinfo{year}{2006}, \bibinfo{journal}{Phys.Rev.}
  \textbf{\bibinfo{volume}{D74}}, \bibinfo{pages}{015010}.

\bibitem[{\citenamefont{Medina} \emph{et~al.}(2007)\citenamefont{Medina, Shah,
  and Wagner}}]{Medina:2007hz}
\bibinfo{author}{\bibnamefont{Medina}, \bibfnamefont{A.~D.}},
  \bibinfo{author}{\bibfnamefont{N.~R.} \bibnamefont{Shah}}, and
  \bibinfo{author}{\bibfnamefont{C.~E.} \bibnamefont{Wagner}},
  \bibinfo{year}{2007}, \bibinfo{journal}{Phys.Rev.}
  \textbf{\bibinfo{volume}{D76}}, \bibinfo{pages}{095010}.

\bibitem[{\citenamefont{Morrissey} \emph{et~al.}(2012)\citenamefont{Morrissey,
  Plehn, and Tait}}]{Morrissey:2009tf}
\bibinfo{author}{\bibnamefont{Morrissey}, \bibfnamefont{D.~E.}},
  \bibinfo{author}{\bibfnamefont{T.}~\bibnamefont{Plehn}}, and
  \bibinfo{author}{\bibfnamefont{T.~M.} \bibnamefont{Tait}},
  \bibinfo{year}{2012}, \bibinfo{journal}{Phys.Rept.}
  \textbf{\bibinfo{volume}{515}}, \bibinfo{pages}{1}.

\bibitem[{\citenamefont{Mrazek} \emph{et~al.}(2011)\citenamefont{Mrazek,
  Pomarol, Rattazzi, Redi, Serra} \emph{et~al.}}]{Mrazek:2011iu}
\bibinfo{author}{\bibnamefont{Mrazek}, \bibfnamefont{J.}},
  \bibinfo{author}{\bibfnamefont{A.}~\bibnamefont{Pomarol}},
  \bibinfo{author}{\bibfnamefont{R.}~\bibnamefont{Rattazzi}},
  \bibinfo{author}{\bibfnamefont{M.}~\bibnamefont{Redi}},
  \bibinfo{author}{\bibfnamefont{J.}~\bibnamefont{Serra}}, \emph{et~al.},
  \bibinfo{year}{2011}, \bibinfo{journal}{Nucl.Phys.}
  \textbf{\bibinfo{volume}{B853}}, \bibinfo{pages}{1}.

\bibitem[{\citenamefont{Mrazek and Wulzer}(2010)}]{Mrazek:2009yu}
\bibinfo{author}{\bibnamefont{Mrazek}, \bibfnamefont{J.}}, and
  \bibinfo{author}{\bibfnamefont{A.}~\bibnamefont{Wulzer}},
  \bibinfo{year}{2010}, \bibinfo{journal}{Phys.Rev.}
  \textbf{\bibinfo{volume}{D81}}, \bibinfo{pages}{075006}.

\bibitem[{\citenamefont{Nambu and
  Jona-Lasinio}(1961{\natexlab{a}})}]{Nambu:1961tp}
\bibinfo{author}{\bibnamefont{Nambu}, \bibfnamefont{Y.}}, and
  \bibinfo{author}{\bibfnamefont{G.}~\bibnamefont{Jona-Lasinio}},
  \bibinfo{year}{1961}{\natexlab{a}}, \bibinfo{journal}{Phys. Rev.}
  \textbf{\bibinfo{volume}{122}}, \bibinfo{pages}{345}.

\bibitem[{\citenamefont{Nambu and
  Jona-Lasinio}(1961{\natexlab{b}})}]{Nambu:1961fr}
\bibinfo{author}{\bibnamefont{Nambu}, \bibfnamefont{Y.}}, and
  \bibinfo{author}{\bibfnamefont{G.}~\bibnamefont{Jona-Lasinio}},
  \bibinfo{year}{1961}{\natexlab{b}}, \bibinfo{journal}{Phys. Rev.}
  \textbf{\bibinfo{volume}{124}}, \bibinfo{pages}{246}.

\bibitem[{\citenamefont{Noble and Perelstein}(2008)}]{Noble:2007kk}
\bibinfo{author}{\bibnamefont{Noble}, \bibfnamefont{A.}}, and
  \bibinfo{author}{\bibfnamefont{M.}~\bibnamefont{Perelstein}},
  \bibinfo{year}{2008}, \bibinfo{journal}{Phys.Rev.}
  \textbf{\bibinfo{volume}{D78}}, \bibinfo{pages}{063518}.

\bibitem[{\citenamefont{Orgogozo and
  Rychkov}(2012{\natexlab{a}})}]{Orgogozo:2011kq}
\bibinfo{author}{\bibnamefont{Orgogozo}, \bibfnamefont{A.}}, and
  \bibinfo{author}{\bibfnamefont{S.}~\bibnamefont{Rychkov}},
  \bibinfo{year}{2012}{\natexlab{a}}, \bibinfo{journal}{JHEP}
  \textbf{\bibinfo{volume}{1203}}, \bibinfo{pages}{046}.

\bibitem[{\citenamefont{Orgogozo and
  Rychkov}(2012{\natexlab{b}})}]{Orgogozo:2012ct}
\bibinfo{author}{\bibnamefont{Orgogozo}, \bibfnamefont{A.}}, and
  \bibinfo{author}{\bibfnamefont{S.}~\bibnamefont{Rychkov}},
  \bibinfo{year}{2012}{\natexlab{b}}, \eprint{1211.5543}.

\bibitem[{\citenamefont{Panico} \emph{et~al.}(2006)\citenamefont{Panico,
  Serone, and Wulzer}}]{Panico:2005dh}
\bibinfo{author}{\bibnamefont{Panico}, \bibfnamefont{G.}},
  \bibinfo{author}{\bibfnamefont{M.}~\bibnamefont{Serone}}, and
  \bibinfo{author}{\bibfnamefont{A.}~\bibnamefont{Wulzer}},
  \bibinfo{year}{2006}, \bibinfo{journal}{Nucl.Phys.}
  \textbf{\bibinfo{volume}{B739}}, \bibinfo{pages}{186}.

\bibitem[{\citenamefont{Panico and Wulzer}(2016)}]{Panico:2015jxa}
\bibinfo{author}{\bibnamefont{Panico}, \bibfnamefont{G.}}, and
  \bibinfo{author}{\bibfnamefont{A.}~\bibnamefont{Wulzer}},
  \bibinfo{year}{2016}, \bibinfo{journal}{Lect. Notes Phys.}
  \textbf{\bibinfo{volume}{913}}, \bibinfo{pages}{pp.}

\bibitem[{\citenamefont{Pappadopulo}
  \emph{et~al.}(2014)\citenamefont{Pappadopulo, Thamm, Torre, and
  Wulzer}}]{Pappadopulo:2014qza}
\bibinfo{author}{\bibnamefont{Pappadopulo}, \bibfnamefont{D.}},
  \bibinfo{author}{\bibfnamefont{A.}~\bibnamefont{Thamm}},
  \bibinfo{author}{\bibfnamefont{R.}~\bibnamefont{Torre}}, and
  \bibinfo{author}{\bibfnamefont{A.}~\bibnamefont{Wulzer}},
  \bibinfo{year}{2014}, \bibinfo{journal}{JHEP} \textbf{\bibinfo{volume}{09}},
  \bibinfo{pages}{060}.

\bibitem[{\citenamefont{Papucci} \emph{et~al.}(2012)\citenamefont{Papucci,
  Ruderman, and Weiler}}]{Papucci:2011wy}
\bibinfo{author}{\bibnamefont{Papucci}, \bibfnamefont{M.}},
  \bibinfo{author}{\bibfnamefont{J.~T.} \bibnamefont{Ruderman}}, and
  \bibinfo{author}{\bibfnamefont{A.}~\bibnamefont{Weiler}},
  \bibinfo{year}{2012}, \bibinfo{journal}{JHEP} \textbf{\bibinfo{volume}{09}},
  \bibinfo{pages}{035}.

\bibitem[{\citenamefont{Park and Song}(2004)}]{Park:2003sq}
\bibinfo{author}{\bibnamefont{Park}, \bibfnamefont{S.~C.}}, and
  \bibinfo{author}{\bibfnamefont{J.-h.} \bibnamefont{Song}},
  \bibinfo{year}{2004}, \bibinfo{journal}{Phys.Rev.}
  \textbf{\bibinfo{volume}{D69}}, \bibinfo{pages}{115010}.

\bibitem[{\citenamefont{Perelstein}(2007)}]{Perelstein:2005ka}
\bibinfo{author}{\bibnamefont{Perelstein}, \bibfnamefont{M.}},
  \bibinfo{year}{2007}, \bibinfo{journal}{Prog.Part.Nucl.Phys.}
  \textbf{\bibinfo{volume}{58}}, \bibinfo{pages}{247}.

\bibitem[{\citenamefont{Perelstein}
  \emph{et~al.}(2004)\citenamefont{Perelstein, Peskin, and
  Pierce}}]{Perelstein:2003wd}
\bibinfo{author}{\bibnamefont{Perelstein}, \bibfnamefont{M.}},
  \bibinfo{author}{\bibfnamefont{M.~E.} \bibnamefont{Peskin}}, and
  \bibinfo{author}{\bibfnamefont{A.}~\bibnamefont{Pierce}},
  \bibinfo{year}{2004}, \bibinfo{journal}{Phys.Rev.}
  \textbf{\bibinfo{volume}{D69}}, \bibinfo{pages}{075002}.

\bibitem[{\citenamefont{Perelstein and Shao}(2011)}]{Perelstein:2011ds}
\bibinfo{author}{\bibnamefont{Perelstein}, \bibfnamefont{M.}}, and
  \bibinfo{author}{\bibfnamefont{J.}~\bibnamefont{Shao}}, \bibinfo{year}{2011},
  \bibinfo{journal}{Phys.Lett.} \textbf{\bibinfo{volume}{B704}},
  \bibinfo{pages}{510}.

\bibitem[{\citenamefont{Perez-Victoria}(2001)}]{PerezVictoria:2001pa}
\bibinfo{author}{\bibnamefont{Perez-Victoria}, \bibfnamefont{M.}},
  \bibinfo{year}{2001}, \bibinfo{journal}{JHEP}
  \textbf{\bibinfo{volume}{0105}}, \bibinfo{pages}{064}.

\bibitem[{\citenamefont{Peskin and Takeuchi}(1990)}]{Peskin:1990zt}
\bibinfo{author}{\bibnamefont{Peskin}, \bibfnamefont{M.~E.}}, and
  \bibinfo{author}{\bibfnamefont{T.}~\bibnamefont{Takeuchi}},
  \bibinfo{year}{1990}, \bibinfo{journal}{Phys. Rev. Lett.}
  \textbf{\bibinfo{volume}{65}}, \bibinfo{pages}{964}.

\bibitem[{\citenamefont{Peskin and Takeuchi}(1992)}]{Peskin:1992sw}
\bibinfo{author}{\bibnamefont{Peskin}, \bibfnamefont{M.~E.}}, and
  \bibinfo{author}{\bibfnamefont{T.}~\bibnamefont{Takeuchi}},
  \bibinfo{year}{1992}, \bibinfo{journal}{Phys. Rev.}
  \textbf{\bibinfo{volume}{D46}}, \bibinfo{pages}{381}.

\bibitem[{\citenamefont{Peskin and Wells}(2001)}]{Peskin:2001rw}
\bibinfo{author}{\bibnamefont{Peskin}, \bibfnamefont{M.~E.}}, and
  \bibinfo{author}{\bibfnamefont{J.~D.} \bibnamefont{Wells}},
  \bibinfo{year}{2001}, \bibinfo{journal}{Phys.Rev.}
  \textbf{\bibinfo{volume}{D64}}, \bibinfo{pages}{093003}.

\bibitem[{\citenamefont{Pich} \emph{et~al.}(2013)\citenamefont{Pich, Rosell,
  and Sanz-Cillero}}]{Pich:2012dv}
\bibinfo{author}{\bibnamefont{Pich}, \bibfnamefont{A.}},
  \bibinfo{author}{\bibfnamefont{I.}~\bibnamefont{Rosell}}, and
  \bibinfo{author}{\bibfnamefont{J.~J.} \bibnamefont{Sanz-Cillero}},
  \bibinfo{year}{2013}, \bibinfo{journal}{Phys.Rev.Lett.}
  \textbf{\bibinfo{volume}{110}}, \bibinfo{pages}{181801}.

\bibitem[{\citenamefont{Plehn and Spannowsky}(2012)}]{Plehn:2011tg}
\bibinfo{author}{\bibnamefont{Plehn}, \bibfnamefont{T.}}, and
  \bibinfo{author}{\bibfnamefont{M.}~\bibnamefont{Spannowsky}},
  \bibinfo{year}{2012}, \bibinfo{journal}{J.Phys.}
  \textbf{\bibinfo{volume}{G39}}, \bibinfo{pages}{083001}.

\bibitem[{\citenamefont{Poland and Simmons-Duffin}(2011)}]{Poland:2010wg}
\bibinfo{author}{\bibnamefont{Poland}, \bibfnamefont{D.}}, and
  \bibinfo{author}{\bibfnamefont{D.}~\bibnamefont{Simmons-Duffin}},
  \bibinfo{year}{2011}, \bibinfo{journal}{JHEP}
  \textbf{\bibinfo{volume}{1105}}, \bibinfo{pages}{017}.

\bibitem[{\citenamefont{Poland} \emph{et~al.}(2012)\citenamefont{Poland,
  Simmons-Duffin, and Vichi}}]{Poland:2011ey}
\bibinfo{author}{\bibnamefont{Poland}, \bibfnamefont{D.}},
  \bibinfo{author}{\bibfnamefont{D.}~\bibnamefont{Simmons-Duffin}}, and
  \bibinfo{author}{\bibfnamefont{A.}~\bibnamefont{Vichi}},
  \bibinfo{year}{2012}, \bibinfo{journal}{JHEP}
  \textbf{\bibinfo{volume}{1205}}, \bibinfo{pages}{110}.

\bibitem[{\citenamefont{Pomarol}(2000)}]{Pomarol:1999ad}
\bibinfo{author}{\bibnamefont{Pomarol}, \bibfnamefont{A.}},
  \bibinfo{year}{2000}, \bibinfo{journal}{Phys.Lett.}
  \textbf{\bibinfo{volume}{B486}}, \bibinfo{pages}{153}.

\bibitem[{\citenamefont{Pomarol and Riva}(2012)}]{Pomarol:2012qf}
\bibinfo{author}{\bibnamefont{Pomarol}, \bibfnamefont{A.}}, and
  \bibinfo{author}{\bibfnamefont{F.}~\bibnamefont{Riva}}, \bibinfo{year}{2012},
  \bibinfo{journal}{JHEP} \textbf{\bibinfo{volume}{1208}},
  \bibinfo{pages}{135}.

\bibitem[{\citenamefont{Ponton}(2012)}]{Ponton:2012bi}
\bibinfo{author}{\bibnamefont{Ponton}, \bibfnamefont{E.}},
  \bibinfo{year}{2012}, \eprint{1207.3827}.

\bibitem[{\citenamefont{Randall and Sundrum}(1999)}]{Randall:1999ee}
\bibinfo{author}{\bibnamefont{Randall}, \bibfnamefont{L.}}, and
  \bibinfo{author}{\bibfnamefont{R.}~\bibnamefont{Sundrum}},
  \bibinfo{year}{1999}, \bibinfo{journal}{Phys.Rev.Lett.}
  \textbf{\bibinfo{volume}{83}}, \bibinfo{pages}{3370}.

\bibitem[{\citenamefont{Rattazzi}(2003)}]{Rattazzi:2003ea}
\bibinfo{author}{\bibnamefont{Rattazzi}, \bibfnamefont{R.}},
  \bibinfo{year}{2003}, in \emph{\bibinfo{booktitle}{{Particle physics and
  cosmology: The interface. Proceedings, NATO Advanced Study Institute, School,
  Cargese, France, August 4-16, 2003}}}, pp. \bibinfo{pages}{461--517},
  \eprint{hep-ph/0607055}.

\bibitem[{\citenamefont{Rattazzi} \emph{et~al.}(2008)\citenamefont{Rattazzi,
  Rychkov, Tonni, and Vichi}}]{Rattazzi:2008pe}
\bibinfo{author}{\bibnamefont{Rattazzi}, \bibfnamefont{R.}},
  \bibinfo{author}{\bibfnamefont{V.~S.} \bibnamefont{Rychkov}},
  \bibinfo{author}{\bibfnamefont{E.}~\bibnamefont{Tonni}}, and
  \bibinfo{author}{\bibfnamefont{A.}~\bibnamefont{Vichi}},
  \bibinfo{year}{2008}, \bibinfo{journal}{JHEP}
  \textbf{\bibinfo{volume}{0812}}, \bibinfo{pages}{031}.

\bibitem[{\citenamefont{Rattazzi and Zaffaroni}(2001)}]{Rattazzi:2000hs}
\bibinfo{author}{\bibnamefont{Rattazzi}, \bibfnamefont{R.}}, and
  \bibinfo{author}{\bibfnamefont{A.}~\bibnamefont{Zaffaroni}},
  \bibinfo{year}{2001}, \bibinfo{journal}{JHEP}
  \textbf{\bibinfo{volume}{0104}}, \bibinfo{pages}{021}.

\bibitem[{\citenamefont{Reuter and Tonini}(2013)}]{Reuter:2012sd}
\bibinfo{author}{\bibnamefont{Reuter}, \bibfnamefont{J.}}, and
  \bibinfo{author}{\bibfnamefont{M.}~\bibnamefont{Tonini}},
  \bibinfo{year}{2013}, \bibinfo{journal}{JHEP}
  \textbf{\bibinfo{volume}{1302}}, \bibinfo{pages}{077}.

\bibitem[{\citenamefont{Reuter} \emph{et~al.}(2014)\citenamefont{Reuter,
  Tonini, and de~Vries}}]{Reuter:2013iya}
\bibinfo{author}{\bibnamefont{Reuter}, \bibfnamefont{J.}},
  \bibinfo{author}{\bibfnamefont{M.}~\bibnamefont{Tonini}}, and
  \bibinfo{author}{\bibfnamefont{M.}~\bibnamefont{de~Vries}},
  \bibinfo{year}{2014}, \bibinfo{journal}{JHEP}
  \textbf{\bibinfo{volume}{1402}}, \bibinfo{pages}{053}.

\bibitem[{\citenamefont{Roy and Schmaltz}(2006)}]{Roy:2005hg}
\bibinfo{author}{\bibnamefont{Roy}, \bibfnamefont{T.~S.}}, and
  \bibinfo{author}{\bibfnamefont{M.}~\bibnamefont{Schmaltz}},
  \bibinfo{year}{2006}, \bibinfo{journal}{JHEP}
  \textbf{\bibinfo{volume}{0601}}, \bibinfo{pages}{149}.

\bibitem[{\citenamefont{Rychkov}(2011)}]{Rychkov:2011br}
\bibinfo{author}{\bibnamefont{Rychkov}, \bibfnamefont{S.}},
  \bibinfo{year}{2011}, \bibinfo{journal}{PoS}
  \textbf{\bibinfo{volume}{EPS-HEP2011}}, \bibinfo{pages}{029}.

\bibitem[{\citenamefont{Rychkov and Vichi}(2009)}]{Rychkov:2009ij}
\bibinfo{author}{\bibnamefont{Rychkov}, \bibfnamefont{V.~S.}}, and
  \bibinfo{author}{\bibfnamefont{A.}~\bibnamefont{Vichi}},
  \bibinfo{year}{2009}, \bibinfo{journal}{Phys.Rev.}
  \textbf{\bibinfo{volume}{D80}}, \bibinfo{pages}{045006}.

\bibitem[{\citenamefont{Ryskin and Shuvaev}(2010)}]{Ryskin:2009kw}
\bibinfo{author}{\bibnamefont{Ryskin}, \bibfnamefont{M.}}, and
  \bibinfo{author}{\bibfnamefont{A.}~\bibnamefont{Shuvaev}},
  \bibinfo{year}{2010}, \bibinfo{journal}{Phys.Atom.Nucl.}
  \textbf{\bibinfo{volume}{73}}, \bibinfo{pages}{965}.

\bibitem[{\citenamefont{Samuel}(1990)}]{Samuel:1990dq}
\bibinfo{author}{\bibnamefont{Samuel}, \bibfnamefont{S.}},
  \bibinfo{year}{1990}, \bibinfo{journal}{Nucl. Phys.}
  \textbf{\bibinfo{volume}{B347}}, \bibinfo{pages}{625}.

\bibitem[{\citenamefont{Sannino}(2009)}]{Sannino:2009za}
\bibinfo{author}{\bibnamefont{Sannino}, \bibfnamefont{F.}},
  \bibinfo{year}{2009}, \bibinfo{journal}{Acta Phys.Polon.}
  \textbf{\bibinfo{volume}{B40}}, \bibinfo{pages}{3533}.

\bibitem[{\citenamefont{Schmaltz}(2003)}]{Schmaltz:2002wx}
\bibinfo{author}{\bibnamefont{Schmaltz}, \bibfnamefont{M.}},
  \bibinfo{year}{2003}, \bibinfo{journal}{Nucl.Phys.Proc.Suppl.}
  \textbf{\bibinfo{volume}{117}}, \bibinfo{pages}{40}.

\bibitem[{\citenamefont{Schmaltz}(2004)}]{Schmaltz:2004de}
\bibinfo{author}{\bibnamefont{Schmaltz}, \bibfnamefont{M.}},
  \bibinfo{year}{2004}, \bibinfo{journal}{JHEP}
  \textbf{\bibinfo{volume}{0408}}, \bibinfo{pages}{056}.

\bibitem[{\citenamefont{Schmaltz} \emph{et~al.}(2010)\citenamefont{Schmaltz,
  Stolarski, and Thaler}}]{Schmaltz:2010ac}
\bibinfo{author}{\bibnamefont{Schmaltz}, \bibfnamefont{M.}},
  \bibinfo{author}{\bibfnamefont{D.}~\bibnamefont{Stolarski}}, and
  \bibinfo{author}{\bibfnamefont{J.}~\bibnamefont{Thaler}},
  \bibinfo{year}{2010}, \bibinfo{journal}{JHEP}
  \textbf{\bibinfo{volume}{1009}}, \bibinfo{pages}{018}.

\bibitem[{\citenamefont{Schmaltz and Thaler}(2009)}]{Schmaltz:2008vd}
\bibinfo{author}{\bibnamefont{Schmaltz}, \bibfnamefont{M.}}, and
  \bibinfo{author}{\bibfnamefont{J.}~\bibnamefont{Thaler}},
  \bibinfo{year}{2009}, \bibinfo{journal}{JHEP}
  \textbf{\bibinfo{volume}{0903}}, \bibinfo{pages}{137}.

\bibitem[{\citenamefont{Schmaltz and Tucker-Smith}(2005)}]{Schmaltz:2005ky}
\bibinfo{author}{\bibnamefont{Schmaltz}, \bibfnamefont{M.}}, and
  \bibinfo{author}{\bibfnamefont{D.}~\bibnamefont{Tucker-Smith}},
  \bibinfo{year}{2005}, \bibinfo{journal}{Ann.Rev.Nucl.Part.Sci.}
  \textbf{\bibinfo{volume}{55}}, \bibinfo{pages}{229}.

\bibitem[{\citenamefont{Seiberg}(1995)}]{Seiberg:1994pq}
\bibinfo{author}{\bibnamefont{Seiberg}, \bibfnamefont{N.}},
  \bibinfo{year}{1995}, \bibinfo{journal}{Nucl.Phys.}
  \textbf{\bibinfo{volume}{B435}}, \bibinfo{pages}{129}.

\bibitem[{\citenamefont{Serone}(2010)}]{Serone:2009kf}
\bibinfo{author}{\bibnamefont{Serone}, \bibfnamefont{M.}},
  \bibinfo{year}{2010}, \bibinfo{journal}{New J.Phys.}
  \textbf{\bibinfo{volume}{12}}, \bibinfo{pages}{075013}.

\bibitem[{\citenamefont{Seymour}(1994)}]{Seymour:1993mx}
\bibinfo{author}{\bibnamefont{Seymour}, \bibfnamefont{M.~H.}},
  \bibinfo{year}{1994}, \bibinfo{journal}{Z.Phys.}
  \textbf{\bibinfo{volume}{C62}}, \bibinfo{pages}{127}.

\bibitem[{\citenamefont{Shelton}(2013)}]{Shelton:2013an}
\bibinfo{author}{\bibnamefont{Shelton}, \bibfnamefont{J.}},
  \bibinfo{year}{2013}, \eprint{1302.0260}.

\bibitem[{\citenamefont{Shifman} \emph{et~al.}(1979)\citenamefont{Shifman,
  Vainshtein, Voloshin, and Zakharov}}]{Shifman:1979eb}
\bibinfo{author}{\bibnamefont{Shifman}, \bibfnamefont{M.~A.}},
  \bibinfo{author}{\bibfnamefont{A.}~\bibnamefont{Vainshtein}},
  \bibinfo{author}{\bibfnamefont{M.}~\bibnamefont{Voloshin}}, and
  \bibinfo{author}{\bibfnamefont{V.~I.} \bibnamefont{Zakharov}},
  \bibinfo{year}{1979}, \bibinfo{journal}{Sov.J.Nucl.Phys.}
  \textbf{\bibinfo{volume}{30}}, \bibinfo{pages}{711}.

\bibitem[{\citenamefont{Sikivie} \emph{et~al.}(1980)\citenamefont{Sikivie,
  Susskind, Voloshin, and Zakharov}}]{Sikivie:1980hm}
\bibinfo{author}{\bibnamefont{Sikivie}, \bibfnamefont{P.}},
  \bibinfo{author}{\bibfnamefont{L.}~\bibnamefont{Susskind}},
  \bibinfo{author}{\bibfnamefont{M.~B.} \bibnamefont{Voloshin}}, and
  \bibinfo{author}{\bibfnamefont{V.~I.} \bibnamefont{Zakharov}},
  \bibinfo{year}{1980}, \bibinfo{journal}{Nucl.Phys.}
  \textbf{\bibinfo{volume}{B173}}, \bibinfo{pages}{189}.

\bibitem[{\citenamefont{Skiba and Terning}(2003)}]{Skiba:2003yf}
\bibinfo{author}{\bibnamefont{Skiba}, \bibfnamefont{W.}}, and
  \bibinfo{author}{\bibfnamefont{J.}~\bibnamefont{Terning}},
  \bibinfo{year}{2003}, \bibinfo{journal}{Phys.Rev.}
  \textbf{\bibinfo{volume}{D68}}, \bibinfo{pages}{075001}.

\bibitem[{\citenamefont{Stancato and Terning}(2009)}]{Stancato:2008mp}
\bibinfo{author}{\bibnamefont{Stancato}, \bibfnamefont{D.}}, and
  \bibinfo{author}{\bibfnamefont{J.}~\bibnamefont{Terning}},
  \bibinfo{year}{2009}, \bibinfo{journal}{JHEP}
  \textbf{\bibinfo{volume}{0911}}, \bibinfo{pages}{101}.

\bibitem[{\citenamefont{Stancato and Terning}(2010)}]{Stancato:2010ay}
\bibinfo{author}{\bibnamefont{Stancato}, \bibfnamefont{D.}}, and
  \bibinfo{author}{\bibfnamefont{J.}~\bibnamefont{Terning}},
  \bibinfo{year}{2010}, \bibinfo{journal}{Phys.Rev.}
  \textbf{\bibinfo{volume}{D81}}, \bibinfo{pages}{115012}.

\bibitem[{\citenamefont{Sullivan}(2003)}]{Sullivan:2003xy}
\bibinfo{author}{\bibnamefont{Sullivan}, \bibfnamefont{Z.}},
  \bibinfo{year}{2003}, in \emph{\bibinfo{booktitle}{{Proceedings, 38th
  Rencontres de Moriond on QCD and High-Energy Hadronic Interactions}}},
  \eprint{hep-ph/0306266}.

\bibitem[{\citenamefont{Sundrum}(2005)}]{Sundrum:2005jf}
\bibinfo{author}{\bibnamefont{Sundrum}, \bibfnamefont{R.}},
  \bibinfo{year}{2005}, in \emph{\bibinfo{booktitle}{{Theoretical Advanced
  Study Institute in Elementary Particle Physics: Many Dimensions of String
  Theory (TASI 2005) Boulder, Colorado, June 5-July 1, 2005}}}, pp.
  \bibinfo{pages}{585--630}, \eprint{hep-th/0508134}.

\bibitem[{\citenamefont{Susskind}(1979)}]{Susskind:1978ms}
\bibinfo{author}{\bibnamefont{Susskind}, \bibfnamefont{L.}},
  \bibinfo{year}{1979}, \bibinfo{journal}{Phys. Rev.}
  \textbf{\bibinfo{volume}{D20}}, \bibinfo{pages}{2619}.

\bibitem[{\citenamefont{Terning}(1991)}]{Terning:1991yt}
\bibinfo{author}{\bibnamefont{Terning}, \bibfnamefont{J.}},
  \bibinfo{year}{1991}, \bibinfo{journal}{Phys.Rev.}
  \textbf{\bibinfo{volume}{D44}}, \bibinfo{pages}{887}.

\bibitem[{\citenamefont{Vecchi}(2010)}]{Vecchi:2010gj}
\bibinfo{author}{\bibnamefont{Vecchi}, \bibfnamefont{L.}},
  \bibinfo{year}{2010}, \bibinfo{journal}{Phys.Rev.}
  \textbf{\bibinfo{volume}{D82}}, \bibinfo{pages}{076009}.

\bibitem[{\citenamefont{Vecchi}(2011{\natexlab{a}})}]{Vecchi:2010em}
\bibinfo{author}{\bibnamefont{Vecchi}, \bibfnamefont{L.}},
  \bibinfo{year}{2011}{\natexlab{a}}, \bibinfo{journal}{JHEP}
  \textbf{\bibinfo{volume}{1111}}, \bibinfo{pages}{102}.

\bibitem[{\citenamefont{Vecchi}(2011{\natexlab{b}})}]{Vecchi:2010aj}
\bibinfo{author}{\bibnamefont{Vecchi}, \bibfnamefont{L.}},
  \bibinfo{year}{2011}{\natexlab{b}}, \bibinfo{journal}{JHEP}
  \textbf{\bibinfo{volume}{1104}}, \bibinfo{pages}{127}.

\bibitem[{\citenamefont{Weinberg}(1967)}]{Weinberg:1967tq}
\bibinfo{author}{\bibnamefont{Weinberg}, \bibfnamefont{S.}},
  \bibinfo{year}{1967}, \bibinfo{journal}{Phys.Rev.Lett.}
  \textbf{\bibinfo{volume}{19}}, \bibinfo{pages}{1264}.

\bibitem[{\citenamefont{Weinberg}(1976)}]{Weinberg:1976gm}
\bibinfo{author}{\bibnamefont{Weinberg}, \bibfnamefont{S.}},
  \bibinfo{year}{1976}, \bibinfo{journal}{Phys. Rev.}
  \textbf{\bibinfo{volume}{D13}}, \bibinfo{pages}{974}.

\bibitem[{\citenamefont{Weinberg}(1979)}]{PhysRevD.19.1277}
\bibinfo{author}{\bibnamefont{Weinberg}, \bibfnamefont{S.}},
  \bibinfo{year}{1979}, \bibinfo{journal}{Phys. Rev. D}
  \textbf{\bibinfo{volume}{19}}, \bibinfo{pages}{1277}.

\bibitem[{\citenamefont{Weisskopf}(1939)}]{Weisskopf:1939zz}
\bibinfo{author}{\bibnamefont{Weisskopf}, \bibfnamefont{V.}},
  \bibinfo{year}{1939}, \bibinfo{journal}{Phys.Rev.}
  \textbf{\bibinfo{volume}{56}}, \bibinfo{pages}{72}.

\bibitem[{\citenamefont{Yamawaki} \emph{et~al.}(1986)\citenamefont{Yamawaki,
  Bando, and iti Matumoto}}]{Yamawaki:1986zg}
\bibinfo{author}{\bibnamefont{Yamawaki}, \bibfnamefont{K.}},
  \bibinfo{author}{\bibfnamefont{M.}~\bibnamefont{Bando}}, and
  \bibinfo{author}{\bibfnamefont{K.}~\bibnamefont{iti Matumoto}},
  \bibinfo{year}{1986}, \bibinfo{journal}{Phys. Rev. Lett.}
  \textbf{\bibinfo{volume}{56}}, \bibinfo{pages}{1335}.

\bibitem[{\citenamefont{Yang} \emph{et~al.}(2014)\citenamefont{Yang, Liu, and
  Han}}]{Yang:2014uza}
\bibinfo{author}{\bibnamefont{Yang}, \bibfnamefont{B.}},
  \bibinfo{author}{\bibfnamefont{N.}~\bibnamefont{Liu}}, and
  \bibinfo{author}{\bibfnamefont{J.}~\bibnamefont{Han}}, \bibinfo{year}{2014},
  \bibinfo{journal}{Phys.Rev.}
  \textbf{\bibinfo{volume}{D89}}(\bibinfo{number}{3}), \bibinfo{pages}{034020}.

\bibitem[{\citenamefont{Yue} \emph{et~al.}(2003)\citenamefont{Yue, Wang, and
  Yu}}]{Yue:2003yk}
\bibinfo{author}{\bibnamefont{Yue}, \bibfnamefont{C.-x.}},
  \bibinfo{author}{\bibfnamefont{S.-z.} \bibnamefont{Wang}}, and
  \bibinfo{author}{\bibfnamefont{D.-q.} \bibnamefont{Yu}},
  \bibinfo{year}{2003}, \bibinfo{journal}{Phys.Rev.}
  \textbf{\bibinfo{volume}{D68}}, \bibinfo{pages}{115004}.

\bibitem[{\citenamefont{Yue and Wang}(2004)}]{Yue:2004xt}
\bibinfo{author}{\bibnamefont{Yue}, \bibfnamefont{C.-x.}}, and
  \bibinfo{author}{\bibfnamefont{W.}~\bibnamefont{Wang}}, \bibinfo{year}{2004},
  \bibinfo{journal}{Nucl.Phys.} \textbf{\bibinfo{volume}{B683}},
  \bibinfo{pages}{48}.

\bibitem[{\citenamefont{Yue and Wang}(2005)}]{Yue:2004fv}
\bibinfo{author}{\bibnamefont{Yue}, \bibfnamefont{C.-x.}}, and
  \bibinfo{author}{\bibfnamefont{W.}~\bibnamefont{Wang}}, \bibinfo{year}{2005},
  \bibinfo{journal}{Phys.Rev.} \textbf{\bibinfo{volume}{D71}},
  \bibinfo{pages}{015002}.

\end{thebibliography}
\newpage

\end{document}